\documentclass[numbib]{comnet2}

\usepackage{etex}

\newcommand{\comment}[1]{}

\usepackage{pifont}
\usepackage{wrapfig}
\usepackage{longtable}
\usepackage{xspace}
\usepackage{url}
\usepackage{afterpage} %% use \afterpage{\clearpage}
\usepackage{amsmath}
\usepackage{amssymb}
\usepackage{nicefrac}
\usepackage{ctable} 
\usepackage{array} 
\usepackage{graphics,subfigure} 
\usepackage{pstricks,pst-plot,pst-coil,pst-text,multido,pst-3d,pst-tree,pst-grad,colortbl,color,multirow,booktabs}
\usepackage{pst-xkey,pst-solides3d} 
\usepackage{pst-3dplot} 

\newcommand{\textfrc}[1]{{\frcseries#1}}
\renewcommand{\thesubfigure}{(\Alph{subfigure})}
\newcommand{\EA}{{\em et al.}\xspace}
\newcommand{\IE}{{\em i.e.}\xspace}
\newcommand{\tx}{^{\rm th}}
\newcommand{\EG}{{\it e.g.}\xspace}
\renewcommand{\cite}{\citep}

\renewcommand\topfraction{.99}        % max fraction of floats at top
\renewcommand\bottomfraction{.99}     % max fraction of floats at bottom
\renewcommand\textfraction{.01}   
\renewcommand{\floatpagefraction}{0.99}
\begin{document}

\title{On Global Stability of Financial Networks}

\shorttitle{Stability of Financial Networks}
\shortauthorlist{DasGupta and Kaligounder} %%% for verso running head

\author{%%%% First author details
{{\sc 
Bhaskar DasGupta$^*$
}}
\address{Department of Computer Science, University of Illinois at Chicago, Chicago, IL 60607, USA\email{$^*$Corresponding author: bdasgup@uic.edu}}
%%%%%%% Second author details
\and
{{\sc 
Lakshmi Kaligounder
}}
\address{Department of Computer Science, University of Illinois at Chicago, Chicago, IL 60607, USA} 
}

\maketitle

\begin{abstract}
{Recent crisis in the global financial world has generated renewed interests in fragilities of global financial networks
among economists and regulatory authorities.
In particular, a potential vulnerability of the financial networks is the ``financial contagion''
process in which insolvencies of individual entities propagate through the ``web of dependencies'' to affect
the entire system. In this paper, we formalize an extension of a financial network model originally proposed by Nier \EA
for scenarios such as the over-the-counter derivatives market,
define a suitable global stability measure for this model, and perform a comprehensive evaluation of this stability measure 
over more than 700,000 combinations of networks types and parameter combinations. Based on our evaluations, we 
discover many interesting implications of our evaluations of this stability measure, and
derive topological properties and parameter combinations that may be used to flag the network as a possible fragile 
network. An interactive software FIN-STAB for computing the stability is available from the website 
\url{www2.cs.uic.edu/~dasgupta/financial-simulator-files.}.}
{
Financial networks, financial contagion, global stability.
}
\\
2000 Math Subject Classification: 91B55, 91G99, 05C82.
\end{abstract}

\section{Introduction}

Recent unprecedented level of global financial crisis has clearly
exposed potential weaknesses of the global economic system, renewing interests in the 
determination of fragilities of various segments of the global economy. 
Since financial institutions governed by borrowing, lending and participation in risky investments
played a crucial role in this crisis, 
they have attracted a major part of the attention of economists; see~\cite{HM11} for a survey.
The issue of instability of free market based financial systems is not new and has been under
discussion among the economists starting with the early works of~\cite{F33,K36} during the 1930's great depression era.
However, the exact causes of such instabilities have {\em not} been unanimously agreed upon yet.
Economists such as Ekelund and Thornton~\cite{ET08} contend that a major reason for the recent financial crisis is 
the enactment of an act that removed several restrictions on mixing investment and consumer banking, 
whereas other economists such as Calabria disagree with such an assertion~\cite{C09}.
Some economists such as Minsky have argued that such instabilities are 
are {\em systemic} for many modern market-based economic systems~\cite{M77}. 

One motivation in this paper to investigate global stabilities of financial 
networks comes from the point of view of a {\em regulatory agency} (as was also the case, for example, 
in~\cite{HM11}). A regulatory agency with 
sufficient knowledge about a part of a global financial network is expected to periodically evaluate
the stability of the network, and flag the network {\em ex ante} for further analysis if it fails some preliminary 
test or exceeds some minimum threshold of vulnerability. In this motivation, 
flagging a network as vulnerable does not necessarily imply 
that such is the case, but that such a network requires further analysis based on other aspects of free market economics 
that are not or simply {\em cannot} be 
modeled\footnote{For example, some such factors are the {\em rumors} and {\em panics} caused by the insolvency
of a large bank and a possible subsequent credit freeze. While fears, panics and rumors are all real aspects in networked economics,
there are hardly any universally agreed upon good way of modelling them.}.
While too many false positives may drain the finite resources of a regulatory agency for further analysis and investigation, 
this motivation assumes that vulnerability is too important an issue to be left for an {\em ex post} analysis.

Similar to prior research works such as~\cite{arxiv,E04,HM11,NYYA07,GK08,MA10},
our study of the vulnerability of financial networks also assumes the absence of government intervention
as banks become insolvent. 
While this is an extreme worst-case situation, the main goal of such type of studies is to see if the network can survive 
a shock even under extreme situations. A further reason for not allowing any intervention is, unlike the case of public health issues such as controlling spread of epidemics, 
government intervention in a capitalist financial system is often not allowed or requires complex political and administrative operatives.

\section{Brief review of related prior works on financial networks}

Although there is a large amount of literature on stability of financial systems 
(\EG, see~\cite{E04,arxiv,FN09,B07,Ace2013,E2013,PYR09,MR13,La13,AG2000,DP83,GK08,XBJ2000,H09,Lie2010,D04,RS98,CM09,NYYA07,CD09,A05}), 
very few prior papers have mathematically defined a global stability measure and 
performed a comprehensive evaluation of such a measure as done in this paper. A most recent research work related to our work is the paper by 
Minoiu and Reyes~\cite{MR13} in which the authors analyzed global banking networks data on cross-border banking flows for 184 countries during 1978-2010
using local connectivity and clustering measures. Below we review other related prior research works. 
Although ordinarily one would expect the risk of contagion to be larger in a highly interconnected banking system, 
prior simulation works indicate that higher connectivity among banks may sometimes lead to lower risk of contagion.
{\em Due to the large volume of prior research works,
we are only able to review a selected subset of related prior research works, leaving many other exciting research results in the bibliographies of the 
cited papers}. 

Allen and Gale~\cite{AG2000} found that when consumers have the liquidity preferences as introduced by Diamond and Dybvig~\cite{DP83} and have 
random liquidity needs, banks {\em perfectly} insure against liquidity fluctuations by exchanging interbank deposits, but the connections created by 
{\em swapping deposits} expose the {\em entire system} to contagion. 
Based on such studies, Allen and Gale~\cite{AG2000} concluded that incomplete networks are {\em more} prone to contagion than networks with maximum connectivity since 
better-connected networks are more resilient via transfer of proportion of the losses in one bank's portfolio to more banks through interbank agreements. 
On the other hand, Gai and Kapadia~\cite{GK08} argued that the higher is the connectivity among banks the more will be the contagion effect during crisis. 
Freixas \EA \cite{XBJ2000} explored the case of banks that face liquidity fluctuations due to the uncertainty about 
consumers withdrawing funds. 
Haldane~\cite{H09} suggested that contagion should be measured based on the interconnectedness of each institution within the financial system.
Liedorp \EA~\cite{Lie2010} argued that both large lending and borrowing shares in interbank markets increase the riskiness of banks active in the {\em dutch} banking market. 

Dasgupta~\cite{D04} explored how linkages between banks, represented by cross-holding of deposits, can be a source of contagious breakdowns by investigating 
how depositors, who receive a private signal about fundamentals of banks, may want to withdraw their deposits if they believe that enough other depositors will do the same. 
Lagunoff and Schreft~\cite{RS98} studies a network model in which the return on an agents' portfolio depends on the portfolio allocations of other agents.  
Iazzetta and Manna~\cite{CM09} used network topology analysis on monthly data on deposits exchange to gain more insight into the way a liquidity crisis spreads. 
Nier \EA~\cite{NYYA07} explored the dependency of systemic risks on the structure of the banking system 
and the resilience (or lack thereof) of such a  system to contagious defaults via graph theoretic approach.
Corbo and Demange~\cite{CD09} explored the relationship of the structure of interbank connections to the contagion risk of defaults
given the exogenous default of set of banks.
Babus~\cite{A05} studied how the trade-off between the benefits and the costs of being linked changes depending on the network structure, and 
observed that, when the network is maximal, liquidity can be redistributed in the system to make the risk of contagion minimal. 

Acemoglu \EA~\cite{Ace2013} and Zawadowski~\cite{Z11} do investigate stability of financial networks, but differently from our study. Both~\cite{Ace2013} and~\cite{Z11}
consider two specific network topologies, namely the ring topology and the complete network topology, as opposed to a more general class of topologies in our study. 
Both~\cite{Ace2013} and~\cite{Z11} consider only the effect of the shock propagation for {\em a few} discrete time steps, as opposed to our study; 
in the terminology of~\cite{arxiv}, this can be thought of as a ``violent death'' of the network as opposed to the ``slow poisoning death'' that our paper investigates. 
The model and structure/terms of bilateral interbank agreements in~\cite{Ace2013}, namely that banks lend to one another through debt contracts with contingency covenants, 
is quite different from ours. As a results, the conclusions in~\cite{Ace2013,Z11} do not directly apply to our model and the corresponding simulation environment.

Attribute propagation models have been investigated in the past in other contexts such as influence maximization in social networks~\cite{kempe1,CWY09,Chen08,borodin}, 
disease spreading in urban networks~\cite{eubank04,coelho08,eubank05}, and percolation models in physics and mathematics~\cite{SA94}. 
However, the shock propagation model in this paper is very different from all these models. For example: 
\begin{itemize}
\item
Almost all of the other models include a trivial solution in which the attribute spreads to the entire network if we inject each node individually with the attribute. 
This is not the case with the shock propagation model.

\item
If shocking a subset of nodes makes $x$ nodes in the network fail, then adding more nodes to this subset may not necessarily lead to the failure of $x$ or more than $x$ nodes of the network.

\item
The complexity of many previous attribute propagation models arises due to the presence of cycles in the graph. 
In contrast, the shock propagation model may be highly complex {\em even when the given network is acyclic}. Instead, 
a key component of the complexity arises due to two or more directed paths sharing a node.
\end{itemize}

\section{Organization of the paper}

The rest of the paper is organized as follows:

\vspace*{0.1in}
\noindent
{\large $\bullet$} $\,$ 
In Section~\ref{model-sec} we describe the network model and the corresponding stability measure. In particular:
\begin{itemize}
\item
In Section~\ref{s1} we define the balance sheet equations, and the two (homogeneous and heterogeneous) versions of our model
that provide an appropriate formalization and extension of the basic prior model of~\cite{E04,NYYA07}.

\item
In Sections~\ref{s2}--\ref{s3}, we provide formalizations of how the initial failures of some nodes in the network (\IE, a shock) 
originate, and how such failures are propagated to other network nodes in successive time steps using a discrete-time {\em shock propagation equation}.

\item
In Section~\ref{stab-def} we define our global network stability measure $\mathcal{K}$.

\item
In Sections~\ref{rat1}~and~\ref{rat2}, we provide rationales for the network model and the global stability measure, respectively.
\end{itemize}

\vspace*{0.1in}
\noindent
{\large $\bullet$} $\,$ 
In Section~\ref{sim-sec} we describe our simulation environment and the combinations of parameters that are being explored. In particular:
\begin{itemize}
\item
In Section~\ref{topo}, we discuss the random network models for generation of network topologies.

\item
In Section~\ref{shock-def} we state and justify the two modes of initial failures ({\em idiosyncratic} and {\em coordinated}) that are being used in the simulation.

\item
In Sections~\ref{alphabeta-sec}--\ref{para}, we describe the combinations of parameters used for homogeneous and heterogeneous networks 
and few other minor details of the simulation environment.
\end{itemize}

\vspace*{0.1in}
\noindent
{\large $\bullet$} $\,$ 
In Section~\ref{empiri-sec} we discuss our findings from the evaluation of the stabilities of the networks. In particular:
\begin{itemize} 
\item
Our six conclusions 
\scalebox{1.5}[1.5]{\ding{172}}--\scalebox{1.5}[1.5]{\ding{177}}
for the stability measure involving various combinations of 
network topology and parameters appear in Sections~\ref{jj1}--\ref{jjj1}.

\item
In Section~\ref{kk2} we discuss two phase transition properties of the stability measure with an intuitive explanation for one of them.
\end{itemize}
Though the issue of stability of financial systems has been discussed by prior 
researchers~\cite{E04,arxiv,FN09,B07,Ace2013,E2013,PYR09,MR13,La13,AG2000,DP83,GK08,XBJ2000,H09,Lie2010,D04,RS98,CM09,NYYA07,CD09,A05}, 
no prior paper has performed a comprehensive evaluation of a global stability measure 
as done in this paper.

\section{Economic policy implications}
\label{econ-motiv}

Returning to our original motivation of flagging financial networks for potential vulnerabilities, 
our results suggest that a network model similar to that used in the paper may be flagged 
for the following cases: 
\begin{itemize}
\item
the equity to asset ratios of most banks are low, 

\item
the network has a highly skewed distribution of external assets and inter-bank exposures among its banks and 
the network is sufficiently sparse, 

\item
the network does not have either a highly skewed distribution of external assets or 
a highly skewed distribution of inter-bank exposures among its banks, but the network is sufficiently dense.
\end{itemize}

%%%%%%%%%%%%%%%%%%%%%%%%%%%%%%%%%%%%%%%%%%%%%%%%%%%%%%%%%%%%%%%%%%%%%%%%%%%%%%%%%%%%%%%%%%%%%%%%%%%%%%%%%%%%%%%%%%%%%%
%%%%%%%%%%%%%%%%%%%%%%%%%%%%%%%%%%%%%%%%%%%%%%%%%%%%%%%%%%%%%%%%%%%%%%%%%%%%%%%%%%%%%%%%%%%%%%%%%%%%%%%%%%%%%%%%%%%%%%
%%% introduction ends; model description starts
%%%%%%%%%%%%%%%%%%%%%%%%%%%%%%%%%%%%%%%%%%%%%%%%%%%%%%%%%%%%%%%%%%%%%%%%%%%%%%%%%%%%%%%%%%%%%%%%%%%%%%%%%%%%%%%%%%%%%%
%%%%%%%%%%%%%%%%%%%%%%%%%%%%%%%%%%%%%%%%%%%%%%%%%%%%%%%%%%%%%%%%%%%%%%%%%%%%%%%%%%%%%%%%%%%%%%%%%%%%%%%%%%%%%%%%%%%%%%

\section{Our financial network model and stability measure}
\label{model-sec}

\newcommand{\din}{{\mathrm{deg^{\,in}}}}
\newcommand{\dout}{{\mathrm{deg^{\,out}}}}
\newcommand{\dave}{{\mathrm{deg_{\,ave}}}}
\newcommand{\vdead}{{V_{\mathrm{\text{\bf\large\ding{34}}}}}}
\newcommand{\bank}{\mathrm{Bank}}
\newcommand{\E}{\mathcal{E}}
\newcommand{\I}{\mathcal{I}}
\newcommand{\R}{{\mathbb R}}
\newcommand{\vs}{{V_{{\text{\footnotesize\ding{54}}}}}}
\newcommand{\ex}{{\mathbb E}}
\newcommand{\Var}{{\mathrm{Var}}}
\newcommand{\A}{{\mathbb A}}
\newcommand{\B}{{\mathbb B}}
\newcommand{\cD}{{\mathcal{D}}}
\newcommand{\e}{{\mathrm{e}}}

Since our model has a large number of parameters, for the benefit of the reader we have included a 
short definition for major parameters at the beginning of each subsection where they are used.

\subsection{Network model and balance sheet}
\label{s1}

\begin{table}[htbp]
\begin{center}
\renewcommand{\tabcolsep}{2pt}
\begin{tabular}{r l  |  r l  |  r l}
\hline
\multicolumn{6}{c}{
A list of major parameters used in this section
}
\\
\hline
$\E$ & total external asset & $\I$ & total inter-bank exposure & $\gamma$ & ratio of equity to asset
\\
\hline
\begin{tabular}{c}
$w(e)$ 
\\
$=$
\\
$w(u,v)$ 
\end{tabular}
& 
\begin{tabular}{c}
weight of edge 
\\
$e=(u,v)$
\end{tabular}
& $\iota_v$ & interbank asset &
$e_v$ & \begin{tabular}{l}
           effective share of
           \\
           total external asset
        \end{tabular}
\\
\hline
$a_v$ & total asset & $b_v$ & 
          \begin{tabular}{l}
            total 
            interbank 
            borrowing
\end{tabular}
&
$c_v$ & net worth (equity)
\\
\hline
\end{tabular}
\renewcommand{\tabcolsep}{6pt}
\end{center}
\end{table}

\vspace*{0.1in}
\noindent
We state a formalization of an {\em ex ante} financial network model
similar to what has been used by researchers from Bank of England and elsewhere~\cite{E04,HM11,NYYA07,GK08,MA10}. 
As was done by these prior researchers, 
we formulate our model in terms of balance-sheet ``insolvency cascades'' in a
network of financial institutions ({\em hereafter simply called ``banks'' and ``banking networks''}) with {\em interlinked} balance sheets, where losses flow into the
asset side of the balance sheets. The same formulation can be used to analyze cascades of
cash-flow insolvency in {\em over-the-counter derivatives markets}. From now on, we will refer to balance-sheet insolvency simply as insolvency.

The banking network is represented by a {\em parameterized} node-weighted and edge-weighted directed graph $G=(V,E,\Gamma)$ in the following 
manner\footnote{The parameters $\E$, $\I$, $A$,  $\gamma$ and $\Phi$ were also used by prior researchers, and the parameters 
$\sigma_v$ and $w(e)$ are generalizations of parameters used by prior researchers.}:  
\begin{itemize}
\item
$\Gamma=\left\{\E,\I,\gamma\right\}$ is the set of parameters where 
\begin{itemize}
\item
$\E\in\R$ is the total external asset,

\item
$\I\in\R$ is the total inter-bank exposure,

\item
$A=\I+\E$ is the total asset, and 

\item
$\gamma\in(0,1)$ is the ratio of equity to asset.
\end{itemize}

\item
$V$ is the set of $n$ banks where 
\begin{itemize}
\item
the node weight $\sigma_v\in[0,1]$ denotes the share of total external asset for each bank $v\in V$ 
$\big(\sum_{v}\sigma_v=1\big)$. 
\end{itemize}

\item
$E$ represents the set of $m$ direct inter-bank exposures where 
\begin{itemize}
\item
$w(e)=w(u,v)>0$ is the weight of a directed edge $e=(u,v)\in E$.
\end{itemize}
\end{itemize}

\renewcommand{\tabcolsep}{6pt}
\begin{table}[htbp]
\caption{Relevant balance sheet details of a node $v$ in the network~\cite{E04,NYYA07,arxiv}.
The total amount of external assets $\E$ is assumed to be large enough such that $b_v-\iota_v+\sigma_v \, \E$ is positive.}
\label{balsheet}
\begin{center}
\begin{tabular}{l | l} 
\hline
\noalign{\smallskip}
\multicolumn{1}{c|}{\bf Assets} & \multicolumn{1}{c}{\bf Liabilities} 
\\
\hline
\noalign{\smallskip}
{
\begin{tabular}{rl}
$\iota_v=\sum_{(v,u)\in E} w(v,u)$ & \hspace*{0.05in} interbank asset 
\\
[0.15in]
$e_v=b_v-\iota_v+\sigma_v \, \E$ & 
            \begin{tabular}{l}
               effective share of
               \\
               total external asset
             \end{tabular}
\\
[0.15in]
$a_v=b_v+\sigma_v \E$ & \hspace*{0.05in} total asset
\end{tabular}
}
&
\begin{tabular}{rl}
$b_v=\sum_{(u,v)\in E} w(u,v)$ & 
\begin{tabular}{l}
            total 
            interbank 
            \\
            borrowing
\end{tabular}
\\
[0.15in]
$c_v=\gamma\,a_v$ & \hspace*{0.05in} net worth (equity)
\end{tabular}
\\
\noalign{\smallskip}
\hline
\end{tabular}
\end{center}
\end{table}
\renewcommand{\tabcolsep}{6pt}

The (interlocked) balance sheet for each node (bank) $v\in V$ is shown in Table~\ref{balsheet}.
Two types of banking network models are considered: 
\begin{description} 
\item[\bf Homogeneous model:]
$\E$ and $\I$ are {\em equally} distributed among the nodes and the edges, respectively, \IE, $\sigma_v=\frac{1}{n}$ for every node $v$, 
and $w(e)=\frac{\I}{m}$ for every edge $e$.

\item[\bf Heterogeneous model:]
$\E$ and $\I$ are not necessarily equally distributed among the nodes and the edges, respectively.
\end{description}
{\em Both} homogeneous and heterogeneous network models are relevant in practice, and have been investigated by prior researchers
such as~\cite{arxiv,HM11,NYYA07,AW12,GK08,MA10}.

\subsection{Initial insolvency via shocks}
\label{s2}

\renewcommand{\tabcolsep}{2pt}
\begin{tabular}{r l  |  r l  |  r l}
\hline
\multicolumn{6}{c}{
A list of major parameters and definitions used in this section
}
\\
\hline
$\vs$ & set of initially shocked nodes & $\Phi$ & severity of initial shock & $t$ & time variable
\\
\hline
\footnotesize shocking mechanism: & \begin{tabular}{l} 
                         \footnotesize rule to select an initial subset
                           \\
                         \footnotesize of nodes to be shocked 
                      \end{tabular}
& 
\\
\hline
\end{tabular}
\renewcommand{\tabcolsep}{6pt}

\vspace*{0.1in}
The initial insolvencies of a banking network at time $t=0$ are caused by ``shocks'' received by a {\em subset} $\emptyset\subset\vs\subseteq V$ of nodes. 
Such shocks can occur, for example, due to {\em operational risks} (\EG, frauds\footnote{Iyer and Peydro~\cite{IP06} show that
fraud is an important cause of bank insolvency.}) or {\em credit risks},
and has the effect of {\em reducing} the external assets of an {\em selected} subset of 
banks\footnote{Nier \EA~\cite{NYYA07} considered shocking only one (or a few) bank and empirically studying the effect of the shock on the entire network.
Berman \EA~\cite{arxiv}, on the other hand, analyzed the computational complexity issues of the 
problem of selecting a subset of nodes such that shocking them will make the network fail.}.
Mathematically, the effect of the initial shock is to {\em simultaneously} decrease the external assets of each shocked node $v\in\vs$ 
from $e_v$ by $s_v=\Phi\,e_v$, thereby reducing the net worth of $v$ from its original value $c_v$ to $c_v-s_v$,
where $(0,1]\ni\Phi>\gamma$ is a parameter denoting the {\em severity} of the initial shock.

In the rest of the paper, by the phrase ``shocking mechanism'', we refer to the {\em rule} that is used to select the initial subset of nodes to be 
shocked\footnote{Shocking mechanisms were not formally defined by prior researchers, but it was often implicit in their discussions.}.

\subsection{Insolvency propagation equation}
\label{s3}

\renewcommand{\tabcolsep}{4pt}
\begin{tabular}{r l  |  r l  }
\hline
\multicolumn{4}{c}{
A list of major parameters used in this section
}
\\
\hline
$\din(v)$ & in-degree of node $v$ & $\vdead\left(t,\vs\right)$ & 
         \begin{tabular}{l}
           set of nodes that became insolvent before time $t$ 
           \\
           when initial shock is provided to nodes in $\vs$ 
         \end{tabular}
\\
\hline
\end{tabular}
\renewcommand{\tabcolsep}{6pt}

\begin{figure*}
\includegraphics[width=1\textwidth]{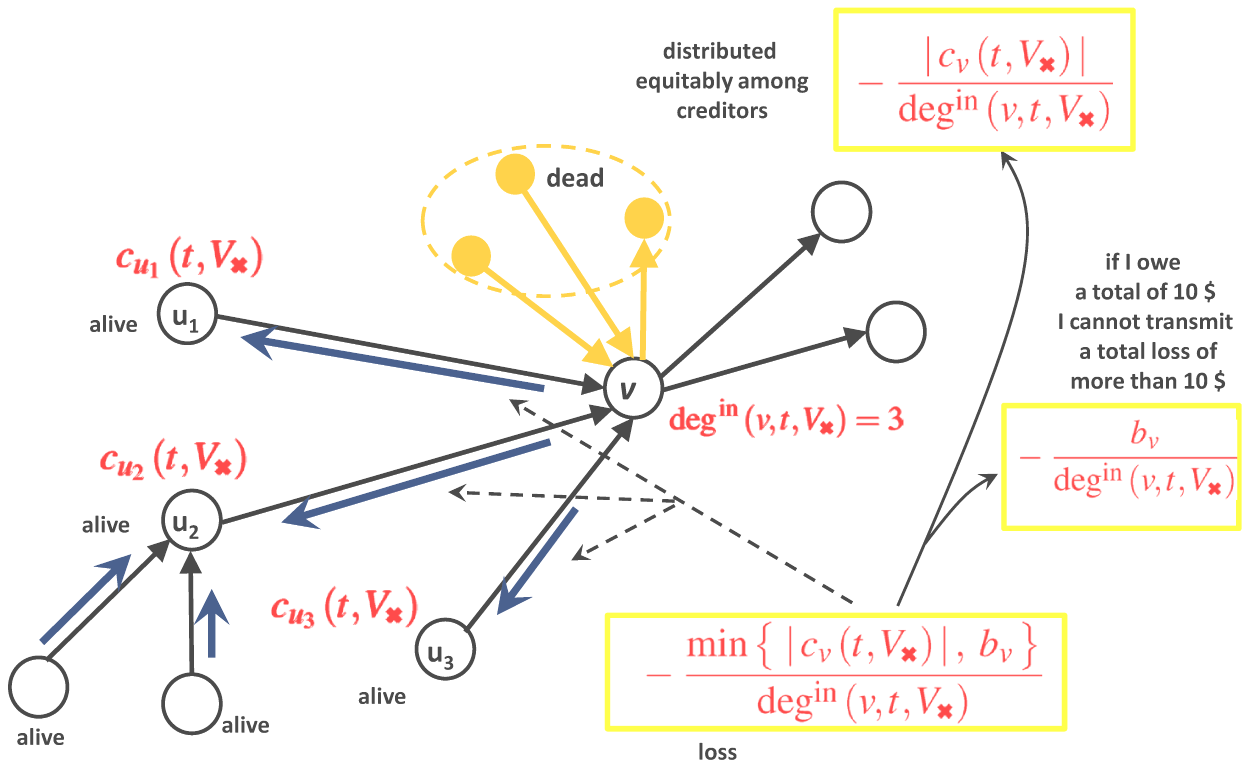}
\caption{Pictorial illustration of the shock transmission equation for a node $v$ from time $t$ to time $t+1$.}
\label{shock-eqn-fig}       
\end{figure*}

\vspace*{0.1in}
Let the notation $\din(v)$ denote the in-degree of node $v$. 
The insolvencies propagate in {\em discrete} time units $t=0,1,2,\dots$; 
we add ``$\left(\dots,t,\vs,\dots\right)$'' to all relevant variables to indicate their dependences on $t$ and on the set $\vs$ of initially shocked 
nodes.
A bank becomes insolvent if its modified net worth becomes {\em negative}, and such a bank is {\em removed} from the network in the next time step. 
Let $\vdead\left(t,\vs\right)\subseteq V$ denote the set of nodes that became insolvent {\em before} time $t$ when an initial shock is 
provided to the nodes
\footnote{Thus, in particular, $\din\left(v,t,\vs\right)$ is the in-degree in the graph induced by the nodes in 
$V\setminus \vdead\left(t,\vs\right)$.}
in $\vs$. The insolvencies of banks at 
time $t$ affect the equity of other banks in the network at the next time step $t+1$ by the following {\em non-linear} ``insolvency propagation 
equation''$^{\,}$\footnote{An equation of same flavor with some simplification and omitted details was described in words by 
Nier \EA~\cite{NYYA07} and Eboli~\cite{E04}.}$^,$\footnote{Equation~\eqref{eq1} is highly non-linear. The results in~\cite{arxiv} indicate that {\em in general} it is 
{\sf NP}-hard to find a subset $\vs$ of initially shocked nodes such that 
$\displaystyle \left|\,\lim_{t\to\infty}\vdead\left(t,\vs\right)\,\right|$ is exactly or approximately maximized.}: 
\begin{gather}
\forall\,u\in V\setminus \vdead\left(t,\vs\right)
\colon
c_u\left(t+1,\vs\right) = c_u\left(t,\vs\right) \,\,\,\scalebox{2.5}[1]{\bf -}
\hspace*{-0.4in}
\sum_{v \, \pmb{\colon} \!\!\!\!\! \substack{\big( c_v\left(t,\vs\right)<0 \big) \\ \,\,\,\,\,\,\,\, \bigwedge \big( v\,\in V\setminus \vdead\left(t,\vs\right) \big) \\ \hspace*{-0.3in} \bigwedge \big( (u,v)\in E \big) } } 
\hspace*{-0.3in}
\frac{\min\Big\{\,\left|\,c_v\left(t,\vs\right)\,\right|\,,\,\,b_v\,\Big\}}{\din\left(v,\,t,\vs\right)}
\label{eq1}
\end{gather}
In Equation~\eqref{eq1}, the term $\dfrac{\left|\,c_v\left(t,\vs\right)\,\right|}{\din\left(v,t,\vs\right)}$ ensures the loss of equity of an insolvent bank to be distributed 
equitably among its creditors that have not become insolvent yet, whereas
the term $\dfrac{b_v}{\din\left(v,t,\vs\right)}$ ensures that the total loss propagated cannot be more than 
the total interbank exposure of the insolvent bank; see {\sc Fig}.~\ref{shock-eqn-fig} for a pictorial illustration.
The insolvency propagation continues until {\em no} new bank becomes insolvent. 
For notational convenience, we may use 
$\left(\dots,t,\dots\right)$
instead of 
$\left(\dots,t,\vs,\dots\right)$
when $\vs$ is clear from the context or is irrelevant.

\subsection{Rationale for the network model and insolvency propagation equation}
\label{rat1}

As prior researchers~\cite{NYYA07,AG08,PYR09} have commented:
\begin{description}
\item
\hspace*{0.05in}
\em 
``conceptual frameworks from the theory of weighted graphs
with additional parameters may provide a powerful tool for analysis of banking network models''.
\end{description}
Several parametric graph-theoretic models, differing in the way
edges are interpreted and additional parameters are used to characterize the contagion,
have been used by prior researchers in 
finance and banking industry to study various research questions involving financial 
systems~\cite{arxiv,E04,HM11,AW12,NYYA07,F03,UW04,M07,ACM11,CMS10,SM12,Z11}.
As noted by researchers in~\cite{NYYA07,ACM11}:
\begin{description}
\item
\hspace*{0.05in}
\em 
``the modelling challenge in studying banking networks lies not so much in analyzing a model that is flexible enough to represent all types of insolvency cascades, but in 
studying a model that can mimic the empirical properties of these different types of networks''.
\end{description}
The insolvency propagation model formalized and evaluated in this paper using a mathematically precise abstraction
is similar to or a generalization of the models in~\cite{arxiv,E04,HM11,NYYA07,AW12,GK08,MA10,ACM11} that represent {\em cascades} of cash-flow insolvencies.
As~\cite{ACM11} observes, over-the-counter (OTC) derivatives and similar markets are {\em prone to this type of cascades}.
In such markets parties deal {\em directly} with one another rather than passing through an exchange, and thus 
each party is subject to the risk that the other party does not fulfill its payment obligations. 
The following example from~\cite{ACM11} illustrates chains of such interactions:
\begin{description}
\item
\hspace*{0.05in}
\cite{ACM11}\em
``Consider two parties A and B, such that A has a
receivable from party B upon the realization of some event. If B does not dispose of enough
liquid reserves, it will default on the payment. Now consider that B has entered an off-setting
contract with another party C, hedging its exposure to the random event. If C is cash-flow
solvent, then the payment will flow through the intermediary B and reach A. However, if C
is cash-flow insolvent and defaults, then the intermediary B might become cash-flow insolvent
if it depends on receivables from C to meet its payment obligations to A''.
\end{description}
The length of such chains of interactions in some OTC markets, like the {\em credit default swap market}, is {\em significant}~\cite{C10,MMM11}, 
thereby {\em increasing} the probability of cascade of cash-flow insolvencies~\cite{AG08}.
As~\cite{NYYA07} observes, an insolvency propagation model such as the one studied here
\begin{description}
\item
\hspace*{0.05in}
\em 
``conceptualises the main characteristics of a financial system using network theory
by relating the cascading behavior of financial networks both to the local properties of the nodes and to the underlying topology of the network, 
allowing us to vary continuously the key parameters of the network''.
\end{description}
Although the cascading effect studied is of somewhat special and simplified nature, as noted by~\cite{HM11}:
\begin{description}
\item
\hspace*{0.05in}
\em
``This is a deliberate oversimplification, aimed at a clearer understanding of how an initial failure can propagate shocks throughout the system''.
\end{description}

\subsection{A measure of global stability\protect\footnote{A mathematically precise definition of global stability measure was omitted by most prior researchers.}}
\label{stab-def}

\begin{table}[htbp]
\begin{center}
\renewcommand{\tabcolsep}{2pt}
\begin{tabular}{r l  |  r l  }
\hline
\multicolumn{4}{c}{
A list of major parameters used in this section
}
\\
\hline
$\Upsilon$ & a shocking mechanism & 
$\xi\left(\mathcal{K},G,\gamma,\Phi,\Upsilon\right)=x$ & 
\begin{tabular}{p{3in}}
on an average $100x\%$ nodes of the network become insolvent with the given values of $\gamma$ and $\Phi$ 
if we provide an initial shock to a random subset of $100\mathcal{K}\%$ of nodes selected using the shocking mechanism $\Upsilon$
\end{tabular}
\\
\hline
\end{tabular}
\renewcommand{\tabcolsep}{6pt}
\end{center}
\end{table}

Consider a banking network model as described in Sections~\ref{s1}--\ref{s3}. Let $\mathcal{K}\in (0,1]$ be a real 
number\footnote{$\mathcal{K}$ is a new parameter not used by prior researchers.}
denoting the fraction of nodes in $V$ that received the initial shock under a shocking mechanism $\Upsilon$ 
and let $S_{\Upsilon,\mathcal{K}}$ be the set of all 
possible $(\mathcal{K}n)$-element subsets of $V$. 
The {\em vulnerability index}\footnote{Although simple topological properties such as clustering coefficients have been used by authors to study 
properties of networks~\cite{OKK04,SOKK}, they are too simplistic for stability analysis of financial networks.}
of the network is then defined as\footnote{In this definition, we implicitly assume that the shocking mechanism $\Upsilon$ allows one to 
select at least one set of $\mathcal{K}n$ nodes for the initial shock. Otherwise, we define $\xi$ to be zero.} 
\[
\xi\left(\mathcal{K},G,\gamma,\Phi,\Upsilon\right)
=\frac{1}{n} \times 
\,
\ex \left[ \,\,\, \left|\,\lim_{t\to\infty}\vdead\left(t,\vs\right)\,\right| \, 
\,\colon \,
\text{
$\vs$ is selected randomly from $S_{\Upsilon,\mathcal{K}}$
}
\right]
\]
In the above definition, the $\frac{1}{n}$ factor is only
for a min-max normalization~\cite{HK00} to ensure that
$0\leq\xi\left(\mathcal{K},G,\gamma,\Phi,\Upsilon\right)\leq 1$.
Noting that no new node in the network may fail at a time $t\geq n$, we may simplify the above expression for $\xi$ as:
\begin{multline*}
\xi\left(\mathcal{K},G,\gamma,\Phi,\Upsilon\right)
=\frac{1}{n} \times 
\,
\ex \left[ \,\,\, \left|\,\vdead\left(n,\vs\right)\,\right| \, 
\,\colon \,
\text{
$\vs$ is selected randomly from $S_{\Upsilon,\mathcal{K}}$
}
\right]
\\
\Longrightarrow \, 
\Pr \left[ 
\,\left|\,\vdead\left(n,\vs\right)\,\right| \, 
\geq 
n\,\xi\left(\mathcal{K},G,\gamma,\Phi,\Upsilon\right)
\,\colon \,
\text{
$\vs$ is selected randomly from $S_{\Upsilon,\mathcal{K}}$
}
\right] > 0 
\end{multline*}
As an example, $\xi\left(0.1,G,0.3,0.5,\mathrm{random}\right)=0.9$ means that with positive probability 
$90\%$ nodes of the network $G$ become insolvent with $\gamma=0.3$ and $\Phi=0.5$ 
if we provide an initial shock to a random subset of $10\%$ of nodes of $G$.
{\em Note that lower values of $\xi$ imply higher global stability of a network}.
For simplicity, we may omit the arguments of $\xi$ when they are clear from the context.
A pseudo-code for calculating $\xi$ is shown in Fig.~\ref{ps1}. 

\begin{figure*}[t]
{
\begin{center}
\begin{tabular}{l}
\hline
\\
[-0.1in]
\hspace*{0.15in}
$(*$ {\sf initialization of parameters} $*)$
\\
\hspace*{0.15in}
$t\leftarrow 0$ ; $\vdead\left(0,\vs\right)\leftarrow\emptyset$ ; continue$\,=${\sf TRUE} ; 
\\
\hspace*{0.15in}
$\vs\leftarrow$ set of $\mathcal{K}\!\!n$ nodes selected for initial insolvency based on shocking mechanism $\Upsilon$
\\
[0.05in]
\hspace*{0.15in}
{\bf for} every node $v\in V$ {\bf do} 
\\
\hspace*{0.25in} $\din\left(v,0,\vs\right)\leftarrow\din(v)$ 
\\
\hspace*{0.15in} 
{\bf endfor}
\\
[0.05in]
\hspace*{0.15in}
$(*$ {\sf provide an initial shock} $*)$
\\
[0.05in]
\hspace*{0.15in}
{\bf for} every node $u\in V'$ {\bf do} 
\\
\hspace*{0.25in} $c_u\left(0,\vs\right)\leftarrow c_u-\Phi\,e_u$ 
\\
\hspace*{0.15in} 
{\bf endfor}
\\
[0.05in]
\hspace*{0.15in}
$(*$ {\sf loop until no new node fails} $*)$
\\
[0.05in]
\hspace*{0.05in}
$
\left[
\begin{array}{l}
\text{{\bf while} $\Big(\big( \,\text{continue$\,=${\sf TRUE}}\, \big) \bigwedge \big(\,\vdead\left(t,\vs\right)\neq V\,\big)\Big)$ {\bf do}} 
\\
\left[
\begin{array}{l}
\text{{\bf for} every node $u\in V\setminus\vdead\left(t,\vs\right)$ {\bf do}}
\\
\hspace*{0.1in}
c_u\left(t+1,\vs\right)\leftarrow c_u\left(t,\vs\right)
\\
[0.05in]
\left[
\begin{array}{l}
\text{{\bf for} every node $v\in V\setminus\vdead\left(t,\vs\right)$ {\bf do}}
\\
\hspace*{0.15in}
\text{{\bf if} $\Big( \big( \, c_v\left(t,\vs\right)<0 \, \big) \bigwedge \big(\,  (u,v)\in E\,  \big) \Big)$ {\bf then}}
\hspace*{0.15in}
\text{
$(*$ {\sf propagate shock to in-neighbors} $*)$
}
\\
\hspace*{0.5in}
c_u\left(t+1,\vs\right)\leftarrow c_u\left(t+1,\vs\right)-\dfrac{\min\big\{\,\big|\,c_v\left(t,\vs\right)\,\big|\,,\,\,b_v\,\big\}}{\din\left(v,t,\vs\right)}
\\
\text{{\bf endfor}} 
\end{array}
\right.
\\
[0.05in]
\hspace*{0.1in}
\vdead\left(t+1,\vs\right) \leftarrow \vdead\left(t,\vs\right) ; 
\\
[0.05in]
\hspace*{0.15in}
\text{{\bf if} $c_u\left(t,\vs\right)<0$ {\bf then}} 
\hspace*{0.1in}
\vdead\left(t+1,\vs\right)\leftarrow\vdead\left(t+1,\vs\right)\cup \big\{u\big\}
\\
\hspace*{2.0in}
\text{
$(*$ {\sf node $u$ fails if its equity becomes negative} $*)$
}
\\
\text{{\bf endfor}} 
\end{array}
\right.
\\
[0.05in]
\hspace*{0.1in}
t\leftarrow t+1
\\
[0.05in]
\hspace*{0.1in}
\text{{\bf if} $\left(\,\vdead\left(t,\vs\right)=\vdead\left(t-1,\vs\right)\,\right)$ {\bf then}} 
\hspace*{0.1in}
\text{continue$\,\leftarrow\,${\sf FALSE}} 
\\
[0.05in]
\left[
\begin{array}{l}
\text{{\bf for} every node $u\in V\setminus\vdead\left(t,\vs\right)$ {\bf do}}
\\
\hspace*{0.1in}
\din\left(u,t,\vs\right)
=
\Big| \Big\{ v \colon \big( \, (v,u)\in E \, \big)  \wedge \big( \, v\in V\setminus\vdead\left(t,\vs\right) \, \big) \Big\} \Big|
\\
\hspace*{0.5in}
\text{
$(*$ {\sf update in-degrees of alive nodes to exclude freshly dead nodes} $*)$
}
\\
\text{{\bf endfor}} 
\end{array}
\right.
\\
\text{{\bf endwhile}} 
\end{array}
\right.
$
\\
[0.05in]
$\xi\left(\mathcal{K},G,\gamma,\Phi,\Upsilon\right)\leftarrow \dfrac{\left|\vdead\left(t,\vs\right)\right|}{n}$
\\
\noalign{\smallskip}
\hline
\end{tabular}
\end{center}
}
\caption{Pseudo-code for calculating $\xi\left(\mathcal{K},G,\gamma,\Phi,\Upsilon\right)$. 
Comments in the pseudo-code are enclosed by $(*$ and $*)$. 
An implementation of the pseudo-code is available
at \protect\url{www2.cs.uic.edu/~dasgupta/financial-simulator-files}.} 
\label{ps1}
\end{figure*}

\subsection{Rationale for the global stability measure}
\label{rat2}

It is possible to think of other alternate measures of global stability than the one quantified above. 
However, the measure introduced above is {\em in tune} with the ideas that references~\cite{E04,NYYA07,HM11,AW12} 
directly (and, some other references such as~\cite{F03,UW04,MA10} implicitly) 
used to empirically study their networks. Thus, in formalizing our global stability measure, 
we have decided to follow the cue provided by other researchers in the banking industry who have studied various insolvency propagation models.
Measures of similar flavor have also been used by prior researchers in social networks in the context of influence
maximization~\cite{Chen08,CWY09}.

\section{Simulation environment and explored parameter space}
\label{sim-sec}

In Table~\ref{sum-par} we provide a summary of our simulation environment and explored parameter space. Individual 
components of the summary are discussed in Sections~\ref{topo}--\ref{para}.

\subsection{Network topology\protect\footnote{One may obviously ask: {\em why not use ``real'' networks}? There are several obstacles however
that make this desirable goal impossible to achieve. For example: {\bf ({\em a})} Due to their highly sensitive nature, such networks with all relevant parameters
are {\em rarely} publicly available. {\bf ({\em b})} For the kind of inferences that we make in this paper, we need hundreds of thousands of large networks to 
have any statistical validity (in this paper, we explore more than 700,000 networks).}}
\label{topo}

\renewcommand{\tabcolsep}{3pt}
\begin{table*}
\begin{tabular}{l l l c}
\cline{1-3}
\noalign{\smallskip}
parameter & \multicolumn{2}{c}{explored values for the parameter} & 
\\
\cline{1-3}
\noalign{\smallskip}
\multirow{3}{*}{\color{red}network type} &                 {homogeneous}                    & 
& \multirow{14}{*}{\scalebox{2}[20]{\bf \}} $\!\!\!\!\!\!$ \begin{tabular}{c} total \\ number of \\ parameter \\ combinations \\ $>\,700,000$ \\[1.0in] \\ \end{tabular} } 
{\smallskip}
\\ 
\cline{2-3}
\noalign{\smallskip}
                                     &  \multirow{2}{*}{$(\alpha,\beta)$-heterogeneous} & $\alpha=0.1,\,\,\beta=0.95$ 
& \\ 
                                     &                                                         & $\alpha=0.2,\,\,\beta=0.6$ 
& \\ \cline{1-3}
\noalign{\smallskip}
\multirow{5}{*}{network topology} &  \multirow{3}{*}{directed scale-free           } & \hspace*{-0.3in}average degree $1$ (in-arborescence) 
& \\ [-0.00in] 
                                         &                                                         & average degree $3$ 
& \\ [-0.01in]
                                         &                                                         & average degree $6$
& \\ [-0.01in] \cline{2-3}
\noalign{\smallskip}
                                         &  \multirow{2}{*}{directed Erd\"{o}s-R\'{e}nyi}   &  average degree $3$ 
& \\ 
                                         &                                                                    &  average degree $6$ 
& \\ \cline{1-3}
\noalign{\smallskip}
\multirow{1}{*}{shocking mechanism           } &  idiosyncratic, coordinated                     &                     
& \\[-0.00in] 
\cline{1-3}
\noalign{\smallskip}
number of nodes                   &  $50,\,100,\,300$                                    &                  
& \\
\cline{1-3}
\noalign{\smallskip}
$\nicefrac{E}{I}$              & \multicolumn{2}{l}{$0.25,0.5,0.75,1,1.25,1.5,1.75,2,2.25,2.5,2.75,3,3.25,3.5$}
& 
\\
\cline{1-3}
\noalign{\smallskip}
$\Phi$                           & $0.5,\,0.6,\,0.7,\,0.8,\,0.9$                             &          
& \\
\cline{1-3}
\noalign{\smallskip}
$\mathcal{K}$                  & \multicolumn{2}{l}{$0.1,\,0.2,\,0.3,\,0.4,\,0.5,\,0.6,\,0.7,\,0.8,\,0.9$}
& \\
\cline{1-3}
\noalign{\smallskip}
$\gamma$                         & $0.05,\,0.1,\,0.15,\dots,\Phi-0.05$                   &               
& \\
\cline{1-3}
\end{tabular}
\caption{\label{sum-par}A summary of simulation environment and explored parameter space.}
\end{table*}
\renewcommand{\tabcolsep}{6pt}

We consider two topology models previously used by economists to generate random financial networks: 
\begin{itemize}
\item
the {\em directed scale-free} ({\sf SF}) network model~\cite{BaAl99} that has been 
used by prior financial network researchers such as~\cite{SR10,M11,ACM11,CMS10}, and 

\item
the {\em directed Erd\"{o}s-R\'{e}nyi} ({\sf ER}) network model~\cite{Bo03} that has been used by prior financial network researchers such as~\cite{A10,GK08,MGS09,CD09,CNSW00}.
\end{itemize}

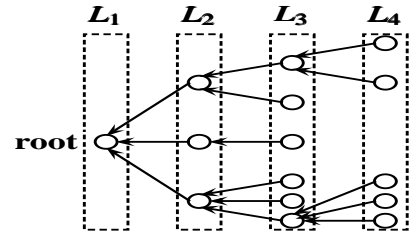
\begin{wrapfigure}[8]{R}{0pt}
%%% in-arborescence
\scalebox{0.55}[0.35]{
\begin{pspicture}(-2.5,-2.2)(7,3.4)
\psset{xunit=0.75cm,yunit=0.75cm}
\rput[mr](-0.9,0){\Huge\bf root}
%%%%%%
\pscircle[origin={0,0},linewidth=2pt](0,0){0.3}
%%%%%%
\pscircle[origin={3,-3},linewidth=2pt](0,0){0.3}
\pscircle[origin={3,0},linewidth=2pt](0,0){0.3}
\pscircle[origin={3,3},linewidth=2pt](0,0){0.3}
%%%%%%
\pscircle[origin={6,-4},linewidth=2pt](0,0){0.3}
\pscircle[origin={6,-3},linewidth=2pt](0,0){0.3}
\pscircle[origin={6,-2},linewidth=2pt](0,0){0.3}
\pscircle[origin={6,0},linewidth=2pt](0,0){0.3}
\pscircle[origin={6,2},linewidth=2pt](0,0){0.3}
\pscircle[origin={6,4},linewidth=2pt](0,0){0.3}
%%%%%%
\pscircle[origin={9,-4},linewidth=2pt](0,0){0.3}
\pscircle[origin={9,-3},linewidth=2pt](0,0){0.3}
\pscircle[origin={9,-2},linewidth=2pt](0,0){0.3}
\pscircle[origin={9,3},linewidth=2pt](0,0){0.3}
\pscircle[origin={9,5},linewidth=2pt](0,0){0.3}
%%%%%%
%%%%%%
%%%%%%
\psline[origin={0,0},linewidth=2pt,linecolor=black,arrowsize=1.5pt 4]{->}(2.7,-3)(0,-0.3)
\psline[origin={0,0},linewidth=2pt,linecolor=black,arrowsize=1.5pt 4]{->}(2.7,0)(0.3,0)
\psline[origin={0,0},linewidth=2pt,linecolor=black,arrowsize=1.5pt 4]{->}(2.7,3)(0,0.3)
%%%%%%
\psline[origin={0,0},linewidth=2pt,linecolor=black,arrowsize=1.5pt 4]{->}(5.7,-3)(3.3,-3)
\psline[origin={0,0},linewidth=2pt,linecolor=black,arrowsize=1.5pt 4]{->}(5.7,-2)(3,-2.7)
\psline[origin={0,0},linewidth=2pt,linecolor=black,arrowsize=1.5pt 4]{->}(5.7,-4)(3,-3.3)
\psline[origin={0,0},linewidth=2pt,linecolor=black,arrowsize=1.5pt 4]{->}(5.7,0)(3.3,0)
\psline[origin={0,0},linewidth=2pt,linecolor=black,arrowsize=1.5pt 4]{->}(5.7,2)(3,2.7)
\psline[origin={0,0},linewidth=2pt,linecolor=black,arrowsize=1.5pt 4]{->}(5.7,4)(3,3.3)
%%%%%%
\psline[origin={0,0},linewidth=2pt,linecolor=black,arrowsize=1.5pt 4]{->}(8.7,-4)(6.3,-4)
\psline[origin={0,0},linewidth=2pt,linecolor=black,arrowsize=1.5pt 4]{->}(8.7,-3)(6.1,-3.9)
\psline[origin={0,0},linewidth=2pt,linecolor=black,arrowsize=1.5pt 4]{->}(8.7,-2)(6,-3.7)
\psline[origin={0,0},linewidth=2pt,linecolor=black,arrowsize=1.5pt 4]{->}(8.7,3)(6,3.7)
\psline[origin={0,0},linewidth=2pt,linecolor=black,arrowsize=1.5pt 4]{->}(8.7,5)(6,4.3)
%
%%%%%%
\psframe[origin={0,0},linewidth=2pt,linestyle=dashed](-0.75,-4.5)(0.75,5.5)
\rput(0,6.4){\Huge$\pmb{L_1}$}
\psframe[origin={3,0},linewidth=2pt,linestyle=dashed](-0.75,-4.5)(0.75,5.5)
\rput(3,6.4){\Huge$\pmb{L_2}$}
\psframe[origin={6,0},linewidth=2pt,linestyle=dashed](-0.75,-4.5)(0.75,5.5)
\rput(6,6.4){\Huge$\pmb{L_3}$}
\psframe[origin={9,0},linewidth=2pt,linestyle=dashed](-0.75,-4.5)(0.75,5.5)
\rput(9,6.4){\Huge$\pmb{L_4}$}
\end{pspicture}
}
\caption{\label{in-arb-fig}An in-arborescence graph.}
\end{wrapfigure}

\noindent
Generation of directed {\sf ER} networks is computationally trivial: given a value $0<p<1$ that parameterizes the {\sf ER} network, for every {\em ordered} pair of distinct nodes $(u,v)$
we let $\Pr\big[(u,v)\in E\big]=\nicefrac{1}{p}$. Letting $p=\nicefrac{d}{n}$ generates a random {\sf ER} network whose average degree is $d$ with high probability.

The directed {\sf SF} networks in this paper are generated using the algorithm outlined by Bollobas \EA~\cite{Bo03}. 
The algorithm works as follows.
Let $a$, $b$, $\eta$, $\delta_{\mathrm{in}}$ (in-degree) and $\delta_{\mathrm{out}}$ (out-degree) 
be non-negative real numbers with $a+b+\eta=1$. The initial graph $G(0)$ at step $\ell=0$ has just one node with no edges. 
At step $\ell>0$ the graph $G(\ell)$ has exactly 
$\ell$ edges and a random number $n_{\ell}$ of nodes. For $\ell\geq 0$, $G(\ell+1)$ is obtained from $G(\ell)$ by using the following rules:
\begin{itemize}
\item 
With probability $a$, add a new node $v$ together with an edge from $v$ to an existing node $w$, where $w$ is chosen randomly 
such that 
\[
\Pr[w=u]=\frac{(d_{\mathrm{in}}(u) + \delta_{\mathrm{in}})}{(\ell + \delta_{\mathrm{in}}\, n_{\ell} )}
\]
for every existing node $u$, where $d_{\mathrm{in}}(u)$ is the in-degree of node $u$ in $G(\ell)$. 

\item 
With probability $b$, add an edge from an existing node $v$ to an existing node $w$, where $v$ and $w$ are chosen independently, such that 
\begin{gather*}
\Pr[v=u]  =  \frac{d_{\mathrm{out}}(u) + \delta_{\mathrm{out}}}{\ell + \delta_{\mathrm{out}} \, n_{\ell} }\,\mbox{ for every existing node $u$}
\\
\Pr[w=u]  =  \frac{d_{\mathrm{in}}(u) + \delta_{\mathrm{in}}}{\ell + \delta_{\mathrm{in}}\,  n_{\ell} }\, \mbox{ for every existing node $u$}
\end{gather*}
where $d_{\mathrm{out}}(u)$ is the out-degree of node $u$ in $G(\ell)$. 

\item 
With probability $\eta$, add a new node $w$ and an edge from an existing node $v$ to $w$, where $v$ is chosen such that 
$\Pr[v\!\!\!=\!u]=\dfrac{d_{\mathrm{out}}(u) + \delta_{\mathrm{out}}}{\ell + \delta_{\mathrm{out}}\,  n_{\ell} }$ for every existing node $u$.
\end{itemize}
To study the {\em effect of connectivity} on network stability, we generated random {\sf SF} and {\sf ER} networks with average 
degrees\footnote{There are many ways to fix the parameters to get the desired average degree. For example, as observed in~\cite{Bo03}, 
letting $\delta_{\mathrm{out}}=0$ and $\alpha>0$, one obtains 
$
{\mathbb E}\big[\mbox{number of nodes in $G(t)$ of in-degree $x$}\big]\propto x^{-\left(1 + \frac{1+\delta_{\mathrm{in}}(\alpha+\eta)}{\alpha+\beta} \right)}  t \,\,
$
and 
$
{\mathbb E}\big[\mbox{number of nodes in $G(t)$ of out-degree $x$}\big]\propto x^{-\left(\frac{2-\alpha}{1-\alpha}\right)}  t \,\,
$.
 }
of $3$ and $6$.

In addition, to study the effect of sparse hierarchical topology on network stability, 
we used the Bar\'{a}basi-Albert {\em preferential-attachment} model~\cite{BaAl99} to generate random {\em in-arborescence} 
networks. 
In-arborescences are directed rooted trees with all edges oriented towards the root (see Fig.~\ref{in-arb-fig}), and have the following well-known topological properties:
\begin{itemize}
\item
They belong to the class of {\em sparsest connected directed acyclic} graphs.

\item
They are {\em hierarchical} networks, \IE, the nodes can be partitioned into levels $L_1,L_2,\dots,L_p$ such that 
$L_1$ has exactly one node (the ``root'') and nodes in any level $L_i$ have directed edges only to nodes in $L_{i-1}$ (see Fig.~\ref{in-arb-fig}). 
The root may model a ``central bank'' that lends to other banks but does not borrow from any bank.
\end{itemize}
The algorithm for generating a random in-arborescence network $G$ using the preferential-attachment model~\cite{BaAl99} is as follows:
\begin{itemize}
\item  
Initialize $G=(V,E)$ to have one node (the root) and no edges.

\item
Repeat the following steps till $G$ has $n$ nodes: 
\begin{itemize}
\item
Randomly select a node $u$ in $G$ such that, for every node $v$ in $G$, 
$
\Pr[u=v]=\frac{\mathrm{deg}(v)}{\sum_{w\in V}\mathrm{deg}(w)}
$
where $\mbox{deg}(y)$ denotes the degree of node $y$ in $G$.

\item
Add a new node $x$ and an undirected edge $\{x,u\}$ in $G$.
\end{itemize}

\item
Orient all the edges towards the root.
\end{itemize}

\subsection{Shocking mechanisms $\Upsilon$}
\label{shock-def}

Recall that a shocking mechanism $\Upsilon$ provides a rule to select the initial subset of nodes to be shocked.
The following two mechanisms are used to select the nodes to receive the initial shock.

\vspace*{0.1in}
\noindent
{\bf Idiosyncratic (random) shocking mechanism}
We select a subset of nodes {\em uniformly at random}. This corresponds to random idiosyncratic initial insolvencies of banks, and 
is a choice that has been used by prior researchers such as~\cite{NYYA07,AW12,HM11,GK08,MA10}.

\vspace*{0.1in}
\noindent
{\bf Coordinated shocking mechanism}\footnote{While correlated shocking mechanisms affecting a correlated subset of banks are relevant in practice,
prior researchers such as~\cite{NYYA07,HM11,AW12,GK08,MA10} have mostly used idiosyncratic shocking mechanisms.
There are at least two reasons for this. Firstly, idiosyncratic shocks are a {\em cleaner} way to study the stability of the topology of the banking network.
Secondly, it is not {\em a priori} clear what kind of correlations in the shocking mechanism would lead to failure of more nodes
than idiosyncratic shocks in a {\em statistically significant way}.
Our coordinated shocking mechanism intuitively corresponds shocks in which banks that are 
``too big to fail'' in terms of their assets are correlated. Our conclusion 
\scalebox{1.5}[1.5]{\ding{177}}
shows that coordinated shocks
do indeed cause more statistically significant damage to the stability of the network as opposed to random shocks.}.
In this type of {\em non-idiosyncratic} correlated shocking mechanism, we seek to play an {\em adversarial} 
role\footnote{Usage of adversarial strategies in measuring the worst-case bounds for network properties are very common in the algorithmic literature; 
see, for example, see the book~\cite{BE98}.} in selecting nodes for the initial shock that may cause more damage to the stability of the network. 
The selection of an adversarial strategy depends on whether the network is homogeneous or heterogeneous.
The coordinated shocking mechanism falls under the general category of {\em correlated} shocks where the nodes with 
high (weighted) in-degrees are correlated. 

For homogeneous networks, recall that all nodes have the same share of the total external asset $\E$. However, the total interbank
exposure $b_v$ of a node $v$ is directly proportional to the in-degree of $v$, and, as per
Equation~\eqref{eq1}, nodes with higher inter-bank exposures are 
more likely to transmit the shock to a larger number of {\em other} nodes.
Thus, we play an adversarial role by selecting a set of $\mathcal{K}\,n$ nodes in {\em non-increasing} order of their in-degrees 
starting from a node with the highest in-degree.

For heterogeneous networks, nodes with higher ``weighted'' in-degrees (\IE, with higher values of the sum of {\em weights} of
incoming edges) represent nodes that have larger external assets than other nodes, 
and have more inter-bank exposures. Thus, for heterogeneous networks
we play an adversarial role by selecting $\mathcal{K}\,n$ nodes in non-increasing order of their {\em weighted in-degrees}
starting from a node with the highest weighted in-degree.

\subsection{Network type: $(\alpha,\beta)$-heterogeneous networks}
\label{alphabeta-sec}

Recall that in a heterogeneous network it is possible to have a few banks whose 
external assets or interbank exposures are {\em significantly larger} than the rest of the banks, \IE, it is possible to 
have a few banks that are ``too big'', and thus heterogeneous networks permit investigation of the effect of such big banks on the global stability of the entire
network. To this end, we define a $(\alpha,\beta)$-heterogeneous network as follows.

\begin{definition}[\bf $\,\pmb{(\alpha,\beta)}$-heterogeneous network]\label{defnet}
Let $\widetilde{V}\subseteq V$ be a random subset $V$ of $\alpha n$ nodes and let $\widetilde{E}$ be the set of edges that have at least one end-point from $\widetilde{V}$.
For $0<\alpha,\beta<1$, a $(\alpha,\beta)$-heterogeneous network $G=(V,E)$ is one in which 
the total external and internal assets are distributed in the following manner:
\begin{quote}
\begin{description}
\item[\bf Distribution of $\E$:]$\,$
\begin{itemize}
\item
distribute $\beta \E$ part of the total external asset $\E$ equally among the $\alpha\,n$ nodes in $\widetilde{V}$, and 

\item
distribute the remaining part $(1-\beta)\E$ of $\E$ equally among the {\em remaining} $(1-\alpha)n$ nodes. 
\end{itemize}

\item[\bf Distribution of $\I$:]$\,$
\begin{itemize}
\item
Distribute $\beta \I$ part of the total interbank exposure $\I$ equally among a random subset of $\alpha|\widetilde{E}|$
of edges from the edges in $\widetilde{E}$, and 

\item
distribute the remaining part $(1-\beta)\I$ of $\I$ equally among the {\em remaining} $|E|-\alpha|\widetilde{E}|$ edges.
\end{itemize}
\end{description}
\end{quote}
\end{definition}

We performed our simulations for $(\alpha,\beta)$-heterogeneous networks for $(\alpha,\beta)=(0.1,0.95)$ and $(\alpha,\beta)=(0.2,0.6)$.
The combination $(\alpha,\beta)=(0.1,0.95)$ corresponds to the extreme situation in which 95\% of the assets and exposures involve 10\% of banks, thus
creating a minority of banks that are significantly larger than the remaining banks. The other combination $(\alpha,\beta)=(0.2,0.6)$ corresponds to a 
less extreme situation in which there are a larger number of moderately large banks.

\subsection{Other minor details}
\label{para}

To correct statistical biases, for each combinations of parameters, shock types and network types, 
we generated $10$ corresponding random networks and computed the average value of the vulnerability index over these $10$ runs.
For {\sf ER} and {\sf SF} random networks, we selected the values of network generation parameters such that the expected number of edges of the network is 
$3n$ or $6n$ depending on whether we require the average degree of the network to be $3$ or $6$, respectively. 

The minimum difference between two non-identical values of the average vulnerability index over $10$ runs for two $n$-node networks is 
$\nicefrac{1}{(10\,n)}$. Thus, to allow for minor statistical biases introduced by any random graph generation method, we consider 
two vulnerability indices to be same (within the margin of statistical error) if their absolute difference is no more than $\nicefrac{1}{(3\,n)}$, 
which is above $\nicefrac{1}{(10\,n)}$ but no more than 0.7\% of the total range of the vulnerability indices.

Finally, we can assume without loss of generality that $\I=m$, since otherwise if $\mu=\frac{\I}{m}\neq 1$ then we can divide each of the 
quantities $\iota_v$, $b_v$ and $\E$ by $\mu$ without changing the
outcome of the insolvency propagation procedure.

\section{Results}
\label{empiri-sec}

In this section, we discuss our uncovering of many interesting relationships of the stability with other 
relevant parameters of the network based on our comprehensive evaluation and analysis of this stability measure. 
It is easy to see that there are many (at least several thousands, but significantly more in most cases) networks in the original sets of networks 
that are compared in two different scenarios in Tables~\ref{table2},\ref{table6},\ref{table8},\ref{table3},\ref{table4and5},\ref{table1},\ref{table10},\ref{table11} and 
related tables in the supplementary documents, thereby assuring the statistical validity of the comparison results.

\subsection{\bf Effect of unequal distribution of total assets $\E$ and $\I$}
\label{jj1}

As our analysis shows, nodes with disproportionately large external assets affect the stability of
the entire network in an {\em adverse} manner, and more uneven distribution of assets among nodes in the network makes the network less stable.

\subsubsection{Effect on global stability}

For the same value of the common parameters $n$, $\frac{\E}{\I}$, $\mathcal{K}$, $\Phi$ and $\gamma$, for the same for network type ({\sf ER}, {\sf SF} or in-arborescence) 
of same average degree (6, 3 or 1) and for the same shocking mechanism $\Upsilon$ (coordinated or idiosyncratic), we compared the value of 
$\xi$ for the homogeneous network with the corresponding values of $\xi$ for $(0.1,0.95)$-heterogeneous and $(0.2,0.6)$-heterogeneous networks.
The comparison results shown in Table~\ref{table2} show most of the entries as being at least 90\%. Thus, we conclude: 

\begin{quote}
{
\scalebox{1.5}[1.5]{\ding{172}}
\em 
networks with all nodes having the same external assets display higher stability 
over similar networks with fewer nodes having disproportionately higher external assets.
}
\end{quote}

%%%%%%%%%%%%%%%%%%%%%%%%%%%%%%%%%%%%%%%%%%%%%%%%%%%%%%%%%%%%%%%%%%%%%%%%%%%%%%%%%%%%%%%%%%%%%%%%%%%%%%%%%%%%%%
%%%%%%%%%%%%%%%%%%%%%%%%%%%%%%%%%%%%%%%%%%%%%%%%%%%%%%%%%%%%%%%%%%%%%%%%%%%%%%%%%%%%%%%%%%%%%%%%%%%%%%%%%%%%%%
%%% Table for xi for homogeneous vs heterogeneous networks
%
\begin{table}[ht]
\renewcommand{\tabcolsep}{2pt}
\caption{Comparison of stabilities of $(\alpha,\beta)$-heterogeneous networks with their homogeneous counter-parts over all parameter ranges.
The numbers are the percentages of data points for which $\xi_{\mathrm{(\alpha,\beta)-heterogeneous}}$ was at least  $\xi_{\,\mathrm{homogeneous}}$.}
\label{table2}
{
\begin{center}
\hspace*{-0.1in}
\begin{tabular}{r c | c | c  | c |c | c |c | c |c | c}
\toprule
%%%%%%%%%%%%%%%%%%%%%%%%%%%%%%%%%%%%%%%%%%%%%%%%%
%% column headings for other than first column
%%%%%%%%%%%%%%%%%%%%%%%%%%%%%%%%%%%%%%%%%%%%%%%%%
& 
       \multicolumn{2}{c|}{In-arborescence} 
       & 
      \multicolumn{2}{c|}{ \begin{tabular}{c} {\sf ER} \\ average degree 3 \end{tabular}} 
      &
      \multicolumn{2}{c|}{\begin{tabular}{c} {\sf ER} \\ average degree 6 \end{tabular}} 
      &
      \multicolumn{2}{c|}{\begin{tabular}{c} {\sf SF} \\ average degree 3 \end{tabular}} 
      &
      \multicolumn{2}{c}{\begin{tabular}{c} {\sf SF} \\ average degree 6 \end{tabular}} 
\\
[0.05in]
&
$\alpha=0.1$ 
&
$\alpha=0.2$ 
&
$\alpha=0.1$ 
&
$\alpha=0.2$ 
&
$\alpha=0.1$ 
&
$\alpha=0.2$ 
&
$\alpha=0.1$ 
&
$\alpha=0.2$ 
&
$\alpha=0.1$ 
&
$\alpha=0.2$ 
\\
&
$\beta=0.95$
&
$\beta=0.6$
&
$\beta=0.95$
&
$\beta=0.6$
&
$\beta=0.95$
&
$\beta=0.6$
&
$\beta=0.95$
&
$\beta=0.6$
&
$\beta=0.95$
&
$\beta=0.6$
\smallskip
\\
\cmidrule{2-11}
\begin{tabular}{c}
coordinated \\ shock
\end{tabular}
%%%%%%%%%%%%%%%%%%%%%%%%%%%%%%%%%%%%%%%%%%%%%%%%%
%% first column ends
%%%%%%%%%%%%%%%%%%%%%%%%%%%%%%%%%%%%%%%%%%%%%%%%%
%
%%%%%%%%%%%%%%%%%%%%%%%%%%%%%%%%%%%%%%%%%%%%%%%%
%% other columns and data points
%%%%%%%%%%%%%%%%%%%%%%%%%%%%%%%%%%%%%%%%%%%%%%%%
          & 66.91\% & 60.22\% & \bf 99.26\% & \bf 98.91\% & \bf 98.46\% & \bf 98.00\% & \bf 98.22\% & \bf 91.68\% & \bf 99.13\% & \bf 97.4\%
\\
& & & & & & & & & &
%%%%%%%%%%%%%%%%%%%%%%%%%%%%%%%%%%%%%%%%%%%%%%%%
%% other columns and data points end
%%%%%%%%%%%%%%%%%%%%%%%%%%%%%%%%%%%%%%%%%%%%%%%%
\\
\begin{tabular}{c}
idiosyncratic \\ shock
\end{tabular}
%%%%%%%%%%%%%%%%%%%%%%%%%%%%%%%%%%%%%%%%%%%%%%%%%
%% first column ends
%%%%%%%%%%%%%%%%%%%%%%%%%%%%%%%%%%%%%%%%%%%%%%%%%
%
%%%%%%%%%%%%%%%%%%%%%%%%%%%%%%%%%%%%%%%%%%%%%%%%
%% other columns and data points
%%%%%%%%%%%%%%%%%%%%%%%%%%%%%%%%%%%%%%%%%%%%%%%%
          & \bf 92.75\% & \bf 81.79\% & \bf 97.76\% & \bf 96.81\% & \bf 98.16\% & \bf 97.61\% & \bf 98.86\% & \bf 94.84\% & \bf 98.83\% & \bf 97.22\%
\\
\bottomrule
\end{tabular}
\end{center}
}
\end{table}

\noindent
{\bf Formal intuition behind the conclusion in \scalebox{1.5}[1.5]{\ding{172}}}

\vspace*{0.1in}
\noindent
In spite of the highly non-linear nature of Equation~\eqref{eq1}, the following formal intuition may help to explain the conclusion in 
\scalebox{1.5}[1.5]{\ding{172}}.

\begin{lemma}[see Section~\ref{g1proof-sec} of the appendix for a proof]\label{g1}
Fix $\gamma$, $\Phi$, $\E$, $\I$ and the graph $G$.
Consider any node $v\in\vs$ and suppose that $v$ fails due to the initial shock. 
For every edge $(u,v)\in E$, let $\Delta_{\,\mathrm{homo}}(u)$ and $\Delta_{\,\mathrm{hetero}}(u)$ be  
the amount of shock received by node $u$ at time $t=1$ if 
$G$ is homogeneous or heterogeneous, respectively. 
Then, 
\[
\ex \left[ \Delta_{\,\mathrm{hetero}}(u)\right] \geq \dfrac{\beta}{\alpha}\,\ex \left[ \Delta_{\,\mathrm{homo}}(u)\right]
=
\begin{array}{rl}
9.5
\,
\ex \left[ \Delta_{\,\mathrm{homo}}(u) \right] 
, & \mbox{ if $(\alpha,\beta)=(0.1,0.95)$}
\\
[0.05in]
3
\,
\ex \left[ \Delta_{\,\mathrm{homo}}(u) \right] 
, & \mbox{ if $(\alpha,\beta)=(0.2,0.6)$}
\end{array}
\]
\end{lemma}

Lemma~\ref{g1} implies that $\ex \left[ \Delta_{\,\mathrm{hetero}}(u)\right]$ is much bigger than $\ex\left[ \Delta_{\,\mathrm{homo}}(u)\right]$, 
and thus more nodes are likely to fail beyond $t>0$ leading to a lower stability 
for heterogeneous networks.

\subsubsection{Effect on residual instability}

For homogeneous networks, if the equity to asset ratio $\gamma$ is close enough to the severity of the shock $\Phi$
then the network almost always tends to be perfectly stable, as one would intuitively expect.  
However, the above property is {\em not} true in general for highly heterogeneous networks in the sense that,
even when $\gamma$ is close to $\Phi$, these networks (irrespective of their topologies and densities) have a {\em minimum} amount of
global instability (which we term as the {\em residual instability})\footnote{For visual illustrations to this phenomena, see {\em supplemental} {\sc Fig}.~\ref{newfig1}--\ref{newfig3}.
For example, in {\em supplemental} {\sc Fig}.~\ref{newfig1}, when $\gamma$ is $45\%$ and $\Phi$ is 
only $5\%$ more than $\gamma$, the vulnerability index $\xi$ is approximately $0$ for {\em all} the $9$ combinations of parameters, but 
in {\em supplemental} {\sc Fig}.~\ref{newfig2}--\ref{newfig3} {\em all} the $18$ networks have a value of $\xi\geq 0.1$ 
even when $\gamma$ is $45\%$ and the severity of the shock is 
only $5\%$ more than $\gamma$.}.

In Table~\ref{table6} and {\em supplemental} Tables~\ref{table7}--\ref{supp4-table7} we tabulated residual instabilities for 
different types of homogeneous and heterogeneous networks under coordinated and idiosyncratic shocks. The numbers in these tables show, for 
each combination of network types, $|V|$, shocking mechanism and values of $\Phi$ and $\gamma$ such that $|\Phi-\gamma|=0.05$, the percentage of 
networks with this combination for which the vulnerability index $\xi$ was less than $0.05$, $0.1$ or $0.2$. As the reader will observe, 
all the numbers for heterogeneous networks are significantly lower than their homogeneous counter-parts. Thus, we conclude: 

\begin{quote}
\scalebox{1.5}[1.5]{\ding{173}}
{\em a heterogeneous network, in contrast to its corresponding homogeneous version, 
has a residual minimum instability even if its equity to asset ratio is very large and close to the severity of the shock}.
\end{quote}

\renewcommand{\tabcolsep}{3pt}
\begin{table}[htbp]
\caption{Residual instabilities of homogeneous versus heterogeneous networks under coordinated shocks. 
The percentages shown are the percentages of networks for which $\xi<0.05$ or $\xi<0.1$ or $\xi<0.2$.
{\bf See also {\em supplementary} Tables~\ref{table7}--\ref{supp4-table7}}.}
\label{table6}
\begin{center}
\scalebox{0.85}[0.75]
{
\begin{tabular}{c c r      r r r | r r r}
\toprule
&
&
&
\multicolumn{6}{c}{\bf coordinated shock}
\\
&
&
&
\multicolumn{3}{c|}{$\Phi=0.5,\gamma=0.45$}
&
\multicolumn{3}{c}{$\Phi=0.5,\gamma=0.40$}
\\
&
&
&
\multicolumn{1}{c}{${\xi<0.05}$}
&
\multicolumn{1}{c}{${\xi<0.1}$}
&
\multicolumn{1}{c|}{${\xi<0.2}$}
&
\multicolumn{1}{c}{${\xi<0.05}$}
&
\multicolumn{1}{c}{${\xi<0.1}$}
&
\multicolumn{1}{c}{${\xi<0.2}$}
\\
\noalign{\smallskip}
\cline{1-9}
\noalign{\smallskip}
\multirow{15}{*}{\begin{tabular}{c}$|V|$ \\ $=$ \\ $50$ \end{tabular} }
&
\multirow{5}{*}{homogeneous} 
&
in-arborescence
&  
\bf 73\%  & \bf 73\%  &  \bf 73\% & \bf 0\% \bf & \bf 31\%  & \bf 59\%
\\
&
&
{\sf ER}, average degree $3$ 
&  
\bf 89\%  & \bf 100\%  &  \bf 100\% & \bf 43\% \bf & \bf 84\%  & \bf 100\%
\\
&
&
{\sf ER}, average degree $6$ 
&  
\bf 100\%  & \bf 100\%  &  \bf 100\% & \bf 100\% \bf & \bf 100\%  & \bf 100\%
\\
&
&
{\sf SF}, average degree $3$ 
&  
\bf 44\%  & \bf 84\%  &  \bf 100\% & \bf 25\% \bf & \bf 57\%  & \bf 88\%
\\
&
&
{\sf SF}, average degree $6$ 
&  
\bf 100\%  & \bf 100\%  &  \bf 100\% & \bf 100\% \bf & \bf 100\%  & \bf 100\%
\\
\noalign{\smallskip}
\cline{2-9}
\noalign{\smallskip}
& 
\multirow{5}{*}{${(0.1,0.95)}$-heterogeneous}
&
in-arborescence
&  
0\%  & 0\%  &  0\% & 0\% \bf & 0\%  & 0\%
\\
&
&
{\sf ER}, average degree $3$ 
&  
0\%  & 0\%  &  1\% & 0\% \bf & 0\%  & 0\%
\\
&
&
{\sf ER}, average degree $6$ 
&  
8\%  & 9\%  &  10\% & 2\% \bf & 6\%  & 6\%
\\
&
&
{\sf SF}, average degree $3$ 
&  
2\%  & 6\%  &  15\% & 0\% \bf & 2\%  & 5\%
\\
&
&
{\sf SF}, average degree $6$ 
&  
18\%  & 23\%  &  30\% & 9\% \bf & 10\%  & 11\%
\\
\noalign{\smallskip}
\cline{2-9}
\noalign{\smallskip}
& 
\multirow{5}{*}{${(0.2,0.6)}$-heterogeneous}
&
in-arborescence
&  
0\%  & 0\%  &  9\% & 0\% \bf & 0\%  & 9\%
\\
&
&
{\sf ER}, average degree $3$ 
&  
4\%  & 7\%  &  19\% & 2\% \bf & 6\%  & 16\%
\\
&
&
{\sf ER}, average degree $6$ 
&  
8\%  & 12\%  &  24\% & 6\% \bf & 7\%  & 16\%
\\
&
&
{\sf SF}, average degree $3$ 
&  
2\%  & 6\%  &  22\% & 0\% \bf & 2\%  & 18\%
\\
&
&
{\sf SF}, average degree $6$ 
&  
8\%  & 12\%  &  24\% & 7\% \bf & 8\%  & 16\%
\\
%
%%%%%%%%%%%%%%%%%%%%%%%%%%%%%%%%%%%%%%%%%%%%%%%%%%%%%%%%%%%%%%%%%%%%%%%%%%%%%%%%%%%%%%%%%%%%%%%%%%%%%%%%%%
\noalign{\smallskip}
\hline
\noalign{\smallskip}
\multirow{15}{*}{ \begin{tabular}{c} $|V|$ \\ $=$ \\ $100$ \end{tabular} }
&
\multirow{5}{*}{homogeneous} 
&
in-arborescence
&  
\bf 73\%  & \bf 73\%  &  \bf 73\% & \bf 0\% \bf & \bf 34\%  & \bf 73\%
\\
&
&
{\sf ER}, average degree $3$ 
&  
\bf 66\%  & \bf 100\%  &  \bf 100\% & \bf 25\% \bf & \bf 64\%  & \bf 100\%
\\
&
&
{\sf ER}, average degree $6$ 
&  
\bf 100\%  & \bf 100\%  &  \bf 100\% & \bf 100\% \bf & \bf 100\%  & \bf 100\%
\\
&
&
{\sf SF}, average degree $3$ 
&  
\bf 29\%  & \bf 61\%  &  \bf 100\% & \bf 20\% \bf & \bf 42\%  & \bf 83\%
\\
&
&
{\sf SF}, average degree $6$ 
&  
\bf 100\%  & \bf 100\%  &  \bf 100\% & \bf 90\% \bf & \bf 100\%  & \bf 100\%
\\
\noalign{\smallskip}
\cline{2-9}
\noalign{\smallskip}
& 
\multirow{5}{*}{${(0.1,0.95)}$-heterogeneous}
&
in-arborescence
&  
0\%  & 0\%  &  0\% & 0\% & 0\%  & 0\%
\\
&
&
{\sf ER}, average degree $3$ 
&  
0\%  & 0\%  &  6\% & 0\% & 0\%  & 1\%
\\
&
&
{\sf ER}, average degree $6$ 
&  
6\%  & 6\%  &  6\% & 4\% & 6\%  & 6\%
\\
&
&
{\sf SF}, average degree $3$ 
&  
0\%  & 0\%  &  7\% & 0\% & 0\%  & 3\%
\\
&
&
{\sf SF}, average degree $6$ 
&  
6\%  & 10\%  &  15\% & 6\% & 6\%  & 6\%
\\
\noalign{\smallskip}
\cline{2-9}
\noalign{\smallskip}
& 
\multirow{5}{*}{${(0.2,0.6)}$-heterogeneous}
&
in-arborescence
&  
0\%  & 0\%  &  9\% & 0\% & 0\%  & 9\%
\\
&
&
{\sf ER}, average degree $3$ 
&  
0\%  & 6\%  &  16\% & 0\% & 4\%  & 16\%
\\
&
&
{\sf ER}, average degree $6$ 
&  
6\%  & 7\%  &  16\% & 6\% & 6\%  & 16\%
\\
&
&
{\sf SF}, average degree $3$ 
&  
0\%  & 2\%  &  14\% & 0\% & 1\%  & 13\%
\\
&
&
{\sf SF}, average degree $6$ 
&  
7\%  & 8\%  &  17\% & 6\% & 7\%  & 16\%
\\
%
%%%%%%%%%%%%%%%%%%%%%%%%%%%%%%%%%%%%%%%%%%%%%%%%%%%%%%%%%%%%%%%%%%%%%%%%%%%%%%%%%%%%%%%%%%%%%%%%%%%%%%%%%%
\noalign{\smallskip}
\hline
\noalign{\smallskip}
\multirow{15}{*}{ \begin{tabular}{c} $|V|$ \\ $=$ \\ $300$ \end{tabular}}
&
\multirow{5}{*}{homogeneous} 
&
in-arborescence
&  
\bf 73\%  & \bf 73\%  &  \bf 73\% & \bf 0\% \bf & \bf 55\%  & \bf 73\%
\\
&
&
{\sf ER}, average degree $3$ 
&  
\bf 71\%  & \bf 97\%  &  \bf 100\% & \bf 22\% \bf & \bf 60\%  & \bf 100\%
\\
&
&
{\sf ER}, average degree $6$ 
&  
\bf 100\%  & \bf 100\%  &  \bf 100\% & \bf 100\% \bf & \bf 100\%  & \bf 100\%
\\
&
&
{\sf SF}, average degree $3$ 
&  
\bf 22\%  & \bf 44\%  &  \bf 86\% & \bf 18\% \bf & \bf 36\%  & \bf 74\%
\\
&
&
{\sf SF}, average degree $6$ 
&  
\bf 100\%  & \bf 100\%  &  \bf 100\% & \bf 88\% \bf & \bf 100\%  & \bf 100\%
\\
\noalign{\smallskip}
\cline{2-9}
\noalign{\smallskip}
& 
\multirow{5}{*}{${(0.1,0.95)}$-heterogeneous}
&
in-arborescence
&  
0\%  & 0\%  &  0\% & 0\% & 0\%  & 0\%
\\
&
&
{\sf ER}, average degree $3$ 
&  
0\%  & 0\%  &  6\% & 0\% & 0\%  & 1\%
\\
&
&
{\sf ER}, average degree $6$ 
&  
6\%  & 6\%  &  6\% & 0\% & 6\%  & 6\%
\\
&
&
{\sf SF}, average degree $3$ 
&  
0\%  & 0\%  &  10\% & 0\% & 0\%  & 2\%
\\
&
&
{\sf SF}, average degree $6$ 
&  
6\%  & 6\%  &  16\% & 6\% & 6\%  & 6\%
\\
\noalign{\smallskip}
\cline{2-9}
\noalign{\smallskip}
& 
\multirow{5}{*}{${(0.2,0.6)}$-heterogeneous}
&
in-arborescence
&  
0\%  & 0\%  &  9\% & 0\% & 0\%  & 9\%
\\
&
&
{\sf ER}, average degree $3$ 
&  
0\%  & 6\%  &  16\% & 0\% & 4\%  & 16\%
\\
&
&
{\sf ER}, average degree $6$ 
&  
6\%  & 6\%  &  16\% & 6\% & 6\%  & 16\%
\\
&
&
{\sf SF}, average degree $3$ 
&  
0\%  & 0\%  &  13\% & 0\% & 0\%  & 12\%
\\
&
&
{\sf SF}, average degree $6$ 
&  
6\%  & 7\%  &  16\% & 6\% & 6\%  & 16\%
\\
\bottomrule
\end{tabular}
}
\end{center}
\end{table}
\renewcommand{\tabcolsep}{6pt}

\subsection{\bf Effect of total external assets}

A controversial belief regarding the cause of the collapse of many major financial institutions around 2007 
asserts that removal of the separation between investment and consumer banking
allowed a ripple effect of insolvencies of individual banks to other banks~\cite{ET08,C09}. 
In our setting, the quantity $\nicefrac{\E}{\I}$ controls the total (normalized) amount of
external investments of all banks in the network. Thus, varying the ratio $\nicefrac{\E}{\I}$ allows us to investigate the role of
the magnitude of total external investments on the stability of our banking network (see Table~\ref{table8}).
All the entries in Table~\ref{table8} are close to $0$,
showing that heterogeneous networks exhibited {\em very small} average changes in the vulnerability index $\xi$
when $\nicefrac{E}{I}$ was varied. Thus, we conclude:

\begin{quote}
\scalebox{1.5}[1.5]{\ding{174}}
{\em for heterogeneous banking networks, global stabilities are affected very little by the amount of the total external asset $\E$ in the system}.
\end{quote}

\noindent
Visual illustrations of \scalebox{1.5}[1.5]{\ding{174}}
are shown in {\em supplemental} {\sc Fig}. \ref{newfig4} and {\sc Fig}. \ref{newfig5} for homogeneous and heterogeneous networks, respectively.

%%%%%%%%%%%%%%%%%%%%%%%%%%%%%%%%%%%%%%%%%%%%%%%%%%%%%%%%%%%%%%%%%%%%%%%%%%%%%%%%%%%%%%%%%%%%%%%%%%%%%%%%%%%%%%
%%% table for average change of E for heterogeneous networks
\begin{table}
{%\normalsize
\caption{Absolute values of the largest change of the vulnerability index $\xi$ in the range $0.25\leq\nicefrac{\E}{\I}\leq 3.5$.}
\label{table8}
\hspace*{-0.0in}
\begin{center}
\begin{tabular}{r c c}
\toprule
& \multicolumn{2}{c}{\hspace*{-0.3in}average values of $\displaystyle\,\,\Big|\,\max_{0.25\leq \frac{\E}{\I}\leq 3.5} \big\{ \xi \big\} {-} \!\!\! \min_{\,0.25\leq \frac{\E}{\I}\leq 3.5} \big\{ \xi \big\} \,\Big|$}
\\[0.1in]
& coordinated shock & idiosyncratic shock 
\\
\noalign{\smallskip}
\cline{2-3}
\noalign{\smallskip}
$(0.1,0.95)$-heterogeneous in-arborescence & \bf 0.017 & \bf 0.045
\\
[0.02in]
$(0.2,0.6)$-heterogeneous in-arborescence & \bf 0.007 & \bf 0.017
\\
[0.02in]
$(0.1,0.95)$-heterogeneous {\sf ER}, average degree 3 & \bf 0.066 & \bf 0.073 
\\
[0.02in]
$(0.2,0.6)$-heterogeneous {\sf ER}, average degree 3 & \bf 0.040 & \bf 0.041
\\
[0.02in]
$(0.1,0.95)$-heterogeneous {\sf ER}, average degree 6 & \bf 0.111 & \bf 0.116
\\
[0.02in]
$(0.2,0.6)$-heterogeneous {\sf ER}, average degree 6 & \bf 0.084 & \bf 0.078
\\
[0.02in]
$(0.1,0.95)$-heterogeneous {\sf SF}, average degree 3 & \bf 0.119 & \bf 0.094
\\
[0.02in]
$(0.2,0.6)$-heterogeneous {\sf SF}, average degree 3 & \bf 0.034 & \bf 0.032
\\
[0.02in]
$(0.1,0.95)$-heterogeneous {\sf SF}, average degree 6 & \bf 0.200 & \bf 0.179 
\\
[0.02in]
$(0.2,0.6)$-heterogeneous {\sf SF}, average degree 6 & \bf 0.054 & \bf 0.054
\\
\bottomrule
\end{tabular}
\end{center}
}
\end{table}
%
%%\renewcommand{\tabcolsep}{6pt}
%%%%%%%%%%%%%%%%%%%%%%%%%%%%%%%%%%%%%%%%%%%%%%%%%%%%%%%%%%%%%%%%%%%%%%%%%%%%%%%%%%%%%%%%%%%%%%%%%%%%%%%%%%%%%%

\subsection{\bf Effect of network connectivity}

Although it is clear that connectivity properties of a banking network 
has a crucial effect on its stability, prior researchers have drawn mixed conclusions on this. For example, 
Allen and Gale~\cite{AG2000} concluded that networks with {\em less} connectivity are {\em more} prone to contagion than networks with higher connectivity due to
improved resilience of banking network topologies with higher connectivity via transfer of proportion of the losses in one bank's portfolio to more banks 
through interbank agreements. 
On the other hand, Babus~\cite{A05} observed that, when the network connectivity is higher, liquidity can be redistributed in the system to make the risk of contagion lower, and 
Gai and Kapadia~\cite{GK08} observed that higher connectivity among banks leads to more contagion effect during a crisis.
The mixed conclusions arise because links between banks have conceptually two {\em conflicting} effects on contagion, namely, 
\begin{itemize}
\item
more interbank links increases the opportunity for spreading insolvencies to other banks,

\item
but, more interbank links also provide banks with co-insurance against fluctuating liquidity flows.
\end{itemize}
As our findings show, these two conflicting effects have different strengths in homogeneous versus highly 
heterogeneous networks, thus justifying the mixed conclusions of past researchers.

%%%%%%%%%%%%%%%%%%%%%%%%%%%%%%%%%%%%%%%%%%%%%%%%%%%%%%%%%%%%%%%%%%%%%%%%%%%%%%%%%%%%%%%%%%%%%%%%%%%%%%%%%%%%%%
%%% Table for xi versus average degree comparison for homogeneous networks
%
\begin{table}[htbp]
\caption{Effect of connectivity on the stability for homogeneous networks under 
coordinated and idiosyncratic shocks. The percentage shown for a comparison of the type ``network A versus network B'' indicates
the percentage of data points for which the stability of networks of type A was at least as much as that of networks of type B.}
\label{table3}
\begin{center}
\begin{tabular}{c | c || c | c}
\toprule
%%%%%%%%%%%%%%%%%%%%%%%%%%%%%%%%%%%%%%%%%%%%%%%
%%% headings
%%%%%%%%%%%%%%%%%%%%%%%%%%%%%%%%%%%%%%%%%%%%%%%
\multicolumn{2}{c||}{{\sf ER} average degree 3 versus {\sf ER} average degree 6} 
         & \multicolumn{2}{c}{{\sf SF}  average degree 3 versus {\sf SF} average degree 6} 
\\
[0.03in]
    $\,\,\,\,\,$ coordinated shock & idiosyncratic shock
        & $\,\,\,\,\,$ coordinated shock & idiosyncratic shock
\\
& & & 
\\
\bf 97.43\% & \bf 97.05\% & \bf 98.89\% & \bf 98.29\% 
\\
\bottomrule
\end{tabular}
\end{center}
\end{table}
%%%%%%%%%%%%%%%%%%%%%%%%%%%%%%%%%%%%%%%%%%%%%%%%%%%%%%%%%%%%%%%%%%%%%%%%%%%%%%%%%%%%%%%%%%%%%%%%%%%%%%%%%%%%%%

\vspace*{0.1in}
\noindent
{\bf Homogeneous networks}
Recall that in a homogeneous network all banks have the same external asset.
Table~\ref{table3} shows sparser homogeneous networks with lower average degrees to be more stable for same values of other parameters.
Thus, we conclude:
\begin{quote}
\scalebox{1.5}[1.5]{\ding{175}}
{\em for homogeneous networks, higher connectivity leads to lower stability}.
\end{quote}

%%%%%%%%%%%%%%%%%%%%%%%%%%%%%%%%%%%%%%%%%%%%%%%%%%%%%%%%%%%%%%%%%%%%%%%%%%%%%%%%%%%%%%%%%%%%%%%%%%%%%%%%%%%%%%%%%%%%%%%%%%%%%%%%%%%%%%%%%%%
%%% Table for xi versus average degree comparison for heterogeneous networks
\begin{table}
\caption{Effect of connectivity on the stability under coordinated and idiosyncratic shocks 
for {\bf (A)} $(\alpha,\beta)$-heterogeneous {\sf ER} and {\sf SF} networks
and {\bf (B)} $(\alpha,\beta)$-heterogeneous in-arborescence versus $(\alpha,\beta)$-heterogeneous {\sf SF} networks.
The percentage shown for a comparison of the type ``network A versus network B'' indicates
the percentage of data points for which the stability of networks of type A was at least as much as that of networks of type B.}
\label{table4and5}
\renewcommand{\tabcolsep}{2pt}
\begin{center}
\scalebox{0.82}[0.82]
{
\begin{tabular}{c c | c c | c c | c c}
\toprule
\multicolumn{2}{c|}{(0,1,0.95) {\sf ER} average degree 6}
     & \multicolumn{2}{c|}{(0.2,0.6) {\sf ER} average degree 6}
           & \multicolumn{2}{c|}{(0,1,0.95) {\sf SF} average degree 6}
              & \multicolumn{2}{c}{(0.2,0.6) {\sf SF} average degree 6}
\\
\multicolumn{2}{c|}{versus}
     & \multicolumn{2}{c|}{versus}
           & \multicolumn{2}{c|}{versus}
              & \multicolumn{2}{c}{versus}
\\
\multicolumn{2}{c|}{(0,1,0.95) {\sf ER} average degree 3}
     & \multicolumn{2}{c|}{(0.2,0.6) {\sf ER} average degree 3}
           & \multicolumn{2}{c|}{(0,1,0.95) {\sf SF} average degree 3}
              & \multicolumn{2}{c}{(0.2,0.6) {\sf SF} average degree 3}
\\
\hline
 coordinated & 
    idiosyncratic & 
         coordinated & 
           idiosyncratic & 
                coordinated & 
                   idiosyncratic & 
                        coordinated & 
                           idiosyncratic 
\\
 shock & 
    shock & 
         shock & 
           shock & 
                shock & 
                   shock & 
                       shock & 
                          shock 
\\
& & & & & & & 
\\
 \bf 89.3\% 
   & \bf 82.39\% 
        & \bf 68.12\% 
           & \bf 61.46\% 
           & \bf 85.51\% 
                & \bf 73.81\% 
                 & \bf 69.29\% 
                     & \bf 73.07\% 
\\
\bottomrule
%
%%%%%%%%%%%%%%%%%%%%%%%%%%%%%%%%%%%%%%%%%%%%%%%%%%%%%%%%%%%%%%%%%%%%%%%%%%%%%%%%%%%%%%%%%%%
%%%%%%%%%%%%%%%%%%%%%%%%%%%%%%%%%%%%%%%%%%%%%%%%%%%%%%%%%%%%%%%%%%%%%%%%%%%%%%%%%%%%%%%%%%%
%%%%%%%%%%%%%%%%%%%%%%%%%%%%%%%%%%%%%%%%%%%%%%%%%%%%%%%%%%%%%%%%%%%%%%%%%%%%%%%%%%%%%%%%%%%
%
\\
\multicolumn{8}{c}{\bf\normalsize (A)}
\\
%%%%%%%%%%%%%%%%%%%%%%%%%%%%%%%%%%%%%%%%%%%%%%%%%%%%%%
%%%%%%%%%%%%%%%%%%%%%%%%%%%%%%%%%%%%%%%%%%%%%%%%%%%%%%
%%%%%%%%%%%%%%%%%%%%%%%%%%%%%%%%%%%%%%%%%%%%%%%%%%%%%%
%%%%%%%%%%%%%%%%%%%%%%%%%%%%%%%%%%%%%%%%%%%%%%%%%%%%%%
\\
\toprule
\multicolumn{4}{c|}{(0,1,0.95) {\sf SF} average degree 3 and average degree 6} & \multicolumn{4}{c}{(0.2,0.6) {\sf SF} average degree 3 and average degree 6}
\\
\multicolumn{4}{c|}{versus} & \multicolumn{4}{c}{versus}
\\
\multicolumn{4}{c|}{(0.1,0.95)-heterogeneous in-arborescence ({\sf SF} ave. degree 1)}
           & \multicolumn{4}{c}{(0.2,0.6)-heterogeneous in-arborescence ({\sf SF} ave. degree 1)}
\\
\hline
               \multicolumn{2}{c}{coordinated shock} & 
                  \multicolumn{2}{c|}{idiosyncratic shock} & 
                       \multicolumn{2}{c}{coordinated shock} & 
                          \multicolumn{2}{c}{idiosyncratic shock} 
\\
\multicolumn{4}{c|}{} & \multicolumn{4}{c}{}
\\
                    \multicolumn{2}{c}{\bf 85.7\%} 
                       & \multicolumn{2}{c|}{\bf 81.86\%} 
                        & \multicolumn{2}{c}{\bf 56.21\%} 
                           & \multicolumn{2}{c}{\bf 51.07\%} 
\\
\bottomrule
%%%%%%%%%%%%%%%%%%%%%%%%%%%%%%%%%%%%%%%%%%%%%%%%%%%%%%%%%%%%%%%%%%%%%%%%%%%%%%%%%%%%%%%%%%
%%%%%%%%%%%%%%%%%%%%%%%%%%%%%%%%%%%%%%%%%%%%%%%%%%%%%%%%%%%%%%%%%%%%%%%%%%%%%%%%%%%%%%%%%%
\\
\multicolumn{8}{c}{\bf\normalsize (B)}
\end{tabular}
}
\end{center}
\renewcommand{\tabcolsep}{6pt}
\end{table}

\vspace*{0.1in}
\noindent
{\bf Heterogeneous networks}
In a heterogeneous network,  
there {\em are} banks that are ``too big'' in the sense that these banks have much larger external assets and inter-bank exposures compared to the remaining banks.
Table~\ref{table4and5} shows that for heterogeneous network models
denser networks with higher average degree are more stable than the corresponding sparser networks for same values of other parameters, 
especially when the heterogeneity of the network is larger (\IE, when $\alpha=0.1,\beta=0.95$).
Thus, we conclude:
\begin{quote}
\scalebox{1.5}[1.5]{\ding{176}}
{\em for heterogeneous networks, higher connectivity leads to higher stability}.
\end{quote}

\vspace*{0.1in}
\noindent
{\bf Formal intuition behind the conclusions in \scalebox{1.5}[1.5]{\ding{175}} and \scalebox{1.5}[1.5]{\ding{176}}}

\vspace*{0.1in}
\noindent
Informally, conclusions 
\scalebox{1.5}[1.5]{\ding{175}} and \scalebox{1.5}[1.5]{\ding{176}}
indicate that 
in homogeneous networks higher connectivity leads to more opportunity for spreading insolvencies to other banks whereas in heterogeneous networks 
higher connectivity provides banks with co-insurance against fluctuating liquidity flows through shared interbank assets. However, a precise 
formal treatment of mechanism that drives such conclusions is complicated due to several reasons such as the random nature of the networks, 
the randomness in asset distribution for heterogeneous networks and the non-linear nature of the insolvency propagation equation. Nevertheless, 
we provide the following, somewhat simplified, formal reasoning.
We will use the following notations and conventions\footnote{In standard algorithmic analysis terminologies, $f\approx g$ implies $\frac{f(r)}{g(r)}=1\pm\mathrm{o}(1)$.}:
\begin{itemize}
\item
$\dave=\frac{|E|}{n}$ will denote the average degree of a graph $G$. It is assumed that $\dave$ is a small positive integer constant independent of $n$ (\EG, in our
simulation work, $\dave\in\{1,3,6\}$).

\item
$\Delta x$ will denote a small change for the value $x$ of a variable.

\item
For two functions $f(r)$ and $g(r)$ of a variable $r$, we will use the notation $f\approx g$ (respectively, $f \lessapprox g$, $f \gtrapprox g$) if 
$\lim\limits_{r \to \infty} \frac{f(r)}{g(r)}=1$ 
(respectively, $\lim\limits_{r \to \infty} \frac{f(r)}{g(r)}\leq 1$,  $\lim\limits_{r \to \infty} \frac{f(r)}{g(r)}\geq 1$).

\item
The standard phrase ``with high probability'' (or {\tt w.h.p.} in short) refers to a probability $p(n)$ such that $\lim_{n\to\infty}p(n)=0$. 

\item
If necessary, we will use the superscripts ``homo'' and ``hetero'' to denote the value of a quantity for homogeneous and heterogeneous networks, respectively.
\end{itemize}
Consider a node $v\in\vs$ with $\din(v)>1$ and suppose that $v$ fails due to the initial shock at $t=0$. By Equation~\eqref{eq1}, for every edge $(u,v)\in E$, the amount of shock 
$u$ receives from $v$ is given by $\B=\min \big\{\A,\,c_1\big\}$ with 
\begin{eqnarray}
\A
&
=
&
\frac
{
\Phi \left( c_1 \, \din(v) - c_1 \, \dout(v) + c_2 \, \E \right) \,-\, \gamma \left( c_1 \, \din(v) + c_2 \, \E \right)
}
{
\din(v)
}
\nonumber
\\
&
=
&
c_1 \big( \Phi - \gamma \big) 
+ 
c_2 \big(\Phi - \gamma \big) \frac{\E}{\din(v)} 
- c_1 \Phi  \frac{\dout(v)}{\din(v)}
\label{eq2}
\end{eqnarray}
for some appropriate positive quantities $c_1$ and $c_2$ that may be estimated as follows:
\begin{itemize}
\item
If $G$ is homogeneous then $c_1^{\mathrm{homo}}=\frac{\I}{n\,\dave}=1$ and $c_2^{\mathrm{homo}}=\frac{1}{n}$. 

\item
If $G$ is $(\alpha,\beta)$-heterogeneous then $c_1^{\mathrm{hetero}}$ and $c_2^{\mathrm{hetero}}$ are random variables independent of $\dave$.
Using the notations in Definition~\ref{defnet} the expected value of $c_2^{\mathrm{hetero}}$ may be estimated as follows: 
\begin{eqnarray*}
\ex\left[c_2^{\mathrm{hetero}}\right] & = & 
\frac{
\Pr\left[v\in\widetilde{V}\right] \frac{\beta\,\E}{\alpha\,n}
+
\Pr\left[v\notin\widetilde{V}\right] 
\frac{(1-\beta)\,\E}{(1-\alpha)\,n}
}{\E}
=
\frac{
\alpha\,\frac{\beta\,\E}{\alpha\,n}
+ (1-\alpha)\, \frac{(1-\beta)\,\E}{(1-\alpha)\,n}
}{\E}
=\frac{1}{n}
\end{eqnarray*}
The expected value of $c_1^{\mathrm{hetero}}$ depend on the nature ({\sf SF} or {\sf ER}) of the random network; its estimation is therefore deferred until later.
\end{itemize}
Our goal is to provide evidence for a claim of the following nature for either random {\sf SF} or random {\sf ER} networks:

\begin{quote}
\em
For many realistic network parameter combinations, {\tt w.h.p.} increasing connectivity from $d$ to $d+\Delta d$ 
decreases the expected amount of shock 
transmitted by a failed node $v$ to $u$ in homogeneous networks (causing improved stability) 
but increases the expected amount of shock 
transmitted by a failed node $v$ to $u$ in heterogeneous networks (leading to worse stability).
\end{quote}

\paragraph{\bf The case of random {\sf SF} networks}
If $G$ is a directed {\sf SF} network, then the discrete probability density function for the degree of any node $v$ in $G$ is given by:
\[
\forall \, k\in \{1,2,\dots,n-1\} \colon 
\Pr\big[ \din(v)=k \big] 
=
\Pr\left[ \dout(v)=k \right] 
=
C \, k^{-\mu} 
\]
where $\mu>2$ is the constant for the exponent of the distribution and $C>0$ is a constant such that 
$\ex\big[ \din(v)\big] = \ex\left[ \dout(v)\right] = \dave$.
For example, for our random in-arborescence networks, the results in~\cite{BaAl99} imply $\mu=3$.
To simplify exposure, in the following we assume that $\mu=3$, though the analysis can be extended in a straightforward manner
for any other $\mu>2$.
$\zeta(s)=\sum_{x=1}^{\infty}x^{-s}$ is the well-known Riemann zeta function; it is well known that 
$\zeta(s)\approx \sum_{x=1}^{n-1}x^{-s}$ for any $s>2$ and for all large $n$ (see~\cite[page 74-75]{Knuth}) and 
$\zeta(s)<\frac{2^{s-1}}{2^{s-1}-1}$ for any $s>1$ (see~\cite[page 489]{Knuth}).
In particular, it is known that $\zeta(2)=\frac{\pi^2}{6}$, $\zeta(3)=1.2020569\cdots$ and $\zeta(4)=\frac{\pi^4}{90}$.
Note that 
\[
\ex\big[ \din(v)\big] = \dave
\,\equiv\,
\sum_{k=1}^{n-1} k \left( C k^{-3} \right) = \dave
\,\Rightarrow\,
C \approx \frac{\dave}{\zeta(2)} = \frac{6\,\dave}{\pi^2}
\]

\begin{lemma}[see Section~\ref{bb1proof-sec} of the appendix for a proof]\label{bb1}
$\displaystyle
\ex\left[\frac{1}{\din(v)} \, \Big| \, \din(v)>0\right]
\approx
\frac{\pi^2}{15}\,\dave
$ 
and 
$\displaystyle
\Var\left[\din(v)\right]
\approx
\frac{6\,\dave}{\pi^2} \,\ln n 
$.
\end{lemma}

We now provide an estimation of $c_1^{\mathrm{hetero}}$ using the notations in Definition~\ref{defnet}.

\begin{lemma}[see Section~\ref{c1het-lemmaproof-sec} of the appendix for a proof]\label{c1het-lemma}
{\rm\tt W.h.p.} 
$\,\,\,1 + \alpha - \beta - \frac{\alpha\,\beta}{2}
\leq 
\ex\left[c_1^{\mathrm{hetero}}\right]  
\leq 
\frac{1 + \alpha\,\beta - \alpha^2 - \beta - \alpha^2\,\beta}{1-2\,\alpha}$.
\end{lemma}

Due to the above lemma, we may assume that 
\[
\mbox{\tt w.h.p. } 
\begin{array}{rl}
0.156875
\leq
\ex\left[c_1^{\mathrm{hetero}}\right]  
\leq
0.16, 
& 
\mbox{ if $G$ is $(0.1,0.95)$-heterogeneous}
\\
[0.05in]
0.54
\leq
\ex\left[c_1^{\mathrm{hetero}}\right]  
\leq
0.76, 
& 
\mbox{ if $G$ is $(0.2,0.6)$-heterogeneous}
\end{array}
\]
We will investigate the sensitivity of the amount of shock $\A$ transmitted from $v$ to $u$ as the average degree $\dave$ is changed 
while keeping all other parameters the same. For notational convenience, the parameters are normalized such that 
$\I=n\,\dave$ at the initial value $d$ of $\dave$. 
As $\dave$ is increased from $d$ to $d+\Delta d$, 
$\I$ is still kept the same.
Thus, 
\[
\I\Big|_{\substack{\mathrm{at} \\ \dave=d+\Delta d}}
=
\I\Big|_{\substack{\mathrm{at} \\ \dave=d}}
\,\Rightarrow \,
c_1^{\mathrm{homo}}
\Big|_{\substack{\mathrm{at} \\ \dave=d+\Delta d}}
=\frac{
\I\Big|_{\substack{\mathrm{at} \\ \dave=d+\Delta d}}
}{n\,\left(d + \Delta d \right)}
=
\frac{n\,d}{n\,\left(d + \Delta d \right)}
=
\frac{d}{d + \Delta d}
\]
Equation~\eqref{eq2} gives the following for homogeneous and heterogeneous networks:
\begin{itemize}
\item
If $G$ is a homogeneous network then
\begin{align*}
& 
\ex\left[\A^{\mathrm{homo}}\,\big|\, \dave=d\right]
\\
\displaybreak[0]
\approx
\,\,\,\,\,
& 
\big( \Phi - \gamma \big) 
+ 
\frac{\E\,\big(\Phi - \gamma \big)}{n} \ex\left[\frac{1}{\din(v)} \, \Big| \, \din(v)>0 \right] 
- \Phi \, \ex\left[\dout(v)\right] \, \ex\left[\frac{1}{\din(v)} \, \Big| \, \din(v)>0 \right]
\\
\displaybreak[0]
=
\,\,\,\,\,
& 
\big( \Phi - \gamma \big) 
+ 
\frac{\pi^2\,\E\,\big(\Phi - \gamma \big)}{15\,n} \,d
- \frac{\pi^2\,\Phi}{15}\,d^2
%%%%%%%%%%%%%%%%%%%%%%%%
\\
\displaybreak[0]
& 
\\
\displaybreak[0]
&
\ex\left[\A^{\mathrm{homo}} \,\big|\, \dave=d+\Delta d \right]
\\
\displaybreak[0]
\approx
\,\,\,\,\,
&
\textstyle
\frac{d}{d+\Delta d}\,
\big( \Phi - \gamma \big) 
+ 
\frac{\E\,\big(\Phi - \gamma \big)}{n} \ex\left[\frac{1}{\din(v)} \, \Big| \, \din(v)>0 \right] 
- 
\frac{d}{d+\Delta d}\,
\Phi \, \ex\left[\dout(v)\right] \, \ex\left[\frac{1}{\din(v)} \, \Big| \, \din(v)>0 \right]
\\
\displaybreak[0]
=
\,\,\,\,\,
&
\frac{d}{d+\Delta d}\,
\big( \Phi - \gamma \big) 
+ 
\frac{\pi^2\,\E\,\big(\Phi - \gamma \big)}{15\,n} \,\big( d + \Delta d \big) 
- 
\frac{d}{d+\Delta d}\,
\frac{\pi^2\,\Phi}{15}\,\big( d + \Delta d \big)^2
%%%%%%%%%%%%%%%%%%%%%%%%
\\
\displaybreak[0]
&
\\
&
\Delta \ex\big[\A^{\mathrm{homo}}\big]
\\
\displaybreak[0]
=
\,\,\,\,\,
&
\ex\left[\A^{\mathrm{homo}} \,\big|\, \dave=d+\Delta d \right]
-
\,
\ex\left[\A^{\mathrm{homo}} \,\big|\, \dave=d \right]
\\
\displaybreak[0]
\approx
\,\,\,\,\,
&
-\,\frac{\Delta d}{d+\Delta d}\,
\big( \Phi - \gamma \big) 
+ 
\frac{\pi^2\,\E\,\big(\Phi - \gamma \big)}{15\,n} \,\Delta d
-
\frac{\pi^2\,\Phi}{15}\,d \,\Delta d
\end{align*}

\item
If $G$ is a heterogeneous network then 
\begin{multline*}
\ex\big[\A^{\mathrm{hetero}}\big]
=
\big( \Phi - \gamma \big) \, 
\ex\left[ c_1^{\mathrm{hetero}} \right]
+ 
\frac{\E\,\big(\Phi - \gamma \big)}{n} \ex\left[\frac{1}{\din(v)}  \, \Big| \, \din(v)>0 \right] 
\\
\displaybreak[0]
- \Phi \,
\ex\left[ c_1^{\mathrm{hetero}} \right]
\, \ex\left[\dout(v)\right] \, \ex\left[\frac{1}{\din(v)}  \, \Big| \, \din(v)>0 \right]
\end{multline*}
which provides the following bounds:
\begin{align*}
\ex\left[\A^{\mathrm{hetero}} \,\big| \, \dave =d \right]
&
\approx 
\big( \Phi - \gamma \big) \, 
\ex\left[ c_1^{\mathrm{hetero}} \right]
+ 
\frac{\pi^2\,\E\,\big(\Phi - \gamma \big)}{15\,n} \,d
- 
\frac{\pi^2\,\Phi}{15}\,d^2 \,
\ex\left[ c_1^{\mathrm{hetero}} \right]
\\
\displaybreak[0]
& 
\\
\displaybreak[0]
%%%%%%%%%%%%%%%%%%%%%%%%
\ex\left[\A^{\mathrm{hetero}} \,\big| \, \dave =d + \Delta d \right]
&
\approx
\big( \Phi - \gamma \big) \, 
\ex\left[ c_1^{\mathrm{hetero}} \right]
+ 
\frac{\pi^2\,\E\,\big(\Phi - \gamma \big)}{15\,n} \,\big( d+\Delta d \big)
- 
\frac{\pi^2\,\Phi}{15}\,d^2 \,
\ex\left[ c_1^{\mathrm{hetero}} \right]
\\
\displaybreak[0]
& 
\\
\displaybreak[0]
%%%%%%%%%%%%%%%%%%%%%%%%
\Delta \ex\big[\A^{\mathrm{hetero}}\big]
&
=
\ex\left[\A^{\mathrm{hetero}} \,\big| \, \dave =d + \Delta d \right]
-
\,\,
\ex\left[\A^{\mathrm{hetero}} \,\big| \, \dave =d \right]
\\
\displaybreak[0]
&
\approx \,
\frac{\pi^2\,\E\,\big(\Phi - \gamma \big)}{15\,n} \,\Delta d
-
\frac{\pi^2\,\Phi}{15\,}\,d \,\Delta d\,
\ex\left[ c_1^{\mathrm{hetero}} \right]
\end{align*}
\end{itemize}
Now note that 
\begin{multline}
\Delta \ex\big[\A^{\mathrm{homo}}\big]
\lessapprox 0
\,\Rightarrow \, 
\frac{ \Phi - \gamma }{d+\Delta d}\,
+
\frac{\pi^2\,\Phi\,d}{15}
>
\frac{\pi^2\,\E\,\big(\Phi - \gamma \big)}{15\,n}
\\
\equiv\,
\left(
\frac{ 1 }{d+\Delta d}
+
\frac{\pi^2\,d}{15}
-
\frac{\pi^2\,\E}{15\,n}
\right)
\, \Phi
> 
\left(
\frac{ 1 }{d+\Delta d}
-
\frac{\pi^2\,\E}{15\,n}
\right)
\, \gamma
\label{eq3}
\end{multline}
\begin{multline}
\Delta \ex\big[\A^{\mathrm{hetero}}\big]
\gtrapprox 0
\,\Rightarrow \, 
\frac{\E\,\big(\Phi - \gamma \big)}{n} 
>
\Phi\,d\,
\ex\left[ c_1^{\mathrm{hetero}} \right]
\equiv\,
\left(
\frac{\E}{n} 
-
d\,
\ex\left[ c_1^{\mathrm{hetero}} \right]
\right)
\, \Phi
>
\left(
\frac{\E}{n} 
\right)
\, \gamma
\\
\Rightarrow \,
\,
\mbox{\tt w.h.p. }
\,
\left(
\frac{\E}{n} 
-
d\,
%%%%%%%%%%%%%%%%%%%%%%%%%%%%%%%%%%%%%%%%%%
\frac{1 + \alpha\,\beta - \alpha^2 - \beta - \alpha^2\,\beta}{1-2\,\alpha}
%%%%%%%%%%%%%%%%%%%%%%%%%%%%%%%%%%%%%%%%%%
\right)
\, \Phi
>
\left(
\frac{\E}{n} 
\right)
\, \gamma
\label{eq4}
\end{multline}
It is easy to verify that constraints~\eqref{eq3}~and~\eqref{eq4} are satisfied for many natural combinations of parameters.
In fact, the constraints~\eqref{eq3}~and~\eqref{eq4} are almost always satisfied when $\E$ grows moderately linearly with $n$.
To see this informally, note that since $\alpha\ll\beta$ and $\alpha$ is small, 
$\dfrac{1 + \alpha\,\beta - \alpha^2 - \beta - \alpha^2\,\beta}{1-2\,\alpha}\approx 1-\beta$ and thus \eqref{eq4} is approximately 
\begin{gather}
\left(
\frac{\E}{n} 
-
d\,
\left( 1 - \beta \right)
\,
\right)
\, \Phi
>
\left(
\frac{\E}{n} 
\right)
\, \gamma
\tag*{$(\ref{eq4}')$}
\label{eq4p}
\end{gather}
Now suppose that $\E<d(1-\beta)n$. Then, \eqref{eq3} and \eqref{eq4p} are always satisfied since $\Phi>\gamma$.
For a numerical example, suppose that $G$ is a $(0.1,0.95)$-heterogeneous network (\IE, $\alpha=0.1$ and $\beta=0.95$), $d=3$, $\Delta d=1$, $\gamma=0.2$ and $\Phi=0.4$. Then 
constraints~\eqref{eq3}~and~\eqref{eq4} reduce to:
\[
0.5+0.4\pi^2-\dfrac{2\,\pi^2\,\E}{15\,n} 
>
0.25-\dfrac{\pi^2\,\E}{15\,n} 
\,\,\,\,\,
\,\,\,\,\,
\text{ and }
\,\,\,\,\,
\,\,\,\,\,
\dfrac{2\,\E}{n}-0.94125
>
\dfrac{\E}{n} 
\]
and these constraints can be satisfied when $\E$ grows moderately linearly with $n$, \IE, when 
$0.94125\,n<\E<6.38\,n$.

\paragraph{\bf The case of random {\sf ER} networks}
In a random {\sf ER} network, the probability of having a particular edge is given by 
the following set of independent Bernoulli trials:
\[
\forall \, u, v\in V \text{ with $u\neq v$ } \colon 
\Pr \left[ (u,v) \in E \right] = p =\frac{\dave}{n-1}
\]
Thus, for every $k\in \{0,1,2,\dots,n-1\}$: 
\begin{align*}
\Pr\big[ \din(v)=k \big] & =\Pr\left[ \dout(v)=k \right]=\binom{n-1}{k} p^k (1-p)^{(n-1)-k}
\\
\displaybreak[0]
\ex\big[\din\big] & =\ex\left[\dout\right]=p\,(n-1)=\dave
\\
\displaybreak[0]
\Var\big[\din(v)\big] & = \Var\left[\dout(v)\right]=p \, (1-p)\,(n-1)=\left(1-\frac{\dave}{n-1} \right)\,\dave
\end{align*}
Since $\dave=p\,(n-1)$ is a constant, one can approximate the above binomial distribution by a Poisson's distribution~\cite[page 72]{rozanov} to obtain 
\[
\Pr\big[ \din(v)=k \big]=\Pr\left[ \dout(v)=k \right] \approx \e^{-\dave} \, \frac{(\dave)^k}{k!}
\]

\begin{lemma}[see Section~\ref{bb3proof-sec} of the appendix for a proof]
\label{bb3}
$%%\displaystyle
\left| \,
\ex\left[\frac{1}{\din(v)}\right]
-
\sum\limits_{k=1}^{\lfloor 3\,\dave + 10 \rfloor } \!\!\!\!\! \!\!\!\!\! \!\!\!\!\! \left(\nicefrac{1}{k}\right) \, \e^{-\dave} \, \frac{(\dave)^k}{k!}
\, \right| 
\lessapprox 10^{-10}
$ 
and 
%%\\
$
\frac{\partial
}{\partial\, d}
\ex\left[\frac{1}{\din(v)} \, \big| \, \dave=d\right]
\approx
\frac{1-\e^{-d}}{d}
$.
\end{lemma}

For notational convenience, let $\Upsilon(x)=\!\!\!\!\sum\limits_{k=1}^{\lfloor 3x + 10 \rfloor} \frac{1}{k} \, \e^{-x} \, \frac{x^k}{k!}$.
Since $10^{-10}$ is an extremely small number for our purposes, we will just use 
$\ex\left[\frac{1}{\din(v)}\right] \approx \Upsilon\left(\dave\right)$ in the sequel. 
Values of $\Upsilon(x)$ and 
$
\frac{\partial \, \Upsilon(x)}{\partial \, x}=\frac{1-\e^{-x}}{x}
$
for a few small integral values of $x$ are shown in Table~\ref{upsi-table}.
It is easy to see that 
$\lim\limits_{x\to\infty}\Upsilon(x)=0$ and, for large $x$, 
$
\frac{\partial \, \Upsilon(x)}{\partial \, x}\approx \nicefrac{1}{x}
$.

\begin{table*}
\caption{Values of $\Upsilon(x)$ and $\frac{\partial \, \Upsilon(x)}{\partial \, x}$ for a few small integral values of $x$ using straightforward calculations.}
\label{upsi-table}
\begin{center}
\begin{tabular}{c | c c c c c c c c c c c }
\toprule
    & \multicolumn{11}{c}{$x=$} 
\\
    & 2 & 3 & 4 & 5 & 6 & 7 & 8 & 9 & 10 & 11 & 12 
\\
\cmidrule{2-12}
$\Upsilon(x)\approx$ & 0.499 & 0.411 & 0.324 & 0.256 & 0.207 & 0.172 & 0.147 & 0.128 & 0.113 & 0.101 & 0.091
\\
\midrule
$\frac{\partial \, \Upsilon(x)}{x} \approx$ & 0.432 & 0.284 & 0.245 & 0.256 & 0.199 & 0.166 & 0.143 & 0.125 & 0.111 & 0.100 & 0.090
\\
\bottomrule
\end{tabular}
\end{center}
%%%
\end{table*}

As before, we first provide an estimation of $c_1^{\mathrm{hetero}}$ using the notations in Definition~\ref{defnet}.

\begin{lemma}[see Section~\ref{bb4proof-sec} of the appendix for a proof]\label{bb4}
{\rm\tt W.h.p.} 
$\,\,\,
1 + \alpha - \beta - \dfrac{\alpha\,\beta}{2}
\leq 
\ex\left[c_1^{\mathrm{hetero}}\right]  
\leq 
\dfrac{1 + \alpha\,\beta - \alpha^2 - \beta - \alpha^2\,\beta}{1-2\,\alpha}
$.
\end{lemma}

Using the above result, the following bounds hold:
\begin{itemize}
%%%%%%
%%%%%%
%%%%%%
\item
If $G$ is a homogeneous network then
\[
\dfrac{\partial}{\partial \, d}
c_1^{\mathrm{homo}}
\Big|_{\substack{\mathrm{at} \\ \dave=d}}
%%%
=
\lim\limits_{\Delta d \to 0}
\dfrac{
\dfrac{ \I\Big|_{\substack{\mathrm{at} \\ \dave=d+\Delta d}} }{n\,\left(d + \Delta d \right)}
\,-\,
\dfrac{ \I\Big|_{\substack{\mathrm{at} \\ \dave=d}} }{n\,d }
}
{ \Delta d}
=
\lim\limits_{\Delta d \to 0}
\dfrac{
\dfrac{n\,d}{n\,\left(d + \Delta d \right)}
\,-\,
1
}
{ \Delta d}
=
\,-\,
\dfrac{1}{d}
\]
%%%%%%%%%%%%%%%%%%%%%%%%%%%%%%%%%%%%%%%
%%%%%%%%%%%%%%%%%%%%%%%%%%%%%%%%%%%%%%%
%%%%%%%%%%%%%%%%%%%%%%%%%%%%%%%%%%%%%%%
%%%%%%%%%%%%%%%%%%%%%%%%%%%%%%%%%%%%%%%
\begin{align*}
&
\ex\left[\A^{\mathrm{homo}} \,\big| \, \dave=d \right]
\\
\displaybreak[0]
=\,\,
&
c_1^{\mathrm{homo}} \,
\big( \Phi - \gamma \big) 
+ 
\frac{\E\,\big(\Phi - \gamma \big)}{n} \ex\left[\frac{1}{\din(v)}  \, \Big| \, \din(v)>0 \right] 
- 
c_1^{\mathrm{homo}} \,
\Phi \, \ex\left[\dout(v)\right] \, \ex\left[\frac{1}{\din(v)}  \, \Big| \, \din(v)>0 \right]
\\
\displaybreak[0]
\approx\,\,
&
c_1^{\mathrm{homo}} \,
\big( \Phi - \gamma \big) 
+ 
\frac{\E\,\big(\Phi - \gamma \big)}{n} \,\Upsilon(d)
- 
c_1^{\mathrm{homo}} \,
\Phi\,d\,\Upsilon(d)
\\
\displaybreak[0]
&
\\
\displaybreak[0]
%%%%%%%%%%%%%%%%%%%%%%%%
%%%%%%%%%%%%%%%%%%%%%%%%%%%%%%%%%%%%%%%
%%%%%%%%%%%%%%%%%%%%%%%%%%%%%%%%%%%%%%%
%%%%%%%%%%%%%%%%%%%%%%%%%%%%%%%%%%%%%%%
&
\frac{\partial}{\partial \, d} \ex\left[\A^{\mathrm{homo}} \,\big| \, \dave=d \right]
\\
\displaybreak[0]
\approx \,\, 
&
\big( \Phi - \gamma \big) 
\,
\frac{\partial}{\partial \, d} 
c_1^{\mathrm{homo}}
\Big|_{\substack{\mathrm{at} \\ \dave=d}}
+
\dfrac{\E\,\big(\Phi - \gamma \big)}{n} \,\frac{\partial}{\partial \, d}\Upsilon(d)
-
\Phi\,
\dfrac{\partial}{\partial \, d} 
\left(
c_1^{\mathrm{homo}}
\Big|_{\substack{\mathrm{at} \\ \dave=d}} \,
d\,\Upsilon(d)
\right)
\\
\displaybreak[0]
= \,\,
&
- \, \dfrac{\Phi - \gamma}{d} \,
+
\left(\dfrac{\E\,\big(\Phi - \gamma \big)}{n}\right) \,\left( \dfrac{1 - \e^{-d}}{d} \right)
-
\Phi \, 
\left(
- \, \Upsilon(d)
+ \, \Upsilon(d)
+ 1 - \e^{-d} 
\right)
\,\,\,\,\,\,\,
\mbox{\begin{tabular}{l} using product rule \\ of derivatives \end{tabular}}
\\
\displaybreak[0]
= \,\,
&
- \, \frac{\Phi - \gamma}{d} \,
+
\left(\frac{\E\,\big(\Phi - \gamma \big)}{n}\right) \,\left( \frac{1 - \e^{-d}}{d} \right)
-
\Phi \, 
\left(
1 - \e^{-d} 
\right)
\end{align*}
%%%%%%%%%%%%%%%%%%%%%%%%

\item
If $G$ is a heterogeneous network then 
\begin{multline*}
\ex\big[\A^{\mathrm{hetero}}\big]
=
\big( \Phi - \gamma \big) \, 
\ex\left[ c_1^{\mathrm{hetero}} \right]
+ 
\frac{\E\,\big(\Phi - \gamma \big)}{n} \ex\left[\frac{1}{\din(v)} \, \Big| \, \din(v)>0 \right] 
\\
\displaybreak[0]
- \Phi \,
\ex\left[ c_1^{\mathrm{hetero}} \right]
\, \ex\left[\dout(v)\right] \, \ex\left[\frac{1}{\din(v)} \, \Big| \, \din(v)>0 \right]
\end{multline*}
which provides the following bounds {\tt w.h.p.}:
\[
\ex\left[\A^{\mathrm{hetero}} \,\big| \, \dave=d \right]
\approx 
\big( \Phi - \gamma \big) \, 
\ex\left[ c_1^{\mathrm{hetero}} \right]
+ 
\frac{\E\,\big(\Phi - \gamma \big)}{n} \,\Upsilon(d)
- 
\Phi\, d \, \Upsilon(d) \,
\ex\left[ c_1^{\mathrm{hetero}} \right]
\]
%%%%%%%%%%%%%%%%%%%%%%%%
\begin{align*}
&
\frac{\partial}{\partial \, d} 
\ex\left[\A^{\mathrm{hetero}} \,\big| \, \dave=d \right]
\\
\displaybreak[0]
\gtrapprox
\,\,
&
\frac{\E\,\big(\Phi - \gamma \big)}{n} \,
\frac{\partial}{\partial \, d} 
\Upsilon(d)
-
\Phi\, \left( \frac{1 + \alpha\,\beta - \alpha^2 - \beta - \alpha^2\,\beta}{1-2\,\alpha} \right) \, 
\frac{\partial}{\partial \, d} 
\big( \, d \, \Upsilon(d) \, \big)
\\
\displaybreak[0]
=
\,\,
&
\left( \frac{\E\,\big(\Phi - \gamma \big)}{n} \right) \,
\left( \frac{1-\e^{-d}}{d} \right) 
-
\Phi\, \left( \frac{1 + \alpha\,\beta - \alpha^2 - \beta - \alpha^2\,\beta}{1-2\,\alpha} \right) \, 
\left(
1 - \e^{-d} + \Upsilon(d)
\right)
\,\,\,\,\,\,\,
\mbox{\begin{tabular}{l} using product \\ rule of \\ derivatives \end{tabular}}
\end{align*}
%%%%%%%%%%%%%%%%%%%%%%%%
%
\end{itemize}
Now note the following:
\begin{multline}
\frac{\partial}{\partial \, d} \ex\left[\A^{\mathrm{homo}} \,\big| \, \dave=d \right]
\lessapprox 0
\\
\displaybreak[0]
\Rightarrow \, 
\frac{\Phi - \gamma}{d} \,
+
\Phi \, 
\left(
1 - \e^{-d} 
\right)
>
\left(\frac{\E\,\big(\Phi - \gamma \big)}{n}\right) \,\left( \frac{1 - \e^{-d}}{d} \right)
%%%%%%%%%%%%%%%%%%%%%%%%%%%%%%%%%%%%%%%%%%%%%%%%%%%%%%%%%%%%%%%%%%%%%%%%%%%%%%%%%%%%%%%%%%%%%%%%%%%%
%%%%%%%%%%%%%%%%%%%%%%%%%%%%%%%%%%%%%%%%%%%%%%%%%%%%%%%%%%%%%%%%%%%%%%%%%%%%%%%%%%%%%%%%%%%%%%%%%%%%
%%%%%%%%%%%%%%%%%%%%%%%%%%%%%%%%%%%%%%%%%%%%%%%%%%%%%%%%%%%%%%%%%%%%%%%%%%%%%%%%%%%%%%%%%%%%%%%%%%%%
%%%%%%%%%%%%%%%%%%%%%%%%%%%%%%%%%%%%%%%%%%%%%%%%%%%%%%%%%%%%%%%%%%%%%%%%%%%%%%%%%%%%%%%%%%%%%%%%%%%%
%%%%%%%%%%%%%%%%%%%%%%%%%%%%%%%%%%%%%%%%%%%%%%%%%%%%%%%%%%%%%%%%%%%%%%%%%%%%%%%%%%%%%%%%%%%%%%%%%%%%
%%%%%%%%%%%%%%%%%%%%%%%%%%%%%%%%%%%%%%%%%%%%%%%%%%%%%%%%%%%%%%%%%%%%%%%%%%%%%%%%%%%%%%%%%%%%%%%%%%%%
\\
\displaybreak[0]
\equiv\,
\left(
\frac{1}{d}
+
1 - \e^{-d}
-
\frac{\E}{n} \,
\left( \frac{1 - \e^{-d}}{d} \right) \,
\right)
\, \Phi
> 
\left(
\frac{ 1}{d}
-
\frac{\E}{n} \,
\left( \frac{1 - \e^{-d}}{d} \right) \,
\right)
\, \gamma
\label{eq5}
\end{multline}
%%%%%%%%%%%%%%%%%%%%%%%%%%%%%%%%%%%%%%%%%%%%%%%%%%%%%%%%%%%%%%%%%%%%%%%%%%%%%%%%%%%%%%%%
\begin{multline}
\frac{\partial}{\partial \, d} \ex\left[\A^{\mathrm{hetero}} \,\big| \, \dave=d \right]
\gtrapprox 0
\\
\displaybreak[0]
\Rightarrow \, 
\left( \frac{\E\,\big(\Phi - \gamma \big)}{n} \right) \,
\left( \frac{1-\e^{-d}}{d} \right) 
>
\Phi\, \left( \frac{1 + \alpha\,\beta - \alpha^2 - \beta - \alpha^2\,\beta}{1-2\,\alpha} \right) \, 
\left(
1 - \e^{-d} + \Upsilon(d)
\right)
%%%%%%%%%%%%%%%%%%%%%%%%%%%%%%%%%%%%%%%%%%%%%%%%%%%%%%%%%%%%%%%%%%%%%%%%%%%%%%%%%%%%%%%%%%%%%%%%%%%%%%%%%%%%%%%%%%%%%%%%%%%%%%%%%%%%%%%%%%%%%%%%%%%%%%%%%
%%%%%%%%%%%%%%%%%%%%%%%%
%%%%%%%%%%%%%%%%%%%%%%%%
%%%%%%%%%%%%%%%%%%%%%%%%
%%%%%%%%%%%%%%%%%%%%%%%%
%%%%%%%%%%%%%%%%%%%%%%%%
%%%%%%%%%%%%%%%%%%%%%%%%
%%%%%%%%%%%%%%%%%%%%%%%%
\\
\displaybreak[0]
\textstyle
\equiv\,
\left(
\frac{\E }{n}
\left( \frac{1-\e^{-d}}{d} \right) 
-
\left( \frac{1 + \alpha\,\beta - \alpha^2 - \beta - \alpha^2\,\beta}{1-2\,\alpha} \right) \, 
\left(
1 - \e^{-d} + \Upsilon(d)
\right)
\,
%%%%%%%%%%%%%%%%
%%%%%%%%%%%%%%%%
%%%%%%%%%%%%%%%%
%%%%%%%%%%%%%%%%
\right)
\, \Phi
>
\frac{\E }{n}
\left( \frac{1-\e^{-d}}{d} \right) 
\, \gamma
%%%%%%%%%%%%%%%%%%%%%%%%%%%%%%%%%%%%%%%%%%%%%
%%%%%%%%%%%%%%%%%%%%%%%%%%%%%%%%%%%%%%%%%%%%%
%%%%%%%%%%%%%%%%%%%%%%%%%%%%%%%%%%%%%%%%%%%%%
%%%%%%%%%%%%%%%%%%%%%%%%%%%%%%%%%%%%%%%%%%%%%
\label{eq6}
\end{multline}
It is easy to verify that constraints~\eqref{eq5}~and~\eqref{eq6} are satisfied for many natural combinations of parameters.
For example, the constraints~\eqref{eq5}~and~\eqref{eq6} are almost always satisfied if $d$ is sufficiently large.
To see this informally, note that if $d$ is large then 
$\frac{1}{d},\e^{-d},\Upsilon(d)\approx 0$.
Moreover, since $\alpha\ll\beta$ and $\alpha$ is small, 
$\frac{1 + \alpha\,\beta - \alpha^2 - \beta - \alpha^2\,\beta}{1-2\,\alpha}\approx 1-\beta$
and then constraints~\eqref{eq5}~and~\eqref{eq6} reduce to 
\begin{gather}
\left(
1 
-
\frac{\E}{n\,d} \,
\right)
\, \Phi
\,
\gtrapprox 
\,
-\,
\frac{\E}{n\,d}
\, \gamma
\,\equiv\,
%%%%%%%%%%%
\Phi
\,
\lessapprox 
\,
\frac{
\frac{\E}{n\,d}
}
{
\frac{\E}{n\,d} \,
-1
}
\, \gamma
\tag*{(\ref{eq5})$^{d\gg 0}$}
\label{eq7}
\end{gather}
%%%%%%%%%%%%%%%%
%%%%%%%%%%%%%%%%
%%%%%%%%%%%%%%%%
%%%%%%%%%%%%%%%%
\begin{gather}
\left(
\,
\frac{\E }{n\,d}
\,
-
\,
1+\beta\,
%%%%%%%%%%%%%%%%
%%%%%%%%%%%%%%%%
%%%%%%%%%%%%%%%%
%%%%%%%%%%%%%%%%
\right)
\, \Phi
\,
\gtrapprox 
\,
\frac{\E }{n\,d}
\, \gamma
\,\equiv\,
\, \Phi
\,
\gtrapprox 
\,
\frac{
\frac{\E }{n\,d}
}
{
\frac{\E }{n\,d}
\,
-
\,
1+\beta\,
}
\,\gamma
\tag*{(\ref{eq6})$^{d\gg 0}$}
\label{eq8}
\end{gather}
If $\E>n\,d$ then 
$\frac{ \frac{\E}{n\,d} } { \frac{\E}{n\,d} -1 }>1$ and 
$\frac{ \frac{\E }{n\,d} } { \frac{\E }{n\,d} - 1+\beta} < \frac{ \frac{\E}{n\,d} } { \frac{\E}{n\,d} -1 }$; 
thus both \ref{eq7} and \ref{eq8} can be satisfied by choosing $\Phi$ appropriately with respect to $\gamma$.
For a numerical example, suppose that $G$ is a $(0.1,0.95)$-heterogeneous network (\IE, $\alpha=0.1$ and $\beta=0.95$), $d=10$, and $\E=12\,n$. Then 
constraints~\eqref{eq5},\eqref{eq6} reduce to:
\[
1.17 < \frac{\Phi}{\gamma} < 11
\]
which corresponds to most settings of $\Phi$ and $\gamma$ used in our simulation.

%%%%%%%%%%%%%%%%%%%%%%%%%%%%%%%%%%%%%%%%%%%%%%%%%%%%%%%%%%%%%%%%%%%%%%%%%%%%%%%%%%%%%%%%%%%%%%%%%%%%%%%%%%%%%%%%%%%%%%%%%%%%%%%%%%%%%%%%%%%%%%%%%%%%%%
%%%%%%%%%%%%%%%%%%%%%%%%%%%%%%%%%%%%%%%%%%%%%%%%%%%%%%%%%%%%%%%%%%%%%%%%%%%%%%%%%%%%%%%%%%%%%%%%%%%%%%%%%%%%%%%%%%%%%%%%%%%%%%%%%%%%%%%%%%%%%%%%%%%%%%
%%%%%%%%%%%%%%%%%%%%%%%%%%%%%%%%%%%%%%%%%%%%%%%%%%%%%%%%%%%%%%%%%%%%%%%%%%%%%%%%%%%%%%%%%%%%%%%%%%%%%%%%%%%%%%%%%%%%%%%%%%%%%%%%%%%%%%%%%%%%%%%%%%%%%%
%%%%%%%%%%%%%%%%%%%%%%%%%%%%%%%%%%%%%%%%%%%%%%%%%%%%%%%%%%%%%%%%%%%%%%%%%%%%%%%%%%%%%%%%%%%%%%%%%%%%%%%%%%%%%%%%%%%%%%%%%%%%%%%%%%%%%%%%%%%%%%%%%%%%%%

%%%%%%%%%%%%%%%%%%%%%%%%%%%%%%%%%%%%%%%%%%%%%%%%%%%%%%%%%%%%%%%%%%%%%%%%%%%%%%%%%%%%%%%%%%%%%%%%%%%%%%%%%%%%%%
%%% Table for random vs coordinated shocks
\renewcommand{\tabcolsep}{4pt}
\begin{table}
\caption{Comparisons of strengths of coordinated versus idiosyncratic shocks.
The percentages indicate the percentage of total number of data points (combinations of parameters $\Phi$, $\gamma$, $\E$ and $\mathcal{K}$) 
for that network type that resulted in $\xi_c\geq\xi_r$, where $\xi_c$ and $\xi_r$ are the vulnerability indices under coordinated and idiosyncratic shocks, respectively.}
\label{table1}
\begin{center}
{\small
\begin{tabular}{c c  | c  c | c c | c c | c c}
\toprule
%%%%%%%%%%%%%%%%%%%%%%%%%%%%%%%%%%%%%%%%%%%%%%%%%
%% column headings for other than first column
%%%%%%%%%%%%%%%%%%%%%%%%%%%%%%%%%%%%%%%%%%%%%%%%%
%
\multicolumn{10}{c}{$(\alpha,\beta)$-heterogeneous networks}
\\
\noalign{\smallskip}
\hline
\noalign{\smallskip}
      \multicolumn{2}{c|}{\footnotesize In-arborescence} 
      &
      \multicolumn{2}{c|}{\footnotesize {\sf ER} average degree 3} 
      &
      \multicolumn{2}{c|}{\footnotesize {\sf ER} average degree 6} 
      &
      \multicolumn{2}{c|}{\footnotesize {\sf SF} average degree 3} 
      &
      \multicolumn{2}{c}{\footnotesize {\sf SF} average degree 6} 
\\
$\alpha=0.1$ 
&
$\alpha=0.2$ 
&
$\alpha=0.1$ 
&
$\alpha=0.2$ 
&
$\alpha=0.1$ 
&
$\alpha=0.2$ 
&
$\alpha=0.1$ 
&
$\alpha=0.2$ 
&
$\alpha=0.1$ 
&
$\alpha=0.2$ 
\\
$\beta=0.95$
&
$\beta=0.6$
&
$\beta=0.95$
&
$\beta=0.6$
&
$\beta=0.95$
&
$\beta=0.6$
&
$\beta=0.95$
&
$\beta=0.6$
&
$\beta=0.95$
&
$\beta=0.6$
\\
 & & & & & & & & & 
\\
%%%%%%%%%%%%%%%%%%%%%%%%%%%%%%%%%%%%%%%%%%%%%%%%%
%% column headings for other than first column end
%%%%%%%%%%%%%%%%%%%%%%%%%%%%%%%%%%%%%%%%%%%%%%%%%
%
%%%%%%%%%%%%%%%%%%%%%%%%%%%%%%%%%%%%%%%%%%%%%%%%
%% other columns and data points
%%%%%%%%%%%%%%%%%%%%%%%%%%%%%%%%%%%%%%%%%%%%%%%%
                      \bf 56.64\% &             \bf 57.27\% &             \bf 89.66\% &             \bf 90.97\% &             \bf 98.99\% 
                     &             \bf 98.04\% &             \bf 93.16\% &             \bf 64.13\% &             \bf 94.44\% &             \bf 66.48\%
\\
\bottomrule
%%%%%%%%%%%%%%%%%%%%%%%%%%%%%%%%%%%%%%%%%%%%%%%%
%% other columns and data points end
%%%%%%%%%%%%%%%%%%%%%%%%%%%%%%%%%%%%%%%%%%%%%%%%
\\
\toprule
\multicolumn{10}{c}{homogeneous networks}
\\
\noalign{\smallskip}
\hline
\noalign{\smallskip}
\multicolumn{2}{c|}{              In-arborescence} 
&
      \multicolumn{2}{c|}{              {\sf ER} average degree 3 } 
      &
      \multicolumn{2}{c|}{              {\sf ER} average degree 6 } 
      &
      \multicolumn{2}{c|}{              {\sf SF} average degree 3 } 
      &
      \multicolumn{2}{c}{              {\sf SF} average degree 6 } 
\\
 & & & & & & & & & 
\\
      \multicolumn{2}{c|}{            \bf 84.62\%} 
      & 
      \multicolumn{2}{c|}{            \bf 74.97\%} 
      &
      \multicolumn{2}{c|}{            \bf 78.59\%} 
      &
      \multicolumn{2}{c|}{            \bf 81.15\%} 
      &
      \multicolumn{2}{c}{            \bf 54.80\%} 
\\
\bottomrule
\end{tabular}
}
\end{center}
\end{table}
\renewcommand{\tabcolsep}{6pt}
%%%%%%%%%%%%%%%%%%%%%%%%%%%%%%%%%%%%%%%%%%%%%%%%%%%%%%%%%%%%%%%%%%%%%%%%%%%%%%%%%%%%%%%%%%%%%%%%%%%%%%%%%%%%%%

\subsection{\bf Random versus correlated initial failures}
\label{jjj1}

For most parameter combinations, our results, tabulated in Table~\ref{table1}, show that coordinated shocks, which are a type of correlated shocks, resulted in insolvencies of higher
number of nodes as opposed to idiosyncratic shocks for the same network with the same parameters, {\em often by a factor of two or more}.
For example, Table~\ref{table1} shows that for $(0.1,0.95)$-heterogeneous {\sf ER} networks of average degree $6$ 
the vulnerability index under coordinated shocks is at least as much as the vulnerability index under idiosyncratic shocks 
$98.99\%$ of the time.
Thus, we conclude:
\begin{quote}
\scalebox{1.5}[1.5]{\ding{177}}
{\em correlated shocking mechanisms are more appropriate to measure the worst-case stability compared to idiosyncratic shocking mechanisms}.
\end{quote}
For visual illustrations of \scalebox{1.5}[1.5]{\ding{177}}, see {\em supplemental} {\sc Fig}.~\ref{newfig1-full}---\ref{supp-newfig5-full}.

\subsection{\bf Phase transition properties of stability}
\label{kk2}

Phase transitions are quite common when one studies various topological properties of graphs~\cite{bolo}.
The stability measure exhibits several sharp {\em phase transitions} for various banking networks and combinations of parameters; 
see {\em supplemental} {\sc Fig}.~\ref{newfig1}---\ref{newfig4} for 
visual illustrations.
We discuss two such interesting phase transitions in the following, with  
an intuitive theoretical explanation for one of them.

\subsubsection{Dense homogeneous networks}

Based on the behavior of $\xi$ with respect to $\left(\Phi-\gamma\right)$,  
we observe that, for smaller value of $\mathcal{K}$ and for denser {\sf ER} and {\sf SF} networks under either coordinated or idiosyncratic shocks,  
there is often a {\em sharp} decrease of stability when $\gamma$ was decreased beyond a particular threshold value.
For example, with $\Phi=0.5$ and $\mathcal{K}=0.1$, $100$-node dense (average degree $6$) {\sf SF} and {\sf ER} homogeneous networks exhibited more than ninefold increase in 
$\xi$ around $\gamma=0.15$ and $\gamma=0.1$, respectively; see {\em supplemental} {\sc Fig}.~\ref{newfig1} for visual illustrations.

To investigate the global extent of such a sharp decrease at a threshold value of $\gamma$ in the range $\big[0.05,0.2\big]$, we 
computed, for each of the five homogeneous network types under coordinated shocks and for each values of the parameters $|V|$, $\Phi$, $\frac{\E}{\I}$ and $\mathcal{K}$, , the ratio 
\[
\Lambda \left( n, \Phi, \frac{\E}{\I}, \mathcal{K} \right)
=
\dfrac
{
\displaystyle
\max_{0.05\leq\gamma\leq 0.2} \big\{ \xi \big\}- \min_{0.05\leq\gamma\leq 0.2} \big\{ \xi \big\}
}
{
\displaystyle
\max_{\text{entire range of } \gamma} \big\{ \xi \big\}- \min_{\text{entire range of } \gamma} \big\{ \xi \big\}
}
\]
that provides the {\em maximum} percentage of the total change of the vulnerability index that occurred within this narrow range of $\gamma$. 
The values of $\Lambda\left( n, \Phi, \frac{\E}{\I}, \mathcal{K} \right)$ for the dense {\sf ER} and {\sf SF} homogeneous networks under coordinated shocks are shown in 
Table~\ref{table10} for
$\Phi=0.5,0.8$ and $\mathcal{K}=0.1,0.2,0.3,0.4,0.5$ (the behaviour of $\xi$ is similar for other intermediate values of $\Phi$).
If the growth of $\xi$ with respect to $\gamma$ was uniform or near uniform over the entire range of $\gamma$, $\Lambda$ would be approximately 
$\lambda=\frac{0.2-0.05}{0.45-0.05}=0.375$; thus, any value of $\Lambda$ significantly higher than $\lambda$ indicates a sharp transition 
within the above-mentioned range of values of $\gamma$. As Table~\ref{table10} shows, 
a {\em significant majority} of the entries for $\Phi\leq 0.8$ and $\kappa\leq 0.2$ are $2\,\lambda$ or more.

%%%%%%%%%%%%%%%%%%%%%%%%%%%%%%%%%%%%%%%%%%%%%%%%%%%%%%%%%%%%%%%%%%%%%%%%%%%%%%%%%%%%%%%%%%%%%%%%%%%%%%%%%%%%%%
%%% table for sharp increase of xi at gamma=0.1 for homogeneous dense ER and SF networks
%%%%%%%%%%%%%%%%%%%%%%%%%%%%%%%%%%%%%%%%%%%%%%%%%%%%%%%%%%%%%%%%%%%%%%%%%%%%%%%%%%%%%%%%%%%%%%%%%%%%%%%%%%%%%%

\renewcommand{\tabcolsep}{4pt}
\begin{table}[htbp]
\caption[]{Values of 
$
\displaystyle
\Lambda \left( n, \Phi, E/I, \mathcal{K} \right)
=
\dfrac
{
\displaystyle
\max_{0.05\leq\gamma\leq 0.2} \big\{ \xi \big\}- \min_{0.05\leq\gamma\leq 0.2} \big\{ \xi \big\}
}
{
\displaystyle
\max_{\text{entire range of } \gamma} \big\{ \xi \big\}- \min_{\text{entire range of } \gamma} \big\{ \xi \big\}
}
$
for homogeneous dense {\sf ER} and {\sf SF} networks under coordinated shocks. Entries that are at least $2\times 0.375$ are shown
in {\bf boldface}.
}
\label{table10}
%%%%
\begin{center}
\scalebox{0.8}[0.8]
{%%\footnotesize
\begin{tabular}{l l |   lllll    lllll ||   lllll    lllll }
%%%\vspace*{-1in}
\toprule
\multicolumn{12}{c||}{\large \hspace*{0.6in} $\pmb{\Phi=0.5}$}
 & \multicolumn{8}{c}{\large \hspace*{0.6in} $\pmb{\Phi=0.8}$} & & 
\\
 & & \multicolumn{5}{c|}{\normalsize {\sf ER} average degree $6$} & \multicolumn{5}{c||}{\normalsize {\sf SF} average degree $6$}
    & \multicolumn{5}{c|}{\normalsize {\sf ER} average degree $6$} & \multicolumn{5}{c}{\normalsize {\sf SF} average degree $6$}
\\
[0.05in]
 & & \multicolumn{5}{c|}{\normalsize $\pmb{\mathcal{K}=}$} & \multicolumn{5}{c||}{\normalsize $\pmb{\mathcal{K}=}$} 
    & \multicolumn{5}{c|}{\normalsize $\pmb{\mathcal{K}=}$} & \multicolumn{5}{c}{\normalsize $\pmb{\mathcal{K}=}$} 
\\
 $\nicefrac{E}{I}$ & \normalsize \hspace*{-0.1in}$|V|$ 
      & \scalebox{0.7}{0.1} & \scalebox{0.7}{0.2} & \scalebox{0.7}{0.3} & \scalebox{0.7}{0.4} & \multicolumn{1}{l|}{\scalebox{0.7}{0.5}}
        & \scalebox{0.7}{0.1} & \scalebox{0.7}{0.2} & \scalebox{0.7}{0.3} & \scalebox{0.7}{0.4} & \multicolumn{1}{l||}{\scalebox{0.7}{0.5}}
          & \scalebox{0.7}{0.1} & \scalebox{0.7}{0.2} & \scalebox{0.7}{0.3} & \scalebox{0.7}{0.4} & \multicolumn{1}{l|}{\scalebox{0.7}{0.5}}
            & \scalebox{0.7}{0.1} & \scalebox{0.7}{0.2} & \scalebox{0.7}{0.3} & \scalebox{0.7}{0.4} & \multicolumn{1}{l}{\scalebox{0.7}{0.5}}
\\
\bottomrule
\end{tabular}
%%%
}
\end{center}
\end{table}
\renewcommand{\tabcolsep}{6pt}
%

%%%%%%%%%%%%%%%%%%%%%%%%%%%%%%%%%%%%%%%%%%%%%%%%%%%%%%%%%%%%%%%%%%%%%%%%%%%%%%%%%%%%%%%%%%%%%%%%%%%%%%%%%%%%%%
%%% END table for sharp increase of xi at gamma=0.1 for homogeneous dense ER and SF networks
%%%%%%%%%%%%%%%%%%%%%%%%%%%%%%%%%%%%%%%%%%%%%%%%%%%%%%%%%%%%%%%%%%%%%%%%%%%%%%%%%%%%%%%%%%%%%%%%%%%%%%%%%%%%%%

\subsubsection{Homogeneous in-arborescence networks}

Homogeneous in-arborescence networks under coordinated shocks exhibited a {\em sharp} increase in stability 
as the ratio $\frac{\E}{\I}$ of the total external asset to the total interbank exposure the system is increased beyond a particular threshold {\em provided the equity to asset ratio $\gamma$ 
was approximately $\nicefrac{\Phi}{2}$}.
For example, 
for a 50-node homogeneous in-arborescence network under coordinated shock, $\xi$ exhibited a sharp decrease from 0.76 to 0.18 
for $\frac{\E}{\I}\in\big[0.75,1\big]$, $\mathcal{K}=0.1$, $\Phi=0.5$ and $\gamma=0.25=\nicefrac{\Phi}{2}$; 
see {\em supplemental} {\sc Fig}.~\ref{newfig4} for a visual illustration.

To investigate the global extent of such a sharp decrease of $\xi$ around a threshold value of $\frac{\E}{\I}$ in the range $\big[0.5,1\big]$ with $\gamma\approx\nicefrac{\Phi}{2}$, we 
computed, for each type of shocking mechanism, and for each values of the parameters $n$, $\Phi$, $\gamma\approx\nicefrac{\Phi}{2}$, and $\kappa$
of the homogeneous in-arborescence network, the ratio 
\[
\Delta(n,\Phi,\gamma,\mathcal{K})
=
\dfrac
{
\displaystyle
\max_{\substack{ 0.5\leq \frac{\E}{\I} \leq 1 \\ } } \big\{ \xi \big\}- \min_{\substack{ 0.5\leq \frac{\E}{\I} \leq 1 } } \big\{ \xi \big\}
}
{
\displaystyle
\max_{\substack{ \text{entire range of } \E } } \big\{ \xi \big\}- \min_{\substack{ \text{entire range of } \E  } } \big\{ \xi \big\}
}
\]
that provides the {\em maximum} percentage of the total change of the vulnerability index that occurred within this range of $\frac{\E}{\I}$. 
If the growth of $\xi$ with respect to $\nicefrac{\E}{\I}$ was uniform or near uniform over the entire range of $\frac{\E}{\I}$, $\Delta$ would be approximately 
$\delta=\frac{1-0.5}{3.5-0.25}\approx 0.16$; thus, any value of $\Delta$ significantly higher than $\delta$ indicates a sharp transition 
within the above-mentioned range of $\nicefrac{\E}{\I}$. As Table~\ref{table11} shows, when $\gamma=\nicefrac{\Phi}{2}$ 
a {\em significant} majority of the entries are coordinated shocks and many entries under idiosyncratic shocks are at least $2\,\delta$.

%%%%%%%%%%%%%%%%%%%%%%%%%%%%%%%%%%%%%%%%%%%%%%%%%%%%%%%%%%%%%%%%%%%%%%%%%%%%%%%%%%%%%%%%%%%%%%%%%%%%%%%%%%%%%%
%%%%%%%%%%%%%%%%%%%%%%%%%%%%%%%%%%%%%%%%%%%%%%%%%%%%%%%%%%%%%%%%%%%%%%%%%%%%%%%%%%%%%%%%%%%%%%%%%%%%%%%%%%%%%%
%%% table for sharp increase of xi at E=0.75 for homogeneous in-arborescence networks
\begin{table}
\caption[]{Values of 
$
\displaystyle
\Delta(n,\Phi,\gamma,\mathcal{K})
=
\dfrac
{
\displaystyle
\max_{\substack{ 0.5\leq E/I \leq 1 \\ } } \big\{ \xi \big\}- \min_{\substack{ 0.5\leq E/I\leq 1 } } \big\{ \xi \big\}
}
{
\displaystyle
\max_{\substack{ \text{entire range of } E } } \big\{ \xi \big\}- \min_{\substack{ \text{entire range of } E  } } \big\{ \xi \big\}
}
$
for homogeneous in-arborescence networks. Entries that are at least $2\times 0.16$ are shown in {\bf boldface black}.
Entries that are at least $\frac{3}{2}\times 0.16$ are shown in {\bf\color{darkgray} boldface gray}.
}
\label{table11}
\begin{center}
\scalebox{0.75}[0.65]
{%%\tiny
\setlength{\extrarowheight}{0ex}
\begin{tabular}{c | c | c | l | l l l l l l l l l }
%%%\vspace*{-1.4in}
\toprule
\multicolumn{1}{c}{} & \multicolumn{1}{c}{} & \multicolumn{1}{c}{} & \multicolumn{1}{c}{} & \multicolumn{9}{c}{\hspace*{-0.1in}\Large \scalebox{8.5}[1.2]{${\leftarrow}$}$\pmb{\mathcal{K}}$\scalebox{8.5}[1.2]{${\rightarrow}$}}
\\
\multicolumn{1}{c}{} & \multicolumn{1}{c}{} & \multicolumn{1}{c}{} & \multicolumn{1}{c}{} 
     & \normalsize 0.1 & \normalsize 0.2 & \normalsize 0.3 & \normalsize 0.4 & \normalsize 0.5 & \normalsize 0.6 & \normalsize 0.7 & \normalsize 0.8 & \normalsize 0.9 
\\
\hline
\multirow{30}{*}{\normalsize \begin{tabular}{c} Coordinated \\ shock \end{tabular}} & 
    \multirow{10}{*}{\normalsize \begin{tabular}{c} $|V|$ \\ $=$ \\ 50 \end{tabular} } & 
      \multirow{2}{*}{\normalsize $\Phi=0.5$} 
         & \scalebox{1   }[1]{$\gamma=0.25$} 
         & \scalebox{1.1}[1.1]{\bf 0.87} & \scalebox{1.1}[1.1]{\bf 0.86} & \scalebox{1.1}[1.1]{\bf 0.84} & \scalebox{1.1}[1.1]{\bf 0.83} & \scalebox{1.1}[1.1]{\bf 0.83} & \scalebox{1.1}[1.1]{\bf 0.83} & \scalebox{1.1}[1.1]{\bf 0.83} & \scalebox{1.1}[1.1]{\bf 0.83} & \scalebox{1.1}[1.1]{\bf 0.83} 
\\
&
&
         & \scalebox{1   }[1]{$\gamma=0.3$}
         & \scalebox{1.1}[1.1]{\bf 1.00} & \scalebox{1.1}[1.1]{\bf 1.00} & \scalebox{1.1}[1.1]{\bf 0.99} & \scalebox{1.1}[1.1]{\bf 1.00} & \scalebox{1.1}[1.1]{\bf 1.00} & \scalebox{1.1}[1.1]{\bf 1.00} & \scalebox{1.1}[1.1]{\bf 1.00} & \scalebox{1.1}[1.1]{\bf 1.00} & \scalebox{1.1}[1.1]{\bf 1.00} 
\\
\cline{3-13}
&
&
      \multirow{2}{*}{\normalsize $\Phi=0.6$} 
         & \scalebox{1   }[1]{$\gamma=0.3$} 
         & \scalebox{1}[1]{\bf\color{darkgray} 0.27} & \scalebox{1.1}[1.1]{\bf 0.49} & \scalebox{1.1}[1.1]{\bf 0.52} & \scalebox{1.1}[1.1]{\bf 0.52} & \scalebox{1.1}[1.1]{\bf 0.52} & \scalebox{1.1}[1.1]{\bf 0.52} & \scalebox{1.1}[1.1]{\bf 0.52} & \scalebox{1.1}[1.1]{\bf 0.52} & \scalebox{1.1}[1.1]{\bf 0.52} 
\\
&
&
         & \scalebox{1   }[1]{$\gamma=0.35$} 
         & \scalebox{1}[1]{\color{darkgray}\bf 0.27} & \scalebox{1.1}[1.1]{\bf 0.51} & \scalebox{1.1}[1.1]{\bf 0.37} & \scalebox{1.1}[1.1]{\bf 0.37} & \scalebox{1.1}[1.1]{\bf 0.37} & \scalebox{1.1}[1.1]{\bf 0.37} & \scalebox{1.1}[1.1]{\bf 0.37} & \scalebox{1.1}[1.1]{\bf 0.37} & \scalebox{1.1}[1.1]{\bf 0.37} 
\\
\cline{3-13}
&
&
      \multirow{2}{*}{\normalsize $\Phi=0.7$} 
         & \scalebox{1   }[1]{$\gamma=0.35$} 
         & \scalebox{1}[1]{\color{darkgray}\bf 0.27} & \scalebox{1.1}[1.1]{\bf 0.48} & \scalebox{1.1}[1.1]{\bf 0.59} & \scalebox{1.1}[1.1]{\bf 0.64} & \scalebox{1.1}[1.1]{\bf 0.65} & \scalebox{1.1}[1.1]{\bf 0.65} & \scalebox{1.1}[1.1]{\bf 0.65} & \scalebox{1.1}[1.1]{\bf 0.65} & \scalebox{1.1}[1.1]{\bf 0.65} 
\\
&
&
         & \scalebox{1   }[1]{$\gamma=0.4$} 
         & \scalebox{1}[1]{\color{darkgray}\bf 0.30} & \scalebox{1.1}[1.1]{\bf 0.34} & \scalebox{1.1}[1.1]{\bf 0.32} & \scalebox{1}[1]{\color{darkgray}\bf 0.29} & \scalebox{1}[1]{\color{darkgray}\bf 0.29} & \scalebox{1}[1]{\color{darkgray}\bf 0.29} & \scalebox{1}[1]{\color{darkgray}\bf 0.29} & \scalebox{1}[1]{\color{darkgray}\bf 0.29} & \scalebox{1}[1]{\color{darkgray}\bf 0.29} 
\\
\cline{3-13}
&
&
      \multirow{2}{*}{\normalsize $\Phi=0.8$} 
         & \scalebox{1   }[1]{$\gamma=0.4$} 
         & \scalebox{1}[1]{\color{darkgray}\bf 0.24} & \scalebox{1.1}[1.1]{\bf 0.42} & \scalebox{1.1}[1.1]{\bf 0.47} & \scalebox{1.1}[1.1]{\bf 0.47} & \scalebox{1.1}[1.1]{\bf 0.47} & \scalebox{1.1}[1.1]{\bf 0.47} & \scalebox{1.1}[1.1]{\bf 0.47} & \scalebox{1.1}[1.1]{\bf 0.47} & \scalebox{1.1}[1.1]{\bf 0.47} 
\\
&
&
         & \scalebox{1   }[1]{$\gamma=0.45$} 
         & \scalebox{1.1}[1.1]{\bf 0.33} & \scalebox{1.1}[1.1]{\bf 0.45} & \scalebox{1.1}[1.1]{\bf 0.43} & \scalebox{1.1}[1.1]{\bf 0.41} & \scalebox{1.1}[1.1]{\bf 0.40} & \scalebox{1.1}[1.1]{\bf 0.40} & \scalebox{1.1}[1.1]{\bf 0.40} & \scalebox{1.1}[1.1]{\bf 0.40} & \scalebox{1.1}[1.1]{\bf 0.40} 
\\
\cline{3-13}
&
&
      \multirow{2}{*}{\normalsize $\Phi=0.9$} 
         & \scalebox{1   }[1]{$\gamma=0.45$} 
         & \scalebox{1}[1]{\color{darkgray}\bf 0.24} & \scalebox{1.1}[1.1]{\bf 0.47} & \scalebox{1.1}[1.1]{\bf 0.54} & \scalebox{1.1}[1.1]{\bf 0.56} & \scalebox{1.1}[1.1]{\bf 0.56} & \scalebox{1.1}[1.1]{\bf 0.56} & \scalebox{1.1}[1.1]{\bf 0.56} & \scalebox{1.1}[1.1]{\bf 0.56} & \scalebox{1.1}[1.1]{\bf 0.56} 
\\
&
&
         & \scalebox{1   }[1]{$\gamma=0.5$} 
         & \scalebox{1}[1]{\color{darkgray}\bf 0.28} & \scalebox{1.1}[1.1]{\bf 0.38} & \scalebox{1.1}[1.1]{\bf 0.37} & \scalebox{1.1}[1.1]{\bf 0.34} & \scalebox{1.1}[1.1]{\bf 0.32} & \scalebox{1.1}[1.1]{\bf 0.32} & \scalebox{1.1}[1.1]{\bf 0.32} & \scalebox{1.1}[1.1]{\bf 0.32} & \scalebox{1.1}[1.1]{\bf 0.32} 
\\
\cline{2-13}
&
    \multirow{10}{*}{\normalsize \begin{tabular}{c} $|V|$ \\ $=$ \\ 100 \end{tabular}}
&
      \multirow{2}{*}{\normalsize $\Phi=0.5$} 
         & \scalebox{1   }[1]{$\gamma=0.25$} 
         & \scalebox{1.1}[1.1]{\bf 0.95} & \scalebox{1.1}[1.1]{\bf 0.94} & \scalebox{1.1}[1.1]{\bf 0.94} & \scalebox{1.1}[1.1]{\bf 0.93} & \scalebox{1.1}[1.1]{\bf 0.93} & \scalebox{1.1}[1.1]{\bf 0.93} & \scalebox{1.1}[1.1]{\bf 0.93} & \scalebox{1.1}[1.1]{\bf 0.93} & \scalebox{1.1}[1.1]{\bf 0.93} 
\\
&
&
         & \scalebox{1   }[1]{$\gamma=0.3$} 
         & \scalebox{1.1}[1.1]{\bf 0.98} & \scalebox{1.1}[1.1]{\bf 1.00} & \scalebox{1.1}[1.1]{\bf 1.00} & \scalebox{1.1}[1.1]{\bf 1.00} & \scalebox{1.1}[1.1]{\bf 1.00} & \scalebox{1.1}[1.1]{\bf 1.00} & \scalebox{1.1}[1.1]{\bf 1.00} & \scalebox{1.1}[1.1]{\bf 1.00} & \scalebox{1.1}[1.1]{\bf 1.00} 
\\
\cline{3-13}
&
&
      \multirow{2}{*}{\normalsize $\Phi=0.6$} 
         & \scalebox{1   }[1]{$\gamma=0.3$} 
         & \scalebox{1.1}[1.1]{\bf 0.39} & \scalebox{1}[1]{\color{darkgray}\bf 0.27} & \scalebox{1.1}[1.1]{\bf 0.33} & \scalebox{1}[1]{\color{darkgray}\bf 0.31} & \scalebox{1}[1]{\color{darkgray}\bf 0.29} & \scalebox{1}[1]{\color{darkgray}\bf 0.29} & \scalebox{1}[1]{\color{darkgray}\bf 0.29} & \scalebox{1}[1]{\color{darkgray}\bf 0.29} & \scalebox{1}[1]{\color{darkgray}\bf 0.29} 
\\
&
&
         & \scalebox{1   }[1]{$\gamma=0.35$} 
         & \scalebox{1.1}[1.1]{\bf 0.32} & \scalebox{1}[1]{\color{darkgray}\bf 0.24} & \scalebox{1}[1]{\color{darkgray}\bf 0.28} & \scalebox{1}[1]{\color{darkgray}\bf 0.30} & \scalebox{1.1}[1.1]{\bf 0.32} & \scalebox{1.1}[1.1]{\bf 0.32} & \scalebox{1.1}[1.1]{\bf 0.32} & \scalebox{1.1}[1.1]{\bf 0.32} & \scalebox{1.1}[1.1]{\bf 0.32} 
\\
\cline{3-13}
&
&
      \multirow{2}{*}{\normalsize $\Phi=0.7$} 
         & \scalebox{1   }[1]{$\gamma=0.35$} 
         & \scalebox{1.1}[1.1]{\bf 0.39} & \scalebox{1.1}[1.1]{\bf 0.57} & \scalebox{1.1}[1.1]{\bf 0.67} & \scalebox{1.1}[1.1]{\bf 0.69} & \scalebox{1.1}[1.1]{\bf 0.69} & \scalebox{1.1}[1.1]{\bf 0.69} & \scalebox{1.1}[1.1]{\bf 0.69} & \scalebox{1.1}[1.1]{\bf 0.69} & \scalebox{1.1}[1.1]{\bf 0.69} 
\\
&
&
         & \scalebox{1   }[1]{$\gamma=0.4$} 
         & \scalebox{1.1}[1.1]{\bf 0.36} & \scalebox{1.1}[1.1]{\bf 0.44} & \scalebox{1.1}[1.1]{\bf 0.42} & \scalebox{1.1}[1.1]{\bf 0.42} & \scalebox{1.1}[1.1]{\bf 0.42} & \scalebox{1.1}[1.1]{\bf 0.42} & \scalebox{1.1}[1.1]{\bf 0.42} & \scalebox{1.1}[1.1]{\bf 0.42} & \scalebox{1.1}[1.1]{\bf 0.42} 
\\
\cline{3-13}
&
&
      \multirow{2}{*}{\normalsize $\Phi=0.8$} 
         & \scalebox{1   }[1]{$\gamma=0.4$} 
         & \scalebox{1.1}[1.1]{\bf 0.36} & \scalebox{1.1}[1.1]{\bf 0.49} & \scalebox{1.1}[1.1]{\bf 0.54} & \scalebox{1.1}[1.1]{\bf 0.53} & \scalebox{1.1}[1.1]{\bf 0.53} & \scalebox{1.1}[1.1]{\bf 0.53} & \scalebox{1.1}[1.1]{\bf 0.53} & \scalebox{1.1}[1.1]{\bf 0.53} & \scalebox{1.1}[1.1]{\bf 0.53} 
\\
&
&
         & \scalebox{1   }[1]{$\gamma=0.45$} 
         & \scalebox{1.1}[1.1]{\bf 0.37} & \scalebox{1.1}[1.1]{\bf 0.46} & \scalebox{1.1}[1.1]{\bf 0.47} & \scalebox{1.1}[1.1]{\bf 0.48} & \scalebox{1.1}[1.1]{\bf 0.48} & \scalebox{1.1}[1.1]{\bf 0.48} & \scalebox{1.1}[1.1]{\bf 0.48} & \scalebox{1.1}[1.1]{\bf 0.48} & \scalebox{1.1}[1.1]{\bf 0.48} 
\\
\cline{3-13}
&
&
      \multirow{2}{*}{\normalsize $\Phi=0.9$} 
         & \scalebox{1   }[1]{$\gamma=0.45$} 
         & \scalebox{1.1}[1.1]{\bf 0.39} & \scalebox{1.1}[1.1]{\bf 0.56} & \scalebox{1.1}[1.1]{\bf 0.66} & \scalebox{1.1}[1.1]{\bf 0.67} & \scalebox{1.1}[1.1]{\bf 0.67} & \scalebox{1.1}[1.1]{\bf 0.67} & \scalebox{1.1}[1.1]{\bf 0.67} & \scalebox{1.1}[1.1]{\bf 0.67} & \scalebox{1.1}[1.1]{\bf 0.67} 
\\
&
&
         & \scalebox{1   }[1]{$\gamma=0.5$} 
         & \scalebox{1.1}[1.1]{\bf 0.37} & \scalebox{1.1}[1.1]{\bf 0.47} & \scalebox{1.1}[1.1]{\bf 0.48} & \scalebox{1.1}[1.1]{\bf 0.47} & \scalebox{1.1}[1.1]{\bf 0.47} & \scalebox{1.1}[1.1]{\bf 0.47} & \scalebox{1.1}[1.1]{\bf 0.47} & \scalebox{1.1}[1.1]{\bf 0.47} & \scalebox{1.1}[1.1]{\bf 0.47} 
\\
\cline{2-13}
&
    \multirow{10}{*}{\normalsize \begin{tabular}{c} $|V|$ \\ $=$ \\ 300 \end{tabular}}
&
      \multirow{2}{*}{\normalsize $\Phi=0.5$} 
         & \scalebox{1   }[1]{$\gamma=0.25$} 
         & \scalebox{1.1}[1.1]{\bf 0.93} & \scalebox{1.1}[1.1]{\bf 0.94} & \scalebox{1.1}[1.1]{\bf 0.94} & \scalebox{1.1}[1.1]{\bf 0.93} & \scalebox{1.1}[1.1]{\bf 0.93} & \scalebox{1.1}[1.1]{\bf 0.93} & \scalebox{1.1}[1.1]{\bf 0.93} & \scalebox{1.1}[1.1]{\bf 0.93} & \scalebox{1.1}[1.1]{\bf 0.93} 
\\
&
&
         & \scalebox{1   }[1]{$\gamma=0.3$} 
         & \scalebox{1.1}[1.1]{\bf 0.99} & \scalebox{1.1}[1.1]{\bf 1.00   } & \scalebox{1.1}[1.1]{\bf 1.00   } & \scalebox{1.1}[1.1]{\bf 1.00   } & \scalebox{1.1}[1.1]{\bf 1.00   } & \scalebox{1.1}[1.1]{\bf 1.00   } & \scalebox{1.1}[1.1]{\bf 1.00   } & \scalebox{1.1}[1.1]{\bf 1.00   } & \scalebox{1.1}[1.1]{\bf 1.00   } 
\\
\cline{3-13}
&
&
      \multirow{2}{*}{\normalsize $\Phi=0.6$} 
         & \scalebox{1   }[1]{$\gamma=0.3$} 
         & \scalebox{1.1}[1.1]{\bf 0.39} & \scalebox{1}[1]{\color{darkgray}\bf 0.27} & \scalebox{1}[1]{\color{darkgray}\bf 0.29} & \scalebox{1}[1]{\color{darkgray}\bf 0.25} & \scalebox{1}[1]{\color{darkgray}\bf 0.25} & \scalebox{1}[1]{\color{darkgray}\bf 0.25} & \scalebox{1}[1]{\color{darkgray}\bf 0.25} & \scalebox{1}[1]{\color{darkgray}\bf 0.25} & \scalebox{1}[1]{\color{darkgray}\bf 0.25} 
\\
&
&
         & \scalebox{1   }[1]{$\gamma=0.35$} 
         & \scalebox{1}[1]{\color{darkgray}\bf 0.31} & \scalebox{1}[1]{\color{darkgray}\bf 0.14} & \scalebox{1}[1]{\color{darkgray}\bf 0.16} & \scalebox{1}[1]{\color{darkgray}\bf 0.14} & \scalebox{1}[1]{\color{darkgray}\bf 0.14} & \scalebox{1}[1]{\color{darkgray}\bf 0.14} & \scalebox{1}[1]{\color{darkgray}\bf 0.14} & \scalebox{1}[1]{\color{darkgray}\bf 0.14} & \scalebox{1}[1]{\color{darkgray}\bf 0.14} 
\\
\cline{3-13}
&
&
      \multirow{2}{*}{\normalsize $\Phi=0.7$} 
         & \scalebox{1   }[1]{$\gamma=0.35$} 
         & \scalebox{1.1}[1.1]{\bf 0.40} & \scalebox{1.1}[1.1]{\bf 0.53} & \scalebox{1.1}[1.1]{\bf 0.61} & \scalebox{1.1}[1.1]{\bf 0.62} & \scalebox{1.1}[1.1]{\bf 0.62} & \scalebox{1.1}[1.1]{\bf 0.62} & \scalebox{1.1}[1.1]{\bf 0.62} & \scalebox{1.1}[1.1]{\bf 0.62} & \scalebox{1.1}[1.1]{\bf 0.62} 
\\
&
&
         & \scalebox{1   }[1]{$\gamma=0.4$} 
         & \scalebox{1.1}[1.1]{\bf 0.34} & \scalebox{1.1}[1.1]{\bf 0.36} & \scalebox{1.1}[1.1]{\bf 0.34} & \scalebox{1.1}[1.1]{\bf 0.33} & \scalebox{1.1}[1.1]{\bf 0.33} & \scalebox{1.1}[1.1]{\bf 0.33} & \scalebox{1.1}[1.1]{\bf 0.33} & \scalebox{1.1}[1.1]{\bf 0.33} & \scalebox{1.1}[1.1]{\bf 0.33} 
\\
\cline{3-13}
&
&
      \multirow{2}{*}{\normalsize $\Phi=0.8$} 
         & \scalebox{1   }[1]{$\gamma=0.4$} 
         & \scalebox{1.1}[1.1]{\bf 0.38} & \scalebox{1.1}[1.1]{\bf 0.47} & \scalebox{1.1}[1.1]{\bf 0.50 } & \scalebox{1.1}[1.1]{\bf 0.50 } & \scalebox{1.1}[1.1]{\bf 0.50 } & \scalebox{1.1}[1.1]{\bf 0.50 } & \scalebox{1.1}[1.1]{\bf 0.50 } & \scalebox{1.1}[1.1]{\bf 0.50 } & \scalebox{1.1}[1.1]{\bf 0.50 } 
\\
&
&
         & \scalebox{1   }[1]{$\gamma=0.45$} 
         & \scalebox{1.1}[1.1]{\bf 0.39} & \scalebox{1.1}[1.1]{\bf 0.43} & \scalebox{1.1}[1.1]{\bf 0.42} & \scalebox{1.1}[1.1]{\bf 0.40 } & \scalebox{1.1}[1.1]{\bf 0.40 } & \scalebox{1.1}[1.1]{\bf 0.40 } & \scalebox{1.1}[1.1]{\bf 0.40 } & \scalebox{1.1}[1.1]{\bf 0.40 } & \scalebox{1.1}[1.1]{\bf 0.40 } 
\\
\cline{3-13}
&
&
      \multirow{2}{*}{\normalsize $\Phi=0.9$} 
         & \scalebox{1   }[1]{$\gamma=0.45$} 
         & \scalebox{1.1}[1.1]{\bf 0.40 } & \scalebox{1.1}[1.1]{\bf 0.53} & \scalebox{1.1}[1.1]{\bf 0.60 } & \scalebox{1.1}[1.1]{\bf 0.61} & \scalebox{1.1}[1.1]{\bf 0.61} & \scalebox{1.1}[1.1]{\bf 0.61} & \scalebox{1.1}[1.1]{\bf 0.61} & \scalebox{1.1}[1.1]{\bf 0.61} & \scalebox{1.1}[1.1]{\bf 0.61} 
\\
&
&
         & \scalebox{1   }[1]{$\gamma=0.5$} 
         & \scalebox{1.1}[1.1]{\bf 0.39} & \scalebox{1.1}[1.1]{\bf 0.44} & \scalebox{1.1}[1.1]{\bf 0.44} & \scalebox{1.1}[1.1]{\bf 0.43} & \scalebox{1.1}[1.1]{\bf 0.43} & \scalebox{1.1}[1.1]{\bf 0.43} & \scalebox{1.1}[1.1]{\bf 0.43} & \scalebox{1.1}[1.1]{\bf 0.43} & \scalebox{1.1}[1.1]{\bf 0.43} 
\\
\hline
%%%%%%%%%%%%%%%%%%%%%%%%%%%%%%
\multirow{30}{*}{\normalsize \begin{tabular}{c} Idiosyncratic \\ shock \end{tabular}} & 
    \multirow{10}{*}{\normalsize \begin{tabular}{c} $|V|$ \\ $=$ \\ 50 \end{tabular}}
&
      \multirow{2}{*}{\normalsize $\Phi=0.5$} 
         & \scalebox{1   }[1]{$\gamma=0.25$} 
         & \scalebox{1.1}[1.1]{\bf 0.95} & \scalebox{1.1}[1.1]{\bf 0.95} & \scalebox{1.1}[1.1]{\bf 0.95} & \scalebox{1.1}[1.1]{\bf 0.76} & \scalebox{1.1}[1.1]{\bf 0.77} & \scalebox{1.1}[1.1]{\bf 0.37} & \scalebox{1}[1]{\color{darkgray}\bf 0.08} & \scalebox{1}[1]{\color{darkgray}\bf 0.11} & \scalebox{1}[1]{\color{darkgray}\bf 0.25} 
\\
&
&
         & \scalebox{1   }[1]{$\gamma=0.3$} 
         & \scalebox{1.1}[1.1]{\bf 0.95} & \scalebox{1.1}[1.1]{\bf 0.95} & \scalebox{1.1}[1.1]{\bf 0.93} & \scalebox{1.1}[1.1]{\bf 0.99} & \scalebox{1.1}[1.1]{\bf 0.76} & \scalebox{1.1}[1.1]{\bf 0.33} & \scalebox{1}[1]{\color{darkgray}\bf 0.07} & \scalebox{1}[1]{\color{darkgray}\bf 0.23} & \scalebox{1}[1]{\color{darkgray}\bf 0.29} 
\\
\cline{3-13}
&
&
      \multirow{2}{*}{\normalsize $\Phi=0.6$} 
         & \scalebox{1   }[1]{$\gamma=0.3$} 
         & \scalebox{1}[1]{\color{darkgray}\bf 0.08} & \scalebox{1.1}[1.1]{\bf 0.37} & \scalebox{1.1}[1.1]{\bf 0.32} & \scalebox{1.1}[1.1]{\bf 0.47} & \scalebox{1.1}[1.1]{\bf 0.48} & \scalebox{1}[1]{\color{darkgray}\bf 0.21} & \scalebox{1.1}[1.1]{\bf 0.35} & \scalebox{1.1}[1.1]{\bf 0.39} & \scalebox{1}[1]{\color{darkgray}\bf 0.25} 
\\
&
&
         & \scalebox{1   }[1]{$\gamma=0.35$} 
         & \scalebox{1}[1]{\color{darkgray}\bf 0.08} & \scalebox{1}[1]{\color{darkgray}\bf 0.12} & \scalebox{1}[1]{\color{darkgray}\bf 0.27} & \scalebox{1}[1]{\color{darkgray}\bf 0.30} & \scalebox{1.1}[1.1]{\bf 0.38} & \scalebox{1}[1]{\color{darkgray}\bf 0.27} & \scalebox{1}[1]{\color{darkgray}\bf 0.24} & \scalebox{1}[1]{\color{darkgray}\bf 0.13} & \scalebox{1}[1]{\color{darkgray}\bf 0.16} 
\\
\cline{3-13}
&
&
      \multirow{2}{*}{\normalsize $\Phi=0.7$} 
         & \scalebox{1   }[1]{$\gamma=0.35$} 
         & \scalebox{1}[1]{\color{darkgray}\bf 0.08} & \scalebox{1}[1]{\color{darkgray}\bf 0.15} & \scalebox{1}[1]{\color{darkgray}\bf 0.12} & \scalebox{1}[1]{\color{darkgray}\bf 0.04} & \scalebox{1}[1]{\color{darkgray}\bf 0.26} & \scalebox{1}[1]{\color{darkgray}\bf 0.13} & \scalebox{1}[1]{\color{darkgray}\bf 0.07} & \scalebox{1}[1]{\color{darkgray}\bf 0.11} & \scalebox{1}[1]{\color{darkgray}\bf 0.15} 
\\
&
&
         & \scalebox{1   }[1]{$\gamma=0.4$} 
         & \scalebox{1}[1]{\color{darkgray}\bf 0.04} & \scalebox{1}[1]{\color{darkgray}\bf 0.06} & \scalebox{1}[1]{\color{darkgray}\bf 0.03} & \scalebox{1}[1]{\color{darkgray}\bf 0.25} & \scalebox{1}[1]{\color{darkgray}\bf 0.17} & \scalebox{1}[1]{\color{darkgray}\bf 0.06} & \scalebox{1}[1]{\color{darkgray}\bf 0.09} & \scalebox{1}[1]{\color{darkgray}\bf 0.06} & \scalebox{1}[1]{\color{darkgray}\bf 0.08} 
\\
\cline{3-13}
&
&
      \multirow{2}{*}{\normalsize $\Phi=0.8$} 
         & \scalebox{1   }[1]{$\gamma=0.4$} 
         & \scalebox{1}[1]{\color{darkgray}\bf 0.05} & \scalebox{1}[1]{\color{darkgray}\bf 0.09} & \scalebox{1}[1]{\color{darkgray}\bf 0.13} & \scalebox{1}[1]{\color{darkgray}\bf 0.21} & \scalebox{1}[1]{\color{darkgray}\bf 0.12} & \scalebox{1}[1]{\color{darkgray}\bf 0.10 } & \scalebox{1}[1]{\color{darkgray}\bf 0.08} & \scalebox{1}[1]{\color{darkgray}\bf 0.14} & \scalebox{1}[1]{\color{darkgray}\bf 0.10 } 
\\
&
&
         & \scalebox{1   }[1]{$\gamma=0.45$} 
         & \scalebox{1}[1]{\color{darkgray}\bf 0.02} & \scalebox{1}[1]{\color{darkgray}\bf 0.17} & \scalebox{1}[1]{\color{darkgray}\bf 0.08} & \scalebox{1}[1]{\color{darkgray}\bf 0.27} & \scalebox{1}[1]{\color{darkgray}\bf 0.17} & \scalebox{1}[1]{\color{darkgray}\bf 0.19} & \scalebox{1}[1]{\color{darkgray}\bf 0.05} & \scalebox{1}[1]{\color{darkgray}\bf 0.04} & \scalebox{1}[1]{\color{darkgray}\bf 0.08} 
\\
\cline{3-13}
&
&
      \multirow{2}{*}{\normalsize $\Phi=0.9$} 
         & \scalebox{1   }[1]{$\gamma=0.45$} 
         & \scalebox{1}[1]{\color{darkgray}\bf 0.09} & \scalebox{1}[1]{\color{darkgray}\bf 0.07} & \scalebox{1}[1]{\color{darkgray}\bf 0.20 } & \scalebox{1}[1]{\color{darkgray}\bf 0.21} & \scalebox{1}[1]{\color{darkgray}\bf 0.03} & \scalebox{1}[1]{\color{darkgray}\bf 0.12} & \scalebox{1}[1]{\color{darkgray}\bf 0.02} & \scalebox{1}[1]{\color{darkgray}\bf 0.17} & \scalebox{1}[1]{\color{darkgray}\bf 0.17} 
\\
&
&
         & \scalebox{1   }[1]{$\gamma=0.5$} 
         & \scalebox{1}[1]{\color{darkgray}\bf 0.08} & \scalebox{1}[1]{\color{darkgray}\bf 0.04} & \scalebox{1}[1]{\color{darkgray}\bf 0.13} & \scalebox{1}[1]{\color{darkgray}\bf 0.24} & \scalebox{1}[1]{\color{darkgray}\bf 0.12} & \scalebox{1}[1]{\color{darkgray}\bf 0.10 } & \scalebox{1}[1]{\color{darkgray}\bf 0.10 } & \scalebox{1}[1]{\color{darkgray}\bf 0.06} & \scalebox{1}[1]{\color{darkgray}\bf 0.10 } 
\\
\cline{2-13}
&
    \multirow{10}{*}{\normalsize \begin{tabular}{c} $|V|$ \\ $=$ \\ 100 \end{tabular}}
&
      \multirow{2}{*}{\normalsize $\Phi=0.5$} 
         & \scalebox{1   }[1]{$\gamma=0.25$} 
         & \scalebox{1.1}[1.1]{\bf 0.98} & \scalebox{1.1}[1.1]{\bf 0.96} & \scalebox{1.1}[1.1]{\bf 0.98} & \scalebox{1.1}[1.1]{\bf 0.96} & \scalebox{1.1}[1.1]{\bf 0.67} & \scalebox{1.1}[1.1]{\bf 0.37} & \scalebox{1}[1]{\color{darkgray}\bf 0.16} & \scalebox{1}[1]{\color{darkgray}\bf 0.03} & \scalebox{1}[1]{\color{darkgray}\bf 0.16} 
\\
&
&
         & \scalebox{1   }[1]{$\gamma=0.3$} 
         & \scalebox{1.1}[1.1]{\bf 0.99} & \scalebox{1.1}[1.1]{\bf 1.00} & \scalebox{1.1}[1.1]{\bf 0.95} & \scalebox{1.1}[1.1]{\bf 0.99} & \scalebox{1.1}[1.1]{\bf 0.71} & \scalebox{1}[1]{\color{darkgray}\bf 0.31} & \scalebox{1}[1]{\color{darkgray}\bf 0.07} & \scalebox{1}[1]{\color{darkgray}\bf 0.11} & \scalebox{1}[1]{\color{darkgray}\bf 0.26} 
\\
\cline{3-13}
&
&
      \multirow{2}{*}{\normalsize $\Phi=0.6$} 
         & \scalebox{1   }[1]{$\gamma=0.3$} 
         & \scalebox{1}[1]{\color{darkgray}\bf 0.04} & \scalebox{1}[1]{\color{darkgray}\bf 0.15} & \scalebox{1}[1]{\color{darkgray}\bf 0.3 } & \scalebox{1.1}[1.1]{\bf 0.44} & \scalebox{1}[1]{\color{darkgray}\bf 0.03} & \scalebox{1}[1]{\color{darkgray}\bf 0.08} & \scalebox{1}[1]{\color{darkgray}\bf 0.22} & \scalebox{1}[1]{\color{darkgray}\bf 0.16} & \scalebox{1}[1]{\color{darkgray}\bf 0.06} 
\\
&
&
         & \scalebox{1   }[1]{$\gamma=0.35$} 
         & \scalebox{1}[1]{\color{darkgray}\bf 0.03} & \scalebox{1}[1]{\color{darkgray}\bf 0.04} & \scalebox{1}[1]{\color{darkgray}\bf 0.29} & \scalebox{1}[1]{\color{darkgray}\bf 0.30} & \scalebox{1}[1]{\color{darkgray}\bf 0.17} & \scalebox{1.1}[1.1]{\bf 0.33} & \scalebox{1}[1]{\color{darkgray}\bf 0.06} & \scalebox{1}[1]{\color{darkgray}\bf 0.10 } & \scalebox{1}[1]{\color{darkgray}\bf 0.10 } 
\\
\cline{3-13}
&
&
      \multirow{2}{*}{\normalsize $\Phi=0.7$} 
         & \scalebox{1   }[1]{$\gamma=0.35$} 
         & \scalebox{1}[1]{\color{darkgray}\bf 0.02} & \scalebox{1}[1]{\color{darkgray}\bf 0.11} & \scalebox{1}[1]{\color{darkgray}\bf 0.09} & \scalebox{1}[1]{\color{darkgray}\bf 0.23} & \scalebox{1}[1]{\color{darkgray}\bf 0.13} & \scalebox{1}[1]{\color{darkgray}\bf 0.01} & \scalebox{1}[1]{\color{darkgray}\bf 0.10 } & \scalebox{1}[1]{\color{darkgray}\bf 0.05} & \scalebox{1}[1]{\color{darkgray}\bf 0.06} 
\\
&
&
         & \scalebox{1   }[1]{$\gamma=0.4$} 
         & \scalebox{1}[1]{\color{darkgray}\bf 0.02} & \scalebox{1}[1]{\color{darkgray}\bf 0.1 } & \scalebox{1}[1]{\color{darkgray}\bf 0.16} & \scalebox{1}[1]{\color{darkgray}\bf 0.11} & \scalebox{1}[1]{\color{darkgray}\bf 0.12} & \scalebox{1}[1]{\color{darkgray}\bf 0.13} & \scalebox{1}[1]{\color{darkgray}\bf 0.05} & \scalebox{1}[1]{\color{darkgray}\bf 0.03} & \scalebox{1}[1]{\color{darkgray}\bf 0.05} 
\\
\cline{3-13}
&
&
      \multirow{2}{*}{\normalsize $\Phi=0.8$} 
         & \scalebox{1   }[1]{$\gamma=0.4$} 
         & \scalebox{1}[1]{\color{darkgray}\bf 0.03} & \scalebox{1}[1]{\color{darkgray}\bf 0.03} & \scalebox{1}[1]{\color{darkgray}\bf 0.07} & \scalebox{1}[1]{\color{darkgray}\bf 0.07} & \scalebox{1}[1]{\color{darkgray}\bf 0.07} & \scalebox{1}[1]{\color{darkgray}\bf 0.04} & \scalebox{1}[1]{\color{darkgray}\bf 0.06} & \scalebox{1}[1]{\color{darkgray}\bf 0.02} & \scalebox{1}[1]{\color{darkgray}\bf 0.08} 
\\
&
&
         & \scalebox{1   }[1]{$\gamma=0.45$} 
         & \scalebox{1}[1]{\color{darkgray}\bf 0.03} & \scalebox{1}[1]{\color{darkgray}\bf 0.06} & \scalebox{1}[1]{\color{darkgray}\bf 0.14} & \scalebox{1}[1]{\color{darkgray}\bf 0.17} & \scalebox{1}[1]{\color{darkgray}\bf 0.09} & \scalebox{1}[1]{\color{darkgray}\bf 0.13} & \scalebox{1}[1]{\color{darkgray}\bf 0.08} & \scalebox{1}[1]{\color{darkgray}\bf 0.03} & \scalebox{1}[1]{\color{darkgray}\bf 0.08} 
\\
\cline{3-13}
&
&
      \multirow{2}{*}{\normalsize $\Phi=0.9$} 
         & \scalebox{1   }[1]{$\gamma=0.45$} 
         & \scalebox{1}[1]{\color{darkgray}\bf 0.04} & \scalebox{1}[1]{\color{darkgray}\bf 0.07} & \scalebox{1}[1]{\color{darkgray}\bf 0.07} & \scalebox{1}[1]{\color{darkgray}\bf 0.10 } & \scalebox{1}[1]{\color{darkgray}\bf 0.14} & \scalebox{1}[1]{\color{darkgray}\bf 0.10 } & \scalebox{1}[1]{\color{darkgray}\bf 0.09} & \scalebox{1}[1]{\color{darkgray}\bf 0.08} & \scalebox{1}[1]{\color{darkgray}\bf 0.07} 
\\
&
&
         & \scalebox{1   }[1]{$\gamma=0.5$} 
         & \scalebox{1}[1]{\color{darkgray}\bf 0.05} & \scalebox{1}[1]{\color{darkgray}\bf 0.15} & \scalebox{1}[1]{\color{darkgray}\bf 0.10 } & \scalebox{1}[1]{\color{darkgray}\bf 0.07} & \scalebox{1}[1]{\color{darkgray}\bf 0.08} & \scalebox{1}[1]{\color{darkgray}\bf 0.13} & \scalebox{1}[1]{\color{darkgray}\bf 0.02} & \scalebox{1}[1]{\color{darkgray}\bf 0.06} & \scalebox{1}[1]{\color{darkgray}\bf 0.04} 
\\
\cline{2-13}
&
    \multirow{10}{*}{\normalsize \begin{tabular}{c} $|V|$ \\ $=$ \\ 300 \end{tabular}}
&
      \multirow{2}{*}{\normalsize $\Phi=0.5$} 
         & \scalebox{1   }[1]{$\gamma=0.25$} 
         & \scalebox{1.1}[1.1]{\bf 1.00   } & \scalebox{1.1}[1.1]{\bf 0.99} & \scalebox{1.1}[1.1]{\bf 0.98} & \scalebox{1.1}[1.1]{\bf 1.00   } & \scalebox{1.1}[1.1]{\bf 0.69} & \scalebox{1.1}[1.1]{\bf 0.36} & \scalebox{1}[1]{\color{darkgray}\bf 0.16} & \scalebox{1}[1]{\color{darkgray}\bf 0.07} & \scalebox{1}[1]{\color{darkgray}\bf 0.16} 
\\
&
&
         & \scalebox{1   }[1]{$\gamma=0.3$} 
         & \scalebox{1.1}[1.1]{\bf 1.00   } & \scalebox{1.1}[1.1]{\bf 0.98} & \scalebox{1.1}[1.1]{\bf 0.96} & \scalebox{1.1}[1.1]{\bf 1.00   } & \scalebox{1.1}[1.1]{\bf 0.69} & \scalebox{1.1}[1.1]{\bf 0.36} & \scalebox{1}[1]{\color{darkgray}\bf 0.12} & \scalebox{1}[1]{\color{darkgray}\bf 0.06} & \scalebox{1}[1]{\color{darkgray}\bf 0.23} 
\\
\cline{3-13}
&
&
      \multirow{2}{*}{\normalsize $\Phi=0.6$} 
         & \scalebox{1   }[1]{$\gamma=0.3$} 
         & \scalebox{1}[1]{\color{darkgray}\bf 0.03} & \scalebox{1}[1]{\color{darkgray}\bf 0.12} & \scalebox{1}[1]{\color{darkgray}\bf 0.03} & \scalebox{1}[1]{\color{darkgray}\bf 0.21} & \scalebox{1}[1]{\color{darkgray}\bf 0.12} & \scalebox{1}[1]{\color{darkgray}\bf 0.14} & \scalebox{1}[1]{\color{darkgray}\bf 0.12} & \scalebox{1}[1]{\color{darkgray}\bf 0.08} & \scalebox{1}[1]{\color{darkgray}\bf 0.14} 
\\
&
&
         & \scalebox{1   }[1]{$\gamma=0.35$} 
         & \scalebox{1}[1]{\color{darkgray}\bf 0.04} & \scalebox{1}[1]{\color{darkgray}\bf 0.03} & \scalebox{1}[1]{\color{darkgray}\bf 0.01} & \scalebox{1}[1]{\color{darkgray}\bf 0.15} & \scalebox{1}[1]{\color{darkgray}\bf 0.08} & \scalebox{1}[1]{\color{darkgray}\bf 0.04} & \scalebox{1}[1]{\color{darkgray}\bf 0.07} & \scalebox{1}[1]{\color{darkgray}\bf 0.09} & \scalebox{1}[1]{\color{darkgray}\bf 0.04} 
\\
\cline{3-13}
&
&
      \multirow{2}{*}{\normalsize $\Phi=0.7$} 
         & \scalebox{1   }[1]{$\gamma=0.35$} 
         & \scalebox{1}[1]{\color{darkgray}\bf 0.06} & \scalebox{1}[1]{\color{darkgray}\bf 0.08} & \scalebox{1   }[1]{\color{darkgray}\bf 0.13} & \scalebox{1   }[1]{\color{darkgray}\bf 0.17} & \scalebox{1   }[1]{\color{darkgray}\bf 0.14} & \scalebox{1   }[1]{\color{darkgray}\bf 0.05} & \scalebox{1   }[1]{\color{darkgray}\bf 0.07} & \scalebox{1   }[1]{\color{darkgray}\bf 0.02} & \scalebox{1   }[1]{\color{darkgray}\bf 0.08} 
\\
&
&
         & \scalebox{1   }[1]{$\gamma=0.4$} 
         & \scalebox{1   }[1]{\color{darkgray}\bf 0.03} & \scalebox{1   }[1]{\color{darkgray}\bf 0.06} & \scalebox{1   }[1]{\color{darkgray}\bf 0.09} & \scalebox{1   }[1]{\color{darkgray}\bf 0.11} & \scalebox{1   }[1]{\color{darkgray}\bf 0.10 } & \scalebox{1   }[1]{\color{darkgray}\bf 0.08} & \scalebox{1   }[1]{\color{darkgray}\bf 0.02} & \scalebox{1   }[1]{\color{darkgray}\bf 0.02} & \scalebox{1   }[1]{\color{darkgray}\bf 0.05} 
\\
\cline{3-13}
&
&
      \multirow{2}{*}{\normalsize $\Phi=0.8$} 
         & \scalebox{1   }[1]{$\gamma=0.4$} 
         & \scalebox{1   }[1]{\color{darkgray}\bf 0.04} & \scalebox{1   }[1]{\color{darkgray}\bf 0.09} & \scalebox{1   }[1]{\color{darkgray}\bf 0.12} & \scalebox{1   }[1]{\color{darkgray}\bf 0.14} & \scalebox{1   }[1]{\color{darkgray}\bf 0.12} & \scalebox{1   }[1]{\color{darkgray}\bf 0.11} & \scalebox{1   }[1]{\color{darkgray}\bf 0.04} & \scalebox{1   }[1]{\color{darkgray}\bf 0.05} & \scalebox{1   }[1]{\color{darkgray}\bf 0.08} 
\\
&
&
         & \scalebox{1   }[1]{$\gamma=0.45$} 
         & \scalebox{1   }[1]{\color{darkgray}\bf 0.04} & \scalebox{1   }[1]{\color{darkgray}\bf 0.17} & \scalebox{1   }[1]{\color{darkgray}\bf 0.11} & \scalebox{1   }[1]{\color{darkgray}\bf 0.15} & \scalebox{1   }[1]{\color{darkgray}\bf 0.17} & \scalebox{1   }[1]{\color{darkgray}\bf 0.09} & \scalebox{1   }[1]{\color{darkgray}\bf 0.04} & \scalebox{1   }[1]{\color{darkgray}\bf 0.04} & \scalebox{1   }[1]{\color{darkgray}\bf 0.05} 
\\
\cline{3-13}
&
&
      \multirow{2}{*}{\normalsize $\Phi=0.9$} 
         & \scalebox{1   }[1]{$\gamma=0.45$} 
         & \scalebox{1   }[1]{\color{darkgray}\bf 0.05} & \scalebox{1   }[1]{\color{darkgray}\bf 0.08} & \scalebox{1   }[1]{\color{darkgray}\bf 0.11} & \scalebox{1   }[1]{\color{darkgray}\bf 0.18} & \scalebox{1   }[1]{\color{darkgray}\bf 0.14} & \scalebox{1   }[1]{\color{darkgray}\bf 0.02} & \scalebox{1   }[1]{\color{darkgray}\bf 0.03} & \scalebox{1   }[1]{\color{darkgray}\bf 0.04} & \scalebox{1   }[1]{\color{darkgray}\bf 0.07} 
\\
&
&
         & \scalebox{1   }[1]{$\gamma=0.5$} 
         & \scalebox{1   }[1]{\color{darkgray}\bf 0.05} & \scalebox{1   }[1]{\color{darkgray}\bf 0.07} & \scalebox{1   }[1]{\color{darkgray}\bf 0.10 } & \scalebox{1   }[1]{\color{darkgray}\bf 0.21} & \scalebox{1   }[1]{\color{darkgray}\bf 0.11} & \scalebox{1   }[1]{\color{darkgray}\bf 0.09} & \scalebox{1   }[1]{\color{darkgray}\bf 0.04} & \scalebox{1   }[1]{\color{darkgray}\bf 0.01} & \scalebox{1   }[1]{\color{darkgray}\bf 0.05} 
\\
\hline
\\
\end{tabular}
\setlength{\extrarowheight}{0ex}
}
\end{center}
\end{table}

%%% END of table for sharp increase of xi at E=0.75 for homogeneous in-arborescence networks
%%%%%%%%%%%%%%%%%%%%%%%%%%%%%%%%%%%%%%%%%%%%%%%%%%%%%%%%%%%%%%%%%%%%%%%%%%%%%%%%%%%%%%%%%%%%
%%%%%%%%%%%%%%%%%%%%%%%%%%%%%%%%%%%%%%%%%%%%%%%%%%%%%%%%%%%%%%%%%%%%%%%%%%%%%%%%%%%%%%%%%%%%
%%%%%%%%%%%%%%%%%%%%%%%%%%%%%%%%%%%%%%%%%%%%%%%%%%%%%%%%%%%%%%%%%%%%%%%%%%%%%%%%%%%%%%%%%%%%

\vspace*{0.1in}
\noindent
{\bf Formal intuition}

\vspace*{0.1in}
\noindent
A formal intuition behind such a sharp decrease of $\xi$ can be provided as follows.

\begin{lemma}[see Section~\ref{g9proof-sec} of the appendix for a proof]\label{g9}
Fix $\gamma$, $\Phi$, $\I$, a homogeneous in-arborescence network $G$ and assume that
$\gamma\approx \nicefrac{\Phi}{2}$. Consider any node $v\in\vs$ with $\din(v)>1$, suppose that $v$ fails due to the initial shock.
Let $u$ be any node such that $u$ is a ``leaf node'' (\IE, $\din(u)=0$), and $(u,v)\in E$.
Then, as the total external asset $\E$ of the network is varied, 
there exists a threshold value $\E_{\tau}(u)$ such that 
\begin{itemize}
\item
if $\E<\E_{\tau}(u)$ then $u$ will become insolvent, but

\item
if $\E>\E_{\tau}(u)$ then 
$u$ will not become insolvent at any time $t\geq 1$, and the shock will not propagate any further through $u$. 
\end{itemize}
\end{lemma}

The next lemma provides a lower 
bound, using the degree distributions of the Bar\'{a}basi-Albert preferential-attachment model~\cite{BaAl99}, 
on the expected value of the number of leaves in a random in-arborescence network for which Lemma~\ref{g9} can be applied.

\begin{lemma}[see Section~\ref{g10proof-sec} of the appendix for a proof]
\label{g10}
Consider a random in-arborescence $G=(V,E)$ generated by the Bar\'{a}basi-Albert preferential-attachment algorithm~{\rm\cite{BaAl99}} as outlined in Section~\ref{topo} and let 
\[
\widehat{V}=\left\{\,u\in V \, \big| \, \big( \din(u)=0 \big) \, \wedge \, \big( \exists\, v \colon \big( \big( \din(v)>1 \big) \, \wedge \, \big( (u,v)\in E \big) \, \big) \, \big) \, \right\}
\]
Then, $\ex\left[\big|\widehat{V}\big|\right]\geq \dfrac{n}{8}-\dfrac{11}{8}$.
\end{lemma}

Let 
$\xi(\E)$ be the value of $\xi$ parameterized by $\E$ (keeping all other parameters unchanged), and let
\begin{align*}
\E_{\tau_{\min}} & = \min \left\{ \E_{\tau}(u) \,\big|\, \din(u)=0,\, (u,v)\in E \text{ and } \din(v)>1\right\}
\\
\displaybreak[0]
\E_{\tau_{\max}} & = \max \left\{ \E_{\tau}(u) \,\big|\, \din(u)=0,\, (u,v)\in E \text{ and } \din(v)>1\right\}
\end{align*}
It then follows that
\[
\ex\left[\xi\left(\E_{\tau_{\min}}\right)\right]
-
\ex\left[\xi\left(\E_{\tau_{\max}}\right)\right]
\geq 
\frac
{
\frac{n}{8}-\frac{11}{8}
}
{n}
\approx \frac{1}{8}
\]
and $\xi\left(\E\right)$ exhibits a sharp decrease around the range $\left[\E_{\tau_{\min}},\,\E_{\tau_{\max}}\right]$.
In practice, the extent of this decrease is expected to be much more than the pessimistic lower bound of $\nicefrac{1}{8}$, as
our simulation results clearly show.

\section{Concluding remarks}
\label{concl-sec}

In this paper, we have initiated a methodology for systematic investigation of the global stabilities of financial networks that arise in 
OTC derivatives market and elsewhere. Our results can be viewed as a much needed beginning of a systematic investigation of these issues, with future
research works concentrating on further improving the network model, the stability measure and parameter choices.

\section*{Acknowledgment}

This work partially supported by NSF grant IIS-1160995.

\appendix

\begin{center}
{\bf APPENDIX}
\end{center}

\section{Proof of Lemma~\ref{g1}}
\label{g1proof-sec}

We will reuse the notations in Definition~\ref{defnet}.
Using Equation~\eqref{eq1}, for every edge $(u,v)\in E$, the amount of shock received by node $u$ at time $t=1$ is as follows:
\begin{itemize}
\item
If $G$ is homogeneous then 
\begin{eqnarray*}
\ex \left[ \Delta_{\,\mathrm{homo}}(u) \right] 
&
=
&
\frac
{
\min \left\{ \Phi \left( \din(v) - \dout(v) + \frac{\E}{n} \right) \,-\, \gamma \left( \din(v) + \frac{\E}{n} \right), \, \din(v) \right\}
}
{
\din(v)
}
\\
&
=
&
\min \left\{
\big( \Phi - \gamma \big) + 
\frac { \frac{\E}{n} (\Phi - \gamma) \,-\, \Phi \, \dout(v) } { \din(v) }
, \, 
1
\right\}
\end{eqnarray*}

\item
If $G$ is $(\alpha,\beta)$-heterogeneous, then 
$\sigma_v=\frac{\beta}{\alpha}$ 
and, using linearity of expectation, we get
%%%%%%%%%%%
%%%%%%%%%%%
%%%%%%%%%%%
\begin{align*}
\ex\left[b_v\right]
&
= 
\ex
\left[
\sum_{(u,v)\in \widetilde{E}}
%%%
\left(
\,
\alpha\,\frac {\beta \I} {\alpha\,|\widetilde{E}|}
+
\big(1-\alpha\big) \, 
\frac {\big(1-\beta\big)\,\I} {|E|-\alpha\,|\widetilde{E}|}
\,
\right)
\right]
\\
\displaybreak[0]
&
=
\din(v) \, 
\ex
\left[
%%%
\left(
\,
\beta\,\frac {\I} {|\widetilde{E}|}
+
\big(1-\alpha\big) \, 
\frac {\big(1-\beta\big)\,\I} {|E|-\alpha\,|\widetilde{E}|}
\,
\right)
\right]
\\
\displaybreak[0]
%%%%%%%%%%%%%%%%%%%%%%%%%%%%%%%%%%%%%%%%%%%%%%%%%%%%%%%%%%%%%%%%%%%%%%%%%%%%%%%%%
%%%%%%%%%%%%%%%%%%%%%%%%%%%%%%%%%%%%%%%%%%%%%%%%%%%%%%%%%%%%%%%%%%%%%%%%%%%%%%%%%
%%%%%%%%%%%%%%%%%%%%%%%%%%%%%%%%%%%%%%%%%%%%%%%%%%%%%%%%%%%%%%%%%%%%%%%%%%%%%%%%%
&
\\
\displaybreak[0]
\ex\left[b_v-\iota_v\right] 
&
=
\ex
\left[
\sum_{(u,v)\in \widetilde{E}}
%%%
\left(
\,
\alpha\,\frac {\beta \I} {\alpha\,|\widetilde{E}|}
+
\big(1-\alpha\big) \, 
\frac {\big(1-\beta\big)\,\I} {|E|-\alpha\,|\widetilde{E}|}
\,
\right)
\,-\,
\sum_{(v,u)\in \widetilde{E}}
%%%
\left(
\,
\alpha\,\frac {\beta \I} {\alpha\,|\widetilde{E}|}
+
\big(1-\alpha\big) \, 
\frac {\big(1-\beta\big)\,\I} {|E|-\alpha\,|\widetilde{E}|}
\,
\right)
\right]
\\
\displaybreak[0]
%%%%%%%%
&
=
\sum_{(u,v)\in \widetilde{E}}
%%%
\ex
\left[
\left(
\,
\beta\, \frac {\I} {|\widetilde{E}|}
+
\big(1-\alpha\big) \, 
\frac {\big(1-\beta\big)\,\I} {|E|-\alpha\,|\widetilde{E}|}
\,
\right)
\right]
\,-\,
\sum_{(v,u)\in \widetilde{E}}
%%%
\ex
\left[
\left(
\,
\beta \, \frac {\I} {|\widetilde{E}|}
+
\big(1-\alpha\big) \, 
\frac {\big(1-\beta\big)\,\I} {|E|-\alpha\,|\widetilde{E}|}
\,
\right)
\right]
\\
\displaybreak[0]
&
=
\big(\,\din(v) - \dout(v) \, \big)
\,
\ex
\left[
\left(
\,
\beta \, \frac {\I} {|\widetilde{E}|}
+
\big(1-\alpha\big) \, 
\frac {\big(1-\beta\big)\,\I} {|E|-\alpha\,|\widetilde{E}|}
\,
\right)
\right]
%%%%%%%%%%%%%%%%
%%%%%%%%%%%%%%%%
%%%%%%%%%%%%%%%%
\\
\displaybreak[0]
&
\\
\displaybreak[0]
\ex \left[ \Delta_{\,\mathrm{hetero}}(u) \right] 
&
= 
\ex \left[
\frac
{
\min \left\{ \Phi \left( b_v - \iota_v + \sigma_v \, \E \right) - \gamma \left( b_v + \sigma_v\,\E \right), \, b_v \right\}
}
{
\din(v)
}
\right]
\\
\displaybreak[0]
&
=
\min 
\left\{
\ex \left[
\frac
{
 \Phi \left( b_v - \iota_v + \sigma_v \, \E \right) - \gamma \left( b_v + \sigma_v\,\E \right)
}
{
\din(v)
}
\right]
,
\,
\ex
\left[
\frac{b_v}{\din(v)}
\right]
\right\}
\\
\displaybreak[0]
&
=
%%%%%%%%%%%%%%%%%%%%%%%%%%%%%%%%%%%%%%%%%%%%%%%%%%%%%%%%%%%%%%%%%%%%%%%%%%%%%%%%%%
\min 
\left\{
\frac{\Phi}{\din(v)} \, \ex \left[ b_v - \iota_v \right]
\,+\,
\frac{\Phi\,\sigma_v\,\E}{\din(v)}
\,-\,
\frac{\gamma}{\din(v)} \, \ex \left[ b_v \right]
\,-\,
\frac{\gamma\,\sigma_v\,\E}{\din(v)}
,
\,\,
\frac{1}{\din(v)}
\,
\ex
\left[
b_v
\right]
\right\}
\\
%%%%%%%%%%%%%%%%%%%%%%%%%%%%%%%%%%%%%%%%%%%%%%%%%%%%%%%%%%%%%%%%%%%%%%%%%%%%%%%%%%
\displaybreak[0]
&
\textstyle
=
\,
\min 
\left\{
\frac{\Phi}{\din(v)} \, 
%%%%%%%%%%
\big(\,\din(v) - \dout(v) \, \big)
\,
\ex
\left[
\left(
\,
\beta\,\frac {\I} {|\widetilde{E}|}
+
\big(1-\alpha\big) \, 
\frac {\big(1-\beta\big)\,\I} {|E|-\alpha\,|\widetilde{E}|}
\,
\right)
\right]
%%%%%%%%%%
\,+\,
\frac{\Phi\,\sigma_v\,\E}{\din(v)}
\right.
\\
\displaybreak[0]
&
\hspace*{0.7in}
\textstyle
\,-\,
\frac{\gamma}{\din(v)} \, 
%%%%%%%
\din(v) \, 
\ex
\left[
%%%
\left(
\,
\beta \,\frac {\I} {|\widetilde{E}|}
+
\big(1-\alpha\big) \, 
\frac {\big(1-\beta\big)\,\I} {|E|-\alpha\,|\widetilde{E}|}
\,
\right)
\right]
%%%%%%%
\,-\,
\frac{\gamma\,\sigma_v\,\E}{\din(v)}
,
\\
\displaybreak[0]
&
\hspace*{1.2in}
\textstyle
\left.
\ex
\left[
%%%
\left(
\,
\beta\,\frac {\I} {|\widetilde{E}|}
+
\big(1-\alpha\big) \, 
\frac {\big(1-\beta\big)\,\I} {|E|-\alpha\,|\widetilde{E}|}
\,
\right)
\right]
%%%%%%%%%%%%%%%%%%
\,
\right\}
\\
%%%%%%%%%%%%%%%%%%
\displaybreak[0]
&
\textstyle
=
\min 
\left\{
\frac{  \Phi \, \big( \din(v) - \dout(v)  \big) - \gamma\,\din(v)}{\din(v)}
\,
\ex
\left[
\left(
\,
\beta\,\frac {\I} {|\widetilde{E}|}
+
\big(1-\alpha\big) \, 
\frac {\big(1-\beta\big)\,\I} {|E|-\alpha\,|\widetilde{E}|}
\,
\right)
\right]
\,+\,
\frac{\big(\Phi - \gamma \big)\,\sigma_v\,\E}{\din(v)}
,
\right.
\\
\displaybreak[0]
&
\hspace*{0.7in}
\textstyle
\left.
\ex
\left[
%%%
\left(
\,
\beta\,\frac {\I} {|\widetilde{E}|}
+
\big(1-\alpha\big) \, 
\frac {\big(1-\beta\big)\,\I} {|E|-\alpha\,|\widetilde{E}|}
\,
\right)
\right]
\,
\right\}
\\
%%%%%%%%%%%%%%%%%%
\displaybreak[0]
&
\textstyle
\geq
\min 
\left\{
\frac{ \Phi \, \big( \din(v) - \dout(v)  \big) - \gamma\,\din(v) }{\din(v)}
\,
\beta
\,
\ex
\left[
\frac {\I} {|\widetilde{E}|}
\right]
\,+\,
\frac{\big(\Phi - \gamma\big)\,\sigma_v\,\E}{\din(v)}
,
\,\,
\beta
\,
\ex
\left[
%%%
\frac {\I} {|\widetilde{E}|}
\right]
\right\}
\\
%%%%%%%%%%%%%%%%%%
\displaybreak[0]
&
\textstyle
\geq
\min 
\left\{
\,
\left(
\frac{  \Phi \, \big( \din(v) - \dout(v)  \big) -          \gamma\,\din(v)  }{\din(v)}
\right)
\,
\left(
\frac{\beta}{\alpha}
\right)
\,+\,
\left(
\frac{\beta}{\alpha}
\right)
\,
\left(
\frac{\big(\Phi - \gamma \big)\,\E}{\din(v)}
\right)
,
\,
\frac{\beta}{\alpha}
\,
\right\}
\\
\displaybreak[0]
&
\hspace*{0.7in}
\textstyle
\text{\small since } \ex \left[ \frac {\I} {|\widetilde{E}|} \right] = \ex \left[ \frac {|E|} {|\widetilde{E}|} \right] = \alpha
\\
\displaybreak[0]
&
\geq
\frac{\beta}{\alpha}
\,
\min 
\left\{
\,
\big(\Phi - \gamma \big)
+
\frac
{
\big(\Phi - \gamma \big)\,\E
-
\Phi\,\dout(v)
}
{
\din(v)
}
,
\,\,
1
\,
\right\}
\\
\displaybreak[0]
&
\geq 
\frac{\beta}{\alpha}
\,
\ex \left[ \Delta_{\,\mathrm{homo}}(u) \right] 
\end{align*}
\end{itemize}

\section{Proof of Lemma~\ref{bb1}}
\label{bb1proof-sec}

Using standard probabilistic calculations, we get
\[
\ex\left[\frac{1}{\din(v)} \, \Big| \, \din(v)>0 \right]
=
\sum_{k=1}^{n-1} \frac{1}{k} \, \Pr\left[ \din(v)=k\right] 
=
C\,\sum_{k=1}^{n-1} k^{-4}
\approx 
C\,\zeta(4)
\approx
\frac{\pi^2}{15}\,\dave
\]
\begin{multline*}
\Var\left[\din(v)\right] =
\ex\left[\,\left(\din(v)\right)^2\right] - \left(\ex\left[\din(v)\right]\right)^2
=
\sum_{k=1}^{n-1} k^2 \left( C k^{-3} \right) - \left( \dave \right)^2
\\
= C \sum_{k=1}^{n-1} \frac{1}{k} - \left( \dave \right)^2
\approx 
\frac{6\,\dave}{\pi^2} \,\ln n - \left( \dave \right)^2
\approx 
\frac{6\,\dave}{\pi^2} \,\ln n 
\end{multline*}

\section{Proof of Lemma~\ref{c1het-lemma}}
\label{c1het-lemmaproof-sec}

Let $\cD=
\sum\limits_{v \in \widetilde{V}} \din(v)
+
\sum\limits_{v \in \widetilde{V}} \dout(v)
$.
By linearity of expectation, we have 
\[
\ex \left[
\cD
\right]
=
2\,
\sum_{v \in \widetilde{V}} \ex \left[\din(v)\right]
=
2\,\alpha\,n\,\dave
\]
and similarly, since $\din(v)$ is independent of any other $\din(u)$ for $u\neq v$, we have 
\[
\Var \left[
\cD
\right]
=
2\,
\sum_{v \in \widetilde{V}} \Var \left[\din(v)\right]
\approx 
\frac{12\,\dave}{\pi^2} \,\alpha\,n\,\ln n 
\]
Thus, via Chebyschev's inequality~\cite[page 37]{AS90}, for any positive $\lambda$ we have 
\[
\textstyle
\Pr \left[
\,
\left|\,
\cD
\,
-
\,
\ex \left[
\cD
\right]
\,\right|
\geq 
\lambda\,
\sqrt{
\Var \left[
\cD
\right]
}
\,
\right]
\leq
\frac{1}{\lambda^2}
%%%%%%%%%%%%%%%%%%%%%%%%%%%%%%%%%%%%
%%%%%%%%%%%%%%%%%%%%%%%%%%%%%%%%%%%%
\,\equiv \,
\Pr \left[
\,
\left|\,
\cD
\,
-
\,
2\,\alpha\,n\,\dave
\,\right|
\gtrapprox
\lambda\,
\sqrt{
\frac{12\,\dave}{\pi^2} \,\alpha\,n\,\ln n 
}
\,
\right]
\leq
\frac{1}{\lambda^2}
\]
Setting $\lambda=
\sqrt{
\frac{\pi^2\,\ln n}{12\,\alpha}
}
$ 
gives 
$
\Pr \left[
\,
\left|\,
\cD
\,
-
\,
2\,\alpha\,n\,\dave
\,\right|
\gtrapprox
\sqrt{n\, \dave}\,\ln n
\,
\right]
\leq
\dfrac {12\,\alpha} {\pi^2\,\ln n}
$
and thus {\tt w.h.p.}  
$\,\,
\cD \approx 
2\,\alpha\,n\,\dave
$.
Since 
$
\frac{\cD}{2}
\leq
|\,\widetilde{E}\,|
\leq 
\cD
$, 
it now follows that
\begin{eqnarray*}
\alpha\,n\,\dave
\leq
&
\ex\left[ | \, \widetilde{E} \, | \right]
&
\leq 
2\,\alpha\,n\,\dave
\\
\mbox{\tt w.h.p.}
\,\,\, \,\,\,
\alpha\,n\,\dave
\leq
&
|\,\widetilde{E}\,|
&
\leq 
2\,\alpha\,n\,\dave
\end{eqnarray*}
Also, note that
$
\Pr\left[ (u,v)\in \widetilde{E} \right]
=
\Pr\left[v\in\widetilde{V}\right]=\alpha
$.
For notational convenience, let $\widetilde{E}_1$ be a random subset of $\alpha|\widetilde{E}|$ of edges from the edges in $\widetilde{E}$
as used in Definition~\ref{defnet}.
This implies that 
\begin{align*}
&
\begin{aligned}
\ex\left[c_1^{\mathrm{hetero}}\right]  
&
=
\Pr\left[ (u,v)\in \widetilde{E}_1 \right] \,\left( \frac{\beta\,\I}{\alpha\,|\,\widetilde{E}\,|} \right)
+ 
\left( 1 - \Pr\left[ (u,v)\in \widetilde{E}_1 \right]\, \right) \, \frac{\I - \beta \,\I}{|E| - \alpha \, |\,\widetilde{E}\,|}
\\
&
=
\alpha \, \Pr\left[ (u,v)\in \widetilde{E} \right] \,\left( \frac{\beta\,\I}{\alpha\,|\,\widetilde{E}\,|} \right)
+ 
\left( 1 - \alpha\,\Pr\left[ (u,v)\in \widetilde{E} \right]\, \right) \, \frac{\I - \beta \,\I}{|E| - \alpha \, |\,\widetilde{E}\,|}
\\
&
=
\alpha^2 \, \left( \frac{\beta\,\I}{\alpha\,|\,\widetilde{E}\,|} \right)
+
\left( 1 - \alpha^2 \,\right) \, \frac{\I - \beta \,\I}{|E| - \alpha \, |\,\widetilde{E}\,|}
\end{aligned}
\\
\displaybreak[0]
\Rightarrow
\,\,
&
\mbox{\tt w.h.p. } 
\,
\alpha^2 \, \frac{\beta\,n\,\dave}{2\, \alpha\,n\,\dave } + \left (1-\alpha^2 \right)\,\frac{(1-\beta)\,n\,\dave}{n\,\dave - \alpha\,n\,\dave}
\\
\displaybreak[0]
&
\hspace*{1.7in}
\leq 
\ex\left[c_1^{\mathrm{hetero}}\right]  
\leq 
\alpha^2 \, \frac{\beta\,n\,\dave}{\alpha\,n\,\dave } + \left(1-\alpha^2\right)\,\frac{(1-\beta)\,n\,\dave}{n\,\dave - 2\,\alpha\,n\,\dave}
\\
\displaybreak[0]
\equiv \,\, 
&
\mbox{\tt w.h.p. } 
\frac{\alpha\,\beta}{2} +
\left( 1 + \alpha \right) \, ( 1 - \beta ) 
\leq 
\ex\left[c_1^{\mathrm{hetero}}\right]  
\leq 
\alpha\,\beta +
\frac{\left( 1 - \alpha^2 \right) \, ( 1 - \beta ) } { 1 - 2\alpha }
\\
\displaybreak[0]
\equiv \,\, 
&
\mbox{\tt w.h.p. } 
1 + \alpha - \beta - \frac{\alpha\,\beta}{2}
\leq 
\ex\left[c_1^{\mathrm{hetero}}\right]  
\leq 
\frac{1 + \alpha\,\beta - \alpha^2 - \beta - \alpha^2\,\beta}{1-2\,\alpha}
\end{align*}
%
%%%%%%%%%%%%%%%%%%%%%%%%%%%%%%%%

\section{Proof of Lemma~\ref{bb3}}
\label{bb3proof-sec}

$\ex\left[\frac{1}{\din(v)}\right] = \sum\limits_{k=1}^{n-1} \frac{1}{k} \, \Pr\left[ \din(v)=k\right]
= \sum\limits_{k=1}^{n-1 } \frac{1}{k} \, \e^{-\dave} \, \frac{(\dave)^k}{k!}$. 
It is easy to see that extending the finite series to an infinite series does not change the asymptotic value of the series since 
\[
\begin{array}{r l l}
&
\sum\limits_{k=1}^{\infty} \dfrac{1}{k} \, \e^{-\dave} \, \dfrac{(\dave)^k}{k!}
\,-\,
\sum\limits_{k=1}^{n-1} \dfrac{1}{k} \, \e^{-\dave} \, \dfrac{(\dave)^k}{k!}
&
\\
=
&
\sum\limits_{k=n}^{\infty} \frac{1}{k} \, \e^{-\dave} \, \dfrac{(\dave)^k}{k!}
&
\\
\leq 
&
\frac{\e^{-\dave}}{n} \, \sum_{k=n}^{\infty} \dfrac{(\dave)^k}{k!}
&
\\
[0.1in]
\leq
&
\left( \dfrac{\e^{-\dave}}{n} \right) \, \left( \dfrac{(\dave)^n}{n!} \right) \, \left( \max\limits_{0\leq x\leq \dave} \left\{ \e^{x} \right\} \right)
&
\mbox{\begin{tabular}{l} using the Lagrange remainder term \\ for the Maclaurin series expansion of $\e^{\dave}$ \end{tabular}}
\\
[0.1in]
=
&
\left( \dfrac{\e^{-\dave}}{n} \right) \, \left( \dfrac{(\dave)^n}{n!} \right) \, \e^{\dave}
&
\\
[0.1in]
=
&
\dfrac{(\dave)^n}{n \, (n!) }
&
\end{array}
\]
and 
$\lim\limits_{n\to\infty} \frac{(\dave)^n}{n \, (n!) }$ because $\dave$ is a constant independent of $n$.
Thus, we can conclude that 
$\sum\limits_{k=1}^{\infty} \frac{1}{k} \, \e^{-\dave} \, \frac{(\dave)^k}{k!} \approx \sum\limits_{k=1}^{n-1} \frac{1}{k} \, \e^{-\dave} \, \frac{(\dave)^k}{k!}$.
It now follows that
\begin{multline*}
\dfrac{\partial }{\partial\, d}
\ex\left[\frac{1}{\din(v)} \, \big| \, \dave=d\right]
\approx
\dfrac{\partial }{\partial\, d}
\sum\limits_{k=1}^{\infty} \frac{1}{k} \, \e^{-d} \, \frac{d^k}{k!}
\\
=
\e^{-d} \, \sum\limits_{k=1}^{\infty} \frac{d^{k-1}}{k!}
=
\frac{\e^{-d}}{d} \, \sum\limits_{k=1}^{\infty} \frac{d^k}{k!}
=
\frac{\e^{-d}}{d} \, \left( \e^d - 1 \right)
=
\frac{1-\e^{-d}}{d}
\end{multline*}
This proves one of the claims in the lemma. To prove the other claim, 
using a well-known approximation on the first inverse moment of Poisson's distribution~\cite[page 173]{jones} we have 
\[
\left| \, 
\sum_{k=1}^{\infty} \, \frac{1}{k} \, \e^{-\dave} \, \frac{(\dave)^k}{k!}
\,-\,
\sum_{k=1}^{\lfloor 3\,\dave + 10 \rfloor } \, \frac{1}{k} \, \e^{-\dave} \, \frac{(\dave)^k}{k!}
\, \right| 
< 10^{-10}
\]
and therefore we obtain 
\[
\left| \, 
\sum_{k=1}^{n-1} \frac{1}{k} \, \e^{-\dave} \, \frac{(\dave)^k}{k!}
\,-\,
\sum_{k=1}^{\lfloor 3\,\dave + 10 \rfloor } \frac{1}{k} \, \e^{-\dave} \, \frac{(\dave)^k}{k!}
\, \right| 
\lessapprox 10^{-10}
\]
%%%%%%%%%%%%%%%%%%%%%%%%%%%%%%%%%%%%%%%%%%%%%%%%%%%%%%%%%
%%%%%%%%%%%%%%%%%%%%%%%%%%%%%%%%%%%%%%%%%%%%%%%%%%%%%%%%%
%%%%%%%%%%%%%%%%%%%%%%%%%%%%%%%%%%%%%%%%%%%%%%%%%%%%%%%%%

\section{Proof of Lemma~\ref{bb4}}
\label{bb4proof-sec}

Let $\cD=
\sum_{v \in \widetilde{V}} \din(v)
+
\sum_{v \in \widetilde{V}} \dout(v)
$.
We can reuse the proof of Lemma~\ref{c1het-lemma} provided we show that 
$\ex \left[\cD \right]=2\,\alpha\,n\,\dave$ and 
{\tt w.h.p.}  $\,\,\cD \approx 2\,\alpha\,n\,\dave$.
By linearity of expectation, we have 
$
\ex \left[
\cD
\right]
=
2\,
\sum_{v \in \widetilde{V}} \ex \left[\din(v)\right]
=
2\,\alpha\,n\,\dave
$, 
and similarly
$
\Var \left[
\cD
\right]
\allowbreak
=
2\,
\sum_{v \in \widetilde{V}} \Var \left[\din(v)\right]
=
2\,\alpha\,n\,
\Big(1-\frac{\dave}{n-1} \Big)\,\dave
\approx 
2\,\alpha\,n\,\dave$.
Thus, via Chebyschev's inequality, for any positive $\lambda$ we have 
\[
\Pr \left[
\,
\left|\,
\cD
\,
-
\,
\ex \left[
\cD
\right]
\,\right|
\geq 
\lambda\,
\sqrt{
\Var \left[
\cD
\right]
}
\,
\right]
\leq
\frac{1}{\lambda^2}
%%%%%%%%%%%%%%%%%%%%%%%%%%%%%%%%%%%%
%%%%%%%%%%%%%%%%%%%%%%%%%%%%%%%%%%%%
\,\equiv \,
\Pr \left[
\,
\left|\,
\cD
\,
-
\,
2\,\alpha\,n\,\dave
\,\right|
\gtrapprox
\lambda\,
\sqrt{
2\,\alpha\,n\,\dave
}
\,
\right]
\leq
\frac{1}{\lambda^2}
\]
Setting $\lambda=
\sqrt{
\frac{\ln n}{2\,\alpha}
}
$ 
gives 
\[
\Pr \left[
\,
\left|\,
\cD
\,
-
\,
2\,\alpha\,n\,\dave
\,\right|
\gtrapprox
\sqrt{n\, \dave\,\ln n}
\,
\right]
\leq
\dfrac {2\,\alpha} {\ln n}
\]
and thus {\tt w.h.p.}  
$\,\,
\cD \approx 
2\,\alpha\,n\,\dave
$.
%%%%%%%%%%%%%%%%%%%%%%%%%%%%%%%%

\section{Proof of Lemma~\ref{g9}}
\label{g9proof-sec}

The amount of shock $\mu$ transmitted from $v$ to $u$ is given by 
\[
\mu=
\min 
\left\{
\left(\Phi-\gamma\right) \, \left( 1 + \frac{\E}{n\,\din(v)} \right)
+
\Phi \, \frac{\dout(v)}{\din(v)}
,\, 1
\right\}
\]
Since $G$ is an in-arborescence, $\dout(v)\leq 1$. First, consider the case of $\dout(v)=0$. In this case,
$
\mu
=
\min 
\left\{
\left(\Phi-\gamma\right) \, \left( 1 + \frac{\E}{n\,\din(v)} \right)
,\, 1
\right\}
$
and thus we have 
\[
c_u(1)=c_u(0) - \mu
=
\gamma \,
\frac{\E}{n}
\,-\,
\min 
\left\{
\left(\Phi-\gamma\right) \, \left( 1 + \frac{\E}{n\,\din(v)} \right)
,\, 1
\right\}
\]
Assuming $\gamma\approx \nicefrac{\Phi}{2}$, we have 
\begin{multline*}
c_u(1)
\approx
\gamma \, 
\frac{\E}{n}+\din(u)
\,-\,
\min 
\left\{
\gamma \, \left( 1 + \frac{\E}{n\,\din(v)} \right)
,\, 1
\right\}
\\
=
\min 
\left\{
\gamma \, \left(
\frac{\E}{n}
- 1 - \frac{\E}{n\,\din(v)}
\right)
,\,\,
\gamma \, 
\frac{\E}{n}
-1
\right\}
\end{multline*}
There are two cases to consider:
\begin{itemize}
\item
If
$
\gamma \, \left(
\frac{\E}{n}
- 1 - \frac{\E}{n\,\din(v)}
\right)
\geq 
\gamma \, 
\frac{\E}{n}
-1
$
then 
\[
c_u(1)
\approx
\gamma \, \left(
\dfrac{\E}{n}
- 1 - \dfrac{\E}{n\,\din(v)}
\right)
=
\gamma \, \left(
\frac{\E}{n} \,
\left(
1 - \frac{1}{\din(v)}
\right)
-1
\right)
\]
Thus, if  $\E> \E_{\tau_1}(u) = \frac{n}{ 1 - \frac{1}{\din(v)} }$ then $c_u(1)$ would be strictly positive, the node 
$u$ will not become insolvent at time $t=1$, but if 
$\E< \E_{\tau_1}(u)$ then $c_u(1)$ would be strictly negative and $u$ would fail.

\item
Otherwise,
$c_u(1) \approx \gamma \, \frac{\E}{n} -1$.
Thus, if  $\E > \E_{\tau_2}(u) = \frac{n}{ \E }$ then $c_u(1)$ would be strictly positive, the node $u$ will not become insolvent
at time $t=1$, but if 
$\E< \E_{\tau_1}(u)$ then $c_u(1)$ would be strictly negative and $u$ would fail.
\end{itemize}
A similar analysis may be carried out if $\dout(v)=1$ leading to {\em slightly} two different threshold values, say $\E_{\tau_3}(u)$ and $\E_{\tau_4}(u)$. 
Since $\dout(u)=1$, if $u$ does not become insolvent at time $t=1$ then it does not become insolvent for any $t>1$ as well.

\section{Proof of Lemma~\ref{g10}}
\label{g10proof-sec}

Let $r$ be the root node of $G$. Note that, for any node $u\in V\setminus \{r\}$, $\dout(u)=1$. Thus, using the results in~\cite{BaAl99}, it follows that for any 
node $u\in V\setminus \{r\}$, $\Pr\left[\din(u)=k-1\right]\propto \nicefrac{1}{k^{3}}$ and in particular 
$\Pr\left[\din(u)=1\right]\leq \nicefrac{1}{4}$.
For $j=0,2,\dots,n$, let $n_j$ be the number of nodes $u$ of $G$ with $\din(u)=j$. 
Thus, $n=\sum\limits_{j=0}^n n_j$, $|E|=n-1=\sum\limits_{j=1}^n j\,n_j$, and 
\begin{multline*}
\sum_{u\in V\setminus \{r\}} \Pr\left[\din(u)=1\right]
\leq
\ex\left[n_1\right]
\leq 
1+ \sum_{u\in V\setminus \{r\}} \Pr\left[\din(u)=1\right]
\\
\displaybreak[0]
\equiv\,
\frac{n-1}{4}
\leq
\ex\left[n_1\right]
\leq 
1+\frac{n-1}{4}
\,\equiv\,
\ex\left[n_1\right]
=
\frac{n-1}{4}+t \,\,\,\,\,\, \text{ for some } t\in[0,1]
\end{multline*}
Letting $n_{>1}=\sum\limits_{j=2}^n n_j$, we have 
\begin{align*}
&
\ex\left[n_0+n_{>1}\right]  = n - \ex\left[n_1\right] = \frac{3n+1}{4} - t 
\\
\displaybreak[0]
&
\\
\displaybreak[0]
&
%%%%%%%%%%%%%%%%%%%%%%%%%%%%%%%%%%%%%%%%%%%%%%%%%%%%%%%%%%%%%%%%%%%%
\ex\left[\sum_{j=1}^n j\,n_j\right]  =  n-1 
\,\equiv\,\, 
\ex\left[n_1\right]  + \ex\left[\sum_{j=2}^n j\,n_j\right]  =  n-1 
\\
\displaybreak[0]
&
\hspace*{0.5in}
\equiv\,\,
\ex\left[n_1\right]  + 2\,\ex\left[n_{>1}\right] \leq  n-1 
\,\Rightarrow\,\,
\ex\left[n_{>1}\right] \leq  \frac{n-1 - \frac{3n+1}{4} + t}{2} 
= \frac{n}{8} +\frac{4t-5}{8}
\\
\displaybreak[0]
&
\\
\displaybreak[0]
&
%%%%%%%%%%%%%%%%%%%%%%%%%%%%%%%%%%%%%%%%%%%%%%%%%%%%%%%%%%%%
\ex\left[n_0\right]=n-\ex\left[n_1\right]-\ex\left[n_{>1}\right]
\geq
n- \left( \frac{n-1}{4}+t \right) - \left( \frac{n}{8} +\frac{4t-5}{8} \right)
=
\frac{n}{8} + \frac{7}{8} - \frac{3\,t}{2}
\end{align*}
and hence we can bound $\ex\left[\big|\widehat{V}\big|\right]$ as 
\begin{multline*}
\ex\left[\big|\widehat{V}\big|\right]
= \ex \left[ n_0 \right] - 
\ex \left[ \, \left| \, \left\{\,u\in V \, \big| \, \big( \din(u)=0 \big) \, \wedge \, \big( \exists\, v \colon \big( \big( \din(v)=1 \big) \, \wedge \, \big( (u,v)\in E \big) \, \big) \, \big) \, \right\} \, \right| \, \right]
\\
\displaybreak[0]
\geq 
\ex \left[ n_0 \right] - \ex \left[ n_1 \right] 
\geq
\left( \frac{n}{8} + \frac{7}{8} - \frac{3\,t}{2} \right)
-
\left( \frac{n-1}{4}+t \right)
=
\frac{n}{8} + \frac{9}{8} - \frac{5\,t}{2}
\geq 
\frac{n}{8} - \frac{11}{8} 
\end{multline*}
%

%%%%%%%%%%%%%%%%%%%%%%%%%%%%%%%%%%%%%%%%%%%%%%%%%%%%%%%%%%%%%%%%%%%%%%%%%%%%%%%%%%%%%%%%%%%%%%%%%%%%%
%%%%%%%%%%%%%%%%%%%%%%%%%%%%%%%%%%%%%%%%%%%%%%%%%%%%%%%%%%%%%%%%%%%%%%%%%%%%%%%%%%%%%%%%%%%%%%%%%%%%%
%%%% Supplemental information follows
%%%%%%%%%%%%%%%%%%%%%%%%%%%%%%%%%%%%%%%%%%%%%%%%%%%%%%%%%%%%%%%%%%%%%%%%%%%%%%%%%%%%%%%%%%%%%%%%%%%%%
%%%%%%%%%%%%%%%%%%%%%%%%%%%%%%%%%%%%%%%%%%%%%%%%%%%%%%%%%%%%%%%%%%%%%%%%%%%%%%%%%%%%%%%%%%%%%%%%%%%%%
\clearpage

\newpage

\renewcommand{\thetable}{S\arabic{table}}
\setcounter{table}{0}

\begin{table}
\hspace*{-0.7in}
\begin{tabular}{c}
\toprule
\\
\\
\\
\bf {\Huge Supplementary documents follow}
\\
\\
\\
\bf {\Huge Supplementary Table~\ref{table7}---Table~\ref{supp4-table7}}
\\
\\
\\
\bf {\Huge Supplementary color figures {\sc Fig}.~\ref{newfig1}---{\sc Fig}.~\ref{supp-newfig5-full}}
\\
\\
\\
\bottomrule
\end{tabular}
\end{table}

\clearpage
\newpage

%%%%%%%%%%%%%%%%%%%%%%%%%%%%%%%%%%%%%%%%%%%%%%%%%%%%%%%%%%%%%%%%%%%%%%%%%%%%%%%%%%%%%%%%%%%%%%%%%%%%%%%%%%%%%%%%%
%%%%%%%%%%%%%%%%%%%%%%%%%%%%%%%%%%%%%%%%%%%%%%%%%%%%%%%%%%%%%%%%%%%%%%%%%%%%%%%%%%%%%%%%%%%%%%%%%%%%%%%%%%%%%%%%%
%%%%%%%%%%%%%%%%%%%%%%%%%%%%%%%%%%%%%%%%%%%%%%%%%%%%%%%%%%%%%%%%%%%%%%%%%%%%%%%%%%%%%%%%%%%%%%%%%%%%%%%%%%%%%%%%%
%%%%%%%%%%%%%%%%%%%%%%%%%%%%%%%%%%%%%%%%%%%%%%%%%%%%%%%%%%%%%%%%%%%%%%%%%%%%%%%%%%%%%%%%%%%%%%%%%%%%%%%%%%%%%%%%%

%%%%%%%%%%%%%%%%%%%%%%%%%%%%%%%%%%%%%%%%%%%%%%%%%%%%%%%%%%%%%%%%%%%%%%%%%%%%%%%%%%%%%%%%%%%%%%%%%%%%%%%%%%%%%%
%%%%%%%%%%%%%%%%%%%%%%%%%%%%%%%%%%%%%%%%%%%%%%%%%%%%%%%%%%%%%%%%%%%%%%%%%%%%%%%%%%%%%%%%%%%%%%%%%%%%%%%%%%%%%%
%%%%%%%%%%%%%%%%%%%%%%%%%%%%%%%%%%%%%%%%%%%%%%%%%%%%%%%%%%%%%%%%%%%%%%%%%%%%%%%%%%%%%%%%%%%%%%%%%%%%%%%%%%%%%%
%%% Table for analysis of residual instability of homogeneous versus heterogeneous networks
%%% coordinated shock
%%%%%%%%%%%%%%%%%%%%%%%%%%%%%%%%%%%%%%%%%%%%%%%%%%%%%%%%%%%%%%%%%%%%%%%%%%%%%%%%%%%%%%%%%%%%%%%%%%%%%%%%%%%%%%
%%%%%%%%%%%%%%%%%%%%%%%%%%%%%%%%%%%%%%%%%%%%%%%%%%%%%%%%%%%%%%%%%%%%%%%%%%%%%%%%%%%%%%%%%%%%%%%%%%%%%%%%%%%%%%
%%%%%%%%%%%%%%%%%%%%%%%%%%%%%%%%%%%%%%%%%%%%%%%%%%%%%%%%%%%%%%%%%%%%%%%%%%%%%%%%%%%%%%%%%%%%%%%%%%%%%%%%%%%%%%
%%% Another table for analysis of residual instability of homogeneous versus heterogeneous networks

\begin{table}
\caption{Residual instabilities of homogeneous versus heterogeneous networks under idiosyncratic shocks. 
The percentages shown are the percentages of networks for which $\xi<0.05$ or $\xi<0.1$ or $\xi<0.2$.}
\label{table7}
\begin{center}
\scalebox{0.75}[0.8]
{%%%\small
\begin{tabular}{c c r      r r r | r r r}
&
&
&
\multicolumn{6}{c}{\bf idiosyncratic shock}
\\
&
&
&
\multicolumn{3}{c|}{$\Phi=0.5,\gamma=0.45$}
&
\multicolumn{3}{c}{$\Phi=0.5,\gamma=0.40$}
\\
&
&
&
\multicolumn{1}{c}{${\xi<0.05}$}
&
\multicolumn{1}{c}{${\xi<0.1}$}
&
\multicolumn{1}{c|}{${\xi<0.2}$}
&
\multicolumn{1}{c}{${\xi<0.05}$}
&
\multicolumn{1}{c}{${\xi<0.1}$}
&
\multicolumn{1}{c}{${\xi<0.2}$}
\\
\cline{1-9}
\multirow{15}{*}{${|V|=50}$}
&
\multirow{5}{*}{homogeneous} 
&
in-arborescence
&  
\bf 76\%  & \bf 78\%  &  \bf 82\% & \bf 26\% \bf & \bf 58\%  & \bf 75\%
\\
&
&
{\sf ER}, average degree $3$ 
&  
\bf 99\%  & \bf 100\%  &  \bf 100\% & \bf 43\% \bf & \bf 84\%  & \bf 100\%
\\
&
&
{\sf ER}, average degree $6$ 
&  
\bf 100\%  & \bf 100\%  &  \bf 100\% & \bf 100\% \bf & \bf 100\%  & \bf 100\%
\\
&
&
{\sf SF}, average degree $3$ 
&  
\bf 42\%  & \bf 100\%  &  \bf 100\% & \bf 23\% \bf & \bf 70\%  & \bf 88\%
\\
&
&
{\sf SF}, average degree $6$ 
&  
\bf 100\%  & \bf 100\%  &  \bf 100\% & \bf 100\% \bf & \bf 100\%  & \bf 100\%
\\
\cline{2-9}
& 
\multirow{5}{*}{${(0.1,0.95)}$-heterogeneous}
&
in-arborescence
&  
1\%  & 2\%  &  15\% & 0\% \bf & 1\%  & 10\%
\\
&
&
{\sf ER}, average degree $3$ 
&  
0\%  & 2\%  &  16\% & 0\% \bf & 1\%  & 8\%
\\
&
&
{\sf ER}, average degree $6$ 
&  
7\%  & 10\%  &  21\% & 2\% \bf & 7\%  & 14\%
\\
&
&
{\sf SF}, average degree $3$ 
&  
0\%  & 6\%  &  22\% & 0\% \bf & 2\%  & 14\%
\\
&
&
{\sf SF}, average degree $6$ 
&  
8\%  & 19\%  &  34\% & 4\% \bf & 12\%  & 21\%
\\
\cline{2-9}
& 
\multirow{5}{*}{${(0.2,0.6)}$-heterogeneous}
&
in-arborescence
&  
0\%  & 2\%  &  12\% & 0\% \bf & 1\%  & 11\%
\\
&
&
{\sf ER}, average degree $3$ 
&  
7\%  & 14\%  &  22\% & 4\% \bf & 9\%  & 18\%
\\
&
&
{\sf ER}, average degree $6$ 
&  
8\%  & 18\%  &  30\% & 6\% \bf & 10\%  & 20\%
\\
&
&
{\sf SF}, average degree $3$ 
&  
0\%  & 9\%  &  19\% & 0\% \bf & 3\%  & 17\%
\\
&
&
{\sf SF}, average degree $6$ 
&  
8\%  & 12\%  &  24\% & 4\% \bf & 7\%  & 18\%
\\
%
%%%%%%%%%%%%%%%%%%%%%%%%%%%%%%%%%%%%%%%%%%%%%%%%%%%%%%%%%%%%%%%%%%%%%%%%%%%%%%%%%%%%%%%%%%%%%%%%%%%%%%%%%%
\hline
\multirow{15}{*}{${|V|=100}$}
&
\multirow{5}{*}{homogeneous} 
&
in-arborescence
&  
\bf 76\%  & \bf 78\%  &  \bf 81\% & \bf 32\% \bf & \bf 62\%  & \bf 81\%
\\
&
&
{\sf ER}, average degree $3$ 
&  
\bf 66\%  & \bf 100\%  &  \bf 100\% & \bf 26\% \bf & \bf 74\%  & \bf 100\%
\\
&
&
{\sf ER}, average degree $6$ 
&  
\bf 100\%  & \bf 100\%  &  \bf 100\% & \bf 100\% \bf & \bf 100\%  & \bf 100\%
\\
&
&
{\sf SF}, average degree $3$ 
&  
\bf 29\%  & \bf 72\%  &  \bf 100\% & \bf 19\% \bf & \bf 53\%  & \bf 88\%
\\
&
&
{\sf SF}, average degree $6$ 
&  
\bf 100\%  & \bf 100\%  &  \bf 100\% & \bf 88\% \bf & \bf 100\%  & \bf 100\%
\\
\cline{2-9}
& 
\multirow{5}{*}{${(0.1,0.95)}$-heterogeneous}
&
in-arborescence
&  
0\%  & 1\%  &  12\% & 0\% & 1\%  & 11\%
\\
&
&
{\sf ER}, average degree $3$ 
&  
0\%  & 0\%  &  15\% & 0\% & 1\%  & 10\%
\\
&
&
{\sf ER}, average degree $6$ 
&  
6\%  & 7\%  &  16\% & 6\% & 6\%  & 10\%
\\
&
&
{\sf SF}, average degree $3$ 
&  
0\%  & 6\%  &  23\% & 0\% & 0\%  & 16\%
\\
&
&
{\sf SF}, average degree $6$ 
&  
8\%  & 16\%  &  30\% & 6\% & 12\%  & 19\%
\\
\cline{2-9}
& 
\multirow{5}{*}{${(0.2,0.6)}$-heterogeneous}
&
in-arborescence
&  
0\%  & 1\%  &  11\% & 0\% & 1\%  & 11\%
\\
&
&
{\sf ER}, average degree $3$ 
&  
6\%  & 10\%  &  18\% & 0\% & 7\%  & 16\%
\\
&
&
{\sf ER}, average degree $6$ 
&  
7\%  & 12\%  &  22\% & 6\% & 8\%  & 17\%
\\
&
&
{\sf SF}, average degree $3$ 
&  
0\%  & 2\%  &  18\% & 0\% & 0\%  & 16\%
\\
&
&
{\sf SF}, average degree $6$ 
&  
5\%  & 9\%  &  18\% & 2\% & 6\%  & 16\%
\\
%
%%%%%%%%%%%%%%%%%%%%%%%%%%%%%%%%%%%%%%%%%%%%%%%%%%%%%%%%%%%%%%%%%%%%%%%%%%%%%%%%%%%%%%%%%%%%%%%%%%%%%%%%%%
\hline
\multirow{15}{*}{${|V|=300}$}
&
\multirow{5}{*}{homogeneous} 
&
in-arborescence
&  
\bf 76\%  & \bf 78\%  &  \bf 81\% & \bf 36\% \bf & \bf 67\%  & \bf 81\%
\\
&
&
{\sf ER}, average degree $3$ 
&  
\bf 76\%  & \bf 100\%  &  \bf 100\% & \bf 22\% \bf & \bf 73\%  & \bf 93\%
\\
&
&
{\sf ER}, average degree $6$ 
&  
\bf 100\%  & \bf 100\%  &  \bf 100\% & \bf 100\% \bf & \bf 100\%  & \bf 100\%
\\
&
&
{\sf SF}, average degree $3$ 
&  
\bf 26\%  & \bf 52\%  &  \bf 100\% & \bf 20\% \bf & \bf 42\%  & \bf 88\%
\\
&
&
{\sf SF}, average degree $6$ 
&  
\bf 100\%  & \bf 100\%  &  \bf 100\% & \bf 87\% \bf & \bf 100\%  & \bf 100\%
\\
\cline{2-9}
& 
\multirow{5}{*}{${(0.1,0.95)}$-heterogeneous}
&
in-arborescence
&  
0\%  & 1\%  &  12\% & 0\% & 1\%  & 12\%
\\
&
&
{\sf ER}, average degree $3$ 
&  
0\%  & 2\%  &  16\% & 0\% & 0\%  & 5\%
\\
&
&
{\sf ER}, average degree $6$ 
&  
0\%  & 6\%  &  12\% & 0\% & 2\%  & 6\%
\\
&
&
{\sf SF}, average degree $3$ 
&  
0\%  & 9\%  &  24\% & 0\% & 1\%  & 16\%
\\
&
&
{\sf SF}, average degree $6$ 
&  
7\%  & 16\%  &  28\% & 6\% & 9\%  & 19\%
\\
\cline{2-9}
& 
\multirow{5}{*}{${(0.2,0.6)}$-heterogeneous}
&
in-arborescence
&  
0\%  & 1\%  &  11\% & 0\% & 1\%  & 11\%
\\
&
&
{\sf ER}, average degree $3$ 
&  
6\%  & 8\%  &  16\% & 0\% & 2\%  & 16\%
\\
&
&
{\sf ER}, average degree $6$ 
&  
6\%  & 8\%  &  17\% & 6\% & 7\%  & 16\%
\\
&
&
{\sf SF}, average degree $3$ 
&  
0\%  & 1\%  &  16\% & 0\% & 0\%  & 16\%
\\
&
&
{\sf SF}, average degree $6$ 
&  
2\%  & 6\%  &  16\% & 2\% & 5\%  & 16\%
\\
\cline{1-9}
\end{tabular}
}
\end{center}
\end{table}

\clearpage

\begin{table}
\caption{Residual instabilities of homogeneous versus heterogeneous networks under coordinated shocks.
The percentages shown are the percentages of networks for which $\xi<0.05$ or $\xi<0.1$ or $\xi<0.2$.}
\label{supp1-table6}
\begin{center}
\scalebox{0.75}[0.8]
{%%\small
\begin{tabular}{c c r      r r r | r r r}
&
&
&
\multicolumn{6}{c}{\bf coordinated shock}
\\
&
&
&
\multicolumn{3}{c|}{$\Phi=0.6,\gamma=0.55$}
&
\multicolumn{3}{c}{$\Phi=0.6,\gamma=0.50$}
\\
&
&
&
\multicolumn{1}{c}{${\xi<0.05}$}
&
\multicolumn{1}{c}{${\xi<0.1}$}
&
\multicolumn{1}{c|}{${\xi<0.2}$}
&
\multicolumn{1}{c}{${\xi<0.05}$}
&
\multicolumn{1}{c}{${\xi<0.1}$}
&
\multicolumn{1}{c}{${\xi<0.2}$}
\\
\cline{1-9}
\multirow{15}{*}{${|V|=50}$}
&
\multirow{5}{*}{homogeneous} 
&
in-arborescence
&  
\bf 93\%  & \bf 93\%  &  \bf 93\% & \bf 2\% \bf & \bf 69\%  & \bf 93\%
\\
&
&
{\sf ER}, average degree $3$ 
&  
\bf 97\%  & \bf 100\%  &  \bf 100\% & \bf 64\% \bf & \bf 95\%  & \bf 100\%
\\
&
&
{\sf ER}, average degree $6$ 
&  
\bf 100\%  & \bf 100\%  &  \bf 100\% & \bf 100\% \bf & \bf 100\%  & \bf 100\%
\\
&
&
{\sf SF}, average degree $3$ 
&  
\bf 56\%  & \bf 96\%  &  \bf 100\% & \bf 44\% \bf & \bf 87\%  & \bf 100\%
\\
&
&
{\sf SF}, average degree $6$ 
&  
\bf 100\%  & \bf 100\%  &  \bf 100\% & \bf 100\% \bf & \bf 100\%  & \bf 100\%
\\
\cline{2-9}
& 
\multirow{5}{*}{${(0.1,0.95)}$-heterogeneous}
&
in-arborescence
&  
0\%  & 0\%  &  1\% & 0\% \bf & 0\%  & 0\%
\\
&
&
{\sf ER}, average degree $3$ 
&  
0\%  & 0\%  &  5\% & 0\% \bf & 0\%  & 1\%
\\
&
&
{\sf ER}, average degree $6$ 
&  
9\%  & 11\%  &  14\% & 5\% \bf & 7\%  & 7\%
\\
&
&
{\sf SF}, average degree $3$ 
&  
5\%  & 9\%  &  19\% & 0\% \bf & 3\%  & 8\%
\\
&
&
{\sf SF}, average degree $6$ 
&  
22\%  & 28\%  &  41\% & 10\% \bf & 15\%  & 17\%
\\
\cline{2-9}
& 
\multirow{5}{*}{${(0.2,0.6)}$-heterogeneous}
&
in-arborescence
&  
0\%  & 0\%  &  9\% & 0\% \bf & 0\%  & 9\%
\\
&
&
{\sf ER}, average degree $3$ 
&  
5\%  & 7\%  &  20\% & 3\% \bf & 7\%  & 16\%
\\
&
&
{\sf ER}, average degree $6$ 
&  
11\%  & 14\%  &  26\% & 7\% \bf & 7\%  & 19\%
\\
&
&
{\sf SF}, average degree $3$ 
&  
2\%  & 8\%  &  23\% & 1\% \bf & 3\%  & 18\%
\\
&
&
{\sf SF}, average degree $6$ 
&  
9\%  & 15\%  &  26\% & 7\% \bf & 8\%  & 18\%
\\
%
%%%%%%%%%%%%%%%%%%%%%%%%%%%%%%%%%%%%%%%%%%%%%%%%%%%%%%%%%%%%%%%%%%%%%%%%%%%%%%%%%%%%%%%%%%%%%%%%%%%%%%%%%%
\hline
\multirow{15}{*}{${|V|=100}$}
&
\multirow{5}{*}{homogeneous} 
&
in-arborescence
&  
\bf 93\%  & \bf 93\%  &  \bf 93\% & \bf 23\% \bf & \bf 84\%  & \bf 93\%
\\
&
&
{\sf ER}, average degree $3$ 
&  
\bf 85\%  & \bf 100\%  &  \bf 100\% & \bf 47\% \bf & \bf 86\%  & \bf 100\%
\\
&
&
{\sf ER}, average degree $6$ 
&  
\bf 100\%  & \bf 100\%  &  \bf 100\% & \bf 100\% \bf & \bf 100\%  & \bf 100\%
\\
&
&
{\sf SF}, average degree $3$ 
&  
\bf 37\%  & \bf 75\%  &  \bf 100\% & \bf 35\% \bf & \bf 71\%  & \bf 100\%
\\
&
&
{\sf SF}, average degree $6$ 
&  
\bf 100\%  & \bf 100\%  &  \bf 100\% & \bf 96\% \bf & \bf 100\%  & \bf 100\%
\\
\cline{2-9}
& 
\multirow{5}{*}{${(0.1,0.95)}$-heterogeneous}
&
in-arborescence
&  
0\%  & 0\%  &  0\% & 0\% & 0\%  & 0\%
\\
&
&
{\sf ER}, average degree $3$ 
&  
0\%  & 1\%  &  7\% & 0\% & 0\%  & 3\%
\\
&
&
{\sf ER}, average degree $6$ 
&  
7\%  & 7\%  &  9\% & 7\% & 7\%  & 7\%
\\
&
&
{\sf SF}, average degree $3$ 
&  
0\%  & 1\%  &  9\% & 0\% & 0\%  & 5\%
\\
&
&
{\sf SF}, average degree $6$ 
&  
10\%  & 15\%  &  21\% & 7\% & 7\%  & 11\%
\\
\cline{2-9}
& 
\multirow{5}{*}{${(0.2,0.6)}$-heterogeneous}
&
in-arborescence
&  
0\%  & 0\%  &  9\% & 0\% & 0\%  & 9\%
\\
&
&
{\sf ER}, average degree $3$ 
&  
0\%  & 7\%  &  17\% & 0\% & 5\%  & 16\%
\\
&
&
{\sf ER}, average degree $6$ 
&  
7\%  & 7\%  &  19\% & 7\% & 7\%  & 16\%
\\
&
&
{\sf SF}, average degree $3$ 
&  
1\%  & 3\%  &  15\% & 0\% & 1\%  & 13\%
\\
&
&
{\sf SF}, average degree $6$ 
&  
7\%  & 9\%  &  18\% & 7\% & 7\%  & 17\%
\\
%
%%%%%%%%%%%%%%%%%%%%%%%%%%%%%%%%%%%%%%%%%%%%%%%%%%%%%%%%%%%%%%%%%%%%%%%%%%%%%%%%%%%%%%%%%%%%%%%%%%%%%%%%%%
\hline
\multirow{15}{*}{${|V|=300}$}
&
\multirow{5}{*}{homogeneous} 
&
in-arborescence
&  
\bf 93\%  & \bf 93\%  &  \bf 93\% & \bf 79\% \bf & \bf 93\%  & \bf 93\%
\\
&
&
{\sf ER}, average degree $3$ 
&  
\bf 93\%  & \bf 100\%  &  \bf 100\% & \bf 48\% \bf & \bf 85\%  & \bf 100\%
\\
&
&
{\sf ER}, average degree $6$ 
&  
\bf 100\%  & \bf 100\%  &  \bf 100\% & \bf 100\% \bf & \bf 100\%  & \bf 100\%
\\
&
&
{\sf SF}, average degree $3$ 
&  
\bf 28\%  & \bf 56\%  &  \bf 97\% & \bf 28\% \bf & \bf 56\%  & \bf 97\%
\\
&
&
{\sf SF}, average degree $6$ 
&  
\bf 100\%  & \bf 100\%  &  \bf 100\% & \bf 93\% \bf & \bf 100\%  & \bf 100\%
\\
\cline{2-9}
& 
\multirow{5}{*}{${(0.1,0.95)}$-heterogeneous}
&
in-arborescence
&  
0\%  & 0\%  &  0\% & 0\% & 0\%  & 0\%
\\
&
&
{\sf ER}, average degree $3$ 
&  
0\%  & 1\%  &  7\% & 0\% & 0\%  & 3\%
\\
&
&
{\sf ER}, average degree $6$ 
&  
7\%  & 7\%  &  7\% & 3\% & 7\%  & 7\%
\\
&
&
{\sf SF}, average degree $3$ 
&  
0\%  & 0\%  &  11\% & 0\% & 0\%  & 3\%
\\
&
&
{\sf SF}, average degree $6$ 
&  
7\%  & 7\%  &  18\% & 7\% & 7\%  & 7\%
\\
\cline{2-9}
& 
\multirow{5}{*}{${(0.2,0.6)}$-heterogeneous}
&
in-arborescence
&  
0\%  & 0\%  &  9\% & 0\% & 0\%  & 9\%
\\
&
&
{\sf ER}, average degree $3$ 
&  
1\%  & 7\%  &  17\% & 1\% & 5\%  & 16\%
\\
&
&
{\sf ER}, average degree $6$ 
&  
7\%  & 7\%  &  16\% & 7\% & 7\%  & 16\%
\\
&
&
{\sf SF}, average degree $3$ 
&  
0\%  & 1\%  &  13\% & 0\% & 0\%  & 13\%
\\
&
&
{\sf SF}, average degree $6$ 
&  
7\%  & 7\%  &  16\% & 7\% & 7\%  & 16\%
\\
\cmidrule{1-9}
\end{tabular}
}
\end{center}
\end{table}

\clearpage

%%%%%%%%%%%%%%%%%%%%%%%%%%%%%%%%%%%%%%%%%%%%%%%%%%%%%%%%%%%%%%%%%%%%%%%%%%%%%%%%%%%%%%%%%%%%%%%%%%%%%%%%%%%%%%
%%%%%%%%%%%%%%%%%%%%%%%%%%%%%%%%%%%%%%%%%%%%%%%%%%%%%%%%%%%%%%%%%%%%%%%%%%%%%%%%%%%%%%%%%%%%%%%%%%%%%%%%%%%%%%
%%%%%%%%%%%%%%%%%%%%%%%%%%%%%%%%%%%%%%%%%%%%%%%%%%%%%%%%%%%%%%%%%%%%%%%%%%%%%%%%%%%%%%%%%%%%%%%%%%%%%%%%%%%%%%
%%% Second table for analysis of residual instability of homogeneous versus heterogeneous networks
%%% coordinated shock

\begin{table}
\caption{Residual instabilities of homogeneous versus heterogeneous networks under coordinated shocks.
The percentages shown are the percentages of networks for which $\xi<0.05$ or $\xi<0.1$ or $\xi<0.2$.} 
\label{supp2-table6}
\begin{center}
\scalebox{0.75}[0.8]
{%%\small
\begin{tabular}{c c r      r r r | r r r}
&
&
&
\multicolumn{6}{c}{\bf coordinated shock}
\\
&
&
&
\multicolumn{3}{c|}{$\Phi=0.7,\gamma=0.65$}
&
\multicolumn{3}{c}{$\Phi=0.7,\gamma=0.60$}
\\
&
&
&
\multicolumn{1}{c}{${\xi<0.05}$}
&
\multicolumn{1}{c}{${\xi<0.1}$}
&
\multicolumn{1}{c|}{${\xi<0.2}$}
&
\multicolumn{1}{c}{${\xi<0.05}$}
&
\multicolumn{1}{c}{${\xi<0.1}$}
&
\multicolumn{1}{c}{${\xi<0.2}$}
\\
\cline{1-9}
\multirow{15}{*}{${|V|=50}$}
&
\multirow{5}{*}{homogeneous} 
&
in-arborescence
&  
\bf 93\%  & \bf 93\%  &  \bf 93\% & \bf 35\% \bf & \bf 88\%  & \bf 93\%
\\
&
&
{\sf ER}, average degree $3$ 
&  
\bf 97\%  & \bf 100\%  &  \bf 100\% & \bf 81\% \bf & \bf 100\%  & \bf 100\%
\\
&
&
{\sf ER}, average degree $6$ 
&  
\bf 100\%  & \bf 100\%  &  \bf 100\% & \bf 100\% \bf & \bf 100\%  & \bf 100\%
\\
&
&
{\sf SF}, average degree $3$ 
&  
\bf 56\%  & \bf 96\%  &  \bf 100\% & \bf 52\% \bf & \bf 96\%  & \bf 100\%
\\
&
&
{\sf SF}, average degree $6$ 
&  
\bf 100\%  & \bf 100\%  &  \bf 100\% & \bf 100\% \bf & \bf 100\%  & \bf 100\%
\\
\cline{2-9}
& 
\multirow{5}{*}{${(0.1,0.95)}$-heterogeneous}
&
in-arborescence
&  
0\%  & 1\%  &  7\% & 0\% \bf & 0\%  & 0\%
\\
&
&
{\sf ER}, average degree $3$ 
&  
0\%  & 0\%  &  5\% & 0\% \bf & 0\%  & 1\%
\\
&
&
{\sf ER}, average degree $6$ 
&  
10\%  & 13\%  &  16\% & 7\% \bf & 7\%  & 9\%
\\
&
&
{\sf SF}, average degree $3$ 
&  
7\%  & 12\%  &  25\% & 1\% \bf & 4\%  & 9\%
\\
&
&
{\sf SF}, average degree $6$ 
&  
26\%  & 32\%  &  49\% & 14\% \bf & 15\%  & 21\%
\\
\cline{2-9}
& 
\multirow{5}{*}{${(0.2,0.6)}$-heterogeneous}
&
in-arborescence
&  
0\%  & 0\%  &  10\% & 0\% \bf & 0\%  & 9\%
\\
&
&
{\sf ER}, average degree $3$ 
&  
5\%  & 8\%  &  21\% & 3\% \bf & 7\%  & 17\%
\\
&
&
{\sf ER}, average degree $6$ 
&  
13\%  & 18\%  &  30\% & 7\% \bf & 7\%  & 20\%
\\
&
&
{\sf SF}, average degree $3$ 
&  
3\%  & 9\%  &  25\% & 1\% \bf & 3\%  & 19\%
\\
&
&
{\sf SF}, average degree $6$ 
&  
9\%  & 16\%  &  28\% & 8\% \bf & 9\%  & 19\%
\\
%
%%%%%%%%%%%%%%%%%%%%%%%%%%%%%%%%%%%%%%%%%%%%%%%%%%%%%%%%%%%%%%%%%%%%%%%%%%%%%%%%%%%%%%%%%%%%%%%%%%%%%%%%%%
\hline
\multirow{15}{*}{${|V|=100}$}
&
\multirow{5}{*}{homogeneous} 
&
in-arborescence
&  
\bf 93\%  & \bf 93\%  &  \bf 93\% & \bf 31\% \bf & \bf 93\%  & \bf 93\%
\\
&
&
{\sf ER}, average degree $3$ 
&  
\bf 85\%  & \bf 100\%  &  \bf 100\% & \bf 62\% \bf & \bf 97\%  & \bf 100\%
\\
&
&
{\sf ER}, average degree $6$ 
&  
\bf 100\%  & \bf 100\%  &  \bf 100\% & \bf 100\% \bf & \bf 100\%  & \bf 100\%
\\
&
&
{\sf SF}, average degree $3$ 
&  
\bf 37\%  & \bf 75\%  &  \bf 100\% & \bf 37\% \bf & \bf 75\%  & \bf 100\%
\\
&
&
{\sf SF}, average degree $6$ 
&  
\bf 100\%  & \bf 100\%  &  \bf 100\% & \bf 100\% \bf & \bf 100\%  & \bf 100\%
\\
\cline{2-9}
& 
\multirow{5}{*}{${(0.1,0.95)}$-heterogeneous}
&
in-arborescence
&  
0\%  & 0\%  &  0\% & 0\% & 0\%  & 0\%
\\
&
&
{\sf ER}, average degree $3$ 
&  
0\%  & 1\%  &  7\% & 0\% & 0\%  & 5\%
\\
&
&
{\sf ER}, average degree $6$ 
&  
7\%  & 7\%  &  9\% & 7\% & 7\%  & 7\%
\\
&
&
{\sf SF}, average degree $3$ 
&  
0\%  & 2\%  &  15\% & 0\% & 0\%  & 5\%
\\
&
&
{\sf SF}, average degree $6$ 
&  
11\%  & 15\%  &  27\% & 7\% & 7\%  & 11\%
\\
\cline{2-9}
& 
\multirow{5}{*}{${(0.2,0.6)}$-heterogeneous}
&
in-arborescence
&  
0\%  & 0\%  &  9\% & 0\% & 0\%  & 9\%
\\
&
&
{\sf ER}, average degree $3$ 
&  
0\%  & 7\%  &  18\% & 0\% & 7\%  & 16\%
\\
&
&
{\sf ER}, average degree $6$ 
&  
7\%  & 7\%  &  19\% & 7\% & 7\%  & 16\%
\\
&
&
{\sf SF}, average degree $3$ 
&  
1\%  & 4\%  &  17\% & 0\% & 2\%  & 13\%
\\
&
&
{\sf SF}, average degree $6$ 
&  
8\%  & 10\%  &  19\% & 7\% & 8\%  & 17\%
\\
%
%%%%%%%%%%%%%%%%%%%%%%%%%%%%%%%%%%%%%%%%%%%%%%%%%%%%%%%%%%%%%%%%%%%%%%%%%%%%%%%%%%%%%%%%%%%%%%%%%%%%%%%%%%
\hline
\multirow{15}{*}{${|V|=300}$}
&
\multirow{5}{*}{homogeneous} 
&
in-arborescence
&  
\bf 93\%  & \bf 93\%  &  \bf 93\% & \bf 56\% \bf & \bf 93\%  & \bf 93\%
\\
&
&
{\sf ER}, average degree $3$ 
&  
\bf 93\%  & \bf 100\%  &  \bf 100\% & \bf 65\% \bf & \bf 97\%  & \bf 100\%
\\
&
&
{\sf ER}, average degree $6$ 
&  
\bf 100\%  & \bf 100\%  &  \bf 100\% & \bf 100\% \bf & \bf 100\%  & \bf 100\%
\\
&
&
{\sf SF}, average degree $3$ 
&  
\bf 28\%  & \bf 56\%  &  \bf 97\% & \bf 28\% \bf & \bf 56\%  & \bf 97\%
\\
&
&
{\sf SF}, average degree $6$ 
&  
\bf 100\%  & \bf 100\%  &  \bf 100\% & \bf 100\% \bf & \bf 100\%  & \bf 100\%
\\
\cline{2-9}
& 
\multirow{5}{*}{${(0.1,0.95)}$-heterogeneous}
&
in-arborescence
&  
0\%  & 0\%  &  0\% & 0\% & 0\%  & 0\%
\\
&
&
{\sf ER}, average degree $3$ 
&  
0\%  & 1\%  &  7\% & 0\% & 0\%  & 4\%
\\
&
&
{\sf ER}, average degree $6$ 
&  
7\%  & 7\%  &  7\% & 6\% & 7\%  & 7\%
\\
&
&
{\sf SF}, average degree $3$ 
&  
0\%  & 0\%  &  15\% & 0\% & 0\%  & 3\%
\\
&
&
{\sf SF}, average degree $6$ 
&  
7\%  & 7\%  &  21\% & 7\% & 7\%  & 7\%
\\
\cline{2-9}
& 
\multirow{5}{*}{${(0.2,0.6)}$-heterogeneous}
&
in-arborescence
&  
0\%  & 0\%  &  9\% & 0\% & 0\%  & 9\%
\\
&
&
{\sf ER}, average degree $3$ 
&  
1\%  & 7\%  &  17\% & 0\% & 7\%  & 16\%
\\
&
&
{\sf ER}, average degree $6$ 
&  
7\%  & 7\%  &  16\% & 7\% & 7\%  & 16\%
\\
&
&
{\sf SF}, average degree $3$ 
&  
0\%  & 1\%  &  14\% & 0\% & 0\%  & 13\%
\\
&
&
{\sf SF}, average degree $6$ 
&  
7\%  & 8\%  &  16\% & 7\% & 7\%  & 16\%
\\
\cline{1-9}
\end{tabular}
}
\end{center}
\end{table}

\clearpage

%%%%%%%%%%%%%%%%%%%%%%%%%%%%%%%%%%%%%%%%%%%%%%%%%%%%%%%%%%%%%%%%%%%%%%%%%%%%%%%%%%%%%%%%%%%%%%%%%%%%%%%%%%%%%%
%%%%%%%%%%%%%%%%%%%%%%%%%%%%%%%%%%%%%%%%%%%%%%%%%%%%%%%%%%%%%%%%%%%%%%%%%%%%%%%%%%%%%%%%%%%%%%%%%%%%%%%%%%%%%%
%%%%%%%%%%%%%%%%%%%%%%%%%%%%%%%%%%%%%%%%%%%%%%%%%%%%%%%%%%%%%%%%%%%%%%%%%%%%%%%%%%%%%%%%%%%%%%%%%%%%%%%%%%%%%%
%%% Third table for analysis of residual instability of homogeneous versus heterogeneous networks
%%% coordinated shock

\begin{table}
\caption{Residual instabilities of homogeneous versus heterogeneous networks under coordinated shocks.
The percentages shown are the percentages of networks for which $\xi<0.05$ or $\xi<0.1$ or $\xi<0.2$.} 
\label{supp3-table6}
\begin{center}
\scalebox{0.75}[0.8]
{%%%\small
\begin{tabular}{c c r      r r r | r r r}
&
&
&
\multicolumn{6}{c}{\bf coordinated shock}
\\
&
&
&
\multicolumn{3}{c|}{$\Phi=0.8,\gamma=0.75$}
&
\multicolumn{3}{c}{$\Phi=0.8,\gamma=0.70$}
\\
&
&
&
\multicolumn{1}{c}{${\xi<0.05}$}
&
\multicolumn{1}{c}{${\xi<0.1}$}
&
\multicolumn{1}{c|}{${\xi<0.2}$}
&
\multicolumn{1}{c}{${\xi<0.05}$}
&
\multicolumn{1}{c}{${\xi<0.1}$}
&
\multicolumn{1}{c}{${\xi<0.2}$}
\\
\cline{1-9}
\multirow{15}{*}{${|V|=50}$}
&
\multirow{5}{*}{homogeneous} 
&
in-arborescence
&  
\bf 0\%  & \bf 0\%  &  \bf 93\% & \bf 0\% \bf & \bf 0\%  & \bf 93\%
\\
&
&
{\sf ER}, average degree $3$ 
&  
\bf 97\%  & \bf 100\%  &  \bf 100\% & \bf 89\% \bf & \bf 100\%  & \bf 100\%
\\
&
&
{\sf ER}, average degree $6$ 
&  
\bf 100\%  & \bf 100\%  &  \bf 100\% & \bf 100\% \bf & \bf 100\%  & \bf 100\%
\\
&
&
{\sf SF}, average degree $3$ 
&  
\bf 56\%  & \bf 96\%  &  \bf 100\% & \bf 54\% \bf & \bf 96\%  & \bf 100\%
\\
&
&
{\sf SF}, average degree $6$ 
&  
\bf 100\%  & \bf 100\%  &  \bf 100\% & \bf 100\% \bf & \bf 100\%  & \bf 100\%
\\
\cline{2-9}
& 
\multirow{5}{*}{${(0.1,0.95)}$-heterogeneous}
&
in-arborescence
&  
1\%  & 7\%  &  7\% & 0\% \bf & 0\%  & 0\%
\\
&
&
{\sf ER}, average degree $3$ 
&  
0\%  & 1\%  &  6\% & 0\% \bf & 0\%  & 1\%
\\
&
&
{\sf ER}, average degree $6$ 
&  
12\%  & 15\%  &  19\% & 7\% \bf & 7\%  & 9\%
\\
&
&
{\sf SF}, average degree $3$ 
&  
9\%  & 15\%  &  32\% & 1\% \bf & 5\%  & 11\%
\\
&
&
{\sf SF}, average degree $6$ 
&  
27\%  & 37\%  &  60\% & 14\% \bf & 19\%  & 21\%
\\
\cline{2-9}
& 
\multirow{5}{*}{${(0.2,0.6)}$-heterogeneous}
&
in-arborescence
&  
0\%  & 0\%  &  10\% & 0\% \bf & 0\%  & 9\%
\\
&
&
{\sf ER}, average degree $3$ 
&  
5\%  & 11\%  &  22\% & 4\% \bf & 7\%  & 17\%
\\
&
&
{\sf ER}, average degree $6$ 
&  
14\%  & 19\%  &  30\% & 7\% \bf & 11\%  & 21\%
\\
&
&
{\sf SF}, average degree $3$ 
&  
3\%  & 11\%  &  27\% & 1\% \bf & 4\%  & 19\%
\\
&
&
{\sf SF}, average degree $6$ 
&  
11\%  & 19\%  &  31\% & 8\% \bf & 9\%  & 21\%
\\
%
%%%%%%%%%%%%%%%%%%%%%%%%%%%%%%%%%%%%%%%%%%%%%%%%%%%%%%%%%%%%%%%%%%%%%%%%%%%%%%%%%%%%%%%%%%%%%%%%%%%%%%%%%%
\hline
\multirow{15}{*}{${|V|=100}$}
&
\multirow{5}{*}{homogeneous} 
&
in-arborescence
&  
\bf 93\%  & \bf 93\%  &  \bf 93\% & \bf 51\% \bf & \bf 93\%  & \bf 93\%
\\
&
&
{\sf ER}, average degree $3$ 
&  
\bf 85\%  & \bf 100\%  &  \bf 100\% & \bf 71\% \bf & \bf 100\%  & \bf 100\%
\\
&
&
{\sf ER}, average degree $6$ 
&  
\bf 100\%  & \bf 100\%  &  \bf 100\% & \bf 100\% \bf & \bf 100\%  & \bf 100\%
\\
&
&
{\sf SF}, average degree $3$ 
&  
\bf 37\%  & \bf 75\%  &  \bf 100\% & \bf 37\% \bf & \bf 75\%  & \bf 100\%
\\
&
&
{\sf SF}, average degree $6$ 
&  
\bf 100\%  & \bf 100\%  &  \bf 100\% & \bf 100\% \bf & \bf 100\%  & \bf 100\%
\\
\cline{2-9}
& 
\multirow{5}{*}{${(0.1,0.95)}$-heterogeneous}
&
in-arborescence
&  
0\%  & 0\%  &  1\% & 0\% & 0\%  & 0\%
\\
&
&
{\sf ER}, average degree $3$ 
&  
0\%  & 1\%  &  7\% & 0\% & 0\%  & 5\%
\\
&
&
{\sf ER}, average degree $6$ 
&  
7\%  & 9\%  &  10\% & 7\% & 7\%  & 7\%
\\
&
&
{\sf SF}, average degree $3$ 
&  
0\%  & 3\%  &  20\% & 0\% & 0\%  & 6\%
\\
&
&
{\sf SF}, average degree $6$ 
&  
11\%  & 19\%  &  33\% & 7\% & 11\%  & 11\%
\\
\cline{2-9}
& 
\multirow{5}{*}{${(0.2,0.6)}$-heterogeneous}
&
in-arborescence
&  
0\%  & 0\%  &  9\% & 0\% & 0\%  & 9\%
\\
&
&
{\sf ER}, average degree $3$ 
&  
0\%  & 7\%  &  19\% & 0\% & 7\%  & 17\%
\\
&
&
{\sf ER}, average degree $6$ 
&  
7\%  & 8\%  &  21\% & 7\% & 7\%  & 17\%
\\
&
&
{\sf SF}, average degree $3$ 
&  
1\%  & 5\%  &  18\% & 0\% & 2\%  & 14\%
\\
&
&
{\sf SF}, average degree $6$ 
&  
8\%  & 10\%  &  23\% & 7\% & 8\%  & 17\%
\\
%
%%%%%%%%%%%%%%%%%%%%%%%%%%%%%%%%%%%%%%%%%%%%%%%%%%%%%%%%%%%%%%%%%%%%%%%%%%%%%%%%%%%%%%%%%%%%%%%%%%%%%%%%%%
\hline
\multirow{15}{*}{${|V|=300}$}
&
\multirow{5}{*}{homogeneous} 
&
in-arborescence
&  
\bf 93\%  & \bf 93\%  &  \bf 93\% & \bf 76\% \bf & \bf 93\%  & \bf 93\%
\\
&
&
{\sf ER}, average degree $3$ 
&  
\bf 93\%  & \bf 100\%  &  \bf 100\% & \bf 76\% \bf & \bf 99\%  & \bf 100\%
\\
&
&
{\sf ER}, average degree $6$ 
&  
\bf 100\%  & \bf 100\%  &  \bf 100\% & \bf 100\% \bf & \bf 100\%  & \bf 100\%
\\
&
&
{\sf SF}, average degree $3$ 
&  
\bf 28\%  & \bf 56\%  &  \bf 97\% & \bf 28\% \bf & \bf 56\%  & \bf 97\%
\\
&
&
{\sf SF}, average degree $6$ 
&  
\bf 100\%  & \bf 100\%  &  \bf 100\% & \bf 99\% \bf & \bf 100\%  & \bf 100\%
\\
\cline{2-9}
& 
\multirow{5}{*}{${(0.1,0.95)}$-heterogeneous}
&
in-arborescence
&  
0\%  & 0\%  &  0\% & 0\% & 0\%  & 0\%
\\
&
&
{\sf ER}, average degree $3$ 
&  
0\%  & 1\%  &  7\% & 0\% & 1\%  & 5\%
\\
&
&
{\sf ER}, average degree $6$ 
&  
7\%  & 7\%  &  7\% & 7\% & 7\%  & 7\%
\\
&
&
{\sf SF}, average degree $3$ 
&  
0\%  & 1\%  &  16\% & 0\% & 0\%  & 6\%
\\
&
&
{\sf SF}, average degree $6$ 
&  
7\%  & 11\%  &  22\% & 7\% & 7\%  & 8\%
\\
\cline{2-9}
& 
\multirow{5}{*}{${(0.2,0.6)}$-heterogeneous}
&
in-arborescence
&  
0\%  & 0\%  &  9\% & 0\% & 0\%  & 9\%
\\
&
&
{\sf ER}, average degree $3$ 
&  
1\%  & 7\%  &  17\% & 1\% & 7\%  & 16\%
\\
&
&
{\sf ER}, average degree $6$ 
&  
7\%  & 7\%  &  16\% & 7\% & 7\%  & 16\%
\\
&
&
{\sf SF}, average degree $3$ 
&  
0\%  & 1\%  &  14\% & 0\% & 1\%  & 13\%
\\
&
&
{\sf SF}, average degree $6$ 
&  
7\%  & 8\%  &  17\% & 7\% & 7\%  & 16\%
\\
\cline{1-9}
\end{tabular}
}
\end{center}
\end{table}

\clearpage

%%%%%%%%%%%%%%%%%%%%%%%%%%%%%%%%%%%%%%%%%%%%%%%%%%%%%%%%%%%%%%%%%%%%%%%%%%%%%%%%%%%%%%%%%%%%%%%%%%%%%%%%%%%%%%
%%%%%%%%%%%%%%%%%%%%%%%%%%%%%%%%%%%%%%%%%%%%%%%%%%%%%%%%%%%%%%%%%%%%%%%%%%%%%%%%%%%%%%%%%%%%%%%%%%%%%%%%%%%%%%
%%%%%%%%%%%%%%%%%%%%%%%%%%%%%%%%%%%%%%%%%%%%%%%%%%%%%%%%%%%%%%%%%%%%%%%%%%%%%%%%%%%%%%%%%%%%%%%%%%%%%%%%%%%%%%
%%% Fourth table for analysis of residual instability of homogeneous versus heterogeneous networks
%%% coordinated shock

\begin{table}
\caption{Residual instabilities of homogeneous versus heterogeneous networks under coordinated shocks.
The percentages shown are the percentages of networks for which $\xi<0.05$ or $\xi<0.1$ or $\xi<0.2$.} 
\label{supp4-table6}
\begin{center}
\scalebox{0.75}[0.8]
{%%%%\small
\begin{tabular}{c c r      r r r | r r r}
&
&
&
\multicolumn{6}{c}{\bf coordinated shock}
\\
&
&
&
\multicolumn{3}{c|}{$\Phi=0.9,\gamma=0.85$}
&
\multicolumn{3}{c}{$\Phi=0.9,\gamma=0.80$}
\\
&
&
&
\multicolumn{1}{c}{${\xi<0.05}$}
&
\multicolumn{1}{c}{${\xi<0.1}$}
&
\multicolumn{1}{c|}{${\xi<0.2}$}
&
\multicolumn{1}{c}{${\xi<0.05}$}
&
\multicolumn{1}{c}{${\xi<0.1}$}
&
\multicolumn{1}{c}{${\xi<0.2}$}
\\
\cline{1-9}
\multirow{15}{*}{${|V|=50}$}
&
\multirow{5}{*}{homogeneous} 
&
in-arborescence
&  
\bf 93\%  & \bf 93\%  &  \bf 93\% & \bf 81\% \bf & \bf 93\%  & \bf 93\%
\\
&
&
{\sf ER}, average degree $3$ 
&  
\bf 98\%  & \bf 100\%  &  \bf 100\% & \bf 95\% \bf & \bf 100\%  & \bf 100\%
\\
&
&
{\sf ER}, average degree $6$ 
&  
\bf 0\%  & \bf 0\%  &  \bf 0\% & \bf 100\% \bf & \bf 100\%  & \bf 100\%
\\
&
&
{\sf SF}, average degree $3$ 
&  
\bf 56\%  & \bf 96\%  &  \bf 100\% & \bf 56\% \bf & \bf 96\%  & \bf 100\%
\\
&
&
{\sf SF}, average degree $6$ 
&  
\bf 100\%  & \bf 100\%  &  \bf 100\% & \bf 100\% \bf & \bf 100\%  & \bf 100\%
\\
\cline{2-9}
& 
\multirow{5}{*}{${(0.1,0.95)}$-heterogeneous}
&
in-arborescence
&  
7\%  & 7\%  &  7\% & 0\% \bf & 0\%  & 0\%
\\
&
&
{\sf ER}, average degree $3$ 
&  
0\%  & 1\%  &  7\% & 0\% \bf & 0\%  & 1\%
\\
&
&
{\sf ER}, average degree $6$ 
&  
13\%  & 18\%  &  22\% & 7\% \bf & 9\%  & 10\%
\\
&
&
{\sf SF}, average degree $3$ 
&  
11\%  & 17\%  &  37\% & 3\% \bf & 5\%  & 13\%
\\
&
&
{\sf SF}, average degree $6$ 
&  
31\%  & 41\%  &  66\% & 18\% \bf & 19\%  & 25\%
\\
\cline{2-9}
& 
\multirow{5}{*}{${(0.2,0.6)}$-heterogeneous}
&
in-arborescence
&  
0\%  & 0\%  &  15\% & 0\% \bf & 0\%  & 9\%
\\
&
&
{\sf ER}, average degree $3$ 
&  
5\%  & 11\%  &  25\% & 5\% \bf & 7\%  & 19\%
\\
&
&
{\sf ER}, average degree $6$ 
&  
14\%  & 19\%  &  34\% & 7\% \bf & 12\%  & 23\%
\\
&
&
{\sf SF}, average degree $3$ 
&  
4\%  & 12\%  &  28\% & 2\% \bf & 6\%  & 21\%
\\
&
&
{\sf SF}, average degree $6$ 
&  
17\%  & 21\%  &  33\% & 8\% \bf & 11\%  & 23\%
\\
%
%%%%%%%%%%%%%%%%%%%%%%%%%%%%%%%%%%%%%%%%%%%%%%%%%%%%%%%%%%%%%%%%%%%%%%%%%%%%%%%%%%%%%%%%%%%%%%%%%%%%%%%%%%
\hline
\multirow{15}{*}{${|V|=100}$}
&
\multirow{5}{*}{homogeneous} 
&
in-arborescence
&  
\bf 93\%  & \bf 93\%  &  \bf 93\% & \bf 75\% \bf & \bf 93\%  & \bf 93\%
\\
&
&
{\sf ER}, average degree $3$ 
&  
\bf 85\%  & \bf 100\%  &  \bf 100\% & \bf 78\% \bf & \bf 100\%  & \bf 100\%
\\
&
&
{\sf ER}, average degree $6$ 
&  
\bf 0\%  & \bf 0\%  &  \bf 0\% & \bf 100\% \bf & \bf 100\%  & \bf 100\%
\\
&
&
{\sf SF}, average degree $3$ 
&  
\bf 37\%  & \bf 75\%  &  \bf 100\% & \bf 37\% \bf & \bf 75\%  & \bf 100\%
\\
&
&
{\sf SF}, average degree $6$ 
&  
\bf 100\%  & \bf 100\%  &  \bf 100\% & \bf 100\% \bf & \bf 100\%  & \bf 100\%
\\
\cline{2-9}
& 
\multirow{5}{*}{${(0.1,0.95)}$-heterogeneous}
&
in-arborescence
&  
0\%  & 0\%  &  1\% & 0\% & 0\%  & 0\%
\\
&
&
{\sf ER}, average degree $3$ 
&  
0\%  & 1\%  &  7\% & 0\% & 0\%  & 6\%
\\
&
&
{\sf ER}, average degree $6$ 
&  
7\%  & 9\%  &  10\% & 7\% & 7\%  & 7\%
\\
&
&
{\sf SF}, average degree $3$ 
&  
1\%  & 4\%  &  25\% & 0\% & 1\%  & 7\%
\\
&
&
{\sf SF}, average degree $6$ 
&  
14\%  & 22\%  &  41\% & 7\% & 11\%  & 15\%
\\
\cline{2-9}
& 
\multirow{5}{*}{${(0.2,0.6)}$-heterogeneous}
&
in-arborescence
&  
0\%  & 0\%  &  9\% & 0\% & 0\%  & 9\%
\\
&
&
{\sf ER}, average degree $3$ 
&  
0\%  & 7\%  &  21\% & 0\% & 7\%  & 17\%
\\
&
&
{\sf ER}, average degree $6$ 
&  
7\%  & 8\%  &  23\% & 7\% & 7\%  & 17\%
\\
&
&
{\sf SF}, average degree $3$ 
&  
1\%  & 5\%  &  19\% & 1\% & 3\%  & 15\%
\\
&
&
{\sf SF}, average degree $6$ 
&  
9\%  & 11\%  &  23\% & 7\% & 9\%  & 17\%
\\
%
%%%%%%%%%%%%%%%%%%%%%%%%%%%%%%%%%%%%%%%%%%%%%%%%%%%%%%%%%%%%%%%%%%%%%%%%%%%%%%%%%%%%%%%%%%%%%%%%%%%%%%%%%%
\hline
\multirow{15}{*}{${|V|=300}$}
&
\multirow{5}{*}{homogeneous} 
&
in-arborescence
&  
\bf 93\%  & \bf 93\%  &  \bf 93\% & \bf 93\% \bf & \bf 93\%  & \bf 93\%
\\
&
&
{\sf ER}, average degree $3$ 
&  
\bf 0\%  & \bf 0\%  &  \bf 100\% & \bf 85\% \bf & \bf 99\%  & \bf 100\%
\\
&
&
{\sf ER}, average degree $6$ 
&  
\bf 0\%  & \bf 0\%  &  \bf 0\% & \bf 100\% \bf & \bf 100\%  & \bf 100\%
\\
&
&
{\sf SF}, average degree $3$ 
&  
\bf 28\%  & \bf 56\%  &  \bf 97\% & \bf 28\% \bf & \bf 56\%  & \bf 97\%
\\
&
&
{\sf SF}, average degree $6$ 
&  
\bf 100\%  & \bf 100\%  &  \bf 100\% & \bf 100\% \bf & \bf 100\%  & \bf 100\%
\\
\cline{2-9}
& 
\multirow{5}{*}{${(0.1,0.95)}$-heterogeneous}
&
in-arborescence
&  
0\%  & 0\%  &  0\% & 0\% & 0\%  & 0\%
\\
&
&
{\sf ER}, average degree $3$ 
&  
0\%  & 1\%  &  7\% & 0\% & 1\%  & 6\%
\\
&
&
{\sf ER}, average degree $6$ 
&  
7\%  & 7\%  &  7\% & 7\% & 7\%  & 7\%
\\
&
&
{\sf SF}, average degree $3$ 
&  
0\%  & 1\%  &  17\% & 0\% & 0\%  & 10\%
\\
&
&
{\sf SF}, average degree $6$ 
&  
7\%  & 11\%  &  24\% & 7\% & 7\%  & 10\%
\\
\cline{2-9}
& 
\multirow{5}{*}{${(0.2,0.6)}$-heterogeneous}
&
in-arborescence
&  
0\%  & 0\%  &  9\% & 0\% & 0\%  & 9\%
\\
&
&
{\sf ER}, average degree $3$ 
&  
1\%  & 7\%  &  17\% & 1\% & 7\%  & 16\%
\\
&
&
{\sf ER}, average degree $6$ 
&  
7\%  & 7\%  &  17\% & 7\% & 7\%  & 16\%
\\
&
&
{\sf SF}, average degree $3$ 
&  
0\%  & 1\%  &  14\% & 0\% & 1\%  & 13\%
\\
&
&
{\sf SF}, average degree $6$ 
&  
7\%  & 8\%  &  17\% & 7\% & 7\%  & 16\%
\\
\cline{1-9}
\end{tabular}
}
\end{center}
\end{table}

\clearpage
%%%%%%%%%%%%%%%%%%%%%%%%%%%%%%%%%%%%%%%%%%%%%%%%%%%%%%%%%%%%%%%%%%%%%%%%%%%%%%%%%%%%%%%%%%%%%%%%%%%%%%%%%%%%%%
%%%%%%%%%%%%%%%%%%%%%%%%%%%%%%%%%%%%%%%%%%%%%%%%%%%%%%%%%%%%%%%%%%%%%%%%%%%%%%%%%%%%%%%%%%%%%%%%%%%%%%%%%%%%%%
%%%%%%%%%%%%%%%%%%%%%%%%%%%%%%%%%%%%%%%%%%%%%%%%%%%%%%%%%%%%%%%%%%%%%%%%%%%%%%%%%%%%%%%%%%%%%%%%%%%%%%%%%%%%%%
%%% Table for analysis of residual instability of homogeneous versus heterogeneous networks
%%% random shock

\begin{table}
\caption{Residual instabilities of homogeneous versus heterogeneous networks under coordinated shocks.
The percentages shown are the percentages of networks for which $\xi<0.05$ or $\xi<0.1$ or $\xi<0.2$.} 
\label{supp1-table7}
\begin{center}
\scalebox{0.75}[0.8]
{%%%%\small
\begin{tabular}{c c r      r r r | r r r}
&
&
&
\multicolumn{6}{c}{\bf idiosyncratic shock}
\\
&
&
&
\multicolumn{3}{c|}{$\Phi=0.6,\gamma=0.55$}
&
\multicolumn{3}{c}{$\Phi=0.6,\gamma=0.50$}
\\
&
&
&
\multicolumn{1}{c}{${\xi<0.05}$}
&
\multicolumn{1}{c}{${\xi<0.1}$}
&
\multicolumn{1}{c|}{${\xi<0.2}$}
&
\multicolumn{1}{c}{${\xi<0.05}$}
&
\multicolumn{1}{c}{${\xi<0.1}$}
&
\multicolumn{1}{c}{${\xi<0.2}$}
\\
\cline{1-9}
\multirow{15}{*}{${|V|=50}$}
&
\multirow{5}{*}{homogeneous} 
&
in-arborescence
&  
\bf 94\%  & \bf 95\%  &  \bf 95\% & \bf 53\% \bf & \bf 87\%  & \bf 95\%
\\
&
&
{\sf ER}, average degree $3$ 
&  
\bf 100\%  & \bf 100\%  &  \bf 100\% & \bf 62\% \bf & \bf 95\%  & \bf 100\%
\\
&
&
{\sf ER}, average degree $6$ 
&  
\bf 57\%  & \bf 58\%  &  \bf 60\% & \bf 57\% \bf & \bf 57\%  & \bf 57\%
\\
&
&
{\sf SF}, average degree $3$ 
&  
\bf 54\%  & \bf 100\%  &  \bf 100\% & \bf 45\% \bf & \bf 90\%  & \bf 100\%
\\
&
&
{\sf SF}, average degree $6$ 
&  
\bf 100\%  & \bf 100\%  &  \bf 100\% & \bf 100\% \bf & \bf 100\%  & \bf 100\%
\\
\cline{2-9}
& 
\multirow{5}{*}{${(0.1,0.95)}$-heterogeneous}
&
in-arborescence
&  
2\%  & 4\%  &  16\% & 1\% \bf & 3\%  & 11\%
\\
&
&
{\sf ER}, average degree $3$ 
&  
0\%  & 1\%  &  14\% & 0\% \bf & 0\%  & 5\%
\\
&
&
{\sf ER}, average degree $6$ 
&  
8\%  & 11\%  &  23\% & 5\% \bf & 7\%  & 12\%
\\
&
&
{\sf SF}, average degree $3$ 
&  
0\%  & 7\%  &  24\% & 0\% \bf & 2\%  & 11\%
\\
&
&
{\sf SF}, average degree $6$ 
&  
9\%  & 21\%  &  39\% & 7\% \bf & 15\%  & 25\%
\\
\cline{2-9}
& 
\multirow{5}{*}{${(0.2,0.6)}$-heterogeneous}
&
in-arborescence
&  
1\%  & 3\%  &  13\% & 1\% \bf & 1\%  & 11\%
\\
&
&
{\sf ER}, average degree $3$ 
&  
7\%  & 14\%  &  25\% & 5\% \bf & 10\%  & 19\%
\\
&
&
{\sf ER}, average degree $6$ 
&  
11\%  & 21\%  &  34\% & 7\% \bf & 11\%  & 23\%
\\
&
&
{\sf SF}, average degree $3$ 
&  
1\%  & 11\%  &  21\% & 0\% \bf & 9\%  & 18\%
\\
&
&
{\sf SF}, average degree $6$ 
&  
10\%  & 15\%  &  27\% & 7\% \bf & 9\%  & 19\%
\\
%
%%%%%%%%%%%%%%%%%%%%%%%%%%%%%%%%%%%%%%%%%%%%%%%%%%%%%%%%%%%%%%%%%%%%%%%%%%%%%%%%%%%%%%%%%%%%%%%%%%%%%%%%%%
\hline
\multirow{15}{*}{${|V|=100}$}
&
\multirow{5}{*}{homogeneous} 
&
in-arborescence
&  
\bf 94\%  & \bf 94\%  &  \bf 95\% & \bf 69\% \bf & \bf 92\%  & \bf 95\%
\\
&
&
{\sf ER}, average degree $3$ 
&  
\bf 77\%  & \bf 100\%  &  \bf 100\% & \bf 45\% \bf & \bf 87\%  & \bf 100\%
\\
&
&
{\sf ER}, average degree $6$ 
&  
\bf 57\%  & \bf 57\%  &  \bf 57\% & \bf 57\% \bf & \bf 57\%  & \bf 57\%
\\
&
&
{\sf SF}, average degree $3$ 
&  
\bf 36\%  & \bf 81\%  &  \bf 100\% & \bf 34\% \bf & \bf 75\%  & \bf 100\%
\\
&
&
{\sf SF}, average degree $6$ 
&  
\bf 100\%  & \bf 100\%  &  \bf 100\% & \bf 95\% \bf & \bf 100\%  & \bf 100\%
\\
\cline{2-9}
& 
\multirow{5}{*}{${(0.1,0.95)}$-heterogeneous}
&
in-arborescence
&  
1\%  & 2\%  &  13\% & 1\% & 1\%  & 11\%
\\
&
&
{\sf ER}, average degree $3$ 
&  
1\%  & 2\%  &  17\% & 0\% & 1\%  & 13\%
\\
&
&
{\sf ER}, average degree $6$ 
&  
7\%  & 7\%  &  18\% & 7\% & 7\%  & 15\%
\\
&
&
{\sf SF}, average degree $3$ 
&  
0\%  & 9\%  &  25\% & 0\% & 3\%  & 17\%
\\
&
&
{\sf SF}, average degree $6$ 
&  
11\%  & 19\%  &  32\% & 7\% & 12\%  & 22\%
\\
\cline{2-9}
& 
\multirow{5}{*}{${(0.2,0.6)}$-heterogeneous}
&
in-arborescence
&  
0\%  & 0\%  &  9\% & 0\% & 0\%  & 9\%
\\
&
&
{\sf ER}, average degree $3$ 
&  
5\%  & 9\%  &  19\% & 1\% & 8\%  & 17\%
\\
&
&
{\sf ER}, average degree $6$ 
&  
8\%  & 13\%  &  23\% & 7\% & 9\%  & 19\%
\\
&
&
{\sf SF}, average degree $3$ 
&  
0\%  & 3\%  &  19\% & 0\% & 1\%  & 17\%
\\
&
&
{\sf SF}, average degree $6$ 
&  
7\%  & 9\%  &  20\% & 7\% & 8\%  & 17\%
\\
%
%%%%%%%%%%%%%%%%%%%%%%%%%%%%%%%%%%%%%%%%%%%%%%%%%%%%%%%%%%%%%%%%%%%%%%%%%%%%%%%%%%%%%%%%%%%%%%%%%%%%%%%%%%
\hline
\multirow{15}{*}{${|V|=300}$}
&
\multirow{5}{*}{homogeneous} 
&
in-arborescence
&  
\bf 94\%  & \bf 95\%  &  \bf 95\% & \bf 93\% \bf & \bf 95\%  & \bf 96\%
\\
&
&
{\sf ER}, average degree $3$ 
&  
\bf 96\%  & \bf 100\%  &  \bf 100\% & \bf 48\% \bf & \bf 89\%  & \bf 100\%
\\
&
&
{\sf ER}, average degree $6$ 
&  
\bf 57\%  & \bf 57\%  &  \bf 57\% & \bf 57\% \bf & \bf 57\%  & \bf 57\%
\\
&
&
{\sf SF}, average degree $3$ 
&  
\bf 34\%  & \bf 70\%  &  \bf 100\% & \bf 31\% \bf & \bf 67\%  & \bf 100\%
\\
&
&
{\sf SF}, average degree $6$ 
&  
\bf 100\%  & \bf 100\%  &  \bf 100\% & \bf 93\% \bf & \bf 100\%  & \bf 100\%
\\
\cline{2-9}
& 
\multirow{5}{*}{${(0.1,0.95)}$-heterogeneous}
&
in-arborescence
&  
1\%  & 1\%  &  12\% & 1\% & 1\%  & 11\%
\\
&
&
{\sf ER}, average degree $3$ 
&  
0\%  & 4\%  &  16\% & 0\% & 1\%  & 10\%
\\
&
&
{\sf ER}, average degree $6$ 
&  
7\%  & 7\%  &  16\% & 5\% & 7\%  & 12\%
\\
&
&
{\sf SF}, average degree $3$ 
&  
0\%  & 8\%  &  25\% & 0\% & 3\%  & 19\%
\\
&
&
{\sf SF}, average degree $6$ 
&  
9\%  & 17\%  &  29\% & 7\% & 11\%  & 21\%
\\
\cline{2-9}
& 
\multirow{5}{*}{${(0.2,0.6)}$-heterogeneous}
&
in-arborescence
&  
0\%  & 0\%  &  9\% & 0\% & 0\%  & 9\%
\\
&
&
{\sf ER}, average degree $3$ 
&  
7\%  & 8\%  &  17\% & 3\% & 7\%  & 16\%
\\
&
&
{\sf ER}, average degree $6$ 
&  
7\%  & 9\%  &  18\% & 7\% & 7\%  & 17\%
\\
&
&
{\sf SF}, average degree $3$ 
&  
0\%  & 1\%  &  17\% & 0\% & 0\%  & 16\%
\\
&
&
{\sf SF}, average degree $6$ 
&  
7\%  & 8\%  &  17\% & 7\% & 7\%  & 16\%
\\
\cline{1-9}
\end{tabular}
}
\end{center}
\end{table}

\clearpage
%%%%%%%%%%%%%%%%%%%%%%%%%%%%%%%%%%%%%%%%%%%%%%%%%%%%%%%%%%%%%%%%%%%%%%%%%%%%%%%%%%%%%%%%%%%%%%%%%%%%%%%%%%%%%%
%%%%%%%%%%%%%%%%%%%%%%%%%%%%%%%%%%%%%%%%%%%%%%%%%%%%%%%%%%%%%%%%%%%%%%%%%%%%%%%%%%%%%%%%%%%%%%%%%%%%%%%%%%%%%%
%%%%%%%%%%%%%%%%%%%%%%%%%%%%%%%%%%%%%%%%%%%%%%%%%%%%%%%%%%%%%%%%%%%%%%%%%%%%%%%%%%%%%%%%%%%%%%%%%%%%%%%%%%%%%%
%%% Second table for analysis of residual instability of homogeneous versus heterogeneous networks
%%% random shock

\begin{table}
\caption{Residual instabilities of homogeneous versus heterogeneous networks under coordinated shocks. 
The percentages shown are the percentages of networks for which $\xi<0.05$ or $\xi<0.1$ or $\xi<0.2$.} 
\label{supp2-table7}
\begin{center}
\scalebox{0.75}[0.8]
{%%%%\small
\begin{tabular}{c c r      r r r | r r r}
&
&
&
\multicolumn{6}{c}{\bf idiosyncratic shock}
\\
&
&
&
\multicolumn{3}{c|}{$\Phi=0.7,\gamma=0.65$}
&
\multicolumn{3}{c}{$\Phi=0.7,\gamma=0.60$}
\\
&
&
&
\multicolumn{1}{c}{${\xi<0.05}$}
&
\multicolumn{1}{c}{${\xi<0.1}$}
&
\multicolumn{1}{c|}{${\xi<0.2}$}
&
\multicolumn{1}{c}{${\xi<0.05}$}
&
\multicolumn{1}{c}{${\xi<0.1}$}
&
\multicolumn{1}{c}{${\xi<0.2}$}
\\
\cline{1-9}
\multirow{15}{*}{${|V|=50}$}
&
\multirow{5}{*}{homogeneous} 
&
in-arborescence
&  
\bf 94\%  & \bf 95\%  &  \bf 96\% & \bf 73\% \bf & \bf 94\%  & \bf 95\%
\\
&
&
{\sf ER}, average degree $3$ 
&  
\bf 100\%  & \bf 100\%  &  \bf 100\% & \bf 81\% \bf & \bf 100\%  & \bf 100\%
\\
&
&
{\sf ER}, average degree $6$ 
&  
\bf 57\%  & \bf 59\%  &  \bf 60\% & \bf 57\% \bf & \bf 57\%  & \bf 57\%
\\
&
&
{\sf SF}, average degree $3$ 
&  
\bf 55\%  & \bf 100\%  &  \bf 100\% & \bf 55\% \bf & \bf 100\%  & \bf 100\%
\\
&
&
{\sf SF}, average degree $6$ 
&  
\bf 100\%  & \bf 100\%  &  \bf 100\% & \bf 100\% \bf & \bf 100\%  & \bf 100\%
\\
\cline{2-9}
& 
\multirow{5}{*}{${(0.1,0.95)}$-heterogeneous}
&
in-arborescence
&  
3\%  & 7\%  &  17\% & 1\% \bf & 2\%  & 12\%
\\
&
&
{\sf ER}, average degree $3$ 
&  
1\%  & 5\%  &  21\% & 0\% \bf & 3\%  & 13\%
\\
&
&
{\sf ER}, average degree $6$ 
&  
10\%  & 18\%  &  28\% & 7\% \bf & 9\%  & 19\%
\\
&
&
{\sf SF}, average degree $3$ 
&  
7\%  & 9\%  &  26\% & 0\% \bf & 3\%  & 19\%
\\
&
&
{\sf SF}, average degree $6$ 
&  
13\%  & 24\%  &  41\% & 9\% \bf & 16\%  & 27\%
\\
\cline{2-9}
& 
\multirow{5}{*}{${(0.2,0.6)}$-heterogeneous}
&
in-arborescence
&  
1\%  & 3\%  &  14\% & 1\% \bf & 1\%  & 11\%
\\
&
&
{\sf ER}, average degree $3$ 
&  
7\%  & 15\%  &  27\% & 7\% \bf & 10\%  & 20\%
\\
&
&
{\sf ER}, average degree $6$ 
&  
13\%  & 23\%  &  35\% & 7\% \bf & 13\%  & 25\%
\\
&
&
{\sf SF}, average degree $3$ 
&  
1\%  & 11\%  &  23\% & 0\% \bf & 6\%  & 19\%
\\
&
&
{\sf SF}, average degree $6$ 
&  
11\%  & 17\%  &  29\% & 7\% \bf & 9\%  & 20\%
\\
%
%%%%%%%%%%%%%%%%%%%%%%%%%%%%%%%%%%%%%%%%%%%%%%%%%%%%%%%%%%%%%%%%%%%%%%%%%%%%%%%%%%%%%%%%%%%%%%%%%%%%%%%%%%
\hline
\multirow{15}{*}{${|V|=100}$}
&
\multirow{5}{*}{homogeneous} 
&
in-arborescence
&  
\bf 94\%  & \bf 95\%  &  \bf 95\% & \bf 79\% \bf & \bf 95\%  & \bf 95\%
\\
&
&
{\sf ER}, average degree $3$ 
&  
\bf 77\%  & \bf 100\%  &  \bf 100\% & \bf 57\% \bf & \bf 97\%  & \bf 100\%
\\
&
&
{\sf ER}, average degree $6$ 
&  
\bf 57\%  & \bf 57\%  &  \bf 57\% & \bf 57\% & \bf 57\%  & \bf 57\%
\\
&
&
{\sf SF}, average degree $3$ 
&  
\bf 37\%  & \bf 78\%  &  \bf 100\% & \bf 35\% \bf & \bf 76\%  & \bf 100\%
\\
&
&
{\sf SF}, average degree $6$ 
&  
\bf 100\%  & \bf 100\%  &  \bf 100\% & \bf 100\% \bf & \bf 100\%  & \bf 100\%
\\
\cline{2-9}
& 
\multirow{5}{*}{${(0.1,0.95)}$-heterogeneous}
&
in-arborescence
&  
1\%  & 3\%  &  13\% & 1\% & 1\%  & 12\%
\\
&
&
{\sf ER}, average degree $3$ 
&  
0\%  & 4\%  &  17\% & 0\% & 1\%  & 14\%
\\
&
&
{\sf ER}, average degree $6$ 
&  
7\%  & 9\%  &  19\% & 7\% & 7\%  & 16\%
\\
&
&
{\sf SF}, average degree $3$ 
&  
0\%  & 10\%  &  25\% & 0\% & 3\%  & 19\%
\\
&
&
{\sf SF}, average degree $6$ 
&  
9\%  & 18\%  &  33\% & 8\% & 15\%  & 25\%
\\
\cline{2-9}
& 
\multirow{5}{*}{${(0.2,0.6)}$-heterogeneous}
&
in-arborescence
&  
1\%  & 1\%  &  11\% & 1\% & 1\%  & 11\%
\\
&
&
{\sf ER}, average degree $3$ 
&  
6\%  & 11\%  &  21\% & 3\% & 9\%  & 17\%
\\
&
&
{\sf ER}, average degree $6$ 
&  
7\%  & 15\%  &  25\% & 7\% & 9\%  & 19\%
\\
&
&
{\sf SF}, average degree $3$ 
&  
0\%  & 3\%  &  19\% & 0\% & 1\%  & 17\%
\\
&
&
{\sf SF}, average degree $6$ 
&  
7\%  & 10\%  &  21\% & 7\% & 8\%  & 17\%
\\
%
%%%%%%%%%%%%%%%%%%%%%%%%%%%%%%%%%%%%%%%%%%%%%%%%%%%%%%%%%%%%%%%%%%%%%%%%%%%%%%%%%%%%%%%%%%%%%%%%%%%%%%%%%%
\hline
\multirow{15}{*}{${|V|=300}$}
&
\multirow{5}{*}{homogeneous} 
&
in-arborescence
&  
\bf 94\%  & \bf 95\%  &  \bf 95\% & \bf 85\% \bf & \bf 95\%  & \bf 95\%
\\
&
&
{\sf ER}, average degree $3$ 
&  
\bf 96\%  & \bf 100\%  &  \bf 100\% & \bf 67\% \bf & \bf 98\%  & \bf 100\%
\\
&
&
{\sf ER}, average degree $6$ 
&  
\bf 57\%  & \bf 57\%  &  \bf 57\% & \bf 57\% \bf & \bf 57\%  & \bf 57\%
\\
&
&
{\sf SF}, average degree $3$ 
&  
\bf 33\%  & \bf 71\%  &  \bf 100\% & \bf 33\% \bf & \bf 69\%  & \bf 100\%
\\
&
&
{\sf SF}, average degree $6$ 
&  
\bf 100\%  & \bf 100\%  &  \bf 100\% & \bf 99\% \bf & \bf 100\%  & \bf 100\%
\\
\cline{2-9}
& 
\multirow{5}{*}{${(0.1,0.95)}$-heterogeneous}
&
in-arborescence
&  
1\%  & 1\%  &  12\% & 1\% & 1\%  & 12\%
\\
&
&
{\sf ER}, average degree $3$ 
&  
1\%  & 5\%  &  16\% & 0\% & 2\%  & 13\%
\\
&
&
{\sf ER}, average degree $6$ 
&  
7\%  & 7\%  &  16\% & 6\% & 7\%  & 15\%
\\
&
&
{\sf SF}, average degree $3$ 
&  
0\%  & 9\%  &  25\% & 0\% & 3\%  & 19\%
\\
&
&
{\sf SF}, average degree $6$ 
&  
7\%  & 17\%  &  31\% & 7\% & 13\%  & 21\%
\\
\cline{2-9}
& 
\multirow{5}{*}{${(0.2,0.6)}$-heterogeneous}
&
in-arborescence
&  
1\%  & 1\%  &  11\% & 1\% & 1\%  & 11\%
\\
&
&
{\sf ER}, average degree $3$ 
&  
7\%  & 8\%  &  17\% & 3\% & 7\%  & 17\%
\\
&
&
{\sf ER}, average degree $6$ 
&  
7\%  & 9\%  &  19\% & 7\% & 8\%  & 17\%
\\
&
&
{\sf SF}, average degree $3$ 
&  
0\%  & 3\%  &  17\% & 0\% & 1\%  & 16\%
\\
&
&
{\sf SF}, average degree $6$ 
&  
7\%  & 8\%  &  17\% & 7\% & 7\%  & 16\%
\\
\cline{1-9}
\end{tabular}
}
\end{center}
\end{table}

\clearpage
%%%%%%%%%%%%%%%%%%%%%%%%%%%%%%%%%%%%%%%%%%%%%%%%%%%%%%%%%%%%%%%%%%%%%%%%%%%%%%%%%%%%%%%%%%%%%%%%%%%%%%%%%%%%%%
%%%%%%%%%%%%%%%%%%%%%%%%%%%%%%%%%%%%%%%%%%%%%%%%%%%%%%%%%%%%%%%%%%%%%%%%%%%%%%%%%%%%%%%%%%%%%%%%%%%%%%%%%%%%%%
%%%%%%%%%%%%%%%%%%%%%%%%%%%%%%%%%%%%%%%%%%%%%%%%%%%%%%%%%%%%%%%%%%%%%%%%%%%%%%%%%%%%%%%%%%%%%%%%%%%%%%%%%%%%%%
%%% Third table for analysis of residual instability of homogeneous versus heterogeneous networks
%%% random shock

\begin{table}
\caption{Residual instabilities of homogeneous versus heterogeneous networks under coordinated shocks. 
The percentages shown are the percentages of networks for which $\xi<0.05$ or $\xi<0.1$ or $\xi<0.2$.} 
\label{supp3-table7}
\begin{center}
\scalebox{0.75}[0.8]
{%%%%%\small
\begin{tabular}{c c r      r r r | r r r}
&
&
&
\multicolumn{6}{c}{\bf idiosyncratic shock}
\\
&
&
&
\multicolumn{3}{c|}{$\Phi=0.8,\gamma=0.75$}
&
\multicolumn{3}{c}{$\Phi=0.8,\gamma=0.70$}
\\
&
&
&
\multicolumn{1}{c}{${\xi<0.05}$}
&
\multicolumn{1}{c}{${\xi<0.1}$}
&
\multicolumn{1}{c|}{${\xi<0.2}$}
&
\multicolumn{1}{c}{${\xi<0.05}$}
&
\multicolumn{1}{c}{${\xi<0.1}$}
&
\multicolumn{1}{c}{${\xi<0.2}$}
\\
\cline{1-9}
\multirow{15}{*}{${|V|=50}$}
&
\multirow{5}{*}{homogeneous} 
&
in-arborescence
&  
\bf 94\%  & \bf 95\%  &  \bf 95\% & \bf 83\% \bf & \bf 94\%  & \bf 95\%
\\
&
&
{\sf ER}, average degree $3$ 
&  
\bf 100\%  & \bf 100\%  &  \bf 100\% & \bf 89\% \bf & \bf 100\%  & \bf 100\%
\\
&
&
{\sf ER}, average degree $6$ 
&  
\bf 57\%  & \bf 60\%  &  \bf 63\% & \bf 57\% \bf & \bf 57\%  & \bf 58\%
\\
&
&
{\sf SF}, average degree $3$ 
&  
\bf 55\%  & \bf 100\%  &  \bf 100\% & \bf 52\% \bf & \bf 100\%  & \bf 100\%
\\
&
&
{\sf SF}, average degree $6$ 
&  
\bf 100\%  & \bf 100\%  &  \bf 100\% & \bf 100\% \bf & \bf 100\%  & \bf 100\%
\\
\cline{2-9}
& 
\multirow{5}{*}{${(0.1,0.95)}$-heterogeneous}
&
in-arborescence
&  
4\%  & 9\%  &  21\% & 1\% \bf & 2\%  & 12\%
\\
&
&
{\sf ER}, average degree $3$ 
&  
0\%  & 7\%  &  21\% & 0\% \bf & 1\%  & 15\%
\\
&
&
{\sf ER}, average degree $6$ 
&  
10\%  & 21\%  &  31\% & 7\% \bf & 9\%  & 17\%
\\
&
&
{\sf SF}, average degree $3$ 
&  
1\%  & 11\%  &  27\% & 0\% \bf & 5\%  & 20\%
\\
&
&
{\sf SF}, average degree $6$ 
&  
9\%  & 23\%  &  43\% & 9\% \bf & 17\%  & 31\%
\\
\cline{2-9}
& 
\multirow{5}{*}{${(0.2,0.6)}$-heterogeneous}
&
in-arborescence
&  
2\%  & 5\%  &  15\% & 1\% \bf & 1\%  & 12\%
\\
&
&
{\sf ER}, average degree $3$ 
&  
8\%  & 18\%  &  29\% & 7\% \bf & 11\%  & 21\%
\\
&
&
{\sf ER}, average degree $6$ 
&  
15\%  & 25\%  &  38\% & 7\% \bf & 15\%  & 27\%
\\
&
&
{\sf SF}, average degree $3$ 
&  
1\%  & 12\%  &  23\% & 0\% \bf & 7\%  & 19\%
\\
&
&
{\sf SF}, average degree $6$ 
&  
12\%  & 21\%  &  33\% & 7\% \bf & 10\%  & 22\%
\\
%
%%%%%%%%%%%%%%%%%%%%%%%%%%%%%%%%%%%%%%%%%%%%%%%%%%%%%%%%%%%%%%%%%%%%%%%%%%%%%%%%%%%%%%%%%%%%%%%%%%%%%%%%%%
\hline
\multirow{15}{*}{${|V|=100}$}
&
\multirow{5}{*}{homogeneous} 
&
in-arborescence
&  
\bf 94\%  & \bf 95\%  &  \bf 95\% & \bf 84\% \bf & \bf 94\%  & \bf 95\%
\\
&
&
{\sf ER}, average degree $3$ 
&  
\bf 78\%  & \bf 100\%  &  \bf 100\% & \bf 68\% \bf & \bf 100\%  & \bf 100\%
\\
&
&
{\sf ER}, average degree $6$ 
&  
\bf 57\%  & \bf 57\%  &  \bf 59\% & \bf 57\% \bf & \bf 57\%  & \bf 57\%
\\
&
&
{\sf SF}, average degree $3$ 
&  
\bf 37\%  & \bf 79\%  &  \bf 100\% & \bf 35\% \bf & \bf 81\%  & \bf 100\%
\\
&
&
{\sf SF}, average degree $6$ 
&  
\bf 100\%  & \bf 100\%  &  \bf 100\% & \bf 100\% \bf & \bf 100\%  & \bf 100\%
\\
\cline{2-9}
& 
\multirow{5}{*}{${(0.1,0.95)}$-heterogeneous}
&
in-arborescence
&  
1\%  & 3\%  &  13\% & 1\% & 1\%  & 13\%
\\
&
&
{\sf ER}, average degree $3$ 
&  
0\%  & 4\%  &  17\% & 0\% & 3\%  & 15\%
\\
&
&
{\sf ER}, average degree $6$ 
&  
7\%  & 9\%  &  21\% & 7\% & 7\%  & 16\%
\\
&
&
{\sf SF}, average degree $3$ 
&  
0\%  & 9\%  &  25\% & 0\% & 4\%  & 19\%
\\
&
&
{\sf SF}, average degree $6$ 
&  
10\%  & 19\%  &  35\% & 7\% & 17\%  & 25\%
\\
\cline{2-9}
& 
\multirow{5}{*}{${(0.2,0.6)}$-heterogeneous}
&
in-arborescence
&  
1\%  & 2\%  &  13\% & 1\% & 1\%  & 11\%
\\
&
&
{\sf ER}, average degree $3$ 
&  
5\%  & 12\%  &  21\% & 5\% & 9\%  & 19\%
\\
&
&
{\sf ER}, average degree $6$ 
&  
9\%  & 16\%  &  28\% & 7\% & 11\%  & 20\%
\\
&
&
{\sf SF}, average degree $3$ 
&  
0\%  & 4\%  &  19\% & 0\% & 1\%  & 17\%
\\
&
&
{\sf SF}, average degree $6$ 
&  
8\%  & 11\%  &  23\% & 7\% & 8\%  & 17\%
\\
%
%%%%%%%%%%%%%%%%%%%%%%%%%%%%%%%%%%%%%%%%%%%%%%%%%%%%%%%%%%%%%%%%%%%%%%%%%%%%%%%%%%%%%%%%%%%%%%%%%%%%%%%%%%
\hline
\multirow{15}{*}{${|V|=300}$}
&
\multirow{5}{*}{homogeneous} 
&
in-arborescence
&  
\bf 94\%  & \bf 95\%  &  \bf 96\% & \bf 92\% \bf & \bf 95\%  & \bf 95\%
\\
&
&
{\sf ER}, average degree $3$ 
&  
\bf 96\%  & \bf 100\%  &  \bf 100\% & \bf 73\% \bf & \bf 100\%  & \bf 100\%
\\
&
&
{\sf ER}, average degree $6$ 
&  
\bf 57\%  & \bf 57\%  &  \bf 57\% & \bf 57\% \bf & \bf 57\%  & \bf 57\%
\\
&
&
{\sf SF}, average degree $3$ 
&  
\bf 33\%  & \bf 67\%  &  \bf 100\% & \bf 31\% \bf & \bf 69\%  & \bf 100\%
\\
&
&
{\sf SF}, average degree $6$ 
&  
\bf 100\%  & \bf 100\%  &  \bf 100\% & \bf 100\% \bf & \bf 100\%  & \bf 100\%
\\
\cline{2-9}
& 
\multirow{5}{*}{${(0.1,0.95)}$-heterogeneous}
&
in-arborescence
&  
1\%  & 1\%  &  12\% & 1\% & 1\%  & 12\%
\\
&
&
{\sf ER}, average degree $3$ 
&  
1\%  & 5\%  &  16\% & 1\% & 3\%  & 16\%
\\
&
&
{\sf ER}, average degree $6$ 
&  
7\%  & 7\%  &  17\% & 7\% & 7\%  & 15\%
\\
&
&
{\sf SF}, average degree $3$ 
&  
0\%  & 9\%  &  25\% & 0\% & 7\%  & 22\%
\\
&
&
{\sf SF}, average degree $6$ 
&  
9\%  & 17\%  &  31\% & 7\% & 13\%  & 25\%
\\
\cline{2-9}
& 
\multirow{5}{*}{${(0.2,0.6)}$-heterogeneous}
&
in-arborescence
&  
1\%  & 1\%  &  11\% & 1\% & 1\%  & 11\%
\\
&
&
{\sf ER}, average degree $3$ 
&  
7\%  & 8\%  &  18\% & 6\% & 7\%  & 17\%
\\
&
&
{\sf ER}, average degree $6$ 
&  
7\%  & 9\%  &  20\% & 7\% & 8\%  & 17\%
\\
&
&
{\sf SF}, average degree $3$ 
&  
0\%  & 2\%  &  17\% & 0\% & 1\%  & 17\%
\\
&
&
{\sf SF}, average degree $6$ 
&  
7\%  & 8\%  &  17\% & 7\% & 7\%  & 17\%
\\
\cline{1-9}
\end{tabular}
}
\end{center}
\end{table}

\clearpage
%%%%%%%%%%%%%%%%%%%%%%%%%%%%%%%%%%%%%%%%%%%%%%%%%%%%%%%%%%%%%%%%%%%%%%%%%%%%%%%%%%%%%%%%%%%%%%%%%%%%%%%%%%%%%%
%%%%%%%%%%%%%%%%%%%%%%%%%%%%%%%%%%%%%%%%%%%%%%%%%%%%%%%%%%%%%%%%%%%%%%%%%%%%%%%%%%%%%%%%%%%%%%%%%%%%%%%%%%%%%%
%%%%%%%%%%%%%%%%%%%%%%%%%%%%%%%%%%%%%%%%%%%%%%%%%%%%%%%%%%%%%%%%%%%%%%%%%%%%%%%%%%%%%%%%%%%%%%%%%%%%%%%%%%%%%%
%%% Fourth table for analysis of residual instability of homogeneous versus heterogeneous networks
%%% random shock

\begin{table}
\caption{Residual instabilities of homogeneous versus heterogeneous networks under coordinated shocks. 
The percentages shown are the percentages of networks for which $\xi<0.05$ or $\xi<0.1$ or $\xi<0.2$.} 
\label{supp4-table7}
\begin{center}
\scalebox{0.75}[0.8]
{%%%%%\small
\begin{tabular}{c c r      r r r | r r r}
&
&
&
\multicolumn{6}{c}{\bf idiosyncratic shock}
\\
&
&
&
\multicolumn{3}{c|}{$\Phi=0.9,\gamma=0.85$}
&
\multicolumn{3}{c}{$\Phi=0.9,\gamma=0.80$}
\\
&
&
&
\multicolumn{1}{c}{${\xi<0.05}$}
&
\multicolumn{1}{c}{${\xi<0.1}$}
&
\multicolumn{1}{c|}{${\xi<0.2}$}
&
\multicolumn{1}{c}{${\xi<0.05}$}
&
\multicolumn{1}{c}{${\xi<0.1}$}
&
\multicolumn{1}{c}{${\xi<0.2}$}
\\
\cline{1-9}
\multirow{15}{*}{${|V|=50}$}
&
\multirow{5}{*}{homogeneous} 
&
in-arborescence
&  
\bf 94\%  & \bf 95\%  &  \bf 95\% & \bf 93\% \bf & \bf 94\%  & \bf 95\%
\\
&
&
{\sf ER}, average degree $3$ 
&  
\bf 100\%  & \bf 100\%  &  \bf 100\% & \bf 96\% \bf & \bf 100\%  & \bf 100\%
\\
&
&
{\sf ER}, average degree $6$ 
&  
\bf 5\%  & \bf 5\%  &  \bf 8\% & \bf 57\% \bf & \bf 57\%  & \bf 59\%
\\
&
&
{\sf SF}, average degree $3$ 
&  
\bf 55\%  & \bf 100\%  &  \bf 100\% & \bf 53\% \bf & \bf 100\%  & \bf 100\%
\\
&
&
{\sf SF}, average degree $6$ 
&  
\bf 100\%  & \bf 100\%  &  \bf 100\% & \bf 100\% \bf & \bf 100\%  & \bf 100\%
\\
\cline{2-9}
& 
\multirow{5}{*}{${(0.1,0.95)}$-heterogeneous}
&
in-arborescence
&  
9\%  & 11\%  &  21\% & 1\% \bf & 3\%  & 14\%
\\
&
&
{\sf ER}, average degree $3$ 
&  
1\%  & 9\%  &  23\% & 0\% \bf & 2\%  & 14\%
\\
&
&
{\sf ER}, average degree $6$ 
&  
11\%  & 21\%  &  31\% & 7\% \bf & 10\%  & 22\%
\\
&
&
{\sf SF}, average degree $3$ 
&  
1\%  & 11\%  &  27\% & 0\% \bf & 7\%  & 20\%
\\
&
&
{\sf SF}, average degree $6$ 
&  
12\%  & 23\%  &  44\% & 9\% \bf & 20\%  & 35\%
\\
\cline{2-9}
& 
\multirow{5}{*}{${(0.2,0.6)}$-heterogeneous}
&
in-arborescence
&  
2\%  & 4\%  &  15\% & 1\% \bf & 2\%  & 12\%
\\
&
&
{\sf ER}, average degree $3$ 
&  
8\%  & 17\%  &  31\% & 7\% \bf & 12\%  & 22\%
\\
&
&
{\sf ER}, average degree $6$ 
&  
17\%  & 25\%  &  39\% & 9\% \bf & 17\%  & 29\%
\\
&
&
{\sf SF}, average degree $3$ 
&  
1\%  & 12\%  &  25\% & 0\% \bf & 8\%  & 19\%
\\
&
&
{\sf SF}, average degree $6$ 
&  
17\%  & 23\%  &  35\% & 9\% \bf & 13\%  & 23\%
\\
%
%%%%%%%%%%%%%%%%%%%%%%%%%%%%%%%%%%%%%%%%%%%%%%%%%%%%%%%%%%%%%%%%%%%%%%%%%%%%%%%%%%%%%%%%%%%%%%%%%%%%%%%%%%
\hline
\multirow{15}{*}{${|V|=100}$}
&
\multirow{5}{*}{homogeneous} 
&
in-arborescence
&  
\bf 94\%  & \bf 95\%  &  \bf 95\% & \bf 92\% \bf & \bf 95\%  & \bf 95\%
\\
&
&
{\sf ER}, average degree $3$ 
&  
\bf 77\%  & \bf 100\%  &  \bf 100\% & \bf 74\% \bf & \bf 100\%  & \bf 100\%
\\
&
&
{\sf ER}, average degree $6$ 
&  
\bf 3\%  & \bf 3\%  &  \bf 5\% & \bf 57\% \bf & \bf 57\%  & \bf 57\%
\\
&
&
{\sf SF}, average degree $3$ 
&  
\bf 34\%  & \bf 81\%  &  \bf 100\% & \bf 37\% \bf & \bf 81\%  & \bf 100\%
\\
&
&
{\sf SF}, average degree $6$ 
&  
\bf 100\%  & \bf 100\%  &  \bf 100\% & \bf 100\% \bf & \bf 100\%  & \bf 100\%
\\
\cline{2-9}
& 
\multirow{5}{*}{${(0.1,0.95)}$-heterogeneous}
&
in-arborescence
&  
1\%  & 3\%  &  14\% & 1\% & 1\%  & 12\%
\\
&
&
{\sf ER}, average degree $3$ 
&  
0\%  & 6\%  &  19\% & 0\% & 1\%  & 16\%
\\
&
&
{\sf ER}, average degree $6$ 
&  
7\%  & 13\%  &  22\% & 7\% & 7\%  & 17\%
\\
&
&
{\sf SF}, average degree $3$ 
&  
0\%  & 9\%  &  25\% & 0\% & 6\%  & 19\%
\\
&
&
{\sf SF}, average degree $6$ 
&  
11\%  & 21\%  &  37\% & 7\% & 17\%  & 29\%
\\
\cline{2-9}
& 
\multirow{5}{*}{${(0.2,0.6)}$-heterogeneous}
&
in-arborescence
&  
1\%  & 2\%  &  13\% & 1\% & 1\%  & 11\%
\\
&
&
{\sf ER}, average degree $3$ 
&  
6\%  & 13\%  &  23\% & 5\% & 9\%  & 19\%
\\
&
&
{\sf ER}, average degree $6$ 
&  
9\%  & 19\%  &  31\% & 7\% & 11\%  & 21\%
\\
&
&
{\sf SF}, average degree $3$ 
&  
0\%  & 5\%  &  21\% & 0\% & 2\%  & 17\%
\\
&
&
{\sf SF}, average degree $6$ 
&  
9\%  & 14\%  &  25\% & 7\% & 8\%  & 18\%
\\
%
%%%%%%%%%%%%%%%%%%%%%%%%%%%%%%%%%%%%%%%%%%%%%%%%%%%%%%%%%%%%%%%%%%%%%%%%%%%%%%%%%%%%%%%%%%%%%%%%%%%%%%%%%%
\hline
\multirow{15}{*}{${|V|=300}$}
&
\multirow{5}{*}{homogeneous} 
&
in-arborescence
&  
\bf 94\%  & \bf 95\%  &  \bf 95\% & \bf 94\% \bf & \bf 95\%  & \bf 95\%
\\
&
&
{\sf ER}, average degree $3$ 
&  
\bf 0\%  & \bf 0\%  &  \bf 100\% & \bf 85\% \bf & \bf 100\%  & \bf 100\%
\\
&
&
{\sf ER}, average degree $6$ 
&  
\bf 3\%  & \bf 3\%  &  \bf 3\% & \bf 57\% \bf & \bf 57\%  & \bf 57\%
\\
&
&
{\sf SF}, average degree $3$ 
&  
\bf 30\%  & \bf 69\%  &  \bf 100\% & \bf 31\% \bf & \bf 69\%  & \bf 100\%
\\
&
&
{\sf SF}, average degree $6$ 
&  
\bf 100\%  & \bf 100\%  &  \bf 100\% & \bf 100\% \bf & \bf 100\%  & \bf 100\%
\\
\cline{2-9}
& 
\multirow{5}{*}{${(0.1,0.95)}$-heterogeneous}
&
in-arborescence
&  
1\%  & 1\%  &  12\% & 1\% & 1\%  & 11\%
\\
&
&
{\sf ER}, average degree $3$ 
&  
0\%  & 5\%  &  17\% & 0\% & 3\%  & 16\%
\\
&
&
{\sf ER}, average degree $6$ 
&  
7\%  & 8\%  &  17\% & 7\% & 7\%  & 16\%
\\
&
&
{\sf SF}, average degree $3$ 
&  
0\%  & 9\%  &  25\% & 0\% & 7\%  & 22\%
\\
&
&
{\sf SF}, average degree $6$ 
&  
10\%  & 17\%  &  33\% & 6\% & 15\%  & 25\%
\\
\cline{2-9}
& 
\multirow{5}{*}{${(0.2,0.6)}$-heterogeneous}
&
in-arborescence
&  
1\%  & 1\%  &  11\% & 1\% & 1\%  & 11\%
\\
&
&
{\sf ER}, average degree $3$ 
&  
7\%  & 9\%  &  18\% & 6\% & 7\%  & 17\%
\\
&
&
{\sf ER}, average degree $6$ 
&  
7\%  & 10\%  &  21\% & 7\% & 9\%  & 17\%
\\
&
&
{\sf SF}, average degree $3$ 
&  
0\%  & 1\%  &  17\% & 0\% & 1\%  & 17\%
\\
&
&
{\sf SF}, average degree $6$ 
&  
7\%  & 9\%  &  17\% & 7\% & 7\%  & 16\%
\\
\cline{1-9}
\end{tabular}
}
\end{center}
\end{table}

\renewcommand{\thefigure}{S\arabic{figure}}
\setcounter{figure}{0}

%%\clearpage
\newpage

%%%%%%%%%%%%%%%%%%%%%%%%%%%%%%%%%%%%%%%%%%%%%%%%%%%%%%%%%%%%%%%%%%%%%%%%%%%%%%%%%%%%%%%%%%%%%%%%%%%%%%%%%%%%%%
%%%%%%%%%%%%%%%%%%%%%%%%%%%%%%%%%%%%%%%%%%%%%%%%%%%%%%%%%%%%%%%%%%%%%%%%%%%%%%%%%%%%%%%%%%%%%%%%%%%%%%%%%%%%%%
%%% newfig1 
%%\input{../tex-figures/newfig1-coordinated-only.tex}

\begin{figure}
\vspace*{0.7in}
%%\hspace*{-0.0in}
\renewcommand{\thesubfigure}{\hspace*{0.6in}\protect\raisebox{0.2in}[0in][0.0in]{(\Alph{subfigure})}}
{
\subfigure[]{
\scalebox{0.7}[0.6]{

}
}
}
%%%%
\vspace*{-0.3in}
\caption{\label{newfig1}Effect of variations of equity to asset ratio (with respect to shock) on the vulnerability index $\xi$ for homogeneous networks.
Lower values of $\xi$ imply higher global stability of a network.
}
\end{figure}

%%
%%%%%%%%%%%%%%%%%%%%%%%%%%%%%%%%%%%%%%%%%%%%%%%%%%%%%%%%%%%%%%%%%%%%%%%%%%%%%%%%%%%%%%%%%%%%%%%%%%%%%%%%%%%%%%
%%%%%%%%%%%%%%%%%%%%%%%%%%%%%%%%%%%%%%%%%%%%%%%%%%%%%%%%%%%%%%%%%%%%%%%%%%%%%%%%%%%%%%%%%%%%%%%%%%%%%%%%%%%%%%

\clearpage
%%%%%%%%%%%%%%%%%%%%%%%%%%%%%%%%%%%%%%%%%%%%%%%%%%%%%%%%%%%%%%%%%%%%%%%%%%%%%%%%%%%%%%%%%%%%%%%%%%%%%%%%%%%%%%
%%%%%%%%%%%%%%%%%%%%%%%%%%%%%%%%%%%%%%%%%%%%%%%%%%%%%%%%%%%%%%%%%%%%%%%%%%%%%%%%%%%%%%%%%%%%%%%%%%%%%%%%%%%%%%
%%% newfig2
%%\input{../tex-figures/newfig2-coordinated-only.tex}

\begin{figure}
\vspace*{0.5in}
%%\hspace*{-0.0in}
%
\subfigure{
\scalebox{0.8}[0.7]{

}
}
%%%%
\vspace*{-0.5in}
\caption{\label{newfig2}Effect of variations of equity to asset ratio (with respect to shock) on the vulnerability index $\xi$ for 
$(\alpha,\beta)$-heterogeneous networks. Lower values of $\xi$ imply higher global stability of a network.
}
\end{figure}

%%%%%%%%%%%%%%%%%%%%%%%%%%%%%%%%%%%%%%%%%%%%%%%%%%%%%%%%%%%%%%%%%%%%%%%%%%%%%%%%%%%%%%%%%%%%%%%%%%%%%%%%%%%%%%
%%%%%%%%%%%%%%%%%%%%%%%%%%%%%%%%%%%%%%%%%%%%%%%%%%%%%%%%%%%%%%%%%%%%%%%%%%%%%%%%%%%%%%%%%%%%%%%%%%%%%%%%%%%%%%

\clearpage
%%%%\thispagestyle{empty}
%%%%%%%%%%%%%%%%%%%%%%%%%%%%%%%%%%%%%%%%%%%%%%%%%%%%%%%%%%%%%%%%%%%%%%%%%%%%%%%%%%%%%%%%%%%%%%%%%%%%%%%%%%%%%%
%%%%%%%%%%%%%%%%%%%%%%%%%%%%%%%%%%%%%%%%%%%%%%%%%%%%%%%%%%%%%%%%%%%%%%%%%%%%%%%%%%%%%%%%%%%%%%%%%%%%%%%%%%%%%%
%%% newfig3
%%\input{../tex-figures/newfig3-coordinated-only.tex}

\begin{figure*}
\vspace*{1.0in}
%
%%\hspace*{-0.0in}
%
\subfigure{
\scalebox{0.7}[0.5]{

}
}
\vspace*{-2.0in}
\caption{\label{newfig3}Effect of variations of equity to asset ratio (with respect to shock) on the vulnerability index $\xi$ for 
$(\alpha,\beta)$-heterogeneous networks. Lower values of $\xi$ imply higher global stability of a network.
}
\end{figure*}

%%%%%%%%%%%%%%%%%%%%%%%%%%%%%%%%%%%%%%%%%%%%%%%%%%%%%%%%%%%%%%%%%%%%%%%%%%%%%%%%%%%%%%%%%%%%%%%%%%%%%%%%%%%%%%
%%%%%%%%%%%%%%%%%%%%%%%%%%%%%%%%%%%%%%%%%%%%%%%%%%%%%%%%%%%%%%%%%%%%%%%%%%%%%%%%%%%%%%%%%%%%%%%%%%%%%%%%%%%%%%

\clearpage
%%%%%%%%%%%%%%%%%%%%%%%%%%%%%%%%%%%%%%%%%%%%%%%%%%%%%%%%%%%%%%%%%%%%%%%%%%%%%%%%%%%%%%%%%%%%%%%%%%%%%%%%%%%%%%
%%%%%%%%%%%%%%%%%%%%%%%%%%%%%%%%%%%%%%%%%%%%%%%%%%%%%%%%%%%%%%%%%%%%%%%%%%%%%%%%%%%%%%%%%%%%%%%%%%%%%%%%%%%%%%
%%%%%%%%%%%%%%%%%%%%%%%%%%%%%%%%%%%%%%%%%%%%%%%%%%%%%%%%%%%%%%%%%%%%%%%%%%%%%%%%%%%%%%%%%%%%%%%%%%%%%%%%%%%%%%
%%% newfig4 
%%\input{../tex-figures/newfig4-coordinated-only.tex}

\begin{figure*}
\vspace*{1.2in}
%%\hspace*{-0.0in}
\begin{minipage}[t]{2in}
\scalebox{0.43}[0.43]{
\begin{pspicture}(-3,-7.5)(-1,9.5)
\psset{xunit=0.4in,yunit=0.4in,runit=0.4in}
\psset{nameX=\LARGE$\pmb{\gamma}$,nameY=\LARGE$\pmb{E}$,nameZ=\LARGE$\pmb{\xi}$}
\psset{coorType=1,Alpha=135,hiddenLine=true}
\def\psxyzlabel#1{\bgroup\large\bf\textsf{#1}\egroup}
%
%%\pstThreeDPlaneGrid[planeGrid=xy,linewidth=0.1pt,ysubticks=14,xsubticks=9,linecolor=darkgray](0,0)(9,14)
%%\pstThreeDPlaneGrid[planeGridOffset=10,linewidth=0.1pt,ysubticks=14,xsubticks=9,linecolor=darkgray](0,0)(9,14)
%
%%\pstThreeDPlaneGrid[planeGrid=xz,linewidth=1pt,xsubticks=9,ysubticks=10,linecolor=darkgray](0,0)(9,10)
%%\pstThreeDPlaneGrid[planeGridOffset=14,planeGrid=xz,linewidth=1pt,xsubticks=9,ysubticks=10,linecolor=darkgray](0,0)(9,10)
%
%\pstThreeDPlaneGrid[planeGrid=yz,linewidth=0.1pt,xsubticks=14,ysubticks=10,linecolor=darkgray](0,0)(14,10)
%%\pstThreeDPlaneGrid[planeGridOffset=1,planeGrid=yz,linewidth=0.1pt,xsubticks=14,ysubticks=10,linecolor=darkgray](0,0)(14,10)
%%\pstThreeDPlaneGrid[planeGridOffset=9,planeGrid=yz,linewidth=0.1pt,xsubticks=14,ysubticks=10,linecolor=darkgray](0,0)(14,10)
%
%%%%%%%%%%%%%%%%%%%%%%%%%%%%%%%%%%%%%%%%%%%%%%%%%%%%%%%%%%%%%%%%%%%%%%%%%%%%%%%%%%%%%%%
%%%% legends
%%%%%%%%%%%%%%%%%%%%%%%%%%%%%%%%%%%%%%%%%%%%%%%%%%%%%%%%%%%%%%%%%%%%%%%%%%%%%%%%%%%%%%%
%
\rput[mr](5,-7.5){\LARGE\bf (A)}
\psline[linewidth=8pt,linecolor=black,arrowsize=1.5pt 2](7.50,16.2)(8.5,16.2)
\rput[mr](7.35,16.2){\LARGE\bf {\sf ER} model (average degree $\pmb{6}$), coordinated shock}
\psline[linewidth=8pt,linecolor=red,arrowsize=1.5pt 2](24.65,16.2)(25.65,16.2)
\rput[mr](24.55,16.2){\LARGE\bf {\sf ER} model (average degree $\pmb{3}$), coordinated shock}
%%
%%\psline[linewidth=2pt,linecolor=black,arrowsize=1.5pt 2](17.20,15.3)(18.20,15.3)
%%\rput[mr](17.05,15.3){\Large\bf {\sf ER} model (average degree $\pmb{6}$), random shock}
%%\psline[linewidth=2pt,linecolor=gray,arrowsize=1.5pt 2](31.55,15.3)(32.55,15.3)
%%\rput[mr](31.05,15.3){\Large\bf {\sf ER} model (average degree $\pmb{3}$), random shock}
%%
\psline[linewidth=8pt,linecolor=yellow,arrowsize=1.5pt 2](7.50,14.4)(8.5,14.4)
\rput[mr](7.35,14.4){\LARGE\bf {\sf SF} model (average degree $\pmb{6}$), coordinated shock}
\psline[linewidth=8pt,linecolor=blue,arrowsize=1.5pt 2](24.65,14.4)(25.65,14.4)
\rput[mr](24.55,14.4){\LARGE\bf {\sf SF} model (average degree $\pmb{3}$), coordinated shock}
%%
%%\psline[linewidth=2pt,linecolor=green,arrowsize=1.5pt 2](17.20,13.5)(18.20,13.5)
%%\rput[mr](17.05,13.3){\Large\bf {\sf SF} model (average degree $\pmb{6}$), random shock}
%%\psline[linewidth=2pt,linecolor=red,arrowsize=1.5pt 2](31.55,13.5)(32.55,13.5)
%%\rput[mr](31.05,13.5){\Large\bf {\sf SF} model (average degree $\pmb{3}$), random shock}
%%
\psline[linewidth=8pt,linecolor=black,linestyle=dashed,arrowsize=1.5pt 2](10,12.5)(11,12.5)
\rput[mr](9.85,12.5){\LARGE\bf in-arborescence (average degree $\pmb{\approx 1}$), coordinated shock}
%%
%%\psline[linewidth=8pt,linecolor=gray,linestyle=dashed,arrowsize=1.5pt 2](25.95,12.5)(26.95,12.5)
%%\rput[mr](25.85,12.5){\Large\bf in-arborescence (average degree $\pmb{\approx 1}$), random shock}
%
\rput[ml](1,11.0){\Large$\pmb{|V|=50}$ \hspace*{0.1in} $\pmb{\mathcal{K}=0.1}$ \hspace*{0.1in}$\pmb{\Phi=0.5}$ \hspace*{0.1in}$\pmb{\gamma=0.05,0.25,0.45}$}
%
%%%%%%%%%%%%%%%% \gamma=0.05,0.25,0.45
%
%%%%%%%%%%%%%%%%%%%%%%%%%%%%%%%%%%%%%%%%%%%%%%%%%%%%%%%%%%%%%%%%%%%%%%%%%%%%%%%%%%%%%%%
%%%%  ER, coordinated, ave degree 6
%%%%%%%%%%%%%%%%%%%%%%%%%%%%%%%%%%%%%%%%%%%%%%%%%%%%%%%%%%%%%%%%%%%%%%%%%%%%%%%%%%%%%%%
%
\pstThreeDLine[linewidth=8pt,linecolor=black,arrowsize=1.5pt 2](1,1,01.06)(1,2,01.06)(1,3,01.06)(1,4,01.36)(1,5,01.36)(1,6,01.36)(1,7,01.36)(1,8,01.64)(1,9,01.64)(1,10,01.64)(1,11,01.64)(1,12,01.88)(1,13,01.88)(1,14,01.88)
\pstThreeDLine[linewidth=8pt,linecolor=black,arrowsize=1.5pt 2](5,1,00.14)(5,2,00.14)(5,3,00.14)(5,4,00.2)(5,5,00.2)(5,6,00.2)(5,7,00.2)(5,8,00.3)(5,9,00.3)(5,10,00.3)(5,11,00.3)(5,12,00.3)(5,13,00.3)(5,14,00.3)
\pstThreeDLine[linewidth=8pt,linecolor=black,arrowsize=1.5pt 2](9,1,00.02)(9,2,00.02)(9,3,00.02)(9,4,00.0)(9,5,00.0)(9,6,00.0)(9,7,00.0)(9,8,00.0)(9,9,00.0)(9,10,00.0)(9,11,00.0)(9,12,00.0)(9,13,00.0)(9,14,00.0)
%
%%%%%%%%%%%%%%%%%%%%%%%%%%%%%%%%%%%%%%%%%%%%%%%%%%%%%%%%%%%%%%%%%%%%%%%%%%%%%%%%%%%%%%%
%%%%  ER, coordinated, ave degree 3
%%%%%%%%%%%%%%%%%%%%%%%%%%%%%%%%%%%%%%%%%%%%%%%%%%%%%%%%%%%%%%%%%%%%%%%%%%%%%%%%%%%%%%%
%
\pstThreeDLine[linewidth=8pt,linecolor=red,arrowsize=1.5pt 2](1,1,02.9)(1,2,02.9)(1,3,02.9)(1,4,03.14)(1,5,03.14)(1,6,03.14)(1,7,03.14)(1,8,03.62)(1,9,03.62)(1,10,03.62)(1,11,03.62)(1,12,04.02)(1,13,04.02)(1,14,04.02)
\pstThreeDLine[linewidth=8pt,linecolor=red,arrowsize=1.5pt 2](5,1,00.58)(5,2,00.58)(5,3,00.58)(5,4,00.4)(5,5,00.4)(5,6,00.4)(5,7,00.4)(5,8,00.52)(5,9,00.52)(5,10,00.52)(5,11,00.52)(5,12,00.58)(5,13,00.58)(5,14,00.58)
\pstThreeDLine[linewidth=8pt,linecolor=red,arrowsize=1.5pt 2](9,1,00.34)(9,2,00.34)(9,3,00.34)(9,4,00.04)(9,5,00.04)(9,6,00.04)(9,7,00.04)(9,8,00.04)(9,9,00.04)(9,10,00.04)(9,11,00.04)(9,12,00.04)(9,13,00.04)(9,14,00.04)
\pstThreeDLine[linewidth=8pt,linecolor=yellow,arrowsize=1.5pt 2](1,1,07.22)(1,2,07.22)(1,3,07.22)(1,4,06.92)(1,5,06.92)(1,6,06.92)(1,7,06.92)(1,8,06.66)(1,9,06.66)(1,10,06.66)(1,11,06.66)(1,12,06.48)(1,13,06.48)(1,14,06.48)
\pstThreeDLine[linewidth=8pt,linecolor=yellow,arrowsize=1.5pt 2](5,1,00.06)(5,2,00.06)(5,3,00.06)(5,4,00.04)(5,5,00.04)(5,6,00.04)(5,7,00.04)(5,8,00.06)(5,9,00.06)(5,10,00.06)(5,11,00.06)(5,12,00.06)(5,13,00.06)(5,14,00.06)
\pstThreeDLine[linewidth=8pt,linecolor=yellow,arrowsize=1.5pt 2](9,1,00.02)(9,2,00.02)(9,3,00.02)(9,4,00.0)(9,5,00.0)(9,6,00.0)(9,7,00.0)(9,8,00.0)(9,9,00.0)(9,10,00.0)(9,11,00.0)(9,12,00.0)(9,13,00.0)(9,14,00.0)
%
%%%%%%%%%%%%%%%%%%%%%%%%%%%%%%%%%%%%%%%%%%%%%%%%%%%%%%%%%%%%%%%%%%%%%%%%%%%%%%%%%%%%%%%
%%%%  SF, coordinated, ave degree 3
%%%%%%%%%%%%%%%%%%%%%%%%%%%%%%%%%%%%%%%%%%%%%%%%%%%%%%%%%%%%%%%%%%%%%%%%%%%%%%%%%%%%%%%
%
\pstThreeDLine[linewidth=8pt,linecolor=blue,arrowsize=1.5pt 2](1,1,08.84)(1,2,08.84)(1,3,08.84)(1,4,08.54)(1,5,08.54)(1,6,08.54)(1,7,08.54)(1,8,08.02)(1,9,08.02)(1,10,08.02)(1,11,08.02)(1,12,07.14)(1,13,07.14)(1,14,07.14)
\pstThreeDLine[linewidth=8pt,linecolor=blue,arrowsize=1.5pt 2](5,1,01.14)(5,2,01.14)(5,3,01.14)(5,4,00.26)(5,5,00.26)(5,6,00.26)(5,7,00.26)(5,8,00.18)(5,9,00.18)(5,10,00.18)(5,11,00.18)(5,12,00.22)(5,13,00.22)(5,14,00.22)
\pstThreeDLine[linewidth=8pt,linecolor=blue,arrowsize=1.5pt 2](9,1,00.72)(9,2,00.72)(9,3,00.72)(9,4,00.06)(9,5,00.06)(9,6,00.06)(9,7,00.06)(9,8,00.06)(9,9,00.06)(9,10,00.06)(9,11,00.06)(9,12,00.06)(9,13,00.06)(9,14,00.06)
\pstThreeDLine[linewidth=8pt,linecolor=black,linestyle=dashed,arrowsize=1.5pt 2](1,1,09.38)(1,2,09.38)(1,3,09.38)(1,4,07.4)(1,5,07.4)(1,6,07.4)(1,7,07.4)(1,8,06.96)(1,9,06.96)(1,10,06.96)(1,11,06.96)(1,12,06.78)(1,13,06.78)(1,14,06.78)
\pstThreeDLine[linewidth=8pt,linecolor=black,linestyle=dashed,arrowsize=1.5pt 2](5,1,07.58)(5,2,07.58)(5,3,07.58)(5,4,01.84)(5,5,01.84)(5,6,01.84)(5,7,01.84)(5,8,01.0)(5,9,01.0)(5,10,01.0)(5,11,01.0)(5,12,01.0)(5,13,01.0)(5,14,01.0)
\pstThreeDLine[linewidth=8pt,linecolor=black,linestyle=dashed,arrowsize=1.5pt 2](9,1,06.62)(9,2,06.62)(9,3,06.62)(9,4,00.3)(9,5,00.3)(9,6,00.3)(9,7,00.3)(9,8,00.3)(9,9,00.3)(9,10,00.3)(9,11,00.3)(9,12,00.34)(9,13,00.34)(9,14,00.34)
\pstThreeDCoor
[IIIDxTicksPlane=xz,IIIDxticksep=-0.6,IIIDyTicksPlane=yz,IIIDyticksep=-0.2,arrowsize=1.5pt 4,linewidth=1.5pt,linecolor=darkgray,IIIDzTicksPlane=yz,yMin=0,yMax=15.5,xMin=0,xMax=10.5,zMin=0,zMax=11,IIIDticks=true,IIIDticksize=0.0,Dy=025,Dx=0.05,Dz=0.1,IIIDlabels=true,IIIDOffset={(0,0,0)}] 
%%%%%%%%%%%%%%%%%
\end{pspicture}
}
\end{minipage}
%
%%%%%%%%%%%%%%%%%%%%%%%%%%%%%%%%%%%%
%%%%%%%%%%%%%%%%%%%%%%%%%%%%%%%%%%%%
%%%%%%%%%%%%%%%%%%%%%%%%%%%%%%%%%%%%
%% part 2 of figure starts
%
\hspace*{1.7in}
%
%%%%%%%%%%%%%%%%%%%%%%%%%%%%%%%%%%%%
%%%%%%%%%%%%%%%%%%%%%%%%%%%%%%%%%%%%
%%%%%%%%%%%%%%%%%%%%%%%%%%%%%%%%%%%%
%
\begin{minipage}[t]{2in}
\scalebox{0.43}[0.43]{
\begin{pspicture}(-4,-7.5)(-2,9.5)
\psset{xunit=0.4in,yunit=0.4in,runit=0.4in}
\psset{nameX=\LARGE$\pmb{\gamma}$,nameY=\LARGE$\pmb{E}$,nameZ=\LARGE$\pmb{\xi}$}
\psset{coorType=1,Alpha=135,hiddenLine=true}

%%\pstThreeDPlaneGrid[planeGrid=xy,linewidth=0.1pt,ysubticks=14,xsubticks=9,linecolor=darkgray](0,0)(9,14)
%%\pstThreeDPlaneGrid[planeGridOffset=10,linewidth=0.1pt,ysubticks=14,xsubticks=9,linecolor=darkgray](0,0)(9,14)
%
%%\pstThreeDPlaneGrid[planeGrid=xz,linewidth=1pt,xsubticks=9,ysubticks=10,linecolor=darkgray](0,0)(9,10)
%%\pstThreeDPlaneGrid[planeGridOffset=14,planeGrid=xz,linewidth=1pt,xsubticks=9,ysubticks=10,linecolor=darkgray](0,0)(9,10)
%
%\pstThreeDPlaneGrid[planeGrid=yz,linewidth=0.1pt,xsubticks=14,ysubticks=10,linecolor=darkgray](0,0)(14,10)
%%\pstThreeDPlaneGrid[planeGridOffset=1,planeGrid=yz,linewidth=0.1pt,xsubticks=14,ysubticks=10,linecolor=darkgray](0,0)(14,10)
%%\pstThreeDPlaneGrid[planeGridOffset=9,planeGrid=yz,linewidth=0.1pt,xsubticks=14,ysubticks=10,linecolor=darkgray](0,0)(14,10)
%
\rput[ml](1,11.0){\Large$\pmb{|V|=50}$ \hspace*{0.1in} $\pmb{\mathcal{K}=0.5}$ \hspace*{0.1in}$\pmb{\Phi=0.5}$ \hspace*{0.1in}$\pmb{\gamma=0.05,0.25,0.45}$}
\rput[mr](5,-7.5){\LARGE\bf (B)}
%
%
%%%%%%%%%%%%%%%% \gamma=0.05,0.25,0.45
%
%%%%%%%%%%%%%%%%%%%%%%%%%%%%%%%%%%%%%%%%%%%%%%%%%%%%%%%%%%%%%%%%%%%%%%%%%%%%%%%%%%%%%%%
%%%%  ER, coordinated, ave degree 6
%%%%%%%%%%%%%%%%%%%%%%%%%%%%%%%%%%%%%%%%%%%%%%%%%%%%%%%%%%%%%%%%%%%%%%%%%%%%%%%%%%%%%%%
%
\pstThreeDLine[linewidth=8pt,linecolor=black,arrowsize=1.5pt 2](1,1,05.03)(1,2,05.03)(1,3,05.03)(1,4,06.70)(1,5,06.70)(1,6,06.70)(1,7,06.70)(1,8,07.90)(1,9,07.90)(1,10,07.90)(1,11,07.90)(1,12,09.00)(1,13,09.00)(1,14,09.00)
\pstThreeDLine[linewidth=8pt,linecolor=black,arrowsize=1.5pt 2](5,1,00.80)(5,2,00.80)(5,3,00.80)(5,4,01.0)(5,5,01.0)(5,6,01.0)(5,7,01.0)(5,8,01.4)(5,9,01.4)(5,10,01.4)(5,11,01.4)(5,12,01.7)(5,13,01.7)(5,14,01.7)
\pstThreeDLine[linewidth=8pt,linecolor=black,arrowsize=1.5pt 2](9,1,00.00)(9,2,00.00)(9,3,00.00)(9,4,00.0)(9,5,00.0)(9,6,00.0)(9,7,00.0)(9,8,00.0)(9,9,00.0)(9,10,00.0)(9,11,00.0)(9,12,00.0)(9,13,00.0)(9,14,00.0)
%
%%%%%%%%%%%%%%%%%%%%%%%%%%%%%%%%%%%%%%%%%%%%%%%%%%%%%%%%%%%%%%%%%%%%%%%%%%%%%%%%%%%%%%%
%%%%  ER, coordinated, ave degree 3
%%%%%%%%%%%%%%%%%%%%%%%%%%%%%%%%%%%%%%%%%%%%%%%%%%%%%%%%%%%%%%%%%%%%%%%%%%%%%%%%%%%%%%%
%
\pstThreeDLine[linewidth=8pt,linecolor=red,arrowsize=1.5pt 2](1,1,05.8)(1,2,05.8)(1,3,05.8)(1,4,07.50)(1,5,07.50)(1,6,07.50)(1,7,07.50)(1,8,08.70)(1,9,08.70)(1,10,08.70)(1,11,08.70)(1,12,09.20)(1,13,09.20)(1,14,09.20)
\pstThreeDLine[linewidth=8pt,linecolor=red,arrowsize=1.5pt 2](5,1,01.60)(5,2,01.60)(5,3,01.60)(5,4,01.7)(5,5,01.7)(5,6,01.7)(5,7,01.7)(5,8,02.50)(5,9,02.50)(5,10,02.50)(5,11,02.50)(5,12,03.60)(5,13,03.60)(5,14,03.60)
\pstThreeDLine[linewidth=8pt,linecolor=red,arrowsize=1.5pt 2](9,1,00.40)(9,2,00.40)(9,3,00.40)(9,4,00.10)(9,5,00.10)(9,6,00.10)(9,7,00.10)(9,8,00.10)(9,9,00.10)(9,10,00.10)(9,11,00.10)(9,12,00.20)(9,13,00.20)(9,14,00.20)
\pstThreeDLine[linewidth=8pt,linecolor=yellow,arrowsize=1.5pt 2](1,1,09.60)(1,2,09.60)(1,3,09.60)(1,4,07.80)(1,5,07.80)(1,6,07.80)(1,7,07.80)(1,8,08.40)(1,9,08.40)(1,10,08.40)(1,11,08.40)(1,12,08.10)(1,13,08.10)(1,14,08.10)
\pstThreeDLine[linewidth=8pt,linecolor=yellow,arrowsize=1.5pt 2](5,1,00.60)(5,2,00.60)(5,3,00.60)(5,4,00.90)(5,5,00.90)(5,6,00.90)(5,7,00.90)(5,8,01.40)(5,9,01.40)(5,10,01.40)(5,11,01.40)(5,12,01.90)(5,13,01.90)(5,14,01.90)
\pstThreeDLine[linewidth=8pt,linecolor=yellow,arrowsize=1.5pt 2](9,1,00.00)(9,2,00.00)(9,3,00.00)(9,4,00.0)(9,5,00.0)(9,6,00.0)(9,7,00.0)(9,8,00.0)(9,9,00.0)(9,10,00.0)(9,11,00.0)(9,12,00.0)(9,13,00.0)(9,14,00.0)
%
%%%%%%%%%%%%%%%%%%%%%%%%%%%%%%%%%%%%%%%%%%%%%%%%%%%%%%%%%%%%%%%%%%%%%%%%%%%%%%%%%%%%%%%
%%%%  SF, coordinated, ave degree 3
%%%%%%%%%%%%%%%%%%%%%%%%%%%%%%%%%%%%%%%%%%%%%%%%%%%%%%%%%%%%%%%%%%%%%%%%%%%%%%%%%%%%%%%
%
\pstThreeDLine[linewidth=8pt,linecolor=blue,arrowsize=1.5pt 2](1,1,08.60)(1,2,08.60)(1,3,08.60)(1,4,08.70)(1,5,08.70)(1,6,08.70)(1,7,08.70)(1,8,08.70)(1,9,08.70)(1,10,08.70)(1,11,08.70)(1,12,08.70)(1,13,08.70)(1,14,08.70)
\pstThreeDLine[linewidth=8pt,linecolor=blue,arrowsize=1.5pt 2](5,1,02.60)(5,2,02.60)(5,3,02.60)(5,4,02.20)(5,5,02.20)(5,6,02.20)(5,7,02.20)(5,8,03.90)(5,9,03.90)(5,10,03.90)(5,11,03.90)(5,12,03.70)(5,13,03.70)(5,14,03.70)
\pstThreeDLine[linewidth=8pt,linecolor=blue,arrowsize=1.5pt 2](9,1,01.10)(9,2,01.10)(9,3,01.10)(9,4,00.40)(9,5,00.40)(9,6,00.40)(9,7,00.40)(9,8,00.40)(9,9,00.40)(9,10,00.40)(9,11,00.40)(9,12,00.40)(9,13,00.40)(9,14,00.40)
\pstThreeDLine[linewidth=8pt,linecolor=black,linestyle=dashed,arrowsize=1.5pt 2](1,1,10.00)(1,2,10.00)(1,3,10.00)(1,4,10.0)(1,5,10.0)(1,6,10.0)(1,7,10.0)(1,8,10.00)(1,9,10.00)(1,10,10.00)(1,11,10.00)(1,12,10.00)(1,13,10.00)(1,14,10.00)
\pstThreeDLine[linewidth=8pt,linecolor=black,linestyle=dashed,arrowsize=1.5pt 2](5,1,08.50)(5,2,08.50)(5,3,08.50)(5,4,04.50)(5,5,04.50)(5,6,04.50)(5,7,04.50)(5,8,03.7)(5,9,03.7)(5,10,03.7)(5,11,03.7)(5,12,03.7)(5,13,03.7)(5,14,03.7)
\pstThreeDLine[linewidth=8pt,linecolor=black,linestyle=dashed,arrowsize=1.5pt 2](9,1,06.60)(9,2,06.60)(9,3,06.60)(9,4,00.3)(9,5,00.3)(9,6,00.3)(9,7,00.3)(9,8,00.3)(9,9,00.3)(9,10,00.3)(9,11,00.3)(9,12,00.40)(9,13,00.40)(9,14,00.40)
%
%%%%%%%%%%%%%%%%%%%%%%%%%%%%%%%%%%%%%%%%%%%%%%%%%%%%%%%%%%%%%%%%%%%%%%%%%%%%%%%%%%%%%%%
%%%%  tree, random
%%%%%%%%%%%%%%%%%%%%%%%%%%%%%%%%%%%%%%%%%%%%%%%%%%%%%%%%%%%%%%%%%%%%%%%%%%%%%%%%%%%%%%%
%
%%\pstThreeDLine[linewidth=8pt,linecolor=gray,linestyle=dashed,arrowsize=1.5pt 2](1,1,05.80)(1,2,06.10)(1,3,05.80)(1,4,07.6)(1,5,06.8)(1,6,07.00)(1,7,07.30)(1,8,07.90)(1,9,08.30)(1,10,07.90)(1,11,08.20)(1,12,07.50)(1,13,08.70)(1,14,08.20)
%%\pstThreeDLine[linewidth=8pt,linecolor=gray,linestyle=dashed,arrowsize=1.5pt 2](5,1,04.30)(5,2,04.10)(5,3,04.40)(5,4,02.10)(5,5,02.20)(5,6,02.50)(5,7,02.00)(5,8,05.0)(5,9,05.0)(5,10,05.0)(5,11,05.0)(5,12,05.0)(5,13,05.0)(5,14,05.0)
%%\pstThreeDLine[linewidth=8pt,linecolor=gray,linestyle=dashed,arrowsize=1.5pt 2](9,1,03.4)(9,2,03.30)(9,3,03.30)(9,4,00.10)(9,5,00.20)(9,6,00.2)(9,7,00.1)(9,8,00.20)(9,9,00.20)(9,10,00.2)(9,11,00.2)(9,12,00.30)(9,13,00.20)(9,14,00.10)
\def\psxyzlabel#1{\bgroup\large\bf\textsf{#1}\egroup}
\pstThreeDCoor
[IIIDxTicksPlane=xz,IIIDxticksep=-0.6,IIIDyTicksPlane=yz,IIIDyticksep=-0.2,arrowsize=1.5pt 4,linewidth=1.5pt,linecolor=darkgray,IIIDzTicksPlane=yz,yMin=0,yMax=15.5,xMin=0,xMax=10.5,zMin=0,zMax=11,IIIDticks=true,IIIDticksize=0.0,Dy=025,Dx=0.05,Dz=0.1,IIIDlabels=true,IIIDOffset={(0,0,0)}] 
\end{pspicture}
}
\end{minipage}
%
%%%%%%%%%%%%%%%%%%%%%%%%%%%%%%%%%%%%
%%%%%%%%%%%%%%%%%%%%%%%%%%%%%%%%%%%%
%%%%%%%%%%%%%%%%%%%%%%%%%%%%%%%%%%%%
%% part 3 of figure starts
%
\\
\hspace*{-0.0in}
%
%%%%%%%%%%%%%%%%%%%%%%%%%%%%%%%%%%%%
%%%%%%%%%%%%%%%%%%%%%%%%%%%%%%%%%%%%
%%%%%%%%%%%%%%%%%%%%%%%%%%%%%%%%%%%%
%
\begin{minipage}[t]{2in}
\scalebox{0.43}[0.43]{
\begin{pspicture}(-3,-7)(-1.95,13)
\psset{xunit=0.4in,yunit=0.4in,runit=0.4in}
\psset{nameX=\LARGE$\pmb{\gamma}$,nameY=\LARGE$\pmb{E}$,nameZ=\LARGE$\pmb{\xi}$}
\psset{coorType=1,Alpha=135,hiddenLine=true}
\def\psxyzlabel#1{\bgroup\large\bf\textsf{#1}\egroup}

%
%%\pstThreeDPlaneGrid[planeGrid=xy,linewidth=0.1pt,ysubticks=14,xsubticks=9,linecolor=darkgray](0,0)(9,14)
%%\pstThreeDPlaneGrid[planeGridOffset=10,linewidth=0.1pt,ysubticks=14,xsubticks=9,linecolor=darkgray](0,0)(9,14)
%
%%\pstThreeDPlaneGrid[planeGrid=xz,linewidth=1pt,xsubticks=9,ysubticks=10,linecolor=darkgray](0,0)(9,10)
%%\pstThreeDPlaneGrid[planeGridOffset=14,planeGrid=xz,linewidth=1pt,xsubticks=9,ysubticks=10,linecolor=darkgray](0,0)(9,10)
%
%\pstThreeDPlaneGrid[planeGrid=yz,linewidth=0.1pt,xsubticks=14,ysubticks=10,linecolor=darkgray](0,0)(14,10)
%%\pstThreeDPlaneGrid[planeGridOffset=1,planeGrid=yz,linewidth=0.1pt,xsubticks=14,ysubticks=10,linecolor=darkgray](0,0)(14,10)
%%\pstThreeDPlaneGrid[planeGridOffset=9,planeGrid=yz,linewidth=0.1pt,xsubticks=14,ysubticks=10,linecolor=darkgray](0,0)(14,10)
%
\rput[ml](1,11.0){\Large$\pmb{|V|=100}$ \hspace*{0.1in} $\pmb{\mathcal{K}=0.1}$ \hspace*{0.1in}$\pmb{\Phi=0.5}$ \hspace*{0.1in}$\pmb{\gamma=0.05,0.25,0.45}$}
\rput[mr](5,-7.5){\LARGE\bf (C)}
%
%
%%%%%%%%%%%%%%%% \gamma=0.05,0.25,0.45
%
%%%%%%%%%%%%%%%%%%%%%%%%%%%%%%%%%%%%%%%%%%%%%%%%%%%%%%%%%%%%%%%%%%%%%%%%%%%%%%%%%%%%%%%
%%%%  ER, coordinated, ave degree 6
%%%%%%%%%%%%%%%%%%%%%%%%%%%%%%%%%%%%%%%%%%%%%%%%%%%%%%%%%%%%%%%%%%%%%%%%%%%%%%%%%%%%%%%
%
\pstThreeDLine[linewidth=8pt,linecolor=black,arrowsize=1.5pt 2](1,1,01.20)(1,2,01.20)(1,3,01.20)(1,4,01.40)(1,5,01.40)(1,6,01.40)(1,7,01.40)(1,8,01.60)(1,9,01.60)(1,10,01.60)(1,11,01.60)(1,12,01.90)(1,13,01.90)(1,14,01.90)
\pstThreeDLine[linewidth=8pt,linecolor=black,arrowsize=1.5pt 2](5,1,00.20)(5,2,00.20)(5,3,00.20)(5,4,00.2)(5,5,00.2)(5,6,00.2)(5,7,00.2)(5,8,00.3)(5,9,00.3)(5,10,00.3)(5,11,00.3)(5,12,00.3)(5,13,00.3)(5,14,00.3)
\pstThreeDLine[linewidth=8pt,linecolor=black,arrowsize=1.5pt 2](9,1,00.00)(9,2,00.00)(9,3,00.00)(9,4,00.0)(9,5,00.0)(9,6,00.0)(9,7,00.0)(9,8,00.0)(9,9,00.0)(9,10,00.0)(9,11,00.0)(9,12,00.0)(9,13,00.0)(9,14,00.0)
%
%%%%%%%%%%%%%%%%%%%%%%%%%%%%%%%%%%%%%%%%%%%%%%%%%%%%%%%%%%%%%%%%%%%%%%%%%%%%%%%%%%%%%%%
%%%%  ER, coordinated, ave degree 3
%%%%%%%%%%%%%%%%%%%%%%%%%%%%%%%%%%%%%%%%%%%%%%%%%%%%%%%%%%%%%%%%%%%%%%%%%%%%%%%%%%%%%%%
%
\pstThreeDLine[linewidth=8pt,linecolor=red,arrowsize=1.5pt 2](1,1,02.3)(1,2,02.3)(1,3,02.3)(1,4,02.30)(1,5,02.30)(1,6,02.30)(1,7,02.30)(1,8,03.00)(1,9,03.00)(1,10,03.00)(1,11,03.00)(1,12,03.30)(1,13,03.30)(1,14,03.30)
\pstThreeDLine[linewidth=8pt,linecolor=red,arrowsize=1.5pt 2](5,1,00.70)(5,2,00.70)(5,3,00.70)(5,4,00.4)(5,5,00.4)(5,6,00.4)(5,7,00.4)(5,8,00.50)(5,9,00.50)(5,10,00.50)(5,11,00.50)(5,12,00.60)(5,13,00.60)(5,14,00.60)
\pstThreeDLine[linewidth=8pt,linecolor=red,arrowsize=1.5pt 2](9,1,00.50)(9,2,00.50)(9,3,00.50)(9,4,00.10)(9,5,00.10)(9,6,00.10)(9,7,00.10)(9,8,00.10)(9,9,00.10)(9,10,00.10)(9,11,00.10)(9,12,00.10)(9,13,00.10)(9,14,00.10)
\pstThreeDLine[linewidth=8pt,linecolor=yellow,arrowsize=1.5pt 2](1,1,09.60)(1,2,09.60)(1,3,09.60)(1,4,07.10)(1,5,07.10)(1,6,07.10)(1,7,07.10)(1,8,06.20)(1,9,06.20)(1,10,06.20)(1,11,06.20)(1,12,07.00)(1,13,07.00)(1,14,07.00)
\pstThreeDLine[linewidth=8pt,linecolor=yellow,arrowsize=1.5pt 2](5,1,00.10)(5,2,00.10)(5,3,00.10)(5,4,00.10)(5,5,00.10)(5,6,00.10)(5,7,00.10)(5,8,00.10)(5,9,00.10)(5,10,00.10)(5,11,00.10)(5,12,00.10)(5,13,00.10)(5,14,00.10)
\pstThreeDLine[linewidth=8pt,linecolor=yellow,arrowsize=1.5pt 2](9,1,00.00)(9,2,00.00)(9,3,00.00)(9,4,00.0)(9,5,00.0)(9,6,00.0)(9,7,00.0)(9,8,00.0)(9,9,00.0)(9,10,00.0)(9,11,00.0)(9,12,00.0)(9,13,00.0)(9,14,00.0)
%
%%%%%%%%%%%%%%%%%%%%%%%%%%%%%%%%%%%%%%%%%%%%%%%%%%%%%%%%%%%%%%%%%%%%%%%%%%%%%%%%%%%%%%%
%%%%  SF, coordinated, ave degree 3
%%%%%%%%%%%%%%%%%%%%%%%%%%%%%%%%%%%%%%%%%%%%%%%%%%%%%%%%%%%%%%%%%%%%%%%%%%%%%%%%%%%%%%%
%
\pstThreeDLine[linewidth=8pt,linecolor=blue,arrowsize=1.5pt 2](1,1,07.60)(1,2,07.60)(1,3,07.60)(1,4,07.00)(1,5,07.00)(1,6,07.00)(1,7,07.00)(1,8,06.30)(1,9,06.30)(1,10,06.30)(1,11,06.30)(1,12,05.80)(1,13,05.80)(1,14,05.80)
\pstThreeDLine[linewidth=8pt,linecolor=blue,arrowsize=1.5pt 2](5,1,01.20)(5,2,01.20)(5,3,01.20)(5,4,00.40)(5,5,00.40)(5,6,00.40)(5,7,00.40)(5,8,00.30)(5,9,00.30)(5,10,00.30)(5,11,00.30)(5,12,00.30)(5,13,00.30)(5,14,00.30)
\pstThreeDLine[linewidth=8pt,linecolor=blue,arrowsize=1.5pt 2](9,1,00.90)(9,2,00.90)(9,3,00.90)(9,4,00.00)(9,5,00.00)(9,6,00.00)(9,7,00.00)(9,8,00.00)(9,9,00.00)(9,10,00.00)(9,11,00.00)(9,12,00.00)(9,13,00.00)(9,14,00.00)
\pstThreeDLine[linewidth=8pt,linecolor=black,linestyle=dashed,arrowsize=1.5pt 2](1,1,09.50)(1,2,09.50)(1,3,09.50)(1,4,07.6)(1,5,07.6)(1,6,07.6)(1,7,07.6)(1,8,07.40)(1,9,07.40)(1,10,07.40)(1,11,07.40)(1,12,07.00)(1,13,07.00)(1,14,07.00)
\pstThreeDLine[linewidth=8pt,linecolor=black,linestyle=dashed,arrowsize=1.5pt 2](5,1,07.60)(5,2,07.60)(5,3,07.60)(5,4,01.40)(5,5,01.40)(5,6,01.40)(5,7,01.40)(5,8,01.0)(5,9,01.0)(5,10,01.0)(5,11,01.0)(5,12,01.0)(5,13,01.0)(5,14,01.0)
\pstThreeDLine[linewidth=8pt,linecolor=black,linestyle=dashed,arrowsize=1.5pt 2](9,1,07.00)(9,2,07.00)(9,3,07.00)(9,4,00.3)(9,5,00.3)(9,6,00.3)(9,7,00.3)(9,8,00.3)(9,9,00.3)(9,10,00.3)(9,11,00.3)(9,12,00.40)(9,13,00.40)(9,14,00.40)
%
%%%%%%%%%%%%%%%%%%%%%%%%%%%%%%%%%%%%%%%%%%%%%%%%%%%%%%%%%%%%%%%%%%%%%%%%%%%%%%%%%%%%%%%
%%%%  tree, random
%%%%%%%%%%%%%%%%%%%%%%%%%%%%%%%%%%%%%%%%%%%%%%%%%%%%%%%%%%%%%%%%%%%%%%%%%%%%%%%%%%%%%%%
%
%%\pstThreeDLine[linewidth=8pt,linecolor=gray,linestyle=dashed,arrowsize=1.5pt 2](1,1,06.40)(1,2,06.70)(1,3,06.80)(1,4,02.6)(1,5,01.8)(1,6,01.50)(1,7,02.10)(1,8,01.80)(1,9,02.40)(1,10,02.30)(1,11,02.50)(1,12,02.10)(1,13,02.40)(1,14,02.40)
%%\pstThreeDLine[linewidth=8pt,linecolor=gray,linestyle=dashed,arrowsize=1.5pt 2](5,1,06.20)(5,2,06.30)(5,3,06.20)(5,4,00.40)(5,5,00.30)(5,6,00.40)(5,7,00.40)(5,8,01.0)(5,9,01.0)(5,10,01.0)(5,11,01.0)(5,12,01.0)(5,13,01.0)(5,14,01.0)
%%\pstThreeDLine[linewidth=8pt,linecolor=gray,linestyle=dashed,arrowsize=1.5pt 2](9,1,06.0)(9,2,06.10)(9,3,06.00)(9,4,00.00)(9,5,00.00)(9,6,00.0)(9,7,00.1)(9,8,00.00)(9,9,00.00)(9,10,00.1)(9,11,00.0)(9,12,00.00)(9,13,00.10)(9,14,00.10)
%
\pstThreeDCoor
[IIIDxTicksPlane=xz,IIIDxticksep=-0.6,IIIDyTicksPlane=yz,IIIDyticksep=-0.2,arrowsize=1.5pt 4,linewidth=1.5pt,linecolor=darkgray,IIIDzTicksPlane=yz,yMin=0,yMax=15.5,xMin=0,xMax=10.5,zMin=0,zMax=11,IIIDticks=true,IIIDticksize=0.0,Dy=025,Dx=0.05,Dz=0.1,IIIDlabels=true,IIIDOffset={(0,0,0)}] 
\end{pspicture}
}
\end{minipage}
%
%%%%%%%%%%%%%%%%%%%%%%%%%%%%%%%%%%%%
%%%%%%%%%%%%%%%%%%%%%%%%%%%%%%%%%%%%
%%%%%%%%%%%%%%%%%%%%%%%%%%%%%%%%%%%%
%% part 4 of figure starts
%
\hspace*{1.7in}
%
%%%%%%%%%%%%%%%%%%%%%%%%%%%%%%%%%%%%
%%%%%%%%%%%%%%%%%%%%%%%%%%%%%%%%%%%%
%%%%%%%%%%%%%%%%%%%%%%%%%%%%%%%%%%%%
%
\begin{minipage}[t]{2in}
\scalebox{0.43}[0.43]{
\begin{pspicture}(-4,-7.5)(-2,9.5)
\psset{xunit=0.4in,yunit=0.4in,runit=0.4in}
\psset{nameX=\LARGE$\pmb{\gamma}$,nameY=\LARGE$\pmb{E}$,nameZ=\LARGE$\pmb{\xi}$}
\psset{coorType=1,Alpha=135,hiddenLine=true}

%%\pstThreeDPlaneGrid[planeGrid=xy,linewidth=0.1pt,ysubticks=14,xsubticks=9,linecolor=darkgray](0,0)(9,14)
%%\pstThreeDPlaneGrid[planeGridOffset=10,linewidth=0.1pt,ysubticks=14,xsubticks=9,linecolor=darkgray](0,0)(9,14)
%
%%\pstThreeDPlaneGrid[planeGrid=xz,linewidth=1pt,xsubticks=9,ysubticks=10,linecolor=darkgray](0,0)(9,10)
%%\pstThreeDPlaneGrid[planeGridOffset=14,planeGrid=xz,linewidth=1pt,xsubticks=9,ysubticks=10,linecolor=darkgray](0,0)(9,10)
%
%\pstThreeDPlaneGrid[planeGrid=yz,linewidth=0.1pt,xsubticks=14,ysubticks=10,linecolor=darkgray](0,0)(14,10)
%%\pstThreeDPlaneGrid[planeGridOffset=1,planeGrid=yz,linewidth=0.1pt,xsubticks=14,ysubticks=10,linecolor=darkgray](0,0)(14,10)
%%\pstThreeDPlaneGrid[planeGridOffset=9,planeGrid=yz,linewidth=0.1pt,xsubticks=14,ysubticks=10,linecolor=darkgray](0,0)(14,10)
%
\rput[ml](1,11.0){\Large$\pmb{|V|=100}$ \hspace*{0.1in} $\pmb{\mathcal{K}=0.5}$ \hspace*{0.1in}$\pmb{\Phi=0.5}$ \hspace*{0.1in}$\pmb{\gamma=0.05,0.25,0.45}$}
\rput[mr](5,-7.5){\LARGE\bf (D)}
%
%
%%%%%%%%%%%%%%%% \gamma=0.05,0.25,0.45
%
%%%%%%%%%%%%%%%%%%%%%%%%%%%%%%%%%%%%%%%%%%%%%%%%%%%%%%%%%%%%%%%%%%%%%%%%%%%%%%%%%%%%%%%
%%%%  ER, coordinated, ave degree 6
%%%%%%%%%%%%%%%%%%%%%%%%%%%%%%%%%%%%%%%%%%%%%%%%%%%%%%%%%%%%%%%%%%%%%%%%%%%%%%%%%%%%%%%
%
\pstThreeDLine[linewidth=8pt,linecolor=black,arrowsize=1.5pt 2](1,1,04.80)(1,2,04.80)(1,3,04.80)(1,4,06.70)(1,5,06.70)(1,6,06.70)(1,7,06.70)(1,8,07.90)(1,9,07.90)(1,10,07.90)(1,11,07.90)(1,12,08.90)(1,13,08.90)(1,14,08.90)
\pstThreeDLine[linewidth=8pt,linecolor=black,arrowsize=1.5pt 2](5,1,00.70)(5,2,00.70)(5,3,00.70)(5,4,01.1)(5,5,01.1)(5,6,01.1)(5,7,01.1)(5,8,01.5)(5,9,01.5)(5,10,01.5)(5,11,01.5)(5,12,01.8)(5,13,01.8)(5,14,01.8)
\pstThreeDLine[linewidth=8pt,linecolor=black,arrowsize=1.5pt 2](9,1,00.00)(9,2,00.00)(9,3,00.00)(9,4,00.0)(9,5,00.0)(9,6,00.0)(9,7,00.0)(9,8,00.0)(9,9,00.0)(9,10,00.0)(9,11,00.0)(9,12,00.0)(9,13,00.0)(9,14,00.0)
%
%%%%%%%%%%%%%%%%%%%%%%%%%%%%%%%%%%%%%%%%%%%%%%%%%%%%%%%%%%%%%%%%%%%%%%%%%%%%%%%%%%%%%%%
%%%%  ER, coordinated, ave degree 3
%%%%%%%%%%%%%%%%%%%%%%%%%%%%%%%%%%%%%%%%%%%%%%%%%%%%%%%%%%%%%%%%%%%%%%%%%%%%%%%%%%%%%%%
%
\pstThreeDLine[linewidth=8pt,linecolor=red,arrowsize=1.5pt 2](1,1,06.3)(1,2,06.3)(1,3,06.3)(1,4,07.60)(1,5,07.60)(1,6,07.60)(1,7,07.60)(1,8,08.80)(1,9,08.80)(1,10,08.80)(1,11,08.80)(1,12,09.30)(1,13,09.30)(1,14,09.30)
\pstThreeDLine[linewidth=8pt,linecolor=red,arrowsize=1.5pt 2](5,1,02.00)(5,2,02.00)(5,3,02.00)(5,4,02.2)(5,5,02.2)(5,6,02.2)(5,7,02.2)(5,8,02.70)(5,9,02.70)(5,10,02.70)(5,11,02.70)(5,12,03.20)(5,13,03.20)(5,14,03.20)
\pstThreeDLine[linewidth=8pt,linecolor=red,arrowsize=1.5pt 2](9,1,00.70)(9,2,00.70)(9,3,00.70)(9,4,00.30)(9,5,00.30)(9,6,00.30)(9,7,00.30)(9,8,00.30)(9,9,00.30)(9,10,00.30)(9,11,00.30)(9,12,00.30)(9,13,00.30)(9,14,00.30)
\pstThreeDLine[linewidth=8pt,linecolor=yellow,arrowsize=1.5pt 2](1,1,09.40)(1,2,09.40)(1,3,09.40)(1,4,07.10)(1,5,07.10)(1,6,07.10)(1,7,07.10)(1,8,07.90)(1,9,07.90)(1,10,07.90)(1,11,07.90)(1,12,08.00)(1,13,08.00)(1,14,08.00)
\pstThreeDLine[linewidth=8pt,linecolor=yellow,arrowsize=1.5pt 2](5,1,00.90)(5,2,00.90)(5,3,00.90)(5,4,01.20)(5,5,01.20)(5,6,01.20)(5,7,01.20)(5,8,01.80)(5,9,01.80)(5,10,01.80)(5,11,01.80)(5,12,02.40)(5,13,02.40)(5,14,02.40)
\pstThreeDLine[linewidth=8pt,linecolor=yellow,arrowsize=1.5pt 2](9,1,00.00)(9,2,00.00)(9,3,00.00)(9,4,00.0)(9,5,00.0)(9,6,00.0)(9,7,00.0)(9,8,00.0)(9,9,00.0)(9,10,00.0)(9,11,00.0)(9,12,00.0)(9,13,00.0)(9,14,00.0)
%
%%%%%%%%%%%%%%%%%%%%%%%%%%%%%%%%%%%%%%%%%%%%%%%%%%%%%%%%%%%%%%%%%%%%%%%%%%%%%%%%%%%%%%%
%%%%  SF, coordinated, ave degree 3
%%%%%%%%%%%%%%%%%%%%%%%%%%%%%%%%%%%%%%%%%%%%%%%%%%%%%%%%%%%%%%%%%%%%%%%%%%%%%%%%%%%%%%%
%
\pstThreeDLine[linewidth=8pt,linecolor=blue,arrowsize=1.5pt 2](1,1,07.60)(1,2,07.60)(1,3,07.60)(1,4,07.80)(1,5,07.80)(1,6,07.80)(1,7,07.80)(1,8,08.10)(1,9,08.10)(1,10,08.10)(1,11,08.10)(1,12,08.30)(1,13,08.30)(1,14,08.30)
\pstThreeDLine[linewidth=8pt,linecolor=blue,arrowsize=1.5pt 2](5,1,03.00)(5,2,03.00)(5,3,03.00)(5,4,02.70)(5,5,02.70)(5,6,02.70)(5,7,02.70)(5,8,03.70)(5,9,03.70)(5,10,03.70)(5,11,03.70)(5,12,04.30)(5,13,04.30)(5,14,04.30)
\pstThreeDLine[linewidth=8pt,linecolor=blue,arrowsize=1.5pt 2](9,1,01.40)(9,2,01.40)(9,3,01.40)(9,4,00.50)(9,5,00.50)(9,6,00.50)(9,7,00.50)(9,8,00.50)(9,9,00.50)(9,10,00.50)(9,11,00.50)(9,12,00.50)(9,13,00.50)(9,14,00.50)
\pstThreeDLine[linewidth=8pt,linecolor=black,linestyle=dashed,arrowsize=1.5pt 2](1,1,10.00)(1,2,10.00)(1,3,10.00)(1,4,10.0)(1,5,10.0)(1,6,10.0)(1,7,10.0)(1,8,10.00)(1,9,10.00)(1,10,10.00)(1,11,10.00)(1,12,10.00)(1,13,10.00)(1,14,10.00)
\pstThreeDLine[linewidth=8pt,linecolor=black,linestyle=dashed,arrowsize=1.5pt 2](5,1,08.50)(5,2,08.50)(5,3,08.50)(5,4,03.70)(5,5,03.70)(5,6,03.70)(5,7,03.70)(5,8,03.3)(5,9,03.3)(5,10,03.3)(5,11,03.3)(5,12,03.3)(5,13,03.3)(5,14,03.3)
\pstThreeDLine[linewidth=8pt,linecolor=black,linestyle=dashed,arrowsize=1.5pt 2](9,1,07.00)(9,2,07.00)(9,3,07.00)(9,4,00.4)(9,5,00.4)(9,6,00.4)(9,7,00.4)(9,8,00.4)(9,9,00.4)(9,10,00.4)(9,11,00.4)(9,12,00.40)(9,13,00.40)(9,14,00.40)
%
%%%%%%%%%%%%%%%%%%%%%%%%%%%%%%%%%%%%%%%%%%%%%%%%%%%%%%%%%%%%%%%%%%%%%%%%%%%%%%%%%%%%%%%
%%%%  tree, random
%%%%%%%%%%%%%%%%%%%%%%%%%%%%%%%%%%%%%%%%%%%%%%%%%%%%%%%%%%%%%%%%%%%%%%%%%%%%%%%%%%%%%%%
%
%%\pstThreeDLine[linewidth=8pt,linecolor=gray,linestyle=dashed,arrowsize=1.5pt 2](1,1,06.30)(1,2,06.30)(1,3,06.00)(1,4,06.3)(1,5,06.3)(1,6,07.30)(1,7,07.30)(1,8,07.70)(1,9,08.00)(1,10,08.60)(1,11,07.80)(1,12,07.50)(1,13,07.90)(1,14,08.20)
%%\pstThreeDLine[linewidth=8pt,linecolor=gray,linestyle=dashed,arrowsize=1.5pt 2](5,1,04.10)(5,2,04.20)(5,3,04.30)(5,4,02.00)(5,5,01.60)(5,6,01.80)(5,7,01.80)(5,8,05.0)(5,9,05.0)(5,10,05.0)(5,11,05.0)(5,12,05.0)(5,13,05.0)(5,14,05.0)
%%\pstThreeDLine[linewidth=8pt,linecolor=gray,linestyle=dashed,arrowsize=1.5pt 2](9,1,03.4)(9,2,03.30)(9,3,03.30)(9,4,00.10)(9,5,00.20)(9,6,00.2)(9,7,00.1)(9,8,00.20)(9,9,00.20)(9,10,00.2)(9,11,00.2)(9,12,00.30)(9,13,00.20)(9,14,00.10)
\def\psxyzlabel#1{\bgroup\large\bf\textsf{#1}\egroup}
\pstThreeDCoor
[IIIDxTicksPlane=xz,IIIDxticksep=-0.6,IIIDyTicksPlane=yz,IIIDyticksep=-0.2,arrowsize=1.5pt 4,linewidth=1.5pt,linecolor=darkgray,IIIDzTicksPlane=yz,yMin=0,yMax=15.5,xMin=0,xMax=10.5,zMin=0,zMax=11,IIIDticks=true,IIIDticksize=0.0,Dy=025,Dx=0.05,Dz=0.1,IIIDlabels=true,IIIDOffset={(0,0,0)}] 
\end{pspicture}
}
\end{minipage}

\vspace*{0.1in}
\caption{\label{newfig4}Effect of variations of the total external to internal asset ratio $E/I$ on the vulnerability index $\xi$ for homogeneous networks.
Lower values of $\xi$ imply higher global stability of a network.}
\end{figure*}

%%%%%%%%%%%%%%%%%%%%%%%%%%%%%%%%%%%%%%%%%%%%%%%%%%%%%%%%%%%%%%%%%%%%%%%%%%%%%%%%%%%%%%%%%%%%%%%%%%%%%%%%%%%%%%
%%%%%%%%%%%%%%%%%%%%%%%%%%%%%%%%%%%%%%%%%%%%%%%%%%%%%%%%%%%%%%%%%%%%%%%%%%%%%%%%%%%%%%%%%%%%%%%%%%%%%%%%%%%%%%

%%%\thispagestyle{empty}

\clearpage
%%%%%%%%%%%%%%%%%%%%%%%%%%%%%%%%%%%%%%%%%%%%%%%%%%%%%%%%%%%%%%%%%%%%%%%%%%%%%%%%%%%%%%%%%%%%%%%%%%%%%%%%%%%%%%
%%%%%%%%%%%%%%%%%%%%%%%%%%%%%%%%%%%%%%%%%%%%%%%%%%%%%%%%%%%%%%%%%%%%%%%%%%%%%%%%%%%%%%%%%%%%%%%%%%%%%%%%%%%%%%
%%% newfig5
%%\input{../tex-figures/newfig5-coordinated-only.tex}

\begin{figure*}
\vspace*{0.5in}
\hspace*{0.0in}
\begin{minipage}[c]{2in}
\scalebox{0.95}[0.7]{

}
\end{minipage}
\vspace*{-3.9in}
\caption{\label{newfig5}Effect of variations of the total external to internal asset ratio $\nicefrac{E}{I}$ on the vulnerability index $\xi$
for $(\alpha,\beta)$-heterogeneous networks. Lower values of $\xi$ imply higher global stability of a network.}
\end{figure*}

%%%%%%%%%%%%%%%%%%%%%%%%%%%%%%%%%%%%%%%%%%%%%%%%%%%%%%%%%%%%%%%%%%%%%%%%%%%%%%%%%%%%%%%%%%%%%%%%%%%%%%%%%%%%%%
%%%%%%%%%%%%%%%%%%%%%%%%%%%%%%%%%%%%%%%%%%%%%%%%%%%%%%%%%%%%%%%%%%%%%%%%%%%%%%%%%%%%%%%%%%%%%%%%%%%%%%%%%%%%%%

%%\thispagestyle{empty}

\clearpage
%%%%%%%%%%%%%%%%%%%%%%%%%%%%%%%%%%%%%%%%%%%%%%%%%%%%%%%%%%%%%%%%%%%%%%%%%%%%%%%%%%%%%%%%%%%%%%%%%%%%%%%%%%%%%%
%%%%%%%%%%%%%%%%%%%%%%%%%%%%%%%%%%%%%%%%%%%%%%%%%%%%%%%%%%%%%%%%%%%%%%%%%%%%%%%%%%%%%%%%%%%%%%%%%%%%%%%%%%%%%%
%newfig1-coordinated-and-random
%%%\input{../tex-figures/newfig1-coordinated-and-random.tex}
%
\begin{figure*}
\renewcommand{\thesubfigure}{\hspace*{0.6in}\protect\raisebox{0.2in}[0in][0.0in]{(\Alph{subfigure})}}
\vspace*{0.8in}
\hspace*{0.0in}
\subfigure[]{
\scalebox{0.8}[0.7]{

}
}
%%%%
\vspace*{-0.3in}
\caption{\label{newfig1-full}Effect of variations of equity to asset ratio (with respect to shock) on the vulnerability index $\xi$ for homogeneous networks.
Lower values of $\xi$ imply higher global stability of a network.
}
\end{figure*}

%%%%%%%%%%%%%%%%%%%%%%%%%%%%%%%%%%%%%%%%%%%%%%%%%%%%%%%%%%%%%%%%%%%%%%%%%%%%%%%%%%%%%%%%%%%%%%%%%%%%%%%%%%%%%%
%%%%%%%%%%%%%%%%%%%%%%%%%%%%%%%%%%%%%%%%%%%%%%%%%%%%%%%%%%%%%%%%%%%%%%%%%%%%%%%%%%%%%%%%%%%%%%%%%%%%%%%%%%%%%%

\clearpage
%%%%%%%%%%%%%%%%%%%%%%%%%%%%%%%%%%%%%%%%%%%%%%%%%%%%%%%%%%%%%%%%%%%%%%%%%%%%%%%%%%%%%%%%%%%%%%%%%%%%%%%%%%%%%%
%%%%%%%%%%%%%%%%%%%%%%%%%%%%%%%%%%%%%%%%%%%%%%%%%%%%%%%%%%%%%%%%%%%%%%%%%%%%%%%%%%%%%%%%%%%%%%%%%%%%%%%%%%%%%%
%newfig2-coordinated-and-random
%%%\input{../tex-figures/newfig2-coordinated-and-random.tex}

\begin{figure*}
\vspace*{0.4in}
\hspace*{-0.0in}
\subfigure{
\scalebox{0.8}[0.7]{

}
}
%%%%
\vspace*{-0.5in}
\caption{\label{newfig2-full}Effect of variations of equity to asset ratio (with respect to shock) on the vulnerability index $\xi$ for 
$(\alpha,\beta)$-heterogeneous networks. Lower values of $\xi$ imply higher global stability of a network.
}
\end{figure*}

%%%%%%%%%%%%%%%%%%%%%%%%%%%%%%%%%%%%%%%%%%%%%%%%%%%%%%%%%%%%%%%%%%%%%%%%%%%%%%%%%%%%%%%%%%%%%%%%%%%%%%%%%%%%%%
%%%%%%%%%%%%%%%%%%%%%%%%%%%%%%%%%%%%%%%%%%%%%%%%%%%%%%%%%%%%%%%%%%%%%%%%%%%%%%%%%%%%%%%%%%%%%%%%%%%%%%%%%%%%%%

%%\thispagestyle{empty}

\clearpage
%%%%%%%%%%%%%%%%%%%%%%%%%%%%%%%%%%%%%%%%%%%%%%%%%%%%%%%%%%%%%%%%%%%%%%%%%%%%%%%%%%%%%%%%%%%%%%%%%%%%%%%%%%%%%%
%%%%%%%%%%%%%%%%%%%%%%%%%%%%%%%%%%%%%%%%%%%%%%%%%%%%%%%%%%%%%%%%%%%%%%%%%%%%%%%%%%%%%%%%%%%%%%%%%%%%%%%%%%%%%%
%%% newfig3-coordinated-and-random
%%\input{../tex-figures/newfig3-coordinated-and-random.tex}

\begin{figure*}
\vspace*{1.2in}
%
%%\hspace*{0.0in}
%
\subfigure{
\scalebox{0.7}[0.50]{

}
}
\vspace*{-2.0in}
\caption{\label{newfig3-full}Effect of variations of equity to asset ratio (with respect to shock) on the vulnerability index $\xi$ for 
$(\alpha,\beta)$-heterogeneous networks. Lower values of $\xi$ imply higher global stability of a network.
}
\end{figure*}

%%%%%%%%%%%%%%%%%%%%%%%%%%%%%%%%%%%%%%%%%%%%%%%%%%%%%%%%%%%%%%%%%%%%%%%%%%%%%%%%%%%%%%%%%%%%%%%%%%%%%%%%%%%%%%
%%%%%%%%%%%%%%%%%%%%%%%%%%%%%%%%%%%%%%%%%%%%%%%%%%%%%%%%%%%%%%%%%%%%%%%%%%%%%%%%%%%%%%%%%%%%%%%%%%%%%%%%%%%%%%

\clearpage
%%%%%%%%%%%%%%%%%%%%%%%%%%%%%%%%%%%%%%%%%%%%%%%%%%%%%%%%%%%%%%%%%%%%%%%%%%%%%%%%%%%%%%%%%%%%%%%%%%%%%%%%%%%%%%
%%%%%%%%%%%%%%%%%%%%%%%%%%%%%%%%%%%%%%%%%%%%%%%%%%%%%%%%%%%%%%%%%%%%%%%%%%%%%%%%%%%%%%%%%%%%%%%%%%%%%%%%%%%%%%
%%%%%%%%%%%%%%%%%%%%%%%%%%%%%%%%%%%%%%%%%%%%%%%%%%%%%%%%%%%%%%%%%%%%%%%%%%%%%%%%%%%%%%%%%%%%%%%%%%%%%%%%%%%%%%
%%% newfig4-coordinated-and-random
%%%\input{../tex-figures/newfig4-coordinated-and-random.tex}

\begin{figure*}
\vspace*{1.2in}
\hspace*{-0.5in}
\begin{minipage}[t]{2in}
\scalebox{0.43}[0.43]{
\begin{pspicture}(-3,-7.5)(-1,9.5)
\psset{xunit=0.4in,yunit=0.4in,runit=0.4in}
\psset{nameX=\LARGE$\pmb{\gamma}$,nameY=\LARGE$\pmb{E}$,nameZ=\LARGE$\pmb{\xi}$}
\psset{coorType=1,Alpha=135,hiddenLine=true}
\def\psxyzlabel#1{\bgroup\large\bf\textsf{#1}\egroup}
%
%%\pstThreeDPlaneGrid[planeGrid=xy,linewidth=0.1pt,ysubticks=14,xsubticks=9,linecolor=darkgray](0,0)(9,14)
%%\pstThreeDPlaneGrid[planeGridOffset=10,linewidth=0.1pt,ysubticks=14,xsubticks=9,linecolor=darkgray](0,0)(9,14)
%
%%\pstThreeDPlaneGrid[planeGrid=xz,linewidth=1pt,xsubticks=9,ysubticks=10,linecolor=darkgray](0,0)(9,10)
%%\pstThreeDPlaneGrid[planeGridOffset=14,planeGrid=xz,linewidth=1pt,xsubticks=9,ysubticks=10,linecolor=darkgray](0,0)(9,10)
%
%\pstThreeDPlaneGrid[planeGrid=yz,linewidth=0.1pt,xsubticks=14,ysubticks=10,linecolor=darkgray](0,0)(14,10)
%%\pstThreeDPlaneGrid[planeGridOffset=1,planeGrid=yz,linewidth=0.1pt,xsubticks=14,ysubticks=10,linecolor=darkgray](0,0)(14,10)
%%\pstThreeDPlaneGrid[planeGridOffset=9,planeGrid=yz,linewidth=0.1pt,xsubticks=14,ysubticks=10,linecolor=darkgray](0,0)(14,10)
%
%%%%%%%%%%%%%%%%%%%%%%%%%%%%%%%%%%%%%%%%%%%%%%%%%%%%%%%%%%%%%%%%%%%%%%%%%%%%%%%%%%%%%%%
%%%% legends
%%%%%%%%%%%%%%%%%%%%%%%%%%%%%%%%%%%%%%%%%%%%%%%%%%%%%%%%%%%%%%%%%%%%%%%%%%%%%%%%%%%%%%%
%
\rput[mr](5,-7.5){\LARGE\bf (A)}
\psline[linewidth=8pt,linecolor=black,arrowsize=1.5pt 2](7.50,16.2)(8.5,16.2)
\rput[mr](7.35,16.2){\Large\bf {\sf ER} model (average degree $\pmb{6}$), coordinated shock}
\psline[linewidth=8pt,linecolor=red,arrowsize=1.5pt 2](25.65,16.2)(26.65,16.2)
\rput[mr](25.55,16.2){\Large\bf {\sf ER} model (average degree $\pmb{3}$), coordinated shock}
\psline[linewidth=2pt,linecolor=black,arrowsize=1.5pt 2](17.20,15.3)(18.20,15.3)
\rput[mr](17.05,15.3){\Large\bf {\sf ER} model (average degree $\pmb{6}$), idiosyncratic shock}
\psline[linewidth=2pt,linecolor=gray,arrowsize=1.5pt 2](35.55,15.3)(36.55,15.3)
\rput[mr](35.05,15.3){\Large\bf {\sf ER} model (average degree $\pmb{3}$), idiosyncratic shock}
\psline[linewidth=8pt,linecolor=yellow,arrowsize=1.5pt 2](7.50,14.4)(8.5,14.4)
\rput[mr](7.35,14.4){\Large\bf {\sf SF} model (average degree $\pmb{6}$), coordinated shock}
\psline[linewidth=8pt,linecolor=blue,arrowsize=1.5pt 2](25.65,14.4)(26.65,14.4)
\rput[mr](25.55,14.4){\Large\bf {\sf SF} model (average degree $\pmb{3}$), coordinated shock}
\psline[linewidth=2pt,linecolor=green,arrowsize=1.5pt 2](17.20,13.5)(18.20,13.5)
\rput[mr](17.05,13.5){\Large\bf {\sf SF} model (average degree $\pmb{6}$), idiosyncratic shock}
\psline[linewidth=2pt,linecolor=red,arrowsize=1.5pt 2](35.55,13.5)(36.55,13.5)
\rput[mr](35.05,13.5){\Large\bf {\sf SF} model (average degree $\pmb{3}$), idiosyncratic shock}
\psline[linewidth=8pt,linecolor=black,linestyle=dashed,arrowsize=1.5pt 2](10,12.5)(11,12.5)
\rput[mr](9.85,12.5){\Large\bf in-arborescence (average degree $\pmb{\approx 1}$), coordinated shock}
\psline[linewidth=8pt,linecolor=gray,linestyle=dashed,arrowsize=1.5pt 2](29.95,12.5)(30.95,12.5)
\rput[mr](29.85,12.5){\Large\bf in-arborescence (average degree $\pmb{\approx 1}$), idiosyncratic shock}
\rput[ml](1,11.0){\Large$\pmb{n=50}$ \hspace*{0.1in} $\pmb{\mathcal{K}=0.1}$ \hspace*{0.1in}$\pmb{\Phi=0.5}$ \hspace*{0.1in}$\pmb{\gamma=0.05,0.25,0.45}$}
%
%%%%%%%%%%%%%%%% \gamma=0.05,0.25,0.45
%
%%%%%%%%%%%%%%%%%%%%%%%%%%%%%%%%%%%%%%%%%%%%%%%%%%%%%%%%%%%%%%%%%%%%%%%%%%%%%%%%%%%%%%%
%%%%  ER, coordinated, ave degree 6
%%%%%%%%%%%%%%%%%%%%%%%%%%%%%%%%%%%%%%%%%%%%%%%%%%%%%%%%%%%%%%%%%%%%%%%%%%%%%%%%%%%%%%%
%
\pstThreeDLine[linewidth=8pt,linecolor=black,arrowsize=1.5pt 2](1,1,01.06)(1,2,01.06)(1,3,01.06)(1,4,01.36)(1,5,01.36)(1,6,01.36)(1,7,01.36)(1,8,01.64)(1,9,01.64)(1,10,01.64)(1,11,01.64)(1,12,01.88)(1,13,01.88)(1,14,01.88)
\pstThreeDLine[linewidth=8pt,linecolor=black,arrowsize=1.5pt 2](5,1,00.14)(5,2,00.14)(5,3,00.14)(5,4,00.2)(5,5,00.2)(5,6,00.2)(5,7,00.2)(5,8,00.3)(5,9,00.3)(5,10,00.3)(5,11,00.3)(5,12,00.3)(5,13,00.3)(5,14,00.3)
\pstThreeDLine[linewidth=8pt,linecolor=black,arrowsize=1.5pt 2](9,1,00.02)(9,2,00.02)(9,3,00.02)(9,4,00.0)(9,5,00.0)(9,6,00.0)(9,7,00.0)(9,8,00.0)(9,9,00.0)(9,10,00.0)(9,11,00.0)(9,12,00.0)(9,13,00.0)(9,14,00.0)
%
%%%%%%%%%%%%%%%%%%%%%%%%%%%%%%%%%%%%%%%%%%%%%%%%%%%%%%%%%%%%%%%%%%%%%%%%%%%%%%%%%%%%%%%
%%%%  ER, coordinated, ave degree 3
%%%%%%%%%%%%%%%%%%%%%%%%%%%%%%%%%%%%%%%%%%%%%%%%%%%%%%%%%%%%%%%%%%%%%%%%%%%%%%%%%%%%%%%
%
\pstThreeDLine[linewidth=8pt,linecolor=red,arrowsize=1.5pt 2](1,1,02.9)(1,2,02.9)(1,3,02.9)(1,4,03.14)(1,5,03.14)(1,6,03.14)(1,7,03.14)(1,8,03.62)(1,9,03.62)(1,10,03.62)(1,11,03.62)(1,12,04.02)(1,13,04.02)(1,14,04.02)
\pstThreeDLine[linewidth=8pt,linecolor=red,arrowsize=1.5pt 2](5,1,00.58)(5,2,00.58)(5,3,00.58)(5,4,00.4)(5,5,00.4)(5,6,00.4)(5,7,00.4)(5,8,00.52)(5,9,00.52)(5,10,00.52)(5,11,00.52)(5,12,00.58)(5,13,00.58)(5,14,00.58)
\pstThreeDLine[linewidth=8pt,linecolor=red,arrowsize=1.5pt 2](9,1,00.34)(9,2,00.34)(9,3,00.34)(9,4,00.04)(9,5,00.04)(9,6,00.04)(9,7,00.04)(9,8,00.04)(9,9,00.04)(9,10,00.04)(9,11,00.04)(9,12,00.04)(9,13,00.04)(9,14,00.04)
%
%%%%%%%%%%%%%%%%%%%%%%%%%%%%%%%%%%%%%%%%%%%%%%%%%%%%%%%%%%%%%%%%%%%%%%%%%%%%%%%%%%%%%%%
%%%%  ER, random, ave degree 6
%%%%%%%%%%%%%%%%%%%%%%%%%%%%%%%%%%%%%%%%%%%%%%%%%%%%%%%%%%%%%%%%%%%%%%%%%%%%%%%%%%%%%%%
%
\pstThreeDLine[linewidth=2pt,linecolor=black,arrowsize=1.5pt 2](1,1,01.82)(1,2,01.76)(1,3,01.12)(1,4,01.18)(1,5,03.56)(1,6,01.86)(1,7,02.84)(1,8,04.3)(1,9,02.0)(1,10,02.88)(1,11,02.16)(1,12,03.96)(1,13,03.56)(1,14,04.66)
\pstThreeDLine[linewidth=2pt,linecolor=black,arrowsize=1.5pt 2](5,1,00.06)(5,2,00.12)(5,3,00.18)(5,4,00.14)(5,5,00.16)(5,6,00.12)(5,7,00.22)(5,8,00.2)(5,9,00.24)(5,10,00.36)(5,11,00.26)(5,12,00.38)(5,13,00.28)(5,14,00.36)
\pstThreeDLine[linewidth=2pt,linecolor=black,arrowsize=1.5pt 2](9,1,00.02)(9,2,00.02)(9,3,00.02)(9,4,00.0)(9,5,00.0)(9,6,00.0)(9,7,00.0)(9,8,00.0)(9,9,00.0)(9,10,00.0)(9,11,00.0)(9,12,00.0)(9,13,00.02)(9,14,00.0)
%
%%%%%%%%%%%%%%%%%%%%%%%%%%%%%%%%%%%%%%%%%%%%%%%%%%%%%%%%%%%%%%%%%%%%%%%%%%%%%%%%%%%%%%%
%%%%  ER, random, ave degree 3
%%%%%%%%%%%%%%%%%%%%%%%%%%%%%%%%%%%%%%%%%%%%%%%%%%%%%%%%%%%%%%%%%%%%%%%%%%%%%%%%%%%%%%%
%
\pstThreeDLine[linewidth=2pt,linecolor=gray,arrowsize=1.5pt 2](1,1,06.8)(1,2,05.72)(1,3,04.44)(1,4,06.76)(1,5,04.66)(1,6,04.12)(1,7,06.28)(1,8,08.18)(1,9,0.624)(1,10,07.04)(1,11,05.3)(1,12,06.32)(1,13,06.84)(1,14,07.28)
\pstThreeDLine[linewidth=2pt,linecolor=gray,arrowsize=1.5pt 2](5,1,00.5)(5,2,00.42)(5,3,00.56)(5,4,00.26)(5,5,00.38)(5,6,00.44)(5,7,00.34)(5,8,01.44)(5,9,00.52)(5,10,00.42)(5,11,01.5)(5,12,00.54)(5,13,00.62)(5,14,00.6)
\pstThreeDLine[linewidth=2pt,linecolor=gray,arrowsize=1.5pt 2](9,1,00.34)(9,2,00.34)(9,3,00.36)(9,4,00.06)(9,5,00.02)(9,6,00.02)(9,7,00.02)(9,8,00.02)(9,9,00.04)(9,10,00.08)(9,11,00.04)(9,12,00.02)(9,13,00.1)(9,14,00.04)
%
%%%%%%%%%%%%%%%%%%%%%%%%%%%%%%%%%%%%%%%%%%%%%%%%%%%%%%%%%%%%%%%%%%%%%%%%%%%%%%%%%%%%%%%
%%%%  SF, coordinated, ave degree 6
%%%%%%%%%%%%%%%%%%%%%%%%%%%%%%%%%%%%%%%%%%%%%%%%%%%%%%%%%%%%%%%%%%%%%%%%%%%%%%%%%%%%%%%
%
\pstThreeDLine[linewidth=8pt,linecolor=yellow,arrowsize=1.5pt 2](1,1,07.22)(1,2,07.22)(1,3,07.22)(1,4,06.92)(1,5,06.92)(1,6,06.92)(1,7,06.92)(1,8,06.66)(1,9,06.66)(1,10,06.66)(1,11,06.66)(1,12,06.48)(1,13,06.48)(1,14,06.48)
\pstThreeDLine[linewidth=8pt,linecolor=yellow,arrowsize=1.5pt 2](5,1,00.06)(5,2,00.06)(5,3,00.06)(5,4,00.04)(5,5,00.04)(5,6,00.04)(5,7,00.04)(5,8,00.06)(5,9,00.06)(5,10,00.06)(5,11,00.06)(5,12,00.06)(5,13,00.06)(5,14,00.06)
\pstThreeDLine[linewidth=8pt,linecolor=yellow,arrowsize=1.5pt 2](9,1,00.02)(9,2,00.02)(9,3,00.02)(9,4,00.0)(9,5,00.0)(9,6,00.0)(9,7,00.0)(9,8,00.0)(9,9,00.0)(9,10,00.0)(9,11,00.0)(9,12,00.0)(9,13,00.0)(9,14,00.0)
%
%%%%%%%%%%%%%%%%%%%%%%%%%%%%%%%%%%%%%%%%%%%%%%%%%%%%%%%%%%%%%%%%%%%%%%%%%%%%%%%%%%%%%%%
%%%%  SF, coordinated, ave degree 3
%%%%%%%%%%%%%%%%%%%%%%%%%%%%%%%%%%%%%%%%%%%%%%%%%%%%%%%%%%%%%%%%%%%%%%%%%%%%%%%%%%%%%%%
%
\pstThreeDLine[linewidth=8pt,linecolor=blue,arrowsize=1.5pt 2](1,1,08.84)(1,2,08.84)(1,3,08.84)(1,4,08.54)(1,5,08.54)(1,6,08.54)(1,7,08.54)(1,8,08.02)(1,9,08.02)(1,10,08.02)(1,11,08.02)(1,12,07.14)(1,13,07.14)(1,14,07.14)
\pstThreeDLine[linewidth=8pt,linecolor=blue,arrowsize=1.5pt 2](5,1,01.14)(5,2,01.14)(5,3,01.14)(5,4,00.26)(5,5,00.26)(5,6,00.26)(5,7,00.26)(5,8,00.18)(5,9,00.18)(5,10,00.18)(5,11,00.18)(5,12,00.22)(5,13,00.22)(5,14,00.22)
\pstThreeDLine[linewidth=8pt,linecolor=blue,arrowsize=1.5pt 2](9,1,00.72)(9,2,00.72)(9,3,00.72)(9,4,00.06)(9,5,00.06)(9,6,00.06)(9,7,00.06)(9,8,00.06)(9,9,00.06)(9,10,00.06)(9,11,00.06)(9,12,00.06)(9,13,00.06)(9,14,00.06)
%
%%%%%%%%%%%%%%%%%%%%%%%%%%%%%%%%%%%%%%%%%%%%%%%%%%%%%%%%%%%%%%%%%%%%%%%%%%%%%%%%%%%%%%%
%%%%  SF, random, ave degree 6
%%%%%%%%%%%%%%%%%%%%%%%%%%%%%%%%%%%%%%%%%%%%%%%%%%%%%%%%%%%%%%%%%%%%%%%%%%%%%%%%%%%%%%%
%
\pstThreeDLine[linewidth=2pt,linecolor=green,arrowsize=1.5pt 2](1,1,02.28)(1,2,01.4)(1,3,01.46)(1,4,01.52)(1,5,00.7)(1,6,02.54)(1,7,01.72)(1,8,01.84)(1,9,01.76)(1,10,00.72)(1,11,01.82)(1,12,02.88)(1,13,02.88)(1,14,00.94)
\pstThreeDLine[linewidth=2pt,linecolor=green,arrowsize=1.5pt 2](5,1,00.14)(5,2,00.14)(5,3,00.18)(5,4,00.2)(5,5,00.16)(5,6,00.18)(5,7,00.22)(5,8,00.36)(5,9,00.36)(5,10,00.36)(5,11,00.3)(5,12,00.5)(5,13,00.4)(5,14,00.4)
\pstThreeDLine[linewidth=2pt,linecolor=green,arrowsize=1.5pt 2](9,1,0.00)(9,2,0.002)(9,3,0.002)(9,4,0.00)(9,5,0.00)(9,6,0.00)(9,7,0.00)(9,8,0.00)(9,9,0.00)(9,10,0.00)(9,11,0.00)(9,12,0.00)(9,13,0.00)(9,14,0.00)
%
%%%%%%%%%%%%%%%%%%%%%%%%%%%%%%%%%%%%%%%%%%%%%%%%%%%%%%%%%%%%%%%%%%%%%%%%%%%%%%%%%%%%%%%
%%%%  SF, random, ave degree 3
%%%%%%%%%%%%%%%%%%%%%%%%%%%%%%%%%%%%%%%%%%%%%%%%%%%%%%%%%%%%%%%%%%%%%%%%%%%%%%%%%%%%%%%
%
\pstThreeDLine[linewidth=2pt,linecolor=red,arrowsize=1.5pt 2](1,1,03.26)(1,2,01.92)(1,3,02.22)(1,4,02.46)(1,5,04.66)(1,6,02.52)(1,7,02.96)(1,8,03.26)(1,9,03.18)(1,10,02.7)(1,11,04.68)(1,12,02.86)(1,13,04.14)(1,14,04.3)
\pstThreeDLine[linewidth=2pt,linecolor=red,arrowsize=1.5pt 2](5,1,01.02)(5,2,00.8)(5,3,00.96)(5,4,00.42)(5,5,00.3)(5,6,00.48)(5,7,00.42)(5,8,00.56)(5,9,00.6)(5,10,00.54)(5,11,00.46)(5,12,00.8)(5,13,00.66)(5,14,00.68)
\pstThreeDLine[linewidth=2pt,linecolor=red,arrowsize=1.5pt 2](9,1,00.66)(9,2,00.68)(9,3,00.74)(9,4,00.02)(9,5,00.04)(9,6,00.08)(9,7,00.1)(9,8,00.04)(9,9,00.16)(9,10,00.06)(9,11,00.16)(9,12,00.06)(9,13,00.08)(9,14,00.08)
%
%%%%%%%%%%%%%%%%%%%%%%%%%%%%%%%%%%%%%%%%%%%%%%%%%%%%%%%%%%%%%%%%%%%%%%%%%%%%%%%%%%%%%%%
%%%%  tree, coordinated
%%%%%%%%%%%%%%%%%%%%%%%%%%%%%%%%%%%%%%%%%%%%%%%%%%%%%%%%%%%%%%%%%%%%%%%%%%%%%%%%%%%%%%%
%
\pstThreeDLine[linewidth=8pt,linecolor=black,linestyle=dashed,arrowsize=1.5pt 2](1,1,09.38)(1,2,09.38)(1,3,09.38)(1,4,07.4)(1,5,07.4)(1,6,07.4)(1,7,07.4)(1,8,06.96)(1,9,06.96)(1,10,06.96)(1,11,06.96)(1,12,06.78)(1,13,06.78)(1,14,06.78)
\pstThreeDLine[linewidth=8pt,linecolor=black,linestyle=dashed,arrowsize=1.5pt 2](5,1,07.58)(5,2,07.58)(5,3,07.58)(5,4,01.84)(5,5,01.84)(5,6,01.84)(5,7,01.84)(5,8,01.0)(5,9,01.0)(5,10,01.0)(5,11,01.0)(5,12,01.0)(5,13,01.0)(5,14,01.0)
\pstThreeDLine[linewidth=8pt,linecolor=black,linestyle=dashed,arrowsize=1.5pt 2](9,1,06.62)(9,2,06.62)(9,3,06.62)(9,4,00.3)(9,5,00.3)(9,6,00.3)(9,7,00.3)(9,8,00.3)(9,9,00.3)(9,10,00.3)(9,11,00.3)(9,12,00.34)(9,13,00.34)(9,14,00.34)
%
%%%%%%%%%%%%%%%%%%%%%%%%%%%%%%%%%%%%%%%%%%%%%%%%%%%%%%%%%%%%%%%%%%%%%%%%%%%%%%%%%%%%%%%
%%%%  tree, random
%%%%%%%%%%%%%%%%%%%%%%%%%%%%%%%%%%%%%%%%%%%%%%%%%%%%%%%%%%%%%%%%%%%%%%%%%%%%%%%%%%%%%%%
%
\pstThreeDLine[linewidth=8pt,linecolor=gray,linestyle=dashed,arrowsize=1.5pt 2](1,1,06.32)(1,2,06.58)(1,3,06.52)(1,4,01.2)(1,5,01.4)(1,6,01.74)(1,7,01.74)(1,8,02.78)(1,9,02.18)(1,10,02.66)(1,11,02.62)(1,12,02.08)(1,13,02.26)(1,14,02.66)
\pstThreeDLine[linewidth=8pt,linecolor=gray,linestyle=dashed,arrowsize=1.5pt 2](5,1,06.12)(5,2,05.84)(5,3,05.78)(5,4,00.26)(5,5,00.36)(5,6,00.68)(5,7,00.48)(5,8,01.0)(5,9,01.0)(5,10,01.0)(5,11,01.0)(5,12,01.0)(5,13,01.0)(5,14,01.0)
\pstThreeDLine[linewidth=8pt,linecolor=gray,linestyle=dashed,arrowsize=1.5pt 2](9,1,05.8)(9,2,05.68)(9,3,05.74)(9,4,00.04)(9,5,00.04)(9,6,00.0)(9,7,00.0)(9,8,00.02)(9,9,00.06)(9,10,00.0)(9,11,00.0)(9,12,00.02)(9,13,00.02)(9,14,00.06)
%
%\pstThreeDPut(1,1,8){haha}
%
%%\savedata{\mydata}[
%%{
%%(1,1,0.00)(1,2,0.02)(1,3,0.06)(1,4,0.12)(1,5,0.2)(1,6,0.3)(1,7,0.34)(1,8,0.56)(1,9,1.36)
%%%(2,1,0.00)(2,2,0.02)(2,3,0.06)(2,4,0.12)(2,5,0.2)(2,6,0.3)(2,7,0.34)(2,8,0.56)(2,9,1.36)
%%}
%%]
%%\dataplotThreeD[plotstyle=line,fillstyle=solid,fillcolor=red,gradbegin=red,gradmidpoint=0.5,linewidth=1pt]{\mydata}
%%%
%%\savedata{\mydata}[
%%{
%%(2,1,0.00)(2,2,0.02)(2,3,0.06)(2,4,0.12)(2,5,0.2)(2,6,0.3)(2,7,0.34)(2,8,0.56)(2,9,1.36)
%%}
%%]
%%\dataplotThreeD[plotstyle=line,fillstyle=solid,fillcolor=blue,gradbegin=red,gradmidpoint=0.5,linewidth=1pt]{\mydata}
%
%\savedata{\newdata}[{2,2,9}, {1,1,1}, {5,5,5}, {6,6,6}, {7,7,7}]
%\dataplotThreeD[plotstyle=curve,fillstyle=solid,fillcolor=blue,gradbegin=blue,gradmidpoint=0.5,linewidth=0pt]{\newdata}
%%%%%%%%%%%%%%%%%
\pstThreeDCoor
[IIIDxTicksPlane=xz,IIIDxticksep=-0.6,IIIDyTicksPlane=yz,IIIDyticksep=-0.2,arrowsize=1.5pt 4,linewidth=1.5pt,linecolor=darkgray,IIIDzTicksPlane=yz,yMin=0,yMax=15.5,xMin=0,xMax=10.5,zMin=0,zMax=11,IIIDticks=true,IIIDticksize=0.0,Dy=025,Dx=0.05,Dz=0.1,IIIDlabels=true,IIIDOffset={(0,0,0)}] 
%%%%%%%%%%%%%%%%%
\end{pspicture}
}
\end{minipage}
%
%%%%%%%%%%%%%%%%%%%%%%%%%%%%%%%%%%%%
%%%%%%%%%%%%%%%%%%%%%%%%%%%%%%%%%%%%
%%%%%%%%%%%%%%%%%%%%%%%%%%%%%%%%%%%%
%% part 2 of figure starts
%
\hspace*{1.7in}
%
%%%%%%%%%%%%%%%%%%%%%%%%%%%%%%%%%%%%
%%%%%%%%%%%%%%%%%%%%%%%%%%%%%%%%%%%%
%%%%%%%%%%%%%%%%%%%%%%%%%%%%%%%%%%%%
%
\begin{minipage}[t]{2in}
\scalebox{0.43}[0.43]{
\begin{pspicture}(-4,-7.5)(-2,9.5)
\psset{xunit=0.4in,yunit=0.4in,runit=0.4in}
\psset{nameX=\LARGE$\pmb{\gamma}$,nameY=\LARGE$\pmb{E}$,nameZ=\LARGE$\pmb{\xi}$}
\psset{coorType=1,Alpha=135,hiddenLine=true}

%%\pstThreeDPlaneGrid[planeGrid=xy,linewidth=0.1pt,ysubticks=14,xsubticks=9,linecolor=darkgray](0,0)(9,14)
%%\pstThreeDPlaneGrid[planeGridOffset=10,linewidth=0.1pt,ysubticks=14,xsubticks=9,linecolor=darkgray](0,0)(9,14)
%
%%\pstThreeDPlaneGrid[planeGrid=xz,linewidth=1pt,xsubticks=9,ysubticks=10,linecolor=darkgray](0,0)(9,10)
%%\pstThreeDPlaneGrid[planeGridOffset=14,planeGrid=xz,linewidth=1pt,xsubticks=9,ysubticks=10,linecolor=darkgray](0,0)(9,10)
%
%\pstThreeDPlaneGrid[planeGrid=yz,linewidth=0.1pt,xsubticks=14,ysubticks=10,linecolor=darkgray](0,0)(14,10)
%%\pstThreeDPlaneGrid[planeGridOffset=1,planeGrid=yz,linewidth=0.1pt,xsubticks=14,ysubticks=10,linecolor=darkgray](0,0)(14,10)
%%\pstThreeDPlaneGrid[planeGridOffset=9,planeGrid=yz,linewidth=0.1pt,xsubticks=14,ysubticks=10,linecolor=darkgray](0,0)(14,10)
%
\rput[ml](1,11.0){\Large$\pmb{n=50}$ \hspace*{0.1in} $\pmb{\mathcal{K}=0.5}$ \hspace*{0.1in}$\pmb{\Phi=0.5}$ \hspace*{0.1in}$\pmb{\gamma=0.05,0.25,0.45}$}
\rput[mr](5,-7.5){\LARGE\bf (B)}
%
%
%%%%%%%%%%%%%%%% \gamma=0.05,0.25,0.45
%
%%%%%%%%%%%%%%%%%%%%%%%%%%%%%%%%%%%%%%%%%%%%%%%%%%%%%%%%%%%%%%%%%%%%%%%%%%%%%%%%%%%%%%%
%%%%  ER, coordinated, ave degree 6
%%%%%%%%%%%%%%%%%%%%%%%%%%%%%%%%%%%%%%%%%%%%%%%%%%%%%%%%%%%%%%%%%%%%%%%%%%%%%%%%%%%%%%%
%
\pstThreeDLine[linewidth=8pt,linecolor=black,arrowsize=1.5pt 2](1,1,05.03)(1,2,05.03)(1,3,05.03)(1,4,06.70)(1,5,06.70)(1,6,06.70)(1,7,06.70)(1,8,07.90)(1,9,07.90)(1,10,07.90)(1,11,07.90)(1,12,09.00)(1,13,09.00)(1,14,09.00)
\pstThreeDLine[linewidth=8pt,linecolor=black,arrowsize=1.5pt 2](5,1,00.80)(5,2,00.80)(5,3,00.80)(5,4,01.0)(5,5,01.0)(5,6,01.0)(5,7,01.0)(5,8,01.4)(5,9,01.4)(5,10,01.4)(5,11,01.4)(5,12,01.7)(5,13,01.7)(5,14,01.7)
\pstThreeDLine[linewidth=8pt,linecolor=black,arrowsize=1.5pt 2](9,1,00.00)(9,2,00.00)(9,3,00.00)(9,4,00.0)(9,5,00.0)(9,6,00.0)(9,7,00.0)(9,8,00.0)(9,9,00.0)(9,10,00.0)(9,11,00.0)(9,12,00.0)(9,13,00.0)(9,14,00.0)
%
%%%%%%%%%%%%%%%%%%%%%%%%%%%%%%%%%%%%%%%%%%%%%%%%%%%%%%%%%%%%%%%%%%%%%%%%%%%%%%%%%%%%%%%
%%%%  ER, coordinated, ave degree 3
%%%%%%%%%%%%%%%%%%%%%%%%%%%%%%%%%%%%%%%%%%%%%%%%%%%%%%%%%%%%%%%%%%%%%%%%%%%%%%%%%%%%%%%
%
\pstThreeDLine[linewidth=8pt,linecolor=red,arrowsize=1.5pt 2](1,1,05.8)(1,2,05.8)(1,3,05.8)(1,4,07.50)(1,5,07.50)(1,6,07.50)(1,7,07.50)(1,8,08.70)(1,9,08.70)(1,10,08.70)(1,11,08.70)(1,12,09.20)(1,13,09.20)(1,14,09.20)
\pstThreeDLine[linewidth=8pt,linecolor=red,arrowsize=1.5pt 2](5,1,01.60)(5,2,01.60)(5,3,01.60)(5,4,01.7)(5,5,01.7)(5,6,01.7)(5,7,01.7)(5,8,02.50)(5,9,02.50)(5,10,02.50)(5,11,02.50)(5,12,03.60)(5,13,03.60)(5,14,03.60)
\pstThreeDLine[linewidth=8pt,linecolor=red,arrowsize=1.5pt 2](9,1,00.40)(9,2,00.40)(9,3,00.40)(9,4,00.10)(9,5,00.10)(9,6,00.10)(9,7,00.10)(9,8,00.10)(9,9,00.10)(9,10,00.10)(9,11,00.10)(9,12,00.20)(9,13,00.20)(9,14,00.20)
%
%%%%%%%%%%%%%%%%%%%%%%%%%%%%%%%%%%%%%%%%%%%%%%%%%%%%%%%%%%%%%%%%%%%%%%%%%%%%%%%%%%%%%%%
%%%%  ER, random, ave degree 6
%%%%%%%%%%%%%%%%%%%%%%%%%%%%%%%%%%%%%%%%%%%%%%%%%%%%%%%%%%%%%%%%%%%%%%%%%%%%%%%%%%%%%%%
%
\pstThreeDLine[linewidth=2pt,linecolor=black,arrowsize=1.5pt 2](1,1,07.20)(1,2,08.20)(1,3,04.60)(1,4,09.00)(1,5,07.80)(1,6,07.40)(1,7,07.20)(1,8,09.0)(1,9,08.8)(1,10,08.00)(1,11,08.60)(1,12,09.20)(1,13,08.60)(1,14,08.90)
\pstThreeDLine[linewidth=2pt,linecolor=black,arrowsize=1.5pt 2](5,1,00.60)(5,2,00.60)(5,3,00.70)(5,4,00.80)(5,5,01.00)(5,6,00.80)(5,7,00.90)(5,8,01.3)(5,9,01.60)(5,10,01.40)(5,11,01.50)(5,12,01.90)(5,13,01.60)(5,14,02.00)
\pstThreeDLine[linewidth=2pt,linecolor=black,arrowsize=1.5pt 2](9,1,00.00)(9,2,00.00)(9,3,00.00)(9,4,00.0)(9,5,00.0)(9,6,00.0)(9,7,00.0)(9,8,00.0)(9,9,00.0)(9,10,00.0)(9,11,00.0)(9,12,00.0)(9,13,00.02)(9,14,00.0)
%
%%%%%%%%%%%%%%%%%%%%%%%%%%%%%%%%%%%%%%%%%%%%%%%%%%%%%%%%%%%%%%%%%%%%%%%%%%%%%%%%%%%%%%%
%%%%  ER, random, ave degree 3
%%%%%%%%%%%%%%%%%%%%%%%%%%%%%%%%%%%%%%%%%%%%%%%%%%%%%%%%%%%%%%%%%%%%%%%%%%%%%%%%%%%%%%%
%
\pstThreeDLine[linewidth=2pt,linecolor=gray,arrowsize=1.5pt 2](1,1,06.7)(1,2,06.70)(1,3,07.60)(1,4,08.30)(1,5,07.20)(1,6,07.80)(1,7,08.10)(1,8,08.90)(1,9,9.000)(1,10,09.10)(1,11,08.8)(1,12,09.00)(1,13,09.00)(1,14,09.20)
\pstThreeDLine[linewidth=2pt,linecolor=gray,arrowsize=1.5pt 2](5,1,01.6)(5,2,01.40)(5,3,01.30)(5,4,01.80)(5,5,01.80)(5,6,02.60)(5,7,01.70)(5,8,02.60)(5,9,03.10)(5,10,02.50)(5,11,02.4)(5,12,03.20)(5,13,03.30)(5,14,03.7)
\pstThreeDLine[linewidth=2pt,linecolor=gray,arrowsize=1.5pt 2](9,1,00.50)(9,2,00.40)(9,3,00.40)(9,4,00.20)(9,5,00.20)(9,6,00.30)(9,7,00.30)(9,8,00.30)(9,9,00.30)(9,10,00.30)(9,11,00.20)(9,12,00.20)(9,13,00.2)(9,14,00.30)
%
%%%%%%%%%%%%%%%%%%%%%%%%%%%%%%%%%%%%%%%%%%%%%%%%%%%%%%%%%%%%%%%%%%%%%%%%%%%%%%%%%%%%%%%
%%%%  SF, coordinated, ave degree 6
%%%%%%%%%%%%%%%%%%%%%%%%%%%%%%%%%%%%%%%%%%%%%%%%%%%%%%%%%%%%%%%%%%%%%%%%%%%%%%%%%%%%%%%
%
\pstThreeDLine[linewidth=8pt,linecolor=yellow,arrowsize=1.5pt 2](1,1,09.60)(1,2,09.60)(1,3,09.60)(1,4,07.80)(1,5,07.80)(1,6,07.80)(1,7,07.80)(1,8,08.40)(1,9,08.40)(1,10,08.40)(1,11,08.40)(1,12,08.10)(1,13,08.10)(1,14,08.10)
\pstThreeDLine[linewidth=8pt,linecolor=yellow,arrowsize=1.5pt 2](5,1,00.60)(5,2,00.60)(5,3,00.60)(5,4,00.90)(5,5,00.90)(5,6,00.90)(5,7,00.90)(5,8,01.40)(5,9,01.40)(5,10,01.40)(5,11,01.40)(5,12,01.90)(5,13,01.90)(5,14,01.90)
\pstThreeDLine[linewidth=8pt,linecolor=yellow,arrowsize=1.5pt 2](9,1,00.00)(9,2,00.00)(9,3,00.00)(9,4,00.0)(9,5,00.0)(9,6,00.0)(9,7,00.0)(9,8,00.0)(9,9,00.0)(9,10,00.0)(9,11,00.0)(9,12,00.0)(9,13,00.0)(9,14,00.0)
%
%%%%%%%%%%%%%%%%%%%%%%%%%%%%%%%%%%%%%%%%%%%%%%%%%%%%%%%%%%%%%%%%%%%%%%%%%%%%%%%%%%%%%%%
%%%%  SF, coordinated, ave degree 3
%%%%%%%%%%%%%%%%%%%%%%%%%%%%%%%%%%%%%%%%%%%%%%%%%%%%%%%%%%%%%%%%%%%%%%%%%%%%%%%%%%%%%%%
%
\pstThreeDLine[linewidth=8pt,linecolor=blue,arrowsize=1.5pt 2](1,1,08.60)(1,2,08.60)(1,3,08.60)(1,4,08.70)(1,5,08.70)(1,6,08.70)(1,7,08.70)(1,8,08.70)(1,9,08.70)(1,10,08.70)(1,11,08.70)(1,12,08.70)(1,13,08.70)(1,14,08.70)
\pstThreeDLine[linewidth=8pt,linecolor=blue,arrowsize=1.5pt 2](5,1,02.60)(5,2,02.60)(5,3,02.60)(5,4,02.20)(5,5,02.20)(5,6,02.20)(5,7,02.20)(5,8,03.90)(5,9,03.90)(5,10,03.90)(5,11,03.90)(5,12,03.70)(5,13,03.70)(5,14,03.70)
\pstThreeDLine[linewidth=8pt,linecolor=blue,arrowsize=1.5pt 2](9,1,01.10)(9,2,01.10)(9,3,01.10)(9,4,00.40)(9,5,00.40)(9,6,00.40)(9,7,00.40)(9,8,00.40)(9,9,00.40)(9,10,00.40)(9,11,00.40)(9,12,00.40)(9,13,00.40)(9,14,00.40)
%
%%%%%%%%%%%%%%%%%%%%%%%%%%%%%%%%%%%%%%%%%%%%%%%%%%%%%%%%%%%%%%%%%%%%%%%%%%%%%%%%%%%%%%%
%%%%  SF, random, ave degree 6
%%%%%%%%%%%%%%%%%%%%%%%%%%%%%%%%%%%%%%%%%%%%%%%%%%%%%%%%%%%%%%%%%%%%%%%%%%%%%%%%%%%%%%%
%
\pstThreeDLine[linewidth=2pt,linecolor=green,arrowsize=1.5pt 2](1,1,08.00)(1,2,08.1)(1,3,06.70)(1,4,05.90)(1,5,05.9)(1,6,06.10)(1,7,07.00)(1,8,07.10)(1,9,07.10)(1,10,06.90)(1,11,06.90)(1,12,08.00)(1,13,08.30)(1,14,07.80)
\pstThreeDLine[linewidth=2pt,linecolor=green,arrowsize=1.5pt 2](5,1,01.00)(5,2,00.80)(5,3,00.70)(5,4,01.2)(5,5,01.00)(5,6,01.30)(5,7,01.30)(5,8,01.70)(5,9,01.60)(5,10,01.70)(5,11,01.7)(5,12,02.2)(5,13,02.1)(5,14,02.2)
\pstThreeDLine[linewidth=2pt,linecolor=green,arrowsize=1.5pt 2](9,1,0.00)(9,2,0.000)(9,3,0.000)(9,4,0.00)(9,5,0.00)(9,6,0.00)(9,7,0.00)(9,8,0.00)(9,9,0.00)(9,10,0.00)(9,11,0.00)(9,12,0.00)(9,13,0.00)(9,14,0.00)
%
%%%%%%%%%%%%%%%%%%%%%%%%%%%%%%%%%%%%%%%%%%%%%%%%%%%%%%%%%%%%%%%%%%%%%%%%%%%%%%%%%%%%%%%
%%%%  SF, random, ave degree 3
%%%%%%%%%%%%%%%%%%%%%%%%%%%%%%%%%%%%%%%%%%%%%%%%%%%%%%%%%%%%%%%%%%%%%%%%%%%%%%%%%%%%%%%
%
\pstThreeDLine[linewidth=2pt,linecolor=red,arrowsize=1.5pt 2](1,1,06.40)(1,2,07.40)(1,3,06.90)(1,4,07.60)(1,5,06.70)(1,6,07.90)(1,7,06.60)(1,8,08.10)(1,9,08.30)(1,10,08.8)(1,11,08.50)(1,12,08.80)(1,13,08.90)(1,14,08.7)
\pstThreeDLine[linewidth=2pt,linecolor=red,arrowsize=1.5pt 2](5,1,02.20)(5,2,03.5)(5,3,02.30)(5,4,01.80)(5,5,02.1)(5,6,02.80)(5,7,02.00)(5,8,03.20)(5,9,03.4)(5,10,04.00)(5,11,03.90)(5,12,03.8)(5,13,03.80)(5,14,03.60)
\pstThreeDLine[linewidth=2pt,linecolor=red,arrowsize=1.5pt 2](9,1,00.70)(9,2,00.70)(9,3,00.70)(9,4,00.30)(9,5,00.30)(9,6,00.50)(9,7,00.4)(9,8,00.30)(9,9,00.40)(9,10,00.30)(9,11,00.40)(9,12,00.30)(9,13,00.40)(9,14,00.50)
%
%%%%%%%%%%%%%%%%%%%%%%%%%%%%%%%%%%%%%%%%%%%%%%%%%%%%%%%%%%%%%%%%%%%%%%%%%%%%%%%%%%%%%%%
%%%%  tree, coordinated
%%%%%%%%%%%%%%%%%%%%%%%%%%%%%%%%%%%%%%%%%%%%%%%%%%%%%%%%%%%%%%%%%%%%%%%%%%%%%%%%%%%%%%%
%
\pstThreeDLine[linewidth=8pt,linecolor=black,linestyle=dashed,arrowsize=1.5pt 2](1,1,10.00)(1,2,10.00)(1,3,10.00)(1,4,10.0)(1,5,10.0)(1,6,10.0)(1,7,10.0)(1,8,10.00)(1,9,10.00)(1,10,10.00)(1,11,10.00)(1,12,10.00)(1,13,10.00)(1,14,10.00)
\pstThreeDLine[linewidth=8pt,linecolor=black,linestyle=dashed,arrowsize=1.5pt 2](5,1,08.50)(5,2,08.50)(5,3,08.50)(5,4,04.50)(5,5,04.50)(5,6,04.50)(5,7,04.50)(5,8,03.7)(5,9,03.7)(5,10,03.7)(5,11,03.7)(5,12,03.7)(5,13,03.7)(5,14,03.7)
\pstThreeDLine[linewidth=8pt,linecolor=black,linestyle=dashed,arrowsize=1.5pt 2](9,1,06.60)(9,2,06.60)(9,3,06.60)(9,4,00.3)(9,5,00.3)(9,6,00.3)(9,7,00.3)(9,8,00.3)(9,9,00.3)(9,10,00.3)(9,11,00.3)(9,12,00.40)(9,13,00.40)(9,14,00.40)
%
%%%%%%%%%%%%%%%%%%%%%%%%%%%%%%%%%%%%%%%%%%%%%%%%%%%%%%%%%%%%%%%%%%%%%%%%%%%%%%%%%%%%%%%
%%%%  tree, random
%%%%%%%%%%%%%%%%%%%%%%%%%%%%%%%%%%%%%%%%%%%%%%%%%%%%%%%%%%%%%%%%%%%%%%%%%%%%%%%%%%%%%%%
%
\pstThreeDLine[linewidth=8pt,linecolor=gray,linestyle=dashed,arrowsize=1.5pt 2](1,1,05.80)(1,2,06.10)(1,3,05.80)(1,4,07.6)(1,5,06.8)(1,6,07.00)(1,7,07.30)(1,8,07.90)(1,9,08.30)(1,10,07.90)(1,11,08.20)(1,12,07.50)(1,13,08.70)(1,14,08.20)
\pstThreeDLine[linewidth=8pt,linecolor=gray,linestyle=dashed,arrowsize=1.5pt 2](5,1,04.30)(5,2,04.10)(5,3,04.40)(5,4,02.10)(5,5,02.20)(5,6,02.50)(5,7,02.00)(5,8,05.0)(5,9,05.0)(5,10,05.0)(5,11,05.0)(5,12,05.0)(5,13,05.0)(5,14,05.0)
\pstThreeDLine[linewidth=8pt,linecolor=gray,linestyle=dashed,arrowsize=1.5pt 2](9,1,03.4)(9,2,03.30)(9,3,03.30)(9,4,00.10)(9,5,00.20)(9,6,00.2)(9,7,00.1)(9,8,00.20)(9,9,00.20)(9,10,00.2)(9,11,00.2)(9,12,00.30)(9,13,00.20)(9,14,00.10)
\def\psxyzlabel#1{\bgroup\large\bf\textsf{#1}\egroup}
\pstThreeDCoor
[IIIDxTicksPlane=xz,IIIDxticksep=-0.6,IIIDyTicksPlane=yz,IIIDyticksep=-0.2,arrowsize=1.5pt 4,linewidth=1.5pt,linecolor=darkgray,IIIDzTicksPlane=yz,yMin=0,yMax=15.5,xMin=0,xMax=10.5,zMin=0,zMax=11,IIIDticks=true,IIIDticksize=0.0,Dy=025,Dx=0.05,Dz=0.1,IIIDlabels=true,IIIDOffset={(0,0,0)}] 
\end{pspicture}
}
\end{minipage}
%
%%%%%%%%%%%%%%%%%%%%%%%%%%%%%%%%%%%%
%%%%%%%%%%%%%%%%%%%%%%%%%%%%%%%%%%%%
%%%%%%%%%%%%%%%%%%%%%%%%%%%%%%%%%%%%
%% part 3 of figure starts
%
\\
%
%%%%%%%%%%%%%%%%%%%%%%%%%%%%%%%%%%%%
%%%%%%%%%%%%%%%%%%%%%%%%%%%%%%%%%%%%
%%%%%%%%%%%%%%%%%%%%%%%%%%%%%%%%%%%%
%
\hspace*{-0.0in}
\begin{minipage}[t]{2in}
\scalebox{0.43}[0.43]{
\begin{pspicture}(-3,-7)(-1.95,13)
\psset{xunit=0.4in,yunit=0.4in,runit=0.4in}
\psset{nameX=\LARGE$\pmb{\gamma}$,nameY=\LARGE$\pmb{E}$,nameZ=\LARGE$\pmb{\xi}$}
\psset{coorType=1,Alpha=135,hiddenLine=true}
\def\psxyzlabel#1{\bgroup\large\bf\textsf{#1}\egroup}

%
%%\pstThreeDPlaneGrid[planeGrid=xy,linewidth=0.1pt,ysubticks=14,xsubticks=9,linecolor=darkgray](0,0)(9,14)
%%\pstThreeDPlaneGrid[planeGridOffset=10,linewidth=0.1pt,ysubticks=14,xsubticks=9,linecolor=darkgray](0,0)(9,14)
%
%%\pstThreeDPlaneGrid[planeGrid=xz,linewidth=1pt,xsubticks=9,ysubticks=10,linecolor=darkgray](0,0)(9,10)
%%\pstThreeDPlaneGrid[planeGridOffset=14,planeGrid=xz,linewidth=1pt,xsubticks=9,ysubticks=10,linecolor=darkgray](0,0)(9,10)
%
%\pstThreeDPlaneGrid[planeGrid=yz,linewidth=0.1pt,xsubticks=14,ysubticks=10,linecolor=darkgray](0,0)(14,10)
%%\pstThreeDPlaneGrid[planeGridOffset=1,planeGrid=yz,linewidth=0.1pt,xsubticks=14,ysubticks=10,linecolor=darkgray](0,0)(14,10)
%%\pstThreeDPlaneGrid[planeGridOffset=9,planeGrid=yz,linewidth=0.1pt,xsubticks=14,ysubticks=10,linecolor=darkgray](0,0)(14,10)
%
\rput[ml](1,11.0){\Large$\pmb{n=100}$ \hspace*{0.1in} $\pmb{\mathcal{K}=0.1}$ \hspace*{0.1in}$\pmb{\Phi=0.5}$ \hspace*{0.1in}$\pmb{\gamma=0.05,0.25,0.45}$}
\rput[mr](5,-7.5){\LARGE\bf (C)}
%
%
%%%%%%%%%%%%%%%% \gamma=0.05,0.25,0.45
%
%%%%%%%%%%%%%%%%%%%%%%%%%%%%%%%%%%%%%%%%%%%%%%%%%%%%%%%%%%%%%%%%%%%%%%%%%%%%%%%%%%%%%%%
%%%%  ER, coordinated, ave degree 6
%%%%%%%%%%%%%%%%%%%%%%%%%%%%%%%%%%%%%%%%%%%%%%%%%%%%%%%%%%%%%%%%%%%%%%%%%%%%%%%%%%%%%%%
%
\pstThreeDLine[linewidth=8pt,linecolor=black,arrowsize=1.5pt 2](1,1,01.20)(1,2,01.20)(1,3,01.20)(1,4,01.40)(1,5,01.40)(1,6,01.40)(1,7,01.40)(1,8,01.60)(1,9,01.60)(1,10,01.60)(1,11,01.60)(1,12,01.90)(1,13,01.90)(1,14,01.90)
\pstThreeDLine[linewidth=8pt,linecolor=black,arrowsize=1.5pt 2](5,1,00.20)(5,2,00.20)(5,3,00.20)(5,4,00.2)(5,5,00.2)(5,6,00.2)(5,7,00.2)(5,8,00.3)(5,9,00.3)(5,10,00.3)(5,11,00.3)(5,12,00.3)(5,13,00.3)(5,14,00.3)
\pstThreeDLine[linewidth=8pt,linecolor=black,arrowsize=1.5pt 2](9,1,00.00)(9,2,00.00)(9,3,00.00)(9,4,00.0)(9,5,00.0)(9,6,00.0)(9,7,00.0)(9,8,00.0)(9,9,00.0)(9,10,00.0)(9,11,00.0)(9,12,00.0)(9,13,00.0)(9,14,00.0)
%
%%%%%%%%%%%%%%%%%%%%%%%%%%%%%%%%%%%%%%%%%%%%%%%%%%%%%%%%%%%%%%%%%%%%%%%%%%%%%%%%%%%%%%%
%%%%  ER, coordinated, ave degree 3
%%%%%%%%%%%%%%%%%%%%%%%%%%%%%%%%%%%%%%%%%%%%%%%%%%%%%%%%%%%%%%%%%%%%%%%%%%%%%%%%%%%%%%%
%
\pstThreeDLine[linewidth=8pt,linecolor=red,arrowsize=1.5pt 2](1,1,02.3)(1,2,02.3)(1,3,02.3)(1,4,02.30)(1,5,02.30)(1,6,02.30)(1,7,02.30)(1,8,03.00)(1,9,03.00)(1,10,03.00)(1,11,03.00)(1,12,03.30)(1,13,03.30)(1,14,03.30)
\pstThreeDLine[linewidth=8pt,linecolor=red,arrowsize=1.5pt 2](5,1,00.70)(5,2,00.70)(5,3,00.70)(5,4,00.4)(5,5,00.4)(5,6,00.4)(5,7,00.4)(5,8,00.50)(5,9,00.50)(5,10,00.50)(5,11,00.50)(5,12,00.60)(5,13,00.60)(5,14,00.60)
\pstThreeDLine[linewidth=8pt,linecolor=red,arrowsize=1.5pt 2](9,1,00.50)(9,2,00.50)(9,3,00.50)(9,4,00.10)(9,5,00.10)(9,6,00.10)(9,7,00.10)(9,8,00.10)(9,9,00.10)(9,10,00.10)(9,11,00.10)(9,12,00.10)(9,13,00.10)(9,14,00.10)
%
%%%%%%%%%%%%%%%%%%%%%%%%%%%%%%%%%%%%%%%%%%%%%%%%%%%%%%%%%%%%%%%%%%%%%%%%%%%%%%%%%%%%%%%
%%%%  ER, random, ave degree 6
%%%%%%%%%%%%%%%%%%%%%%%%%%%%%%%%%%%%%%%%%%%%%%%%%%%%%%%%%%%%%%%%%%%%%%%%%%%%%%%%%%%%%%%
%
\pstThreeDLine[linewidth=2pt,linecolor=black,arrowsize=1.5pt 2](1,1,02.90)(1,2,01.00)(1,3,01.90)(1,4,01.50)(1,5,03.00)(1,6,02.20)(1,7,01.20)(1,8,02.9)(1,9,03.4)(1,10,02.40)(1,11,01.40)(1,12,03.30)(1,13,04.00)(1,14,03.00)
\pstThreeDLine[linewidth=2pt,linecolor=black,arrowsize=1.5pt 2](5,1,00.10)(5,2,00.10)(5,3,00.10)(5,4,00.20)(5,5,00.20)(5,6,00.20)(5,7,00.20)(5,8,00.2)(5,9,00.30)(5,10,00.30)(5,11,00.30)(5,12,00.30)(5,13,00.40)(5,14,00.40)
\pstThreeDLine[linewidth=2pt,linecolor=black,arrowsize=1.5pt 2](9,1,00.00)(9,2,00.00)(9,3,00.00)(9,4,00.0)(9,5,00.0)(9,6,00.0)(9,7,00.0)(9,8,00.0)(9,9,00.0)(9,10,00.0)(9,11,00.0)(9,12,00.0)(9,13,00.02)(9,14,00.0)
%
%%%%%%%%%%%%%%%%%%%%%%%%%%%%%%%%%%%%%%%%%%%%%%%%%%%%%%%%%%%%%%%%%%%%%%%%%%%%%%%%%%%%%%%
%%%%  ER, random, ave degree 3
%%%%%%%%%%%%%%%%%%%%%%%%%%%%%%%%%%%%%%%%%%%%%%%%%%%%%%%%%%%%%%%%%%%%%%%%%%%%%%%%%%%%%%%
%
\pstThreeDLine[linewidth=2pt,linecolor=gray,arrowsize=1.5pt 2](1,1,07.4)(1,2,08.10)(1,3,09.00)(1,4,06.80)(1,5,06.70)(1,6,05.90)(1,7,06.80)(1,8,05.80)(1,9,7.300)(1,10,07.20)(1,11,07.7)(1,12,08.20)(1,13,07.00)(1,14,07.90)
\pstThreeDLine[linewidth=2pt,linecolor=gray,arrowsize=1.5pt 2](5,1,00.7)(5,2,00.80)(5,3,00.60)(5,4,00.40)(5,5,00.30)(5,6,00.30)(5,7,00.40)(5,8,00.50)(5,9,00.40)(5,10,00.50)(5,11,00.5)(5,12,01.50)(5,13,00.60)(5,14,00.6)
\pstThreeDLine[linewidth=2pt,linecolor=gray,arrowsize=1.5pt 2](9,1,00.40)(9,2,00.50)(9,3,00.40)(9,4,00.00)(9,5,00.10)(9,6,00.00)(9,7,00.00)(9,8,00.00)(9,9,00.10)(9,10,00.10)(9,11,00.10)(9,12,00.10)(9,13,00.1)(9,14,00.10)
%
%%%%%%%%%%%%%%%%%%%%%%%%%%%%%%%%%%%%%%%%%%%%%%%%%%%%%%%%%%%%%%%%%%%%%%%%%%%%%%%%%%%%%%%
%%%%  SF, coordinated, ave degree 6
%%%%%%%%%%%%%%%%%%%%%%%%%%%%%%%%%%%%%%%%%%%%%%%%%%%%%%%%%%%%%%%%%%%%%%%%%%%%%%%%%%%%%%%
%
\pstThreeDLine[linewidth=8pt,linecolor=yellow,arrowsize=1.5pt 2](1,1,09.60)(1,2,09.60)(1,3,09.60)(1,4,07.10)(1,5,07.10)(1,6,07.10)(1,7,07.10)(1,8,06.20)(1,9,06.20)(1,10,06.20)(1,11,06.20)(1,12,07.00)(1,13,07.00)(1,14,07.00)
\pstThreeDLine[linewidth=8pt,linecolor=yellow,arrowsize=1.5pt 2](5,1,00.10)(5,2,00.10)(5,3,00.10)(5,4,00.10)(5,5,00.10)(5,6,00.10)(5,7,00.10)(5,8,00.10)(5,9,00.10)(5,10,00.10)(5,11,00.10)(5,12,00.10)(5,13,00.10)(5,14,00.10)
\pstThreeDLine[linewidth=8pt,linecolor=yellow,arrowsize=1.5pt 2](9,1,00.00)(9,2,00.00)(9,3,00.00)(9,4,00.0)(9,5,00.0)(9,6,00.0)(9,7,00.0)(9,8,00.0)(9,9,00.0)(9,10,00.0)(9,11,00.0)(9,12,00.0)(9,13,00.0)(9,14,00.0)
%
%%%%%%%%%%%%%%%%%%%%%%%%%%%%%%%%%%%%%%%%%%%%%%%%%%%%%%%%%%%%%%%%%%%%%%%%%%%%%%%%%%%%%%%
%%%%  SF, coordinated, ave degree 3
%%%%%%%%%%%%%%%%%%%%%%%%%%%%%%%%%%%%%%%%%%%%%%%%%%%%%%%%%%%%%%%%%%%%%%%%%%%%%%%%%%%%%%%
%
\pstThreeDLine[linewidth=8pt,linecolor=blue,arrowsize=1.5pt 2](1,1,07.60)(1,2,07.60)(1,3,07.60)(1,4,07.00)(1,5,07.00)(1,6,07.00)(1,7,07.00)(1,8,06.30)(1,9,06.30)(1,10,06.30)(1,11,06.30)(1,12,05.80)(1,13,05.80)(1,14,05.80)
\pstThreeDLine[linewidth=8pt,linecolor=blue,arrowsize=1.5pt 2](5,1,01.20)(5,2,01.20)(5,3,01.20)(5,4,00.40)(5,5,00.40)(5,6,00.40)(5,7,00.40)(5,8,00.30)(5,9,00.30)(5,10,00.30)(5,11,00.30)(5,12,00.30)(5,13,00.30)(5,14,00.30)
\pstThreeDLine[linewidth=8pt,linecolor=blue,arrowsize=1.5pt 2](9,1,00.90)(9,2,00.90)(9,3,00.90)(9,4,00.00)(9,5,00.00)(9,6,00.00)(9,7,00.00)(9,8,00.00)(9,9,00.00)(9,10,00.00)(9,11,00.00)(9,12,00.00)(9,13,00.00)(9,14,00.00)
%
%%%%%%%%%%%%%%%%%%%%%%%%%%%%%%%%%%%%%%%%%%%%%%%%%%%%%%%%%%%%%%%%%%%%%%%%%%%%%%%%%%%%%%%
%%%%  SF, random, ave degree 6
%%%%%%%%%%%%%%%%%%%%%%%%%%%%%%%%%%%%%%%%%%%%%%%%%%%%%%%%%%%%%%%%%%%%%%%%%%%%%%%%%%%%%%%
%
\pstThreeDLine[linewidth=2pt,linecolor=green,arrowsize=1.5pt 2](1,1,03.10)(1,2,04.2)(1,3,01.50)(1,4,03.40)(1,5,00.8)(1,6,01.60)(1,7,02.40)(1,8,01.70)(1,9,02.70)(1,10,02.40)(1,11,00.90)(1,12,01.00)(1,13,01.00)(1,14,01.10)
\pstThreeDLine[linewidth=2pt,linecolor=green,arrowsize=1.5pt 2](5,1,00.20)(5,2,00.30)(5,3,00.20)(5,4,00.2)(5,5,00.30)(5,6,00.30)(5,7,00.30)(5,8,00.40)(5,9,00.40)(5,10,00.40)(5,11,00.4)(5,12,00.5)(5,13,00.5)(5,14,00.5)
\pstThreeDLine[linewidth=2pt,linecolor=green,arrowsize=1.5pt 2](9,1,0.00)(9,2,0.000)(9,3,0.000)(9,4,0.00)(9,5,0.00)(9,6,0.00)(9,7,0.00)(9,8,0.00)(9,9,0.00)(9,10,0.00)(9,11,0.00)(9,12,0.00)(9,13,0.00)(9,14,0.00)
%
%%%%%%%%%%%%%%%%%%%%%%%%%%%%%%%%%%%%%%%%%%%%%%%%%%%%%%%%%%%%%%%%%%%%%%%%%%%%%%%%%%%%%%%
%%%%  SF, random, ave degree 3
%%%%%%%%%%%%%%%%%%%%%%%%%%%%%%%%%%%%%%%%%%%%%%%%%%%%%%%%%%%%%%%%%%%%%%%%%%%%%%%%%%%%%%%
%
\pstThreeDLine[linewidth=2pt,linecolor=red,arrowsize=1.5pt 2](1,1,02.70)(1,2,02.90)(1,3,03.90)(1,4,03.40)(1,5,03.80)(1,6,01.60)(1,7,03.50)(1,8,01.60)(1,9,04.10)(1,10,01.9)(1,11,02.60)(1,12,02.90)(1,13,01.90)(1,14,04.2)
\pstThreeDLine[linewidth=2pt,linecolor=red,arrowsize=1.5pt 2](5,1,01.20)(5,2,01.0)(5,3,01.90)(5,4,00.50)(5,5,00.5)(5,6,00.60)(5,7,00.60)(5,8,01.20)(5,9,00.5)(5,10,00.60)(5,11,01.40)(5,12,00.9)(5,13,00.90)(5,14,00.80)
\pstThreeDLine[linewidth=2pt,linecolor=red,arrowsize=1.5pt 2](9,1,00.90)(9,2,00.90)(9,3,00.80)(9,4,00.10)(9,5,00.10)(9,6,00.10)(9,7,00.1)(9,8,00.20)(9,9,00.10)(9,10,00.10)(9,11,00.10)(9,12,00.10)(9,13,00.10)(9,14,00.10)
%
%%%%%%%%%%%%%%%%%%%%%%%%%%%%%%%%%%%%%%%%%%%%%%%%%%%%%%%%%%%%%%%%%%%%%%%%%%%%%%%%%%%%%%%
%%%%  tree, coordinated
%%%%%%%%%%%%%%%%%%%%%%%%%%%%%%%%%%%%%%%%%%%%%%%%%%%%%%%%%%%%%%%%%%%%%%%%%%%%%%%%%%%%%%%
%
\pstThreeDLine[linewidth=8pt,linecolor=black,linestyle=dashed,arrowsize=1.5pt 2](1,1,09.50)(1,2,09.50)(1,3,09.50)(1,4,07.6)(1,5,07.6)(1,6,07.6)(1,7,07.6)(1,8,07.40)(1,9,07.40)(1,10,07.40)(1,11,07.40)(1,12,07.00)(1,13,07.00)(1,14,07.00)
\pstThreeDLine[linewidth=8pt,linecolor=black,linestyle=dashed,arrowsize=1.5pt 2](5,1,07.60)(5,2,07.60)(5,3,07.60)(5,4,01.40)(5,5,01.40)(5,6,01.40)(5,7,01.40)(5,8,01.0)(5,9,01.0)(5,10,01.0)(5,11,01.0)(5,12,01.0)(5,13,01.0)(5,14,01.0)
\pstThreeDLine[linewidth=8pt,linecolor=black,linestyle=dashed,arrowsize=1.5pt 2](9,1,07.00)(9,2,07.00)(9,3,07.00)(9,4,00.3)(9,5,00.3)(9,6,00.3)(9,7,00.3)(9,8,00.3)(9,9,00.3)(9,10,00.3)(9,11,00.3)(9,12,00.40)(9,13,00.40)(9,14,00.40)
%
%%%%%%%%%%%%%%%%%%%%%%%%%%%%%%%%%%%%%%%%%%%%%%%%%%%%%%%%%%%%%%%%%%%%%%%%%%%%%%%%%%%%%%%
%%%%  tree, random
%%%%%%%%%%%%%%%%%%%%%%%%%%%%%%%%%%%%%%%%%%%%%%%%%%%%%%%%%%%%%%%%%%%%%%%%%%%%%%%%%%%%%%%
%
\pstThreeDLine[linewidth=8pt,linecolor=gray,linestyle=dashed,arrowsize=1.5pt 2](1,1,06.40)(1,2,06.70)(1,3,06.80)(1,4,02.6)(1,5,01.8)(1,6,01.50)(1,7,02.10)(1,8,01.80)(1,9,02.40)(1,10,02.30)(1,11,02.50)(1,12,02.10)(1,13,02.40)(1,14,02.40)
\pstThreeDLine[linewidth=8pt,linecolor=gray,linestyle=dashed,arrowsize=1.5pt 2](5,1,06.20)(5,2,06.30)(5,3,06.20)(5,4,00.40)(5,5,00.30)(5,6,00.40)(5,7,00.40)(5,8,01.0)(5,9,01.0)(5,10,01.0)(5,11,01.0)(5,12,01.0)(5,13,01.0)(5,14,01.0)
\pstThreeDLine[linewidth=8pt,linecolor=gray,linestyle=dashed,arrowsize=1.5pt 2](9,1,06.0)(9,2,06.10)(9,3,06.00)(9,4,00.00)(9,5,00.00)(9,6,00.0)(9,7,00.1)(9,8,00.00)(9,9,00.00)(9,10,00.1)(9,11,00.0)(9,12,00.00)(9,13,00.10)(9,14,00.10)
\pstThreeDCoor
[IIIDxTicksPlane=xz,IIIDxticksep=-0.6,IIIDyTicksPlane=yz,IIIDyticksep=-0.2,arrowsize=1.5pt 4,linewidth=1.5pt,linecolor=darkgray,IIIDzTicksPlane=yz,yMin=0,yMax=15.5,xMin=0,xMax=10.5,zMin=0,zMax=11,IIIDticks=true,IIIDticksize=0.0,Dy=025,Dx=0.05,Dz=0.1,IIIDlabels=true,IIIDOffset={(0,0,0)}] 
\end{pspicture}
}
\end{minipage}
%
%%%%%%%%%%%%%%%%%%%%%%%%%%%%%%%%%%%%
%%%%%%%%%%%%%%%%%%%%%%%%%%%%%%%%%%%%
%%%%%%%%%%%%%%%%%%%%%%%%%%%%%%%%%%%%
%% part 4 of figure starts
%
\hspace*{1.7in}
%
%%%%%%%%%%%%%%%%%%%%%%%%%%%%%%%%%%%%
%%%%%%%%%%%%%%%%%%%%%%%%%%%%%%%%%%%%
%%%%%%%%%%%%%%%%%%%%%%%%%%%%%%%%%%%%
%
\begin{minipage}[t]{2in}
\scalebox{0.43}[0.43]{
\begin{pspicture}(-4,-7.5)(-2,9.5)
\psset{xunit=0.4in,yunit=0.4in,runit=0.4in}
\psset{nameX=\LARGE$\pmb{\gamma}$,nameY=\LARGE$\pmb{E}$,nameZ=\LARGE$\pmb{\xi}$}
\psset{coorType=1,Alpha=135,hiddenLine=true}

%%\pstThreeDPlaneGrid[planeGrid=xy,linewidth=0.1pt,ysubticks=14,xsubticks=9,linecolor=darkgray](0,0)(9,14)
%%\pstThreeDPlaneGrid[planeGridOffset=10,linewidth=0.1pt,ysubticks=14,xsubticks=9,linecolor=darkgray](0,0)(9,14)
%
%%\pstThreeDPlaneGrid[planeGrid=xz,linewidth=1pt,xsubticks=9,ysubticks=10,linecolor=darkgray](0,0)(9,10)
%%\pstThreeDPlaneGrid[planeGridOffset=14,planeGrid=xz,linewidth=1pt,xsubticks=9,ysubticks=10,linecolor=darkgray](0,0)(9,10)
%
%\pstThreeDPlaneGrid[planeGrid=yz,linewidth=0.1pt,xsubticks=14,ysubticks=10,linecolor=darkgray](0,0)(14,10)
%%\pstThreeDPlaneGrid[planeGridOffset=1,planeGrid=yz,linewidth=0.1pt,xsubticks=14,ysubticks=10,linecolor=darkgray](0,0)(14,10)
%%\pstThreeDPlaneGrid[planeGridOffset=9,planeGrid=yz,linewidth=0.1pt,xsubticks=14,ysubticks=10,linecolor=darkgray](0,0)(14,10)
%
\rput[ml](1,11.0){\Large$\pmb{n=100}$ \hspace*{0.1in} $\pmb{\mathcal{K}=0.5}$ \hspace*{0.1in}$\pmb{\Phi=0.5}$ \hspace*{0.1in}$\pmb{\gamma=0.05,0.25,0.45}$}
\rput[mr](5,-7.5){\LARGE\bf (D)}
%
%
%%%%%%%%%%%%%%%% \gamma=0.05,0.25,0.45
%
%%%%%%%%%%%%%%%%%%%%%%%%%%%%%%%%%%%%%%%%%%%%%%%%%%%%%%%%%%%%%%%%%%%%%%%%%%%%%%%%%%%%%%%
%%%%  ER, coordinated, ave degree 6
%%%%%%%%%%%%%%%%%%%%%%%%%%%%%%%%%%%%%%%%%%%%%%%%%%%%%%%%%%%%%%%%%%%%%%%%%%%%%%%%%%%%%%%
%
\pstThreeDLine[linewidth=8pt,linecolor=black,arrowsize=1.5pt 2](1,1,04.80)(1,2,04.80)(1,3,04.80)(1,4,06.70)(1,5,06.70)(1,6,06.70)(1,7,06.70)(1,8,07.90)(1,9,07.90)(1,10,07.90)(1,11,07.90)(1,12,08.90)(1,13,08.90)(1,14,08.90)
\pstThreeDLine[linewidth=8pt,linecolor=black,arrowsize=1.5pt 2](5,1,00.70)(5,2,00.70)(5,3,00.70)(5,4,01.1)(5,5,01.1)(5,6,01.1)(5,7,01.1)(5,8,01.5)(5,9,01.5)(5,10,01.5)(5,11,01.5)(5,12,01.8)(5,13,01.8)(5,14,01.8)
\pstThreeDLine[linewidth=8pt,linecolor=black,arrowsize=1.5pt 2](9,1,00.00)(9,2,00.00)(9,3,00.00)(9,4,00.0)(9,5,00.0)(9,6,00.0)(9,7,00.0)(9,8,00.0)(9,9,00.0)(9,10,00.0)(9,11,00.0)(9,12,00.0)(9,13,00.0)(9,14,00.0)
%
%%%%%%%%%%%%%%%%%%%%%%%%%%%%%%%%%%%%%%%%%%%%%%%%%%%%%%%%%%%%%%%%%%%%%%%%%%%%%%%%%%%%%%%
%%%%  ER, coordinated, ave degree 3
%%%%%%%%%%%%%%%%%%%%%%%%%%%%%%%%%%%%%%%%%%%%%%%%%%%%%%%%%%%%%%%%%%%%%%%%%%%%%%%%%%%%%%%
%
\pstThreeDLine[linewidth=8pt,linecolor=red,arrowsize=1.5pt 2](1,1,06.3)(1,2,06.3)(1,3,06.3)(1,4,07.60)(1,5,07.60)(1,6,07.60)(1,7,07.60)(1,8,08.80)(1,9,08.80)(1,10,08.80)(1,11,08.80)(1,12,09.30)(1,13,09.30)(1,14,09.30)
\pstThreeDLine[linewidth=8pt,linecolor=red,arrowsize=1.5pt 2](5,1,02.00)(5,2,02.00)(5,3,02.00)(5,4,02.2)(5,5,02.2)(5,6,02.2)(5,7,02.2)(5,8,02.70)(5,9,02.70)(5,10,02.70)(5,11,02.70)(5,12,03.20)(5,13,03.20)(5,14,03.20)
\pstThreeDLine[linewidth=8pt,linecolor=red,arrowsize=1.5pt 2](9,1,00.70)(9,2,00.70)(9,3,00.70)(9,4,00.30)(9,5,00.30)(9,6,00.30)(9,7,00.30)(9,8,00.30)(9,9,00.30)(9,10,00.30)(9,11,00.30)(9,12,00.30)(9,13,00.30)(9,14,00.30)
%
%%%%%%%%%%%%%%%%%%%%%%%%%%%%%%%%%%%%%%%%%%%%%%%%%%%%%%%%%%%%%%%%%%%%%%%%%%%%%%%%%%%%%%%
%%%%  ER, random, ave degree 6
%%%%%%%%%%%%%%%%%%%%%%%%%%%%%%%%%%%%%%%%%%%%%%%%%%%%%%%%%%%%%%%%%%%%%%%%%%%%%%%%%%%%%%%
%
\pstThreeDLine[linewidth=2pt,linecolor=black,arrowsize=1.5pt 2](1,1,07.60)(1,2,08.00)(1,3,06.50)(1,4,06.90)(1,5,08.30)(1,6,08.50)(1,7,08.60)(1,8,08.7)(1,9,08.1)(1,10,08.40)(1,11,08.70)(1,12,09.10)(1,13,09.00)(1,14,09.20)
\pstThreeDLine[linewidth=2pt,linecolor=black,arrowsize=1.5pt 2](5,1,00.70)(5,2,00.70)(5,3,00.70)(5,4,01.00)(5,5,01.00)(5,6,01.00)(5,7,01.20)(5,8,01.4)(5,9,01.40)(5,10,01.40)(5,11,01.30)(5,12,01.70)(5,13,01.80)(5,14,01.70)
\pstThreeDLine[linewidth=2pt,linecolor=black,arrowsize=1.5pt 2](9,1,00.00)(9,2,00.00)(9,3,00.00)(9,4,00.0)(9,5,00.0)(9,6,00.0)(9,7,00.0)(9,8,00.0)(9,9,00.0)(9,10,00.0)(9,11,00.0)(9,12,00.0)(9,13,00.02)(9,14,00.0)
%
%%%%%%%%%%%%%%%%%%%%%%%%%%%%%%%%%%%%%%%%%%%%%%%%%%%%%%%%%%%%%%%%%%%%%%%%%%%%%%%%%%%%%%%
%%%%  ER, random, ave degree 3
%%%%%%%%%%%%%%%%%%%%%%%%%%%%%%%%%%%%%%%%%%%%%%%%%%%%%%%%%%%%%%%%%%%%%%%%%%%%%%%%%%%%%%%
%
\pstThreeDLine[linewidth=2pt,linecolor=gray,arrowsize=1.5pt 2](1,1,07.2)(1,2,07.30)(1,3,07.80)(1,4,08.40)(1,5,08.40)(1,6,07.80)(1,7,08.20)(1,8,08.60)(1,9,8.800)(1,10,09.00)(1,11,08.7)(1,12,09.20)(1,13,09.00)(1,14,09.00)
\pstThreeDLine[linewidth=2pt,linecolor=gray,arrowsize=1.5pt 2](5,1,01.7)(5,2,01.60)(5,3,02.40)(5,4,02.00)(5,5,01.90)(5,6,01.90)(5,7,01.90)(5,8,03.90)(5,9,02.50)(5,10,02.50)(5,11,04.0)(5,12,03.00)(5,13,04.40)(5,14,03.6)
\pstThreeDLine[linewidth=2pt,linecolor=gray,arrowsize=1.5pt 2](9,1,00.50)(9,2,00.60)(9,3,00.50)(9,4,00.30)(9,5,00.30)(9,6,00.30)(9,7,00.30)(9,8,00.30)(9,9,00.30)(9,10,00.30)(9,11,00.20)(9,12,00.30)(9,13,00.3)(9,14,00.30)
%
%%%%%%%%%%%%%%%%%%%%%%%%%%%%%%%%%%%%%%%%%%%%%%%%%%%%%%%%%%%%%%%%%%%%%%%%%%%%%%%%%%%%%%%
%%%%  SF, coordinated, ave degree 6
%%%%%%%%%%%%%%%%%%%%%%%%%%%%%%%%%%%%%%%%%%%%%%%%%%%%%%%%%%%%%%%%%%%%%%%%%%%%%%%%%%%%%%%
%
\pstThreeDLine[linewidth=8pt,linecolor=yellow,arrowsize=1.5pt 2](1,1,09.40)(1,2,09.40)(1,3,09.40)(1,4,07.10)(1,5,07.10)(1,6,07.10)(1,7,07.10)(1,8,07.90)(1,9,07.90)(1,10,07.90)(1,11,07.90)(1,12,08.00)(1,13,08.00)(1,14,08.00)
\pstThreeDLine[linewidth=8pt,linecolor=yellow,arrowsize=1.5pt 2](5,1,00.90)(5,2,00.90)(5,3,00.90)(5,4,01.20)(5,5,01.20)(5,6,01.20)(5,7,01.20)(5,8,01.80)(5,9,01.80)(5,10,01.80)(5,11,01.80)(5,12,02.40)(5,13,02.40)(5,14,02.40)
\pstThreeDLine[linewidth=8pt,linecolor=yellow,arrowsize=1.5pt 2](9,1,00.00)(9,2,00.00)(9,3,00.00)(9,4,00.0)(9,5,00.0)(9,6,00.0)(9,7,00.0)(9,8,00.0)(9,9,00.0)(9,10,00.0)(9,11,00.0)(9,12,00.0)(9,13,00.0)(9,14,00.0)
%
%%%%%%%%%%%%%%%%%%%%%%%%%%%%%%%%%%%%%%%%%%%%%%%%%%%%%%%%%%%%%%%%%%%%%%%%%%%%%%%%%%%%%%%
%%%%  SF, coordinated, ave degree 3
%%%%%%%%%%%%%%%%%%%%%%%%%%%%%%%%%%%%%%%%%%%%%%%%%%%%%%%%%%%%%%%%%%%%%%%%%%%%%%%%%%%%%%%
%
\pstThreeDLine[linewidth=8pt,linecolor=blue,arrowsize=1.5pt 2](1,1,07.60)(1,2,07.60)(1,3,07.60)(1,4,07.80)(1,5,07.80)(1,6,07.80)(1,7,07.80)(1,8,08.10)(1,9,08.10)(1,10,08.10)(1,11,08.10)(1,12,08.30)(1,13,08.30)(1,14,08.30)
\pstThreeDLine[linewidth=8pt,linecolor=blue,arrowsize=1.5pt 2](5,1,03.00)(5,2,03.00)(5,3,03.00)(5,4,02.70)(5,5,02.70)(5,6,02.70)(5,7,02.70)(5,8,03.70)(5,9,03.70)(5,10,03.70)(5,11,03.70)(5,12,04.30)(5,13,04.30)(5,14,04.30)
\pstThreeDLine[linewidth=8pt,linecolor=blue,arrowsize=1.5pt 2](9,1,01.40)(9,2,01.40)(9,3,01.40)(9,4,00.50)(9,5,00.50)(9,6,00.50)(9,7,00.50)(9,8,00.50)(9,9,00.50)(9,10,00.50)(9,11,00.50)(9,12,00.50)(9,13,00.50)(9,14,00.50)
%
%%%%%%%%%%%%%%%%%%%%%%%%%%%%%%%%%%%%%%%%%%%%%%%%%%%%%%%%%%%%%%%%%%%%%%%%%%%%%%%%%%%%%%%
%%%%  SF, random, ave degree 6
%%%%%%%%%%%%%%%%%%%%%%%%%%%%%%%%%%%%%%%%%%%%%%%%%%%%%%%%%%%%%%%%%%%%%%%%%%%%%%%%%%%%%%%
%
\pstThreeDLine[linewidth=2pt,linecolor=green,arrowsize=1.5pt 2](1,1,07.10)(1,2,05.9)(1,3,05.80)(1,4,05.80)(1,5,05.2)(1,6,06.70)(1,7,05.70)(1,8,07.50)(1,9,07.30)(1,10,07.20)(1,11,07.60)(1,12,08.40)(1,13,08.70)(1,14,08.10)
\pstThreeDLine[linewidth=2pt,linecolor=green,arrowsize=1.5pt 2](5,1,01.00)(5,2,01.00)(5,3,01.10)(5,4,01.4)(5,5,01.50)(5,6,01.50)(5,7,01.30)(5,8,02.10)(5,9,01.90)(5,10,01.80)(5,11,02.0)(5,12,02.7)(5,13,02.6)(5,14,02.8)
\pstThreeDLine[linewidth=2pt,linecolor=green,arrowsize=1.5pt 2](9,1,0.00)(9,2,0.100)(9,3,0.000)(9,4,0.00)(9,5,0.00)(9,6,0.00)(9,7,0.00)(9,8,0.00)(9,9,0.00)(9,10,0.00)(9,11,0.00)(9,12,0.00)(9,13,0.00)(9,14,0.00)
%
%%%%%%%%%%%%%%%%%%%%%%%%%%%%%%%%%%%%%%%%%%%%%%%%%%%%%%%%%%%%%%%%%%%%%%%%%%%%%%%%%%%%%%%
%%%%  SF, random, ave degree 3
%%%%%%%%%%%%%%%%%%%%%%%%%%%%%%%%%%%%%%%%%%%%%%%%%%%%%%%%%%%%%%%%%%%%%%%%%%%%%%%%%%%%%%%
%
\pstThreeDLine[linewidth=2pt,linecolor=red,arrowsize=1.5pt 2](1,1,06.80)(1,2,06.30)(1,3,06.70)(1,4,07.80)(1,5,06.80)(1,6,07.50)(1,7,07.80)(1,8,08.00)(1,9,08.60)(1,10,08.20)(1,11,08.20)(1,12,08.60)(1,13,08.50)(1,14,08.3)
\pstThreeDLine[linewidth=2pt,linecolor=red,arrowsize=1.5pt 2](5,1,02.40)(5,2,02.6)(5,3,02.50)(5,4,02.40)(5,5,02.5)(5,6,02.60)(5,7,02.60)(5,8,03.60)(5,9,03.3)(5,10,02.80)(5,11,02.90)(5,12,04.2)(5,13,04.20)(5,14,04.20)
\pstThreeDLine[linewidth=2pt,linecolor=red,arrowsize=1.5pt 2](9,1,01.10)(9,2,00.90)(9,3,01.00)(9,4,00.60)(9,5,00.50)(9,6,00.50)(9,7,00.5)(9,8,00.50)(9,9,00.60)(9,10,00.50)(9,11,00.60)(9,12,00.50)(9,13,00.60)(9,14,00.70)
%
%%%%%%%%%%%%%%%%%%%%%%%%%%%%%%%%%%%%%%%%%%%%%%%%%%%%%%%%%%%%%%%%%%%%%%%%%%%%%%%%%%%%%%%
%%%%  tree, coordinated
%%%%%%%%%%%%%%%%%%%%%%%%%%%%%%%%%%%%%%%%%%%%%%%%%%%%%%%%%%%%%%%%%%%%%%%%%%%%%%%%%%%%%%%
%
\pstThreeDLine[linewidth=8pt,linecolor=black,linestyle=dashed,arrowsize=1.5pt 2](1,1,10.00)(1,2,10.00)(1,3,10.00)(1,4,10.0)(1,5,10.0)(1,6,10.0)(1,7,10.0)(1,8,10.00)(1,9,10.00)(1,10,10.00)(1,11,10.00)(1,12,10.00)(1,13,10.00)(1,14,10.00)
\pstThreeDLine[linewidth=8pt,linecolor=black,linestyle=dashed,arrowsize=1.5pt 2](5,1,08.50)(5,2,08.50)(5,3,08.50)(5,4,03.70)(5,5,03.70)(5,6,03.70)(5,7,03.70)(5,8,03.3)(5,9,03.3)(5,10,03.3)(5,11,03.3)(5,12,03.3)(5,13,03.3)(5,14,03.3)
\pstThreeDLine[linewidth=8pt,linecolor=black,linestyle=dashed,arrowsize=1.5pt 2](9,1,07.00)(9,2,07.00)(9,3,07.00)(9,4,00.4)(9,5,00.4)(9,6,00.4)(9,7,00.4)(9,8,00.4)(9,9,00.4)(9,10,00.4)(9,11,00.4)(9,12,00.40)(9,13,00.40)(9,14,00.40)
%
%%%%%%%%%%%%%%%%%%%%%%%%%%%%%%%%%%%%%%%%%%%%%%%%%%%%%%%%%%%%%%%%%%%%%%%%%%%%%%%%%%%%%%%
%%%%  tree, random
%%%%%%%%%%%%%%%%%%%%%%%%%%%%%%%%%%%%%%%%%%%%%%%%%%%%%%%%%%%%%%%%%%%%%%%%%%%%%%%%%%%%%%%
%
\pstThreeDLine[linewidth=8pt,linecolor=gray,linestyle=dashed,arrowsize=1.5pt 2](1,1,06.30)(1,2,06.30)(1,3,06.00)(1,4,06.3)(1,5,06.3)(1,6,07.30)(1,7,07.30)(1,8,07.70)(1,9,08.00)(1,10,08.60)(1,11,07.80)(1,12,07.50)(1,13,07.90)(1,14,08.20)
\pstThreeDLine[linewidth=8pt,linecolor=gray,linestyle=dashed,arrowsize=1.5pt 2](5,1,04.10)(5,2,04.20)(5,3,04.30)(5,4,02.00)(5,5,01.60)(5,6,01.80)(5,7,01.80)(5,8,05.0)(5,9,05.0)(5,10,05.0)(5,11,05.0)(5,12,05.0)(5,13,05.0)(5,14,05.0)
\pstThreeDLine[linewidth=8pt,linecolor=gray,linestyle=dashed,arrowsize=1.5pt 2](9,1,03.4)(9,2,03.30)(9,3,03.30)(9,4,00.10)(9,5,00.20)(9,6,00.2)(9,7,00.1)(9,8,00.20)(9,9,00.20)(9,10,00.2)(9,11,00.2)(9,12,00.30)(9,13,00.20)(9,14,00.10)
\def\psxyzlabel#1{\bgroup\large\bf\textsf{#1}\egroup}
\pstThreeDCoor
[IIIDxTicksPlane=xz,IIIDxticksep=-0.6,IIIDyTicksPlane=yz,IIIDyticksep=-0.2,arrowsize=1.5pt 4,linewidth=1.5pt,linecolor=darkgray,IIIDzTicksPlane=yz,yMin=0,yMax=15.5,xMin=0,xMax=10.5,zMin=0,zMax=11,IIIDticks=true,IIIDticksize=0.0,Dy=025,Dx=0.05,Dz=0.1,IIIDlabels=true,IIIDOffset={(0,0,0)}] 
\end{pspicture}
}
\end{minipage}
\vspace*{0.0in}
\caption{\label{newfig4-full}Effect of variations of the total external to internal asset ratio $\nicefrac{E}{I}$ on the vulnerability index $\xi$ for homogeneous networks.
Lower values of $\xi$ imply higher global stability of a network.}
\end{figure*}

%%%%%%%%%%%%%%%%%%%%%%%%%%%%%%%%%%%%%%%%%%%%%%%%%%%%%%%%%%%%%%%%%%%%%%%%%%%%%%%%%%%%%%%%%%%%%%%%%%%%%%%%%%%%%%
%%%%%%%%%%%%%%%%%%%%%%%%%%%%%%%%%%%%%%%%%%%%%%%%%%%%%%%%%%%%%%%%%%%%%%%%%%%%%%%%%%%%%%%%%%%%%%%%%%%%%%%%%%%%%%
%%%%%%%%%%%%%%%%%%%%%%%%%%%%%%%%%%%%%%%%%%%%%%%%%%%%%%%%%%%%%%%%%%%%%%%%%%%%%%%%%%%%%%%%%%%%%%%%%%%%%%%%%%%%%%

\clearpage
%%%%%%%%%%%%%%%%%%%%%%%%%%%%%%%%%%%%%%%%%%%%%%%%%%%%%%%%%%%%%%%%%%%%%%%%%%%%%%%%%%%%%%%%%%%%%%%%%%%%%%%%%%%%%%
%%%%%%%%%%%%%%%%%%%%%%%%%%%%%%%%%%%%%%%%%%%%%%%%%%%%%%%%%%%%%%%%%%%%%%%%%%%%%%%%%%%%%%%%%%%%%%%%%%%%%%%%%%%%%%
%%%%%%%%%%%%%%%%%%%%%%%%%%%%%%%%%%%%%%%%%%%%%%%%%%%%%%%%%%%%%%%%%%%%%%%%%%%%%%%%%%%%%%%%%%%%%%%%%%%%%%%%%%%%%%
%%% newfig5-coordinated-and-random
%%%\input{../tex-figures/newfig5-coordinated-and-random.tex}

\begin{figure*}
\vspace*{0.5in}
\hspace*{-0.0in}
\subfigure{
\scalebox{0.95}[0.7]{

}

\vspace*{-5.3in}
\caption{\label{newfig5-full}Effect of variations of the total external to internal asset ratio $\nicefrac{E}{I}$ on the vulnerability index $\xi$
for $(\alpha,\beta)$-heterogeneous networks. Lower values of $\xi$ imply higher global stability of a network.}
\end{figure*}

%%%%%%%%%%%%%%%%%%%%%%%%%%%%%%%%%%%%%%%%%%%%%%%%%%%%%%%%%%%%%%%%%%%%%%%%%%%%%%%%%%%%%%%%%%%%%%%%%%%%%%%%%%%%%%
%%%%%%%%%%%%%%%%%%%%%%%%%%%%%%%%%%%%%%%%%%%%%%%%%%%%%%%%%%%%%%%%%%%%%%%%%%%%%%%%%%%%%%%%%%%%%%%%%%%%%%%%%%%%%%
%%%%%%%%%%%%%%%%%%%%%%%%%%%%%%%%%%%%%%%%%%%%%%%%%%%%%%%%%%%%%%%%%%%%%%%%%%%%%%%%%%%%%%%%%%%%%%%%%%%%%%%%%%%%%%

\clearpage
%%%%%%%%%%%%%%%%%%%%%%%%%%%%%%%%%%%%%%%%%%%%%%%%%%%%%%%%%%%%%%%%%%%%%%%%%%%%%%%%%%%%%%%%%%%%%%%%%%%%%%%%%%%%%%
%%%%%%%%%%%%%%%%%%%%%%%%%%%%%%%%%%%%%%%%%%%%%%%%%%%%%%%%%%%%%%%%%%%%%%%%%%%%%%%%%%%%%%%%%%%%%%%%%%%%%%%%%%%%%%
%%%%%%%%%%%%%%%%%%%%%%%%%%%%%%%%%%%%%%%%%%%%%%%%%%%%%%%%%%%%%%%%%%%%%%%%%%%%%%%%%%%%%%%%%%%%%%%%%%%%%%%%%%%%%%
%%% supp-newfig5-coordinated-and-random
%%%\input{../tex-figures/supp-newfig5-coordinated-and-random.tex}

\begin{figure*}
\vspace*{0.7in}
%%%\hspace*{0.2in}
%
\subfigure{
\scalebox{1}[0.8]{

}
}
\vspace*{-4.4in}
\caption{\label{supp-newfig5-full}Effect of variations of the total external to internal asset ratio $\frac{\E}{\I}$ on the vulnerability index $\xi$
for $(\alpha,\beta)$-heterogeneous networks. Lower values of $\xi$ imply higher global stability of a network.}
\end{figure*}
%%%%%%%%%%%%%%%%%

\end{document}